\documentclass[aps,twocolumn,showpacs, amssymb, longbibliography,
nofootinbib]{revtex4-2}
\usepackage{amsmath} 
\usepackage{graphicx}
\usepackage{dcolumn}
\usepackage{bm}
\usepackage{caption} \usepackage{subcaption}
\usepackage{comment} 
\usepackage[dvipsnames]{xcolor} 
\usepackage{tikz} 
\usepackage{pgfplots} 
\usepackage{hyperref}
\usepackage[nameinlink, english]{cleveref} 
\usepackage{enumerate}
\usepackage[f]{esvect} 
\usepackage{siunitx}
\DeclareSIUnit\dyne{dyn}
\usepackage{mathtools}

\newcommand*{\parens}[1]{\left( #1 \right)}

\makeatletter \renewcommand*\env@matrix[1][*\c@MaxMatrixCols c]{%
  \hskip -\arraycolsep \let\@ifnextchar\new@ifnextchar \array{#1}} \makeatother

\begin{document}
\title{The Co-Moving
  Velocity and Projective Transformations}

\author{Håkon Pedersen\email{hakon.pedersen@ntnu.no}}  \affiliation{PoreLab, Department of
  Physics, Norwegian University of Science and Technology, NO--7491 Trondheim,
  Norway}
\date{\today {}}
\begin{abstract}
  In a string of recent papers starting with (Transport in Porous Media, 125,
  565 (2018)), a theory of immiscible two-phase flow in porous media based on
  Euler homogeneity of the total volumetric flow rate has been investigated. The
  thermodynamic-like theory has an associated statistical mechanics based on a
  maximum entropy principle. A quantity called the co-moving velocity connects
  the equations of state of the intensive thermodynamic velocities and the
  physical seepage velocities of two flowing fluids. The obtained relations have
  a structure that can be interpreted using affine- and projective geometry. The
  co-moving velocity can be expressed as a transformation of the saturation
  using projective duality of points and lines. One obtains an exact
  constitutive relation depending on a projective invariant, the cross-ratio,
  which allows the co-moving velocity to be expressed in terms of a simple
  steady-state advection equation. A kinematic view of the
  velocity relations is presented, modeled by a well-known non-trivial geometry which turns
  out to be pseudo-Euclidean. The cross-ratio determines a hyperbolic angle in
  this space, and can be parametrized in terms of three numbers using a linear
  fractional transformation. Knowing these parameters, the pore velocity and
  the derivative of the pore velocity with respect to the saturation, an
  approximation for the co-moving velocity can be obtained for a range of applied
  pressures, viscosity ratios and surface tensions. The parametrization is
  demonstrated using data from a dynamic pore network model and
  relative permeability data from the literature. This paper only considers the
  pore areas as extensive variables, however, the geometric principles are 
  general, and the same idea could potentially be used in other systems.

\end{abstract}
\maketitle
\section{Introduction}
\label{sec:intro}

One of the greatest unsolved challenges of modern day research on two-phase flow
in porous media is to determine an effective continuum-level theory for the flow
based in rigorous physical principles. Even in the case of immiscible and
incompressible flow in a homogeneous porous medium, the system exhibit a range
of phenomena on several time- and length scales due to the complicated
interaction between the fluids and the porous medium
\cite{bearDynamicsFluidsPorous1988, bluntMultiphaseFlowPermeable2017,
  federPhysicsFlowPorous2022}. In the limit of large length scales, the tracking
of individual particles is unviable, and even discretized fluid elements at the
mesoscopic level must often be substituted for a continuum theory as one goes up
in scale to extract useful results. The details of how one performs this scaling
and how this respects the physics at the smaller length scales is called the
\textit{upscaling problem}. This problem is a manifestation of the general
difficulties in obtaining macroscale properties from smaller constituents, which
is prevalent in all physics \cite{andersonMoreDifferent1972}.

To date, there is no agreed-upon solution to the upscaling problem in porous
media, even though it is practically as old as the field of fluid dynamics
itself. This stems from the interactions between the fluids in the random
network formed by the solid matrix of the porous medium, which is complicated
further by factors such as driving forces, fluid properties, wetting conditions
and long range correlations due to dominant flow paths
\cite{bluntMultiphaseFlowPermeable2017,federPhysicsFlowPorous2022}. A central
difficulty lies in determining the correct variables for describing the combined
flow of the two fluids given the individual fluid properties, with the
additional complication that local properties of the porous medium may be
intractable or unavailable, especially for large systems. The unification of the
many complicated phenomena at several time- and length scales leads to
complicated relations that often lend themselves better to approximations which
are sometimes justified, other times not.

There are many proposed solutions to this problem, all with different strengths
and drawbacks. The most well-known example is \textit{relative permeability
  theory} \cite{wyckoffFlowGasLiquid1936, leverettCapillaryBehaviorPorous1941,
  leverettFlowOilWaterMixtures1939}, which assumes that the fluids flowing in
the medium has their available pore space reduced because they share it with
other fluids. This gives a reduction in the fluid conductance, and the factor by
which it is reduced is the relative permeability. Other examples of approaches
to the upscaling problem are \textit{Thermodynamically Constrained Averaging
  Theory} (TCAT) \cite{grayIntroductionThermodynamicallyConstrained2014,
  hassanizadehMechanicsThermodynamicsMultiphase1990}, non-equilibrium
statistical mechanics \cite{kjelstrupNonisothermalTransportMultiphase2018,
  kjelstrupNonisothermalTransportMultiphase2019}, \textit{decomposition of
  Prototype Flow} \cite{valavanidesMechanisticModelSteadyState1998,
  valavanidesReviewSteadyStateTwoPhase2018,
  valavanidesOilFragmentationInterfacial2018}, and using Minkowski functionals
to characterize the flowing fluids in the porous medium
\cite{mcclureGeometricStateFunction2018,
  armstrongPorousMediaCharacterization2019,
  schroder-turkMinkowskiTensorsAnisotropic2013}.

Hansen et al. \cite{hansenRelationsSeepageVelocities2018} introduced a
pseudo-thermodynamic theory based on an effective description of the fluid flow
using Euler homogeneity, which has been shown to correspond well with both
numerical and experimental results over a large parameter-space
\cite{royFlowAreaRelationsImmiscible2020,royCoMovingVelocityImmiscible2022,
  hansenStatisticalMechanicsFramework2023}. In this theory, the two fluids are
mapped to a single ``effective'' fluid measured at the continuum-scale. By
introducing a quantity called the \textit{co-moving velocity}, which has no
transport associated with it, the mapping becomes a bijection, and no
information is lost in the transformation between the two sets of velocities.
In most cases, the co-moving velocity has a simple, linear behaviour. The pseudo-thermodynamics enters when steady-state is defined as the
situation where the chosen macroscopic variables fluctuate around well-defined
average values. One can define this steady-state as an equilibrium state of the
system, which allows for a formulation of associated statistical mechanics based
on information theory \cite{jaynesInformationTheoryStatistical1957}.

In \cite{hansenRelationsSeepageVelocities2018}, Euler homogeneity was used to
derive equations relating the seepage velocities of two immiscible
fluids $v_w$ and $v_n$, respectively more wetting and the other less wetting
with respect to the porous medium, in steady-state flow through a representative
area element (REA), with an average seepage, or ``pore''-, velocity $v$. The
co-moving velocity $v_m$ and $v$ contain the necessary information to determine
the seepage velocities. The derived equations allowed for the transformation
$(v,v_m) \to (v_w,v_n)$, given as
\begin{align}
  v_w=v+S_n\left(\frac{dv}{dS_w}-v_m\right)\ \label{eq:e1} \;, \\
  v_n=v-S_w\left(\frac{dv}{dS_w}-v_m\right) \label{eq:e2} \;,
\end{align}
where $S_w$ and $S_n$ are the wetting and non-wetting saturations respectively.
The derivation of \cref{eq:e1,eq:e2} necessitates the introduction of a set of
\textit{thermodynamic velocities} $\left(\hat{v}_w , \hat{v}_n \right)$, defined
as the partial derivatives of the volumetric flow rate $Q$ with respect to the
wetting- and non-wetting areas $A_w, A_n $ of the REA respectively.

\subsection{Choice of geometric context}
\label{sec:geometric-context}

To classify the transformation of \cref{eq:e1,eq:e2}, it is desirable to obtain
a geometric interpretation of the mappings
\begin{align}
  (v,v_m) \ \leftrightarrows& \ (v_w,v_n) \ \label{eq:primary-transf} \;, \\
  (\hat{v}_w,\hat{v}_n) \ \leftrightarrows& \ (v_w,v_n) \ \label{eq:secondary-transf} \;.
\end{align}
Such an interpretation can aid in further generalization and incorporation of
more variables, in addition to leading the way to constructing equations of
state. At the same time, there are a number of geometric contexts one could
consider. A convenient structure that allows for a distinction between a set of
variables viewed as points of a space and associated quantities parametrized by
these points are \textit{fiber bundles}
\cite{leeIntroductionSmoothManifolds2012, milnorCharacteristicClasses1974}
\footnote{Not to be confused with the \textit{capillary fiber bundle} of porous
  media research \cite{hansenFiberBundleModel2015}.}, which can be understood as
consisting of a base space and some space, the fiber, ``attached'' at each point
of the base space. A common example of a fiber bundle is a surface, potentially
defined in terms of a single function, along with the spaces of tangent vectors
to this surface, called a \textit{tangent vector bundle}. Relations between
objects on the tangent vector bundle and the ``dual'' \textit{cotangent vector
  bundle} \cite{leeIntroductionSmoothManifolds2012} are central in defining
thermodynamic relations and geometric equivalents of equilibrium states. In the
context of (co-)tangent vector bundles, a choice of vector in the fiber for each
point of the base space is called a \textit{(co-)tangent vector field}, which
are just vector fields in the ordinary physical sense \footnote{One calls such a
  choice of element in the fiber at each point of the base space a
  \textit{section} of the bundle.}. These bundle structures are fundamental in
\textit{contact geometry} \cite{arnoldMathematicalMethodsClassical1978}, which
is often recognised as the appropriate geometric setting for thermodynamics
\cite{bravettiContactGeometryThermodynamics2019}, with close ties to statistical
mechanics \cite{jaynesInformationTheoryStatistical1957}.

This article will apply a geometrical model based on classical \textit{affine} and
\textit{projective} geometry \cite{bergerGeometry1994}. An affine space is intuitively a
vector space without a choice of origin. A projective space
\footnote{Projective- and affine spaces are quite general, and can be introduced
  in more ways than presented here.} contains as points the lines through the
origin of a vector space, i.e. the points are the one dimensional vector
subspaces themselves. Affine spaces are contained in projective spaces; they are
the complements of projective subspaces. The points of the projective subspaces
are called \textit{``improper''} or \textit{infinite} points with respect to
the affine space, while the points of the affine space are \textit{finite}
points \footnote{In fact, projective- and affine spaces are examples of
  \textit{homogeneous spaces}. These spaces are ``symmetric'' in the sense that
  there are no preferred points such as in a vector space. The separation into
  finite and infinite points is a result of our choice of hyperplane in the
  projective space.}

In \cite{pedersenParameterizationsImmiscibleTwophase2023b}, different coordinate
systems were considered on a vector space of extensive variables. Both extensive
(pore areas) and intensive (velocities) quantities were taken as components of
vectors in the same vector space. In this paper, the extensive variables simply
represent points of $\mathbb{R}^{n}$. The intensive quantities will then appear
as ``dual'' coordinates of the extensive coordinates, which can be inferred from
the definition of the total volumetric flow rate $Q$. The duality is to be
interpreted in a \textit{projective} sense, which is a duality between
\textit{points and lines} \footnote{This duality is closely related to the
  duality between so-called \textit{conjugate variables} in thermodynamics, see
  \cref{sec:slip-ratio}.}. Both sets of coordinates can be parametrized in terms
of a saturation, $S_w$ or $S_n$, where primarily $S_w$ will be
used. 

This paper represents a \textit{kinematic} view of the two-phase flow problem.
This could, broadly speaking, be viewed in contrast to the \textit{differential
  geometric} point of view that contact structures represent. In particular, the
extensivity required from the volumetric flow rates is not straightforwardly
generalized to manifolds \cite{garcia-arizaGeometricApproachConcept2019}, but
baked into the projective description considered here. Applying differential
geometry is not the topic of this paper, see e.g. \cite{pedersen_geometric_2025}
\footnote{Note that projective spaces can appear in different ways in the
  differential geometric analysis as well, see e.g.
  \cite{vanderschaftLiouvilleGeometryClassical2021}.}. The geometry presented in
this paper can give insight into the relations initially presented in
\cite{hansenRelationsSeepageVelocities2018}, in the case where one is only
considering two pore areas as variables. The relations can later be extended to
a more general geometric framework when more variables are added. The
interpretation of the co-moving velocity $v_m$ might change in this
generalization process. In this paper, $v_m$ is a function that scales the
length of a displacement vector between two points. In a
differential-geometric context, $v_m$ is better treated in terms of
\textit{connection forms}\cite{pedersenParameterizationsImmiscibleTwophase2023b,
  pedersen_geometric_2025} , which is equivalent to the existence of a
\textit{covariant derivative} \cite{misnerGravitation1973} on the manifold.
Connection forms and covariant derivatives are both methods of relating the
fibers at different points of the manifolds, which means that they relate how
the velocities change from point to point on the manifold.

Another possible interpretation which appears in contact geometry is that $v_m$
describes the departure from the manifold of equilibrium states
\cite{balianHamiltonianStructureThermodynamics2001}. This is in fact quite
sensible, since one is effectively redefining what is meant by
equilibrium in a porous media two-phase flow system where the theory contains
the pore areas only \footnote{This deviation might be resolved if one were to
  include more thermodynamic variables relevant for the system.}. The deviation
can in a sense be ``eliminated'' by requiring that the physical velocities are
parametrized by $S_w$ only and imposing degree-1 homogeneity in $Q$, reducing
the number of variables to one. Note that it is important to keep in mind that
$S_w$ is a parameter on a subspace of the space of extensive areas; it is an
\textit{affine} coordinate on an affine space. This is a useful distinction from
viewing it as the coordinate of a vector in a vector space
\cite{pedersenParameterizationsImmiscibleTwophase2023b}.

No matter what geometric model one applies, some legwork is required in
interpreting the areas and velocities and how they are embedded in a potentially
larger ambient space. In the present case, projective space makes these
identifications relatively natural. The requirement that the wetting- and
non-wetting pore areas sum to the total pore area can be used to define a set of
convenient infinite points of our projective space representing physical
configurations of the areas. The finite points then represent the situation
where there is a ``unphysical'' excess area in the pore space. This situation
was already considered in \cite{royFlowAreaRelationsImmiscible2020}, where this
excess area was required to be zero. In this paper, results are obtained from
sending and intersecting objects in a designated finite part of a projective
space with the infinite hyperplane where this physical area conservation is
fulfilled. Hence, no spurious areas enter into the physical predictions.

To arrive at the results, it is necessary to introduce \textit{projective
  frames} or \textit{projective references}
\cite{casas-alveroAnalyticProjectiveGeometry2014} on the projective space, which
allow for the introduction of \textit{homogeneous coordinates} which are
straightforward to work with. Using these coordinates, one can obtain an
expression for the most important invariant in projective geometry, the
\textit{cross-ratio} or \textit{anharmonic ratio} \cite{bergerGeometry1994,
  richtergebertPerspectivesProjectiveGeometry2011}. The co-moving velocity
appears as a scaling function of a displacement vector between two points
defined by the thermodynamic- and seepage velocities. Since one in projective
spaces consider all vectors up to scaling, the co-moving velocity drops out,
which can be seen as a geometric analogue of the cancellation of $v_m$ in the
average pore velocity \cite{hansenRelationsSeepageVelocities2018}. By
considering a particular choice of quadruple of lines, one can define an angle
between two of them. It turns out that this angle is hyperbolic, with the
effective theory having a pseudo-Euclidean structure. The most well-known
pseudo-Euclidean space is \textit{Minkowski Space} \cite{misnerGravitation1973},
used in special relativity. This paper shows how one is naturally led to this
geometry by the simple fact that the saturation is bounded between zero and one.
By using the hyperbolic angle, one can define a modified saturation which
parametrizes a subspace of projective space. This can be used to parametrize
$v_{m}$.

A projective map is used to relate the parameter $S_{w}$ to the modified
saturation. This map will depend on three parameters. In most cases one can
neglect one of them, leaving only two parameters. If one plugs this relation
back into a general expression for the co-moving velocity where the cross-ratio
itself is a parameter, one can compare this to the co-moving velocity calculated
from the definition. This will demonstrate how one can parametrize $v_m$ in
terms of the macroscopic extensive variables, or equivalently $S_w$, using a
mathematical framework that is very closely tied to thermodynamics, and
generalizes more easily than other phenomenological approaches in the case of
more variables, for instance the configurational entropy
\cite{hansenStatisticalMechanicsFramework2023}.

The preliminaries of the pseudo-thermodynamic description of two-phase flow is
given in \cref{sec:rea}. The important assumption is degree-$1$ homogeneity of
the total volumetric flow rate in the areas.\cref{sec:affine-transf-vm,sec:vm-as-affine-transf,sec:vm-from-proj-transf} 

The projective geometry gives a way
of assigning parameters to this description. By letting the extensive areas
represent homogeneous coordinates of a projective line, a ``flexibility'' in the
definition of the saturation is introduced, or rather, the same saturation is
viewed in different frames. This is reflected via a set of modified saturations
that sum to unity, but is ``distributed'' differently along the number line
\cite{stillwellFourPillarsGeometry2005}. To demonstrate the above, the co-moving
velocity will be computed using data obtained from simulations with a 2d dynamic
pore network model \cite{sinhaFluidMeniscusAlgorithms2021}, and computed from
relative permeability data \cite{bennionRelativePermeabilityCharacteristics2005,
  virnovskyImplementationMultirateTechnique1998,
  oakThreePhaseRelativePermeability1990, reynoldsCharacterizingFlowBehavior2015,
  fulcherEffectCapillaryNumber1985}.


\section{Preliminaries \& Euler's Homogeneous Function Theorem}
\label{sec:rea}

What follows is review of the necessary fundamental quantities of the two-phase
flow theory of the co-moving velocity. More details can be found elsewhere
\cite{hansenThermodynamicslikeFormalismImmiscible2025}.

\begin{figure}[tb]
  \centering \includegraphics[width=\linewidth]{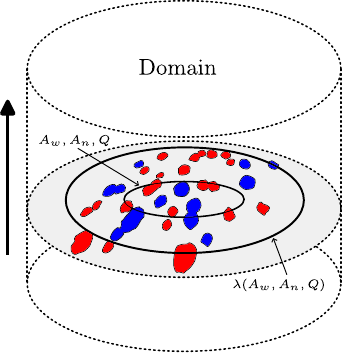}
  \caption[Euler scaling]{A porous media domain in which there is a well-defined
    overall direction of the volumetric flow rate $Q$ through $A$. Scaling the
    area $A$ by a factor $\lambda$ scales the volumetric flow rate $Q$ and the
    wetting- and non-wetting areas $A_w, A_n$ (denoted by red and blue) in the
    same manner.}
  \label{fig:euler-scaling}
\end{figure}

Take a cross-sectional area $A$ perpendicular to the total flow at each point of
a porous media domain, see \cref{fig:euler-scaling}. The area is assumed to be
large enough so that the quantities of interest have well-defined averages. This
volume is the \textit{representative elementary area} (REA). There is a time
averaged volumetric flow rate $Q$ passing through $A$ at each instant. Using the
REA, define the \textit{porosity} $\phi$ of the porous medium as
\begin{equation}
  \label{e4}
  \phi \ \equiv \ \frac{A_p}{A} \ \;,
\end{equation}
where $A_p$ is the cross-sectional pore area, the area of $A$ covered by pores.
The solid matrix $A_s$ of the porous medium is given by $A_s= A \left(1-\phi
\right)$. $\phi$ is constant, meaning the porous medium is considered homogeneous.

The pore area $A_p$ is an extensive variable as it scales by a factor $\lambda$ when
$A$ is scaled by $\lambda$. The porosity $\phi$ is unaltered, and is therefore intensive.
In some volume of the domain, $A_p$ is taken as the average over all
cross-sections orthogonal to the flow. The REA contains an area $A_w$ of (more)
wetting fluid and an area $A_n$ of (less) non-wetting fluid. Define the
\textit{wetting and non-wetting saturations} $S_w=A_w/A_p$ and $S_n=A_n/A_p$ so
that one has
\begin{align}
  A_w+A_n \ =& \ A_p \label{e11} \;, \\
  S_w+S_n \ =& \ 1 \label{e10} \ \;.
\end{align}
The saturations are intensive variables. As $A_p$ is an average over all
cross-sectional planes orthogonal to the flow direction, the same is true for
$A_w$ and $A_n$.

$Q$ can be decomposed as a sum of volumetric flow rates of the individual
fluids, denoted $Q_{w}$ and $Q_{n}$,
\begin{equation}
  \label{eq:e6}
  Q=Q_{w}+Q_{n}\;.
\end{equation}
The {\it seepage velocities\/} are defined as
\begin{align}
  v_{p}=&\frac{Q}{A_p} \label{e7}\;,\\
  v_w=&\frac{Q_w}{A_w} \label{eq:vw-seepage} \;,\\
  v_n=&\frac{Q_n}{A_n} \label{eq:vn-seepage}\;.
\end{align}
These are the measured velocities of the individual fluids passing thought the
REA. Using equation \cref{eq:e6}-\cref{eq:vn-seepage}, one finds
\begin{align}
  v =& \ \frac{Q}{A_p}=\frac{Q_w}{A_p}+\frac{Q_n}{A_p}\nonumber\\
  =& \ \frac{A_w}{A_p}\ \frac{Q_w}{A_w}+\frac{A_n}{A_p}\ \frac{Q_n}{A_n}\nonumber\\
  =& \ S_w v_w+S_n v_n\;. \label{eq:e14}
\end{align}

\Cref{eq:e6} can then be written as
\begin{equation}
  \label{eq:Q-seepage}
  Q=A_wv_{w}+A_nv_{n}\;.  
\end{equation}

The volumetric flow rates $Q$, $Q_w$ and $Q_n$ are functions of the driving
forces across the REA, for instance an applied pressure gradient or a gradient
in saturation or wettability. These are taken to be intensive quantities. All
driving forces are assumed to be constant, and dropped from the notation.


The volumetric flow rate $Q$ is extensive in the variables $A_w$ and $A_n$,
\begin{equation}
  \label{eq:Q-homogeneity}
  Q(\lambda A_w,\lambda A_n) 
  =\lambda Q(A_w,A_n) \;.
\end{equation} 
$Q$ is a degree-1 Euler homogeneous function in the areas
\cite{hansenRelationsSeepageVelocities2018}.
By taking the derivative with respect to $\lambda$ on both sides of
\cref{eq:Q-homogeneity} and setting $\lambda =1$, one obtains
\begin{equation}
  \label{eq:Q-euler-hom}
  Q(A_{w},A_{n})=A_{w}\left( \frac{\partial Q}{\partial A_{w}}\right)
  _{A_{n}}+A_{n}\left( \frac{\partial Q}{\partial A_{n}}\right)
  _{A_{w}}\;.
\end{equation}
Dividing ~\cref{eq:Q-euler-hom} by $A_{p}$, one gets
\begin{equation}
  \label{eqn10-1}
  v=S_{w}\left( \frac{\partial Q}{\partial A_w}\right)
  _{A_n}+S_{n}\left( \frac{\partial Q}{\partial A_n}\right)_{A_w}\;.  
\end{equation}
The partial derivatives acting on $Q$ have units of velocity. These define the
{\it thermodynamic velocities\/} $\hat{v}_w$ and $\hat{v}_n$,
\begin{align}
  \hat{v}_{w}=&\left( \frac{\partial Q}{\partial A_{w}}\right)_{A_{n}} \;,  
                \label{eq:vw-therm} \\
  \hat{v}_{n}=&\left( \frac{\partial Q}{\partial A_{n}}\right)_{A_{w}}  \;,  
                \label{eq:vn-therm}
\end{align}
so one may write \cref{eqn10-1} as
\begin{equation}
  \label{eqn10-5}
  v=S_w \hat{v}_w+S_n\hat{v}_n\;.
\end{equation}
The notation $ \left\{ \hat{v}_i \right\}$, where $i = w, n$, will be used for
the set of thermodynamic velocities $\left( \hat{v}_w, \hat{v}_{n} \right)$, and
the same (unhatted) notation for the set of seepage velocities.

The thermodynamic velocities $\left\{ \hat{v}_i \right\}$ are not the same as
the physical velocities $\left\{ v_{i} \right\}$. Rather, the most general
relation between $\left\{ v_i \right\}$ and $\left\{ \hat{v}_i \right\}$ that
fulfills \cref{eq:e14,eqn10-5}, namely
\begin{equation}
  \label{eq:v-both-definitions}
  v=S_w \hat{v}_w +S_n\hat{v}_n = S_wv_w+S_nv_n \ \;,
\end{equation} 
is given by
\begin{align}
  \hat{v}_w \ =& \ v_w + S_n v_m \label{eq:vw-transf} \;, \\
  \hat{v}_n \ =& \ v_n - S_w v_m \label{eq:vn-transf} \;,
\end{align} 
which defines the \textit{co-moving velocity} $v_m$. Hence, $v_m$ relates the
thermodynamic and physical seepage velocities. It can be shown that
\cite{hansenRelationsSeepageVelocities2018}
\begin{equation}
  \label{eq:therm-vel-differences}
  \left( \frac{\partial Q}{\partial S_w} \right)_{A_p} \ = \ A_p \left( \hat{v}_w - \hat{v}_n \right) \;,
\end{equation}
from which one can to show that
\begin{align}
  \label{eq:vm-derivative-form}
  v_m + v_w - v_n \ =& \ \hat{v}_w - \hat{v}_{n} \ \nonumber \\
  =& \ \frac{d v}{d S_{w}} \ \;.
\end{align}
Moreover, $\left\{ \hat{v}_i \right\}$ satisfy an analogue of the
\textit{Gibbs-Duhem} relation,
\begin{equation}
  \label{eq:GD}
  S_w \left( \frac{d \hat{v}_{w}}{d S_w} \right) + S_n\left( \frac{d \hat{v}_{n}}{d S_w} \right) \ = \ 0 \ \;.
\end{equation}
$\left\{ v_i \right\}$ do not satisfy the Gibbs-Duhem relation, which can be
formulated as
\begin{equation}
  \label{eq:co-moving-convex-comb}
  v_m \ = \ S_w \left( \frac{d v_{w}}{d S_w} \right) + S_n \left( \frac{d v_{n}}{d S_w} \right) \ \;.
\end{equation}

Numerical and experimental data show that $v_m$ can be described by a simple
constitutive relation in $v^{\prime} = dv/d S_w$, expressible as an affine function
of the form
\begin{equation}
  \label{eq:vm-constitutive}
  v_m \ = \ b v^{\prime} + a v_0 \ \;,
\end{equation}
where $a$ and $b$ are dependent on other macroscopic variables than the areas,
viscosity ratio and driving forces, and $v_0$ is a velocity scale. The
interpretation is that the co-moving velocity is an effective description of the
porous medium and the interactions between the two fluids
\cite{royCoMovingVelocityImmiscible2022, hansen2024linearity}.

\subsection{Requirement for affine \& projective spaces}
\label{sec:affine-transf-vm}

A central question posed in several recent works
\cite{hansenRelationsSeepageVelocities2018, royCoMovingVelocityImmiscible2022,
  pedersenParameterizationsImmiscibleTwophase2023b} is how to classify the map
between the velocities $\left\{ \hat{v}_i \right\}$ and $\left\{ v_i \right\}$
in \cref{eq:v-both-definitions}. The relation describes going from one set of
equations of state to another, with the equilibrium assumption being
steady-state flow \cite{hansenStatisticalMechanicsFramework2023}, and a
degree-$1$ homogeneous volumetric flow rate is assumed. Linear, homogeneous
equations, where the coefficients of the linear equations here are given by the
velocities, are fundamental objects in projective geometry, of which affine
spaces are a part of.

One can connect the relations in \cref{sec:rea} to affine spaces
\cite{pedersen_geometric_2025}. An \textit{affine space}
\cite{gallierGeometricMethodsApplications2011,
  crampinApplicableDifferentialGeometry1987, bergerGeometry1994} is a space $\mathcal{A}$
of points with an accompanying vector space $\mathcal{V}$, where the space $\mathcal{V}$ can be seen
the space of displacements of points $p \in \mathcal{A}$. A vector $v \in \mathcal{V}$ maps a point $p$
to another point $p' \in \mathcal{A}$ by translation \footnote{ The difference between two
  points $p, p' \in \mathcal{A}$ can in the same manner be identified with an element $v \in
  \mathcal{V}$, which intuitively just means that the difference between two points can be
  identified with a vector between them.}. An \textit{affine transformation} $f$
is a composition of a linear transformation and a translation
\cite{bergerGeometry1994}, and transforms affine spaces to themselves
\footnote{In other words, one has added the ability to translate the origin to
  the ordinary linear transformations of vectors.}.

By fixing any point $o \in \mathcal{A}$ as the origin, where $\mathrm{dim}\left( \mathcal{A} \right)
=n$, one obtains a vector space at $o$. On $\mathcal{A}$, one can define an \textit{affine
  frame} \cite{crampinApplicableDifferentialGeometry1987}, which is just a
choice of origin $o$ and basis vectors $\left( e_1, e_2, \ldots, e_n \right) \equiv
\left\{ e_i \right\}$ of the vector space at $o$. Then, any point $p \in \mathcal{A}$ can be
expressed with respect to this affine frame as
\begin{align}
  \label{eq:affine-coordinates}
  p \ =& \ o + \left( p - o \right) \\
  =& \  o + p^ie_i \ \;,
\end{align}
where $\left( p-o \right)$ is a vector at $o$ and a summation convention in $i$
is implied. This vector can be expanded in the basis $\left\{ e_i \right\}$,
with components $\left\{ p^{i} \right\}$ called the \textit{affine coordinates}
of the point $p$ with
respect to this basis.  \\

General linear combinations of vectors are not well-defined on affine spaces,
since they depend on the choice of origin. However, linear combinations where
the coefficients $\alpha_i$ sum to $1$ are still well-defined. Such a linear
combination is called an \textit{affine combination} \footnote{ If one adds the
  condition that $\left\{ \alpha_i \right\} \geq 0 $, it is called a
  \textit{convex combination} \cite{rockafellarConvexAnalysis1997}. In fact,
  affine/convex combinations are defined for both vector and \textit{points},
  since the choice of origin always cancel.}. An affine combination of two
points results in a point on the line through these points, and a convex
combination of two points yields a point on the line segment between these
points. Thus, \cref{eq:v-both-definitions} implies that one can view the pore
velocity $v$ as a point on some one-dimensional line $L$, where the position of
the point is just the value of the velocity function. $v \in L$ is on a line
segment with endpoints $\left\{ \hat{v}_i \right\} $, or between $\left\{ v_i
\right\}$.

An expression where affine points are defined as a weighted combinations of a
number of reference- or basis points are called \textit{barycentric
  coordinates}, where the basis points define a \textit{barycentric frame}
\cite{gallierGeometricMethodsApplications2011}. In our case, the barycentric
coordinates or \textit{weights} $S_w, S_n$ with respect to the basis $\left\{
  \hat{v}_i \right\}$ or $\left\{ v_i \right\}$ are normalized such that they
sum to unity \footnote{In this case one often calls them \textit{homogeneous
    barycentric coordinates}. A slight abuse of terminology will be used, as
  they will be called either barycentric coordinates or just affine coordinates.
  The definition of barycentric coordinates do not require that the weights are
  bounded between zero and one. This is only the case if the points being
  described are within the line segment defined by the basis points.} This paper
will use barycentric coordinates, as they can be completed to coordinates on a
projective space. Before considering barycentric coordinates exclusively, the
relation between affine- and barycentric coordinates is shown in the next
paragraph.

\Cref{eq:vw-transf,eq:vn-transf,eq:affine-coordinates} can be used to write
\cref{eqn10-5} as
\begin{align}
  \label{eq:affine-basis-and-coordinates}
  Q \ =& \ A_w \left(\hat{v}_w - S_n v_m  \right) + A_n \left( \hat{v}_n + S_w v_m \right) \nonumber \\
  =& \ A_pv_m + A_w \left( \hat{v}_w - v_m \right) + A_n \left( \hat{v}_n - v_m \right) \nonumber \\
  =& \ A_pv_m+ A_w \left( v_w - v_m \right) + A_n \left( v_n - v_m \right) \nonumber \\
  =& \ A_{p}v_m + p^i e_{i} \ \;.
\end{align}
Note that this is not related to the definition of $v_m$, as it holds for any
quantity. Instead, note that if $v, v_m, \left\{ \hat{v}_i \right\}, \left\{ v_i
\right\} \in L$, the velocity \textit{differences} corresponds to the vectors
$\left\{ e_i \right\}$ of an affine frame. By setting $o = A_pv_m$ in eq.
\eqref{eq:affine-coordinates}, one can identify $A_pv_m$ with a choice of origin
on $L$. Up to scaling by $A_p$, the affine frame would then be $\left\{v_m, (v_w
  - v_m), (v_n - v_m) \right\}$ (or equivalently, with hatted velocities), i.e.
an origin $v_m$ and two vectors with respect to this origin.

By imposing $A_p = A_w + A_n$, which gives $S_w + S_n = 1$, and scaling
eq.~\eqref{eq:affine-basis-and-coordinates} by $A_p$, one can rewrite the above
as a relation on a one-dimensional affine space $\mathbb{A}^1$ \footnote{It is
  important that one remembers that this space is embedded in a higher
  dimensional space.},
\begin{align}
  \label{eq:two-to-one-dim-affine-frame}
  A_pv \ = & \ A_p v_m + A_w \left( \hat{v}_w - v_m  \right) + A_n \left( \hat{v}_n - v_m \right) \nonumber \\
  \mapsto & \ v_m + S_w \left( \hat{v}_w -v_m \right) + S_n \left( \hat{v}_n - v_m \right) \;, 
\end{align}
which means that
\begin{align}
  \label{eq:more-rewrite}
  v \ =& \ \hat{v}_n + S_w \left( \hat{v}_w- \hat{v}_n \right) \nonumber \\
  =& \ \hat{v}_n + S_w v' \;.
\end{align}
Equivalently, one can replace the thermodynamic velocities by seepage
velocities. If for instance $v_n$ was the starting point and $v_w$ the endpoint
on the line segment $\subset \mathbb{A}^{1}$, $S_w$ determines where $v$ is
located on the line segment of length $( v_w - v_n )$.



\subsection{$v_m$ via affine transformation}
\label{sec:vm-as-affine-transf}

For an affine combination with coefficients $\alpha_i$ and points (or vectors) $
x_{i}$, an affine transformation $f$ satisfies
\begin{align}
  \label{eq:affine-transformation}
  f\left(  \sum\displaylimits_{i=1}^n \alpha_i x_{i}\right) =&   \sum\displaylimits_{i=1}^n \alpha_i f\parens{x_{i}}  \ \;.
\end{align}

Comparing with \cref{eq:v-both-definitions}, one is motivated to define an
affine transformation $f$ such that $f (v) = v$ \cite{pedersen_geometric_2025},
and
\begin{equation}
  \label{eq:f-v-fixed-point}
  f\parens{v} \ = \ S_w f\left( \hat{v}_w \right) + S_n f \left( \hat{v}_{n} \right) \ = \ S_w v_w + S_n v_n \;,
\end{equation}
which means that
\begin{align}
  \label{eq:affine-map-f-w}
  f \left( \hat{v}_w \right) \ = \ v_w \;, \\
  f \left( \hat{v}_n \right) \ = \ v_n \label{eq:affine-map-f-n} \;.
\end{align}
Using \cref{eq:vw-transf,eq:vn-transf}, one has
\begin{align}
  \label{eq:affine-map-f-w-vm}
  \hat{v}_w - f \left( \hat{v}_w \right) \ \equiv \ S_nv_m \;, \\
  \hat{v}_n - f \left( \hat{v}_n \right) \ \equiv \ -S_wv_m \label{eq:affine-map-f-n-vm} \;.
\end{align}
By the definition of an affine transformation, by setting $\hat{v}_i = f \left(
  v_{0,i} \right)$ for a reference velocity $v_{0,i}$ (which can be a function
of $S_w$), the right-hand side of
\cref{eq:affine-map-f-w-vm,eq:affine-map-f-n-vm} is a linear transformation of
the vector $(\hat{v}_i - \hat{v}_{0,i})$, which in one dimension is just
multiplication by a scalar $\lambda$. Intuitively, these linear transformations
``move'' the endpoints $\left\{ \hat{v}_i \right\}$ or $\left\{ v_i \right\}$
such that the position of $v$ on $L$ relative to these endpoints is preserved.
If one is only interested in the position of $v$, the value of $v_m$ is
irrelevant since only the relative position of $v$ matters.

\Cref{eq:affine-map-f-w-vm,eq:affine-map-f-n-vm} looks like a non-autonomous
system, where the hatted velocities are the points and $S_w$ is analogous to a
time-parameter \cite{arnoldOrdinaryDifferentialEquations1992}. One needs two
different reference functions $\hat{v}_{0,i}$ to get two thermodynamic
velocities if one only has a single map $f$. However, knowing the solution to
either one of \cref{eq:affine-map-f-w-vm,eq:affine-map-f-n-vm} determines the
other, so only one of the relations are needed, with $v_i = \hat{v}_{i}$ two
fixed points of $f$. Defining the map $f$ in this way represents another
formulation of the problem which will not be pursued further, as it is too
general and not clear on how $f$ is computed.

\subsection{Volumetric flow rate scaling \& line at infinity}
\label{sec:vm-from-proj-transf}

Physically, the map $\left( v_w, v_n \right) \leftrightarrow \left( \hat{v}_w, \hat{v}_n
\right)$ implies making different physical assumptions about the fluids through
constitutive relations. In a porous two-phase flow system, the saturation range
itself depends on the properties of the system. For the seepage velocities, due
to the presence of irreducible wetting- and non-wetting saturations, one often
scales $S_w$ to get a normalized saturation $\in \left[ 0,1 \right]$,
\begin{equation}
  \label{eq:saturation-scaling}
  S_w \mapsto a_{r}S_w + b_{r} \equiv S_{w,r} \;,
\end{equation}
where the constants $a_{r},b_{r}$ depend on the irreducible saturations.
Therefore, one is implicitly using different saturation ranges depending on if
the thermodynamic- or seepage- velocities are used. In eq.
\eqref{eq:saturation-scaling}, the relation between the normalized and
non-normalized saturation ranges is an \textit{affine transformation} in the
parameter $S_w$.

More generally, one could imagine defining other transformations between two
saturation ranges, which would imply that one is describing the saturation(s) of
a pair of abstract fluids which may or may not be measurable. Then, $S_{w,r}$
could possibly be a more complicated function of $S_w$ than in
\cref{eq:saturation-scaling}. Call the wetting saturation in this case
$\tilde{S}_w$. It is only when the saturation ranges $S_w$, $\tilde{S}_{w}$ are
compared that one can see the non-trivial relation that could be different from
\cref{eq:saturation-scaling}. The functional relation between $\tilde{S}_{w}
\left( S_{w} \right)$ also encodes a relation between the velocities related to
the two saturation ranges.

By requiring extensivity of $Q$ in the areas, the velocities themselves do not
interfere with the homogeneous structure, meaning there is a degree of freedom
in which fluid phase is attributed to which velocity. Homogeneity is a central
concept in projective geometry, and the connection between the transformation
alluded to in the above and projective geometry will is possible due to the
homogeneity-assumption in $Q$. A crucial point will be the following: through
the requirement $A_w + A_n = A_p$, one does not require that $A_p$ is constant.
One may pick two independent variables among $\left( A_w, A_n, A_p \right)$. One
can then scale everything by one of the variables in the above set, say $A_p$
\footnote{There is a difference in whether one divides all relations by a
  \textit{fixed} value or consider all relations up to arbitrary scaling, i.e.
  modulo $A_p$. The latter is a non-trivial scaling behaviour that can scale the
  relations differently depending on the value of the parameter $S_w$, so one
  must treat $A_p$ as an arbitrary scaling factor.}.

A possible way of viewing the relation between the projective formalism and the
extensivity and additivity of $Q$ is now presented. It is \textit{only} assumed
that $Q$ is extensive in the pore areas (for a suitable REA), $Q(\lambda A_w,
\lambda A_n) = \lambda Q \left( A_w, A_n \right) $. Due to interactions between
the fluid phases and the porous medium, the same is \textit{not} automatically
assumed for $Q_i$, depending on our external control variables. By definition of
a \textit{thermodynamic subsystem}
\cite{callenThermodynamicsIntroductionThermostatistics1985,
  touchetteWhenQuantityAdditive2002}, $Q_i$ should only be functions of $A_i$,
i.e. it is completely determined by the a subset of the variables, which can
only be one of the areas in $\left( A_w, A_n\right)$. Only in this case is the
system \textit{additive} \footnote{Even though that extensivity is taken here,
  note that additivity do not imply extensivity in general
  \cite{touchetteWhenQuantityAdditive2002}.}, meaning here that
\begin{equation}
  \label{eq:additive-Q}
  Q \left( A_w, A_n \right) \ \approx \ Q_w \left( A_w \right) + Q_n \left( A_n \right) \;.
\end{equation}
$Q_i \left( A_i \right)$ in the above equation describes the situation where the
volumetric flow rate of fluid $i$ is completely determined by the amount of
wetting area, i.e. the ``state'' of the subsystem $Q_{i}$ is completely
determined by an appropriate subset of the variables. The physical situation
during two-phase flow in a porous medium is, in general, too complicated to
justify \cref{eq:additive-Q} in all cases. There are interactions between the
two phases, and the phases and the porous medium itself. It will also depend on
the ensemble and boundary conditions on the flow domain. To state the general
non-additivity more clearly, one can add a term $Q_{I}$ in \cref{eq:additive-Q}
analogous to an interaction energy which takes into account the correlations
between the flow rates \cite{oppenheimThermodynamicsLongrangeInteractions},
\begin{equation}
  \label{eq:interaction-flow}
  Q \left( A_w, A_n \right) \ = \ Q_w \left( A_w \right) + Q_n \left( A_n \right) + Q_I \left(A_w, A_n  \right) \;,
\end{equation}
where it is assumed that the ``interaction flow'' $Q_{I}$ must be a function of
both areas. For some fixed $Q$, one can imagine $Q_I$ as assigning a part of the
flow rate to contain the interaction between the phases.
Given some physical interpretation of the mechanism of this interaction, one can
justify partitioning $Q_I$ and absorbing it into the $Q_i$´s, implying a
redistribution of the flow rate between the phases. Note that this is closely
related to the interpretation of the co-moving velocity
\cite{hansenRelationsSeepageVelocities2018, royFlowAreaRelationsImmiscible2020,
  royCoMovingVelocityImmiscible2022}.

Again, even though one assumes extensivity in the total $Q$, how each term
scales on the left-hand side of \cref{eq:interaction-flow} under $A \mapsto
\lambda A$ depends on the particular system and imposed restrictions. Moreover,
even though one selects a suitable REA and restrict $Q_{I}$ to be degree-1
homogeneous, $\left\{ Q_i \right\}$ in \cref{eq:interaction-flow} can enter a
non-trivial scaling regime. This is because as the REA is scaled, the fluids
will at a certain scale interact with the pore structure itself. When
considering the fluids in terms of the saturation $S_w$, one cannot get away
from this effect, since one or the other fluid will, depending on the saturation
range, enter this regime when the saturation is varied.

In the case that the areas are the control variables, the REA contains more or
less pores as one scales $A \mapsto \lambda A$. When the porous medium is
assumed to be homogeneous, this will not alter the extensivity in $Q$ until one
reaches the scale where the REA is not applicable
\cite{fyhnLocalStatisticsImmiscible2023}. The situation considered in sec.
\ref{sec:rea} is equivalent to defining
\begin{equation}
  \label{eq:Q-rewritten-partition}
  Q \left( A_w, A_n \right) \ = \ Q_w \left( A_w, A_n \right) + Q_{n} \left( A_w, A_n \right) \;,
\end{equation}
where $\left\{ Q_i \right\}$ are different functions than the $\left\{ Q_i
\right\}$ in \cref{eq:additive-Q,eq:interaction-flow}. In
\cref{eq:Q-rewritten-partition}, the interaction is absorbed into the phase flow
rates. These $Q_i$'s are those that define $\left\{ v_i \right\}$ in
\cref{eq:vw-seepage,eq:vn-seepage}. However, the thermodynamic formulation in
terms of subsystems is different, \cref{eq:interaction-flow}. Here, the
subsystems are only functions of a single area, so one has no way of formulating
the fact that the flow also depends on the other area in $\left\{ Q_i \right\}$.
Passing to saturations, this makes it clear that there is non-trivial scaling
present from the definition of the saturation, $S_w = A_w/(A_w + A_n)$. The area
$A_i$ in $Q_i$ in \cref{eq:interaction-flow} scales as if the other area was
constant. Therefore, this scaling is present even though one is outside the
non-trivial scaling regime where $\lambda A$ is close to the pore scale.

One can express this scaling a bit more explicitly. Sending $A \mapsto \lambda
A$ in \cref{eq:interaction-flow}, one has
\begin{align}
  \lambda^{r} Q \ =& \ Q \left( \lambda A_w, \lambda A_{n}\right)    \nonumber \\
  =& \ Q_w ( \lambda A_w) + Q_n (  \lambda A_n) + Q_I \left( \lambda A_{w}, \lambda A_n \right)  \nonumber \\
  =& \ \lambda^{s} Q_w \left( A_w \right) + \lambda^{t} Q_n \left(  A_n \right) + \lambda^u Q_I \left( A_w, A_n \right) \label{eq:quasi-homogeneity-areas} \;,
\end{align}
where $s, t, u$ are in general functions, with $s \neq t \neq u$ being assumed.
One can write \cref{eq:quasi-homogeneity-areas} in terms of different control
variables as
\begin{align}
  \label{eq:quasi-homogeneity-flow-rates}
  \lambda Q \left( Q_w, Q_n, Q_{I} \right) \ = \ Q \left( \lambda^s Q_{w}, \lambda^t Q_n, \lambda^u Q_I \right) \;,
\end{align}
in other words, $Q$ is a \textit{quasi-homogeneous function}
\cite{belgiornoQuasiHomogeneousThermodynamicsBlack2005,
  bravettiZerothLawQuasihomogeneous2017} of degree $r = 1$ and type $\left( s,t
  , u\right)$ in the volumetric flow rates $\left\{ Q_i \right\}, Q_{I}$.
$\lambda^s, \lambda^{t}, \lambda^{u}$ are dependent functions of the areas, and
reflect the fact that there is an indeterminacy in which volumetric flow rate
corresponds to which phase. This is due to mutual co-carrying transport of the
fluids due to their interactions in the pore space, and with the pore space
itself. Since there are two independent variables $A_w, A_n$ which determines
the state of the system, there must be dependency among $Q_w, Q_n, Q_I$. One is
free to absorb $Q_I$ into one of the other flow rates, for instance $\lambda Q$,
to define $Q_{0} = \lambda Q - \lambda^u Q_{I}$.

The above discussion has interesting implications. Analogous to
\cref{e7,eq:vw-seepage,eq:vn-seepage}, define the new velocities (assume for now
that $A_p$ is independent of $A_w, A_n$)
\begin{align}
  \tilde{v}_w \ \equiv& \ \lambda^{s}\frac{Q_{w} \left( A_w \right)}{A_w} \label{eq:pseudo-velocities-vw} \;, \\
  \tilde{v}_n \ \equiv& \ \lambda^{t}\frac{Q_{n} \left( A_n \right)}{A_n} \label{eq:pseudo-velocities-vn} \;, \\
  \tilde{v} \ \equiv& \ \frac{Q_{0} \left( A_w, A_n \right)}{A_p} \label{eq:pseudo-velocities-v} \;,
\end{align}
so that \cref{eq:quasi-homogeneity-areas} can be written as
\begin{equation}
  \label{eq:dot-product-lambda-scaling}
  0 = \vec{A}^{\,T} \cdot \vec{\tilde{v}} 
\end{equation}
with vectors
\begin{equation}
  \label{eq:tilde-vectors}
  \vec{A} \equiv
  \begin{pmatrix}
    A_{w} \\
    A_{n} \\
    A_{p}
  \end{pmatrix} \;, \ \ 
  \vec{\tilde{v}} \equiv
  \begin{pmatrix}
    \tilde{v}_{w} \\
    \tilde{v}_{n} \\
    -\tilde{v}
  \end{pmatrix} \;.
\end{equation}
\Cref{eq:dot-product-lambda-scaling} determines a linear equation in the
components of $\vec{A}$, or equivalently, a linear equation in the components of
$\vec{\tilde{v}}$. The ``dual'' components in $\vec{\tilde{v}}$ consist of some
combination of areas, volumetric flow rates and scaling parameters, and are not
straightforwardly interpreted as velocities, because they do not scale as
intensive variables in general. Note that the equation is invariant under
scaling by a common factor in the components in either representation.

The relations in terms of the scaling parameters in this section are far too
general and complicated, and gives no clear way towards obtaining a
parametrization of $v_m$. Rather, this section should be seen as a demonstration
of how projective geometry can be introduced into the problem. In projective
geometry, a scalar product such as \cref{eq:dot-product-lambda-scaling}
determines projective hyperplanes, embedded in some higher dimensional
projective space. In fact, one can write
\begin{equation}
  \label{eq:ordinary-dot-product}
  0 \ = \ \vec{A}^{\,T} \cdot \vec{v} \ = \  \vec{A}^{\,T} \cdot \vec{\hat{v}}  \;,
\end{equation}
with
\begin{align}
  \label{eq:v-vec}
  \vec{v} \ \equiv & \ \left( v_w, v_n, -v \right)^{T} \;, \\
  \vec{\hat{v}} \ \equiv & \ \left( \hat{v}_w, \hat{v}_n, -v \right)^{T}\label{eq:v-hat-vec} \;.
\end{align}
$0 = \vec{A}^{\,T} \cdot \vec{v} = \vec{A}^{\,T} \cdot \vec{\hat{v}}$ is only
satisfied if $\vec{\hat{v}}$ and $\vec{v}$ (and $\vec{\tilde{v}}$ for that
matter) are related by rotation in a plane perpendicular to $\vec{A}$, or are
scalar multiples of each other (this is clearly not the case in general). If one
imposes $A_w + A_n - A_p = 0 $, the vectors $\vec{A}^{\,T}$ with different $A_w,
A_n$ form a plane. However, in this case, there is no freedom for the vectors
$\vec{v}, \vec{\hat{v}}$ to rotate: they must both be normal to the plane
determined by $\vec{A}^{\,T}$, hence they are parallel and can only differ by a
scalar multiple (and possibly a translation).

In working with projective geometry, instead of imposing $A_w + A_n - A_p = 0$
in the equations from the start, the three areas are kept independent. Then, one
can enforce the relation between the areas in the velocities by intersecting the
lines given by the velocity vectors with a chosen improper line. This line is
the one defined by $A_w + A_n - A_p = 0$, which is seen by writing
\begin{align}
  A_w + A_n - A_p \ =& \ A_w \cdot \left( 1 \right) + A_n \cdot \left( 1 \right) - A_p \cdot \left( -1 \right) \nonumber \\
  \equiv& \ \vec{A}^{\,T} \cdot  l_{\infty} = 0 \label{eq:area-relation-as-line} \;,
\end{align}
which defines the line at infinity $l_{\infty} = (1,1,-1)^{T}$ \footnote{ The
  term ``velocity vector'' as the dual vector to $\vec{A}^{\,T}$ in the scalar
  product can be misleading, because transformations can in fact mix areas and
  velocities. Hence, this term will be avoided.}. This allows for a more
flexible geometric framework: one can work with finite points and lines where
the area conservation of \cref{e11} is not fulfilled, and then send the obtained
configurations to infinity so that they can be interpreted physically. One sees
that the vector $\vec{\tilde{v}}$ determines an arbitrary line in this space,
which is a projective plane, see \cref{sec:projective-space}.

\section{Algebraic viewpoint, affine subspaces}
\label{sec:spaces}

Affine and projective relations will now be introduced from the relations in
\cref{sec:rea}. \Cref{eq:Q-seepage} can be interpreted algebraically by writing
\begin{align}
  \label{eq:Q-alg-line}
  0 = \left( A_w v_w + A_n v_{n} - Q \right) =& \left( A_w v_w + A_n v_{n} - A_p v \right) \nonumber \\
  =& \left( A_w \left( v_w - v \right) + A_n \left( v_{n} - v \right)  \right) \;.
\end{align}
From \cref{eq:Q-euler-hom}, eq. \eqref{eq:Q-alg-line} holds with $\left\{ v_{i}
\right\}$ replaced by $\left\{ \hat{v}_i \right\}$. In general, there are
several possibilities of identifying the extensive coordinates. For instance,
one could consider either $\left( A_w, A_n, Q\right)$, $\left( A_w, A_n, A_p
\right)$, or $\left( A_w, A_n \right)$. Which set of variables is used depends
on the framework, if the space of extensive variables is viewed as embedded in
some ambient space, or if the space of variables is viewed as intrinsic
\footnote{ In equilibrium thermodynamics, one would take the set $\left( A_w,
    A_n, Q \right)$ as extensive variables, and define a thermodynamic system as
  a surface $Q = Q \left( A_w, A_n \right)$ via a constitutive relation.}. Here,
$\left( A_w, A_n, A_p \right)$ and the homogeneity of $Q$ is used to write $Q =
A_p v$. $v$ is then viewed as a variable associated to the extensive variable
$A_p$. The dependency between the areas is given by the condition that the
homogeneous equation
\begin{equation}
  \label{eq:linear-form-F}
  F = A_w + A_n - A_p
\end{equation}
equals $0$.

For the purposes of demonstration, consider only the set $\left( A_w, A_n
\right) \equiv x$ as independent variables
\cite{pedersenParameterizationsImmiscibleTwophase2023b}. $x$ determines a point
in a two-dimensional space $\mathcal{A}$. Eq. \eqref{eq:Q-alg-line} then
determines a curve $C \subset \mathcal{A}$ as $\left( A_w \left( v_w - v \right)
  + A_n \left( v_{n} - v \right) \right) \equiv G \left(x \right) = 0$. The
curve determined by $G$ is the line where all points $x$ satisfy $G \left( x
\right) = 0$. The velocities $v_w,v_{n}, v$ are functions of the parameter
$S_w$.
For each point $x$, eq. \eqref{eq:Q-alg-line} determines a new line $G_{S_w}
\left( x \left( S_w \right) \right) =0$, i.e. $G$ is parametrized by $S_w$.
Dually, one can consider all lines through a fixed point $x$ by taking the
velocities $\left\{ \left( v_i - v \right) \right\}$ as coordinates. The
velocities are the dual coordinates of the areas, defined on the dual space
$\mathcal{A}^{\ast}$ of $\mathcal{A}$. The ``dual'' is interpreted in the
projective sense, which means $\mathcal{A}^{*}$ is the space of lines in
$\mathcal{A}$ \cite{richtergebertPerspectivesProjectiveGeometry2011}. Taking
either $\left\{ \left( v_i - v\right) \right\}$ or $\left\{ \left( \hat{v}_i - v
  \right) \right\}$ as coordinates correspond to different choices of
coordinates on $\mathcal{A}^{\ast}$. $S_w$ is called a parameter because it
determines the position along the line $A_w + A_n - A_p =0$, similar to a
time-parameter describing the path of a particle.

The generalization is straightforward when $A_p$ is added as an additional
independent variable. Let $\left( A_w, A_n, A_p \right) \in E = \mathbb{R}^3$
\cite{pedersenParameterizationsImmiscibleTwophase2023b}. An affine space can be
obtained by setting $A_p = c$, where $c$ is a some constant. The plane $A_p = c$
is an affine plane $\mathcal{H}$, written algebraically as $Z = A_p - c = 0$. It
is just the linear hyperplane \footnote{A \textit{linear} hyperplane goes
  through the origin $\left( A_w, A_n, A_{p} \right) = \left( 0,0,0 \right)$. }
spanned by $A_w$, $A_n$ translated to $\left( 0,0,c \right)$. This space is a 2d
affine space, see \cref{sec:affine-transf-vm}. \Cref{eq:linear-form-F} set to
zero, $F = 0$, defines a linear hyperplane $H$ in $E$. The intersection between
$H$ and the affine plane $Z=0$ is just an affine line $L$ in $\mathcal{H}$. Setting $A_p =
c$ in $F$ and dividing by $c$, one can write $S_w + S_n - 1 = 0$, where $S_w \equiv
A_w/A_p$, $S_n \equiv A_n/A_p$. When $S_w, S_n \in \left[ 0,1 \right]$, these
quantities are the ordinary saturations.

$A_p$ is arbitrary in general, and not necessarily constant. The relation $S_w +
S_n = 1$ is still valid in this case \footnote{ However, changing $A_p$ changes
  the overall scaling, so the amount of total pore area one is defining the
  saturations with respect to might change.}, as long as if one considers the
points up to arbitrary scaling. $\left( S_w, S_n \right)$ are the coordinates of
the points of $L$ in the plane $\mathcal{H}$, scaled by $A_p$. Since $S_n = \left( 1 - S_w
\right)$, one only has a single independent parameter $S_w$ determining the
position on $L$. In this context, $S_w$ is called a \textit{affine parameter}
along the one-dimensional space $L$, and is the geometric interpretation of the
saturation.



\subsection{Projective spaces}
\label{sec:projective-space}

Consider the vector space $E \cong \mathbb{R}^{3}$ with independent coordinates $\left( A_w ,
  A_n , A_{p} \right)$. A \textit{(real) projective space}
\cite{anderssonComplexConvexityAnalytic2004} of dimension $n$, denoted by $\mathbb{R}\mathbb{P}^n$
or just $\mathbb{P}^{n}$, can be constructed as the set of lines through the origin in
$\mathbb{R}^{n+1}$. Alternatively, it can be defined by identifying any two points $x,y$
in $\mathbb{R}^{n+1} \backslash \left\{ 0 \right\}$ that are related by multiplication by a
non-zero scalar $\lambda$, meaning that $x \sim y$ if $x = \lambda y$.

A projective space can be constructed from any vector space. If this space is
$E$, the \textit{projectivization of $E$} \cite{perrinAlgebraicGeometry2008,
  gallierGeometricMethodsApplications2011}, denoted $\mathbb{P}\left( E \right)$, is
defined by taking the one-dimensional subspaces of $E$ to be the points of
$\mathbb{P}\left( E \right)$. A one-dimensional vector subspace $N$ of $E$ is simply a
line through the origin in $E$ \footnote{In other words, the $\mathbb{P}\left( E \right)$
  is quotient of $E - \left\{ 0 \right\}$ (all vectors except the zero vector)
  by the action of multiplication of scalars. This means that one can introduce
  an equivalence relation $x \sim \lambda x$ for all points $x \in E$ and scalars $\lambda \neq
  0$.}. Applying this construction to $E$, $\mathrm{dim}(E) = 3$, one obtains
the \textit{ projective plane} $\mathbb{P}^2$ \cite{stillwellFourPillarsGeometry2005}
\footnote{ In general, $\mathbb{P}^n_{M}$ denotes the projective space of dimension $n$
  obtained from the projectivization of a general vector space $M$ of dimension
  $n+1$.}. In this article, the notation $p: E \ \left\{ 0 \right\} \mapsto \mathbb{P} \left( E
\right)$ is used for the map that sends a point in $E$ to the line through the
origin in $E$, which is a point of $\mathbb{P} \left( E \right)$.

As mentioned in \cref{sec:spaces}, linear equations in $\left( A_w, A_n, A_{p}
\right)$ specifies linear hyperplanes in $E$. Analogous to \cref{sec:spaces},
one can define $W(A_{w}, A_n, A_{p}) = A_w v_w + A_nv_n - A_pv = 0$, which gives
a hyperplane $H \subset E$. $\mathbb{P} \left( H \right)$ is a projective subspace of $\mathbb{P}(E)$, a
projective space of dimension $1$, called a (real) \textit{projective line} $ L
= \mathbb{P}^{1}$ sitting inside $\mathbb{P}^{2}$ \footnote{One should really be careful to denote
  when $A_w$, $A_{n}$ are used as coordinates in $\left( A_w, A_n, A_{p}
  \right)$ or in the hyperplane $H \in E$ with coordinates $(A_w, A_n)$. This
  article will (a bit sloppily) not make this distinction in the
  notation. \label{fn:sloppily} }. Since the velocities are functions of $S_w$,
each value of $S_w$ defines a new hyperplane. Thus, $W = W_{S_w} = 0$ describes
a family of hyperplanes parametrized by $S_w$.

\textit{Homogeneous coordinates} \cite{gallierGeometricMethodsApplications2011,
  stillwellFourPillarsGeometry2005} are useful when working with projective
spaces. These are just the components of the points in the vector space $E$ up
to an arbitrary scalar. A point (vector) $\vec{x} \in E$ given by $\left( A_w,
  A_n, A_{p} \right) \equiv \vec{x}$, where in $\mathbb{P}^{2}$ one has $\vec{x} \sim \lambda \vec{x}$
for scalars $\lambda \neq 0$, has homogeneous coordinates $\left[ A_w:A_n:A_p \right]$.
Since the homogeneous coordinates are defined up to a scalar, then for $F=0$ one
can write
\begin{align}
  \label{eq:hom-coord-equivalence}
  \left[ A_{w}:A_n:A_{p}
  \right] \ = \ \left[ \frac{A_{w}}{A_p}:\frac{A_n}{A_p}:\frac{A_{p}}{A_p} \right] =&
                                                                                      \left[ S_w: S_n: 1 \right] \nonumber \\
  \equiv& \ \left( S_w, S_n \right) \;,
\end{align}
where $A_w, A_n$ and $A_p$ are dependent. Thus, $\left[ A_w:A_n:\left( A_w + A_n
  \right) \right]$ describes a projective subspace of $\mathbb{P}^{2}$. This is subspace
is a projective line $\mathbb{P}^1$ since the defining equation $F$ is linear in the
areas.

$\left( S_w, S_n \right)$ are dependent coordinates on an affine space of
dimension two \footnote{In eq. \eqref{eq:hom-coord-equivalence}, $S_w$ and $S_n$
  are dependent, so even though there are three coordinates, one describes a
  two-dimensional affine plane. }, which describe a affine line $\mathbb{A}^{1}$
embedded in the affine plane. Any vector line through the origin except for
those where $A_p = 0$ (which are parallel to the plane) hits this affine plane.
Since everything is considered up to scaling, the value of $A_p$ does not
matter, so this is simply the affine plane defined by $A_p \neq 0$. Note that one
gets the saturations (which sum to one) by additionally setting $A_w + A_n
=A_p$, otherwise, the affine coordinates do not necessarily lie on a line. In
general, one can choose any affine plane in $E$, and consider the points of
intersections of the vector lines and this plane \footnote{Letting any
  homogeneous coordinate be non-zero gives an open subset of projective space,
  which is always an affine space \cite{perrinAlgebraicGeometry2008}. One then
  divides the other coordinates by this non-zero coordinate. The exact value of
  the non-zero homogeneous coordinate is not important, and it could in general
  be a function of some parameter. }.


A useful view of the projective line $\mathbb{P}^{1}$ is as the ordinary number line
together with all possible projections onto itself. After embedding the line in
a plane, one can pick a point of projection and a second line, and project the
points from the first line onto the second via lines through the point of
projection. The procedure can be repeated by picking an arbitrary finite number
of new lines and points of projections. The number line together with all such
projections is the real projective line $\mathbb{P}^{1}$
\cite{stillwellFourPillarsGeometry2005}. An arbitrary finite composition of
projections is in general called a \textit{projective map} or a
\textit{homography}, to be treated in sec. \ref{sec:rp-homographies} and
\cref{app:rp1}.
\subsection{Duality}
\label{sec:affine-space-points}

A consequence of the projective space construction is that there is a duality
between points and lines. This is so because hyperplanes in the underlying
vector space (here $E$) \cite{richtergebertPerspectivesProjectiveGeometry2011,
  bergerGeometry1994}, which correspond to projective lines $\mathbb{P}^{1} \subset \mathbb{P}^{2}$, has
a unique projective point associated to them, namely the projective point
corresponding to the normal vector of the hyperplane. Any statement about points
in projective space therefore has a corresponding dual description
\footnote{Note that this form of duality is the basis for the perhaps more
  well-known (at least among physicists) \textit{Legendre-Fenchel transform}
  \cite{touchetteLegendreFenchelTransformsNutshell}.} in terms of lines.

Consider $\mathbb{P}^{2}_{E}$ with independent homogeneous coordinates $\left[
  A_w:A_{n}:A_{p} \right]$. As per \cref{sec:projective-space}, a projective
subspace is associated to any vector subspace of $E$. A general hyperplane in
$E$ is given by an equation of the form
\begin{equation}
  \label{eq:hyperplane-coordinates}
  \alpha A_w+ \beta A_n + \gamma A_{p} \equiv  \left( \vec{u} \right)^{\,T} \cdot \vec{A} = 0 \;,
\end{equation}
where $\vec{u} = \left( \alpha, \beta, \gamma \right)$. For some point $[ A_w:A_n:A_{p}] \in
\mathbb{P}^{2}_{E}$, changing $\vec{u}$ gives different lines through this point. Dually,
for a fixed line $\vec{u} = \left( \alpha, \beta, \gamma \right)$, varying $\vec{A}$ gives
different points along this line. $\vec{A} = \left( A_w, A_n,A_{p} \right)$ are
called \textit{point coordinates}, and $\vec{u} = \left( \alpha, \beta, \gamma \right)$
\textit{hyperplane coordinates}, or, in the special case of $\mathrm{dim}\left(
  E \right) = 3$, $\mathrm{dim}\left( \mathbb{P}^{2}_E \right) = 2$, $\vec{u}$ are often
called \textit{line coordinates}. A general hyperplane is given either by all
points $\left( A_{w}, A_n, A_{p} \right)$ that satisfies
\eqref{eq:hyperplane-coordinates} for fixed $\vec{u}$, or dually, by all the
lines $\left( \alpha, \beta, \gamma \right)$ satisfying \cref{eq:hyperplane-coordinates} for
fixed $\vec{A}$. It is arbitrary whether one specifies a hyperplane $H \subset E$ (or
equivalently a projective line inside $\mathbb{P}^2_E$) using point-coordinates or line
coordinates \footnote{This is equivalent to specifying surfaces in terms of its
  supporting hyperplanes.}. For instance, let $\vec{u}$ in
eq.~\eqref{eq:hyperplane-coordinates} be given by the fluid velocities. Thus,
the areas and velocities are projective duals in the sense of eq.
\eqref{eq:hyperplane-coordinates} \footnote{However, recall from
  \cref{sec:vm-from-proj-transf} that the components of a projective line,
  specifically $\vec{\tilde{v}}$, need not scale as degree-$0$ homogeneous
  functions in $A_{w}, A_n, A_p$. Thus, they are strictly speaking not
  necessarily ``velocities''.}. As already mentioned, line coordinates are the
homogeneous coordinates on the dual projective space $\left( \mathbb{P}^{n}
\right)^{\ast}$, obtained by the projectivization of the dual vector space
$E^{\ast}$ of hyperplanes in $E$, $\mathbb{P}\left( E^{\ast} \right) \cong
\left( \mathbb{P}^{n} \right)^{\ast}$.

An important consequence of duality is that \textbf{there really are no
  distinction between areas and velocities}. A homogeneous point given by the
areas can be scaled by a parameter of arbitrary physical units, and the same
holds for a homogeneous point expressed using the velocities. Moreover, a point
described by areas can equally well describe a line via duality, so one cannot
make the statement that `` areas describe points'' and ``velocities describe
lines''. All quantities can be seen as dimensionless by using the scaling
property in \cref{sec:projective-space}. This principle is important in the
analysis that follows, since points will frequently be described by velocities.


Duality has an important implication for $\mathbb{P}^1$, which is self-dual,
meaning homogeneous coordinates of lines and points gives point on the same
curve.
Consider the homogeneous coordinates $\left[ A_w:A_n \right]$ on
$\mathbb{P}^{1}$. Rewrite \cref{eq:Q-alg-line} as
\begin{equation}
  \label{eq:projective-point-2D-Q}
  (v_w -v)A_w + (v_n - v)A_n  = \vec{u}^{\,T} \cdot \vec{A} = 0\;,
\end{equation}
now with $\vec{A} = (A_w, A_n)$, $\vec{u} = \left( (v_w - v), (v_n - v)
\right)$, where $F=0$ was use to eliminate $A_p$. One can use either $\left[
  A_w:A_n \right]$ or $\left[\left( v_w - v \right):\left( v_n - v \right)
\right]$ to specify points on $\mathbb{P}^{1}$. Observe that $\left[ A_w:A_n
\right] = \left[ A_w/A_n : 1 \right] \sim A_w/A_n \equiv S_w / S_n$. From
eq.~\eqref{eq:projective-point-2D-Q}, one has
\begin{equation}
  \label{eq:simple-ratio}
  \frac{A_w}{A_n} \ = \ -\frac{v_n - v}{v_w - v} \ = \ -\frac{\hat{v}_n - v}{\hat{v}_w - v} \;.
\end{equation}
Thus, ratios of areas are equivalent to ratios of velocity differences. The
ratio in \cref{eq:simple-ratio} is often called a \textit{simple ratio}, the
ratio of two lengths \footnote{Note that under an affine transformation of the
  velocities, which would correspond to a transformation $\left\{ u_{i} \right\}
  \mapsto a \left\{u_{i} \right\} + b$ where $\left\{ u_{i} \right\}$ are the
  velocities $\left\{ v, v_w, v_n \right\}$ or $\left\{ v, \hat{v}_w, \hat{v}_n
  \right\}$ and $a,b$ are scalars, the simple ratio is invariant.}. When $v_w =
v$, one has $v_n = v$. The same holds for the hatted velocities. The ratio
$S_w/S_n$ is an affine parameter in the same way as $S_w$, but in a different
set of homogeneous coordinates on $\mathbb{P}^{1}$. If one instead considered
$\left[ A_w:A_p \right]$ by substituting $A_n = A_p - A_w$ in
\cref{eq:Q-alg-line}, $S_w$ would be the affine parameter instead.
\Cref{eq:simple-ratio} can simply be rewritten as
\begin{equation}
  \label{eq:Sw-simple-ratio}
  S_w \ = \ \frac{v - v_n}{v_w - v_n}
\end{equation}
to make this clearer, which also holds for the hatted velocities.

\subsection{Intersection and joins in $\mathbb{P}^2$}
\label{sec:lines-points-p2}

Using homogeneous coordinates, one can easily compute the line joining two
points, or dually, the intersection point between two lines
\cite{richtergebertPerspectivesProjectiveGeometry2011} in $\mathbb{P}^{2}$. The
line $l$ defined by joining two projective points $p_{1}$, $p_2$ is given by the
ordinary cross product of their homogeneous point coordinates. The resulting
coordinates are then the line coordinates of the line $l$. Dually, the
intersection point $p$ of two lines $l_1$, $l_2$ is the cross-product of the
line coordinates of these lines, which results in the homogeneous point
coordinates of $p$ \footnote{Moreover, one can check if three points are on a
  common line or if three lines are incident to a common point by computing the
  determinant of the point- or line coordinates of the three objects, which
  should work out to zero in the two cases. These are possibly the simplest
  geometric statements that can be made using linear algebra and homogeneous
  coordinates. See e.g.\cite{richtergebertPerspectivesProjectiveGeometry2011}
  for a much more detailed treatment of this subject. }.

To compute the intersection of the two lines defined by the linear form
\begin{align}
  G \ =& \ \alpha A_w + \beta A_n + \gamma A_p = 0 \label{eq:linear-form-G} \;,
\end{align}
and the line $l_{\infty}$ from \cref{sec:vm-from-proj-transf}, one forms the
cross-product
\begin{equation}
  \label{eq:cross-product-lines}
  \begin{pmatrix}
    1 \\
    1 \\
    -1 
  \end{pmatrix}
  \times
  \begin{pmatrix}
    \alpha \\
    \beta \\
    \gamma 
  \end{pmatrix} \ = \ 
  \begin{pmatrix}
    \beta + \gamma \\
    - \left( \alpha + \gamma \right) \\
    \beta - \alpha
  \end{pmatrix} \;,
\end{equation}
which are homogeneous coordinates of a point of $\mathbb{P}^2_{E}$. Consider
either $\left( \alpha, \beta, \gamma \right) = \left( v_w , v_n , -v \right) =
\vec{v}$ or $ \left( \alpha, \beta, \gamma \right) = \left( \hat{v}_w ,
  \hat{v}_n , -v \right) = \vec{\hat{v}}$. The former gives
\begin{align}
  \label{eq:cross-prod-homogeneous}
  & \ \left[ \left( v_n - v \right):\left(v - v_w  \right):\left( v_{n } - v_w\right) \right] \nonumber \\
  =& \ \left[ \frac{\left( v_n - v \right)}{\left( v_{n } - v_w\right)}: \frac{\left(v - v_w  \right)}{\left( v_{n } - v_w\right)}:1 \right] = \left[ S_w : S_n :1 \right] \;,
\end{align}
and similarily for $\vec{\hat{v}}$. The last equality
in~\cref{eq:cross-prod-homogeneous} follows from \cref{eq:e14}. The points of
intersection between $F=0$ and $G=0$ is given by the affine points $\left( S_w,
  S_n \right)$. This is the case for both $\vec{v}$ and $\vec{\hat{v}}$, so the
two lines defined by these two vectors are by definition parallel, with
direction given by $\left( S_w, S_n \right)$. Declaring $l_{\infty}$ to contain
improper points, the two lines $l = p \left( \vec{v} \right)$ and $\hat{l} = p
\left( \vec{\hat{v}} \right)$ has the same intersection point at infinity, which
is the definition of parallelism in an affine plane embedded in
$\mathbb{P}^{2}$.

Since $l$, $\hat{l}$ are parallel, the intersection of these lines gives points
at infinity,
\begin{align}
  \label{eq:cross-product-hatted-seepage-lines}
  l \times \hat{l} \ = \ 
  \begin{pmatrix}
    v \left( \hat{v}_n - v_{n} \right) \\
    v \left( v_w - \hat{v}_{w} \right) \\
    v_w \hat{v}_n - v_n \hat{v}_{w}
  \end{pmatrix} \ =& \ 
                     \begin{pmatrix}
                       -S_w v_m \\
                       -S_n v_m \\
                       - v_m
                     \end{pmatrix} 
                     \sim 
                     \begin{pmatrix}
                       S_w \\
                       S_n  \\
                       1
                     \end{pmatrix} \;,
\end{align}
where \cref{eq:vw-transf,eq:vn-transf} and the relation
\begin{equation}
  \label{eq:vm-quad}
  v_m \ = \ \frac{v_n\hat{v}_w - v_w\hat{v}_n}{v} 
\end{equation}
has been used. Scaling out $-v_m$ from eq.~\eqref{eq:vm-quad} gives the points
$\left[ S_w:S_n:1 \right]$ on $l_{\infty}$. From this, one sees that $v_m$
``dilates'' the coordinates, and cancels up to scaling. This dilation relates
the hyperplanes given by the line coordinates $\left( v_w, v_n, -v \right)$ and
$\left( \hat{v}_w, \hat{v}_n, -v \right)$. \textbf{This is a concrete geometric
  interpretation of the co-moving velocity.} The identity in
eq.~\eqref{eq:vm-quad} appears naturally in the projective setting, but can in
fact be obtained directly from the relations in \cref{eq:vw-transf,eq:vn-transf}
by rewriting the ratio $v_n/v_w$ and solve for $v_m$. The quantity $v_n/v_w
\equiv \mathcal{S}$ is often called the \textit{slip ratio} in the two-phase
flow literature \cite{zuberAverageVolumetricConcentration1965}. It has an
important role in later sections.

The choice of $l_{\infty}$ as an infinite line ties nicely into the expressions
in \cref{sec:affine-transf-vm}. The barycentric coordinates in that section are
by definition barycentric coordinates on the line $l_{\infty}$.

\section{Homographies \& cross-ratio}
\label{sec:rp-homographies}

\textit{Homographies} or \textit{projective transformations} are the
transformations of projective spaces
\cite{richtergebertPerspectivesProjectiveGeometry2011,
  stillwellFourPillarsGeometry2005}.
Homographies map lines to lines, but do not preserve parallel lines like affine
transformations do. A homography can be expressed as an injective linear map
\footnote{ If the map is not injective, one only obtains a \textit{partial}
  map.} of the vector space underlying the projective space. It acts by
multiplying the vector of homogeneous coordinates of \cref{sec:projective-space}
by a matrix.

Homographies of $\mathbb{P}^{1}$ was introduced in \cref{sec:projective-space}
as a composition of a finite number of projections of an affine line onto
itself. If one initially has an affine parameter $S_w$ on the line, a series of
projections result in an affine parameter $\tilde{S}_w = \tilde{S}_w (S_w)$
\footnote{Note that a series of arbitrary projections can not only translate and
  scale the points, but also change their order.}. The relation $\tilde{S}_w =
\tilde{S}_w (S_w)$ is particularly simple in the case of homographies of
$\mathbb{P}^{1}$; they are simply linear fractional transformations in $S_w$.
One often denotes the group of such homographies by $\mathrm{PGL}(2,
\mathbb{R})$, where $2$ is the dimension of the underlying linear space. These
transformation are often called \textit{Möbius transformations} in the context
of complex numbers. Homographies of $\mathbb{P}^2$ (and $\mathbb{P}^n$, for that
matter) work in the same manner, generalized to higher dimensions, see
\cref{app:rep-maps,app:rp1}. The interest in projective transformations,
especially for $\mathbb{P}^1$, is that they are described by a matrix of numbers
which can be used in the parametrization of $v_{m}$. For $\mathbb{P}^1$, this
matrix has four parameters, which can be reduced to three upon scaling.

\subsection{Projective frames }
\label{sec:frames}

In \cref{sec:affine-transf-vm}, the location of a point $v$ on a affine line was
defined with respect to an affine frame, with $v_m$ as a choice of origin. On a
projective line, any three distinct points can serve as a \textit{projective
  frame} or \textit{projective reference}, and \footnote{ Since projective
  points are defined up to a scalar, it turns out that a projective frame of
  $\mathbb{P}^{1}$ has to contain three points. Since each basis vector is
  defined up to a scalar, one gets an extra degree of freedom that has to be
  fixed when considering projective transformations between the frames, see
  \cite{bergerGeometry1994}.}. These can be used to obtain homogeneous
coordinates of a fourth point with respect to the frame
\cite{gallierGeometricMethodsApplications2011,
  bergerGeometry1994,casas-alveroAnalyticProjectiveGeometry2014}. One can view
this projective frame as the completion of a barycentric frame, see sec.
\cref{sec:affine-transf-vm,sec:vm-as-affine-transf}, in which one adds a
``unit''-point to the frame, defined as the projectivization of the sum of our
chosen basis vectors for the underlying linear space.

The idea of a projective frame generalizes to higher dimensions. For a
projective space of dimension $n$, with underlying vector space of dimension
$n+1$, one requires $n+2$ points for defining a projective frame. As a frame,
one can take an arbitrary basis of the vector space and the sum of all the
vectors in the basis. Upon projectivization, one has a frame of $\mathbb{P}^n$,
denoted by $\Delta$. A point $X \in \mathbb{P}^{n}$ can be written as the
projectivization of a linear combination of the basis vectors $\left( e_0,
  \ldots, e_{n} \right)$ of the underlying vector space as
\begin{equation}
  \label{eq:projective-frame-point}
  X = p\left( x_0 e_0 + \ldots + x_n e_n \right) \;.
\end{equation}
The components $\left( x_0, \ldots, x_n \right)$ are the homogeneous coordinates
of $X$ with respect to the frame $\Delta$.

A frame $\Delta$ for $\mathbb{P}^1$ can by definition be expressed as
\begin{equation}
  \label{eq:projective-frame}
  \left( p \left( e_1 \right), p \left( e_2 \right), p \left( e_1 + e_2 \right) \right) \equiv \left( m_1, m_2, m_{0} \right)  \;,
\end{equation}
where $e_1, e_2$ forms a basis of the underlying vector space \footnote{ Note
  that one can pick \textit{any} three points of $\mathbb{P}^1_{H}$ as a
  projective frame: the crucial element is that there exists some basis of $H$
  such that one can write the frame as in \cref{eq:projective-frame}. Two
  projective frames $m_i, m_i^{\prime}$ are equivalent if one can find a single
  scalar $\lambda \neq 0$ such that the associated bases $\left\{ e_i \right\},
  \left\{ e_i^{\prime} \right\}$ of $\mathcal{H}$ are related as $e_i = \lambda
  e_i^{\prime}$.}. The point $p \left( e_1 + e_2 \right)$ is the unit-point.
With a projective frame $\Delta$, the homogeneous coordinates of a projective
point $q = p(\vec{q})$ is just the vector coordinates of the vector $\vec{q}$
with respect to the basis $e_i$. Using this description, one can parametrize a
projective line $L \subset \mathbb{P}^{2}$ spanned by two projective points
$p(e_1), p(e_{2}) \in \mathbb{P}^{2}$, as
\cite{richtergebertPerspectivesProjectiveGeometry2011}
\begin{equation}
  \label{eq:parametric-projective-line}
  L =  p \left(  \lambda e_{1} + \mu e_2 \right) \;,
\end{equation}
which holds up to scaling and arbitrary $\lambda, \mu$, the homogeneous
coordinates of $L$ relative to the given frame. Both points $p \left( e_1
\right), p \left( e_2 \right)$ satisfy an implicit equation of the form $ax + by
+ cz = 0$, so the parameters $\lambda, \mu$ are related to the coefficients
$a,b,c$.

Any two triples of points on the projective line can be related to each other
via a unique homography $h$
\cite{richtergebertPerspectivesProjectiveGeometry2011}. Thus, for two frames
$\Delta$, $\Delta^{\prime}$, which means any two triples, one has
\begin{equation}
  \label{eq:homography-frames}
  h \left( \Delta \right) \ = \ \Delta^{\prime}  \;.
\end{equation}
This property means that any triple is essentially equivalent to any other
triple, which is the reason for needing three points to form a frame of
$\mathbb{P}^1$. Since the position of these three reference points is arbitrary,
one might place them in a ``canonical'' position where one point is taken to be
at infinity, one point is the origin and one point defines a unit length scale
on the line. The homogeneous coordinates of a fourth point is defined to be the
cross-ratio of this point with respect to a frame consisting of three points.
The cross-ratio of four points is in fact an invariant under projective
transformations, which will be elaborated on in the next section.

\subsection{Cross-ratio}
\label{sec:cross-ratio}

The simple ratio in~\cref{eq:simple-ratio} is invariant under affine
transformations of the velocities, but not invariant under projective
transformations, since a homography can send any three points to any three
points. The corresponding projective invariant, the \textit{cross-ratio}
\cite{bergerGeometry1994, richtergebertPerspectivesProjectiveGeometry2011,
  gallierGeometricMethodsApplications2011}, must then be defined from four
points.

The cross-ratio \cite{bergerGeometry1994} of four points $a,b,c,d$ on a
projective line $L$, where three of them must be distinct, is denoted
$\left[a,b,c,d \right]$ or simply $\left[a_i\right]$. It is the element $k \in
\mathbb{R} \cup \infty $ defined as
\begin{equation}
  \label{eq:cross-ratio-formal}
  k \ \equiv \ \left[a,b,c,d \right] \ = \ f_{a,b,c} \left(d \right)
\end{equation}
such that $f \left( a \right) = \infty$, $f \left( b \right) = 0$ and $f \left(
  c \right) = 1$. Thus, the cross-ratio ``is'' the homography $f_{a,b,c}$ acting
on the point $d$, and is equivalent to the homogeneous coordinates of $d$ with
respect to the chosen frame as explained at the end of section \ref{sec:frames}.
The cross-ratio with respect to a frame $m_i$ and a arbitrary fourth point $d$
is denoted by $\left[ m_i , d \right]$.

In general, if one expresses the homogeneous coordinates of the points by $a =
\left( a_1, a_2 \right)^{T}$ etc., and the determinant of two points $a,b$ as
\begin{equation}
  \label{eq:determinant}
  \mathrm{det}\left( a,b \right) \ = \
  \begin{vmatrix}
    a_1 & b_1 \\
    a_2 & b_2
  \end{vmatrix} \ = \ a_1 b_2 - b_1 a_{2} \;,
\end{equation}
then the cross-ratio is computed as
\begin{equation}
  \label{eq:cross-ratio-determinant}
  \left[ a,b,c,d \right] \ = \ \frac{ \mathrm{det}\left( a,c \right) \mathrm{det}\left( b,d \right)}{  \mathrm{det}\left( a,d \right) \mathrm{det}\left( b,c \right)} \;.
\end{equation}
One can compute the cross ratio of four points on a line or a conic (see
\cref{sec:five-point-conic}) by picking a point $O$ \footnote{In the case of
  four points on a line, the point cannot be on this line. In the case of a
  conic, the point has to be on the conic itself.} and computing the cross-ratio
``as seen'' from this point. In fact, the cross-ratio is independent of the
choice of $O$. Define the determinant of three points $\mathrm{det}\left( a,b,c
\right)$ analogous to \cref{eq:determinant}. The cross ratio is then
\begin{equation}
  \label{eq:cross-ratio-projective-plane}
  \left[ a,b,c,d \right]_{O} \ = \ \frac{ \mathrm{det}\left(O, a,c \right) \mathrm{det}\left(O, b,d \right)}{  \mathrm{det}\left(O, a,d \right) \mathrm{det}\left(O, b,c \right)} \;.
\end{equation}

If the four points of an affine line has affine coordinates $x_1, x_2, x_3,
x_4$, the cross ratio can be written as
\begin{equation}
  \label{eq:coss-ratio-distance}
  \left[ x_1, x_2, x_3, x_{4} \right] \ = \ \frac{\left( x_3 - x_1 \right)\left( x_4 - x_2 \right)}{\left( x_3 - x_2 \right) \left( x_4 - x_1 \right)} \;.
\end{equation}
This is related to eq. \eqref{eq:cross-ratio-determinant} by using that $a =
\left[a_1:a_2 \right] = \left[ a_1/a_2 : 1 \right] \mapsto a_1/a_2$, so one
simply sets $x_1 = a_1/a_2$ as the affine coordinate of the point $a$ and so
forth \footnote{ The cross-ratio of four points on a line is constructed as a
  ``ratio of ratios'' of distances on the line, in contrast to the single ratio
  in \cref{eq:simple-ratio}.}.

The cross-ratio can become infinite if either $x_3 = x_2$ or $x_4 = x_1$, which
represents an infinite point on $L$. Changing the order of the points in the
cross-ratio alters the value of the cross ratio in a predictable manner
\cite{bergerGeometry1994}. Using the identities
\begin{gather}
  \label{eq:cross-ratio-identities-permute}
  \left[ a,b,c,d \right] = \left[ b,a,c,d \right]^{-1} = \left[ a,b,d,c \right]^{-1} \\
  \left[ a,b,c,d \right] + \left[ a,c,b,d \right] = 1 \label{eq:cross-ratio-identities-permute} \;,
\end{gather}
one can show that there are only \textit{six possible} values of the cross-ratio
upon permutation of points. Setting $k = \left[ a,b,c,d \right]$, these values
are
\begin{equation}
  \label{eq:six-cross-ratios}
  k \;, \ \frac{1}{k} \;, \ 1 - k \;, \ 1 - \frac{1}{k}\;, \ \frac{1}{1 - k} \;, \ \frac{k}{k-1}\;.
\end{equation}
These permutations correspond to permuting the location of $\infty$, $0$, $1$
and the
point $f_{a,b,c} \left( d \right)$. \\

Lastly, the cross-ratio can be computed from the homogeneous coordinates of the
parametric representation in \cref{sec:frames}. If one has the points
\begin{align*}
  a = p \left( e_1 \right)  \;,\\
  b = p \left( e_2 \right)  \;, \\
  c = p \left(\lambda e_1 + \mu e_2 \right) \;, \\
  d = p \left(\gamma e_1 + \delta e_2 \right) \;,
\end{align*}
then the cross-ratio $\left[ a_i \right]$ is the homogeneous coordinates of the
fourth point $d$ with respect to the frame $\Delta = \left( a,b,c \right)$.
Since
\begin{equation}
  \label{eq:cross-ratio-hom-coords}
  d = p \left( \gamma e_1 + \delta e_{2} \right) = p \left( \frac{ \gamma}{\lambda} (\lambda e_1) + \frac{\delta}{\mu} (\mu e_{2}) \right) \;,
\end{equation}
the homogeneous coordinates of $d$ with respect to the frame $\Delta$ is simply
\begin{equation}
  \label{eq:cross-ratio-hom-coords-demonstration}
  d = p \left( \frac{ \gamma}{\lambda} (\lambda e_1) + \frac{\delta}{\mu} (\mu e_{2}) \right) \sim \left[ \frac{\gamma}{\lambda}:\frac{\delta}{\mu} \right] \mapsto \frac{\gamma}{\lambda} \frac{\mu}{\delta} \;.
\end{equation}

To relate $v_m$ to the cross-ratio, one must define a sensible set of points on
a line to compute the cross-ratio of. Due to area-scaling, it is the homogeneous
expressions in the areas that determine the structure of the problem. One of
these homogeneous expressions describe a physical limitation of the system, $A_w
+ A_n - A_p \equiv F = 0$, which was occasionally ignored in the general
developments of \cref{sec:spaces}. It is clear that $F = 0$ signifies a
privileged line $l_{\infty}$ of $\mathbb{P}^2$. Only the case $F=0$ makes
physical sense, however, the relations of \cref{sec:rea} have a structure that
defines objects in $\mathbb{P}^2$. One therefore needs to transfer these finite
objects where $F \neq 0$ to the line $l_{\infty}$. Moreover, the limitation that
the saturations are bounded $\in \left[ 0,1 \right]$ will have to be enforced.
How this is encoded is shown in the next section.

\section{Degenerate quadratic form}
\label{sec:degenerate-quadratic-form}

Upon selecting $l_{\infty} = \left( 1,1,-1 \right)$ as an infinite line,
intersection points of $l_{\infty}$ and the lines $l, \hat{l}$ was given by
$\left( S_w, S_n \right) = \left( S_w, 1 - S_w \right)$. One can pick out two
special points on $l_{\infty}$, the ones corresponding to $S_w = 1$ and $S_{w} =
0$.
In $\mathbb{P}^{2}$, these are
\begin{align}
  \label{eq:I-J}
  I \ \equiv \
  \begin{pmatrix}
    1 \\
    0 \\
    1
  \end{pmatrix} \ \;, \
  J \ \equiv \
  \begin{pmatrix}
    0 \\
    1 \\
    1
  \end{pmatrix} \;,
\end{align}
on $l_{\infty}$, meaning both satisfy
\begin{equation}
  \label{eq:infinite-points-on-line}
  l_{\infty}^T \cdot I = l_{\infty}^T \cdot J = 0 \;.
\end{equation}
Thus, a distinguished line $l_{\infty}$ with two designated real points on it
has been picked out. In classical projective geometry, one can use such a
configuration to perform measurements, meaning one can define notions of
distance and angle on the affine spaces embedded in the projective space. One
defines a reference object in $\mathbb{P}^{2}$ \footnote{One can in fact
  generalize this idea to higher dimensional projective spaces
  \cite{richtergebertPerspectivesProjectiveGeometry2011}.}, which induces this
metric-structure on the complement of this reference object. The reference
object is taken to be a \textit{projective conic} $\mathcal{C}$, and the
obtained geometry from the choice of conic $\mathcal{C}$ and scaling parameters
for distances and angles is called a \textit{Cayley-Klein geometry}
\cite{richtergebertPerspectivesProjectiveGeometry2011,
  casas-alveroAnalyticProjectiveGeometry2014}. Cayley-Klein geometries unify
several well-known classical geometries: elliptic, hyperbolic, Euclidean,
Galilean and pseudo-Euclidean \footnote{Pseudo-Euclidean geometry is more
  commonly known among physicists as the geometry of special relativity
  \cite{misnerGravitation1973}. } geometries are all describable in this
framework. Only the strictly necessary parts of this framework is presented in
this paper, see e.g. \cite{richtergebertPerspectivesProjectiveGeometry2011} for
a more exhaustive treatment.

A \textit{projective conic} $\mathcal{C}$
\cite{richtergebertPerspectivesProjectiveGeometry2011,
  casas-alveroAnalyticProjectiveGeometry2014} is defined by a quadratic equation
of the type
\begin{equation}
  \label{eq:quadratic-form}
  \left(x,y,z \right)
  \begin{pmatrix}
    a & b & c \\
    b & d & e \\
    c & e & f
  \end{pmatrix}
  \begin{pmatrix}
    x \\
    y \\
    z
  \end{pmatrix} \ \equiv \ p^T \mathcal{M} p \ = \ 0
\end{equation}
for constants $a, \ldots , f$ and $p = \left( x,y,z \right)^{T}$ \footnote{ One
  can always assume that the matrix of the quadratic form is symmetric, which
  can be seen by writing out the quadratic form with a general matrix and
  observe that only the diagonals and sums of the off-diagonals matter. One can
  always create a symmetric matrix from a non-symmetric matrix $M$ by computing
  $\left( M + M^{T} \right)/2$. }. $\mathcal{M}$ is the \textit{matrix of the
  quadratic form}, or just the \textit{quadratic form} for simplicity. The set
of points $p \in \mathbb{P}^{2}$ that satisfies \cref{eq:quadratic-form} defines
a conic in the projective plane, since the resulting equation one gets from
multiplying out \cref{eq:quadratic-form} is degree-$2$ homogeneous in $p$, and
\cref{eq:quadratic-form} is invariant under the scaling $p \mapsto \lambda p$.

The type of conic defined by \cref{eq:quadratic-form} depends on the matrix
$\mathcal{M}$. It turns out that one can classify the types of possible conics
up to projective equivalence, i.e. up to a projective transformation of $p$.
Since a projective transformation of $p$ is performed by multiplying $p$ by a
non-singular matrix $M$, one can see from \cref{eq:quadratic-form} that such a
transformation can alter the eigenvalues of $\mathcal{M}$ but not the signature
of the eigenvalues, the number of positive, negative and zero eigenvalues of
$\mathcal{M}$. For instance, if the signature of $\mathcal{M}$ is $\left( +1,
  +1, -1 \right)$, a projective transformation will not change the type of conic
defined by $\mathcal{M}$ \footnote{This is due to \textit{Sylvester's law of
    inertia} \cite{richtergebertPerspectivesProjectiveGeometry2011}.}.

A projective conic $\mathcal{C}$ can naturally be introduced from the relations
in the previous sections. Letting $p = \left( A_w, A_n, A_p \right)^{T}$ with
independent areas defined a linear equation $ l_{\infty}^{T} \cdot p = A_w + A_n
- A_p \equiv F = 0$. After squaring this relation, $\left( A_w +A_n - A_p
\right)^2 = 0$ and writing out the terms, one can compare the coefficients with
\eqref{eq:quadratic-form} to see that $F^{2}=0$ can be written as the quadratic
form
\begin{equation}
  \label{eq:our-quad-form}
  \left(A_w,A_n,A_p \right)
  \begin{pmatrix*}[r]
    1 & 1 & -1 \\
    1 & 1 & -1 \\
    -1 & -1 & 1
  \end{pmatrix*}
  \begin{pmatrix}
    A_{w} \\
    A_{n} \\
    A_{p}
  \end{pmatrix} \ \equiv \ p^TA p = 0 \;.
\end{equation}
The rank of $A$ is one, and can therefore be written as an outer product $u
\otimes u$ for a vector $u$, here abbreviated as $uu^{T}$. The matrix $A$ is
just $l_{\infty}l_{\infty}^{T}$, which encodes a double-cover of the line
$l_{\infty}$. This signifies that the reference conic defined by $A$ is
degenerate, with signature equal to $\left( +,0,0 \right)$. This conic is
defined from the points satisfying \cref{eq:quadratic-form}, called the
\textit{primal} conic. Since quadratic forms encode conics in the way described
above, we will henceforth refer to the conics by the matrices representing them.

Due to projective duality, see \cref{sec:affine-space-points}, one has a
\textit{dual conic} defined by all lines $l$ satisfying an equation
\begin{equation}
  \label{eq:dual-quad-form}
  l^{T}B l \ = \ 0
\end{equation}
for some matrix $B$ \footnote{$B$ is often written as the adjugate matrix of of
  $A$, $\mathrm{adj}\left( A \right) = \mathrm{det}\left( A \right) A^{-1} $.}.
In the case of a non-degenerate primal conic, $B$ is uniquely determined from
$A$, but this is not the case for degenerate conics. If $A$ is degenerate and
does not have full rank, $B$ cannot be computed directly from $A$. It is still
possible to assign a dual object to $A$ in the degenerate case
\cite{richtergebertPerspectivesProjectiveGeometry2011}, as it turns out that for
a primal/dual pair $\left( A, B \right)$ where $A = ll^{T}$ for some line $l$
(as in our case), the dual $B$ is of the form
\begin{equation}
  \label{eq:B-form-dual}
  B \ = \ xy^T + y x^{T}
\end{equation}
for a choice of two points $x,y$ (which could be real or complex), where
\begin{equation}
  \label{eq:IJ-on-l}
  l^T x = 0 \ \;, \ l^T y = 0 \;,
\end{equation}
meaning $x,y \in l$. Let the line be given by $l_{\infty}$, such that $A$ is the
matrix in \cref{eq:our-quad-form}. The two points $I,J$ in \cref{eq:I-J} lies on
$l_{\infty}$, and can be used to construct a dual conic.
From~\cref{eq:B-form-dual} with $x=I, y=J$, one obtains from
\cref{eq:B-form-dual} the dual matrix
\begin{equation}
  \label{eq:dual-B}
  B \ = \
  \begin{pmatrix}
    0 & 1 & 1 \\
    1 & 0 & 1 \\
    1 & 1 & 2
  \end{pmatrix} \;.
\end{equation}
Thus, there is a sensible choice of primal and dual objects, namely the double
covered infinite line $l_{\infty}$ and the two points $I$, $J$, represented by
matrices $(A, B)$ corresponding to degenerate primal and dual conics
respectively. The pair $\left( A,B \right)$ not only describe the primal/dual
conics. The operation $p \mapsto Ap$ for some point $p$ produces a line tangent
to the conic, while $l \mapsto Bl$ for some line $l$ produces a point on the
conic. The maps $p \mapsto Ap$ and $l \mapsto Bl$ are called \textit{polarities}
\cite{richtergebertPerspectivesProjectiveGeometry2011}, and will be used in the
next section.

To sum up, a choice of primal/dual objects corresponds to a Cayley-Klein
geometry \cite{richtergebertPerspectivesProjectiveGeometry2011} which allow for
a definition of measurement in the projective plane, on the complement of the
reference object. The complement of $l_{\infty}$ in $\mathbb{P}^{2}$ is a space
of finite objects, which are not subject to the condition $F=0$. With the choice
$\left( A, B \right)$ as in this section, the geometry of these finite objects
is \textit{pseudo-Euclidean}. This is the same geometry as the more well-known
Minkowski space of special relativity. This geometry has an hyperbolic
angle-measure, which means that a series of lines with of equal angles between
them will tend toward a limiting line. This is the same principle as having the
limiting asymptotes of a hyperbola in Minkowski space define a light cone, which
defines an upper limit of velocities of objects.


\subsection{Transforming the degenerate conic}
\label{sec:transf-degenerate-conic}

The matrices $\left( A, B \right)$ defining the double line and the two points
on this line are not in their simplest possible form. One can simplify the
matrices $\left( A, B \right)$ by suitable projective transformations, which
transforms the conics $A,B$ to $A_0, B_0$. Such a projective transformation can
map $l_{\infty}$ to a new line $\tilde{l}_{\infty}$, and possibly send $I$, $J$
or both to new points $I_0, J_{0}$. One can view projective transformation as
acting on the points themselves instead of transforming the conic.

Consider first $B$, the least degenerate object in the primal/dual pair. In what
follows, all matrices are $3\times3$ matrices. The simplest possible form of
dual matrix $B$ for this geometry has the form $\mathrm{diag}\left( +1,-1,0
\right)$, up to permutation of the columns. Call this matrix $B_0$. The
transformation to define is then $B_0 \mapsto B$ \footnote{One can formulate the
  mapping in the other direction instead, if one wishes. }. A dual conic
represented by a symmetric matrix $B$ transforms as
\cite{richtergebertPerspectivesProjectiveGeometry2011}
\begin{equation}
  \label{eq:inner-automorphism}
  B \mapsto M B M^{T} \;,
\end{equation}
for some invertible matrix $M$ representing the projective transformation
\footnote{ If $B = M B_0 M^{T}$ for some $M$, $B$ and $B_{0}$ are said to be
  \textit{congruent}.}. Note that the primal conic $A$ transforms as
\begin{equation}
  \label{eq:primal-conic-transf}
  A \mapsto \left( M^{-1} \right)^T A \left( M^{-1} \right) \;.
\end{equation}
Since $B$ is symmetric, one can consider its eigendecomposition
\begin{equation}
  \label{eq:eigendecomposition}
  B \ = \ Q D_B Q \ = \ Q S B_0 S^T Q^{T} \equiv Q_{0}B_0 Q_{0}^{T} \;,
\end{equation}
where $Q$ is a orthonormal matrix, $D_B$ is a diagonal matrix containing the
eigenvalues of $B$, $Q_{0} \equiv QS$ and $S$ is a matrix of the form
\begin{equation}
  \label{eq:S-scaling-matrix}
  S_{ii} =
  \begin{cases}
    \sqrt{\left| \lambda_{i} \right|},& \text{if} \ \lambda_i \neq 0 \\
    1 ,& \text{if} \ \lambda_i = 0 \;,
  \end{cases}
\end{equation}
where $\left\{ \lambda_i \right\}$ are the eigenvalues of $B$. This matrix
in~\cref{eq:eigendecomposition} ensures that the diagonal only contains $1,-1$
and $0$.

From $Q_{0}$, the transformation $A \mapsto A_{0}$ can be found from
\cref{eq:primal-conic-transf}, with $M = Q_{0}$. As explained in the previous
section, the fact that $\mathrm{rank} \left( A \right)=1$ means one can write
\begin{equation}
  \label{eq:A-tilde-decomp}
  A_{0} \ = \ \tilde{l}_{\infty} \tilde{l}_{\infty}^{T} \;,
\end{equation}
where $\tilde{l}_{\infty}$ is interpreted as the line at infinity $l_{\infty}$
after the transformation. To find $\tilde{l}_{\infty}$ explicitly, take a single
column of $A_{0}$ and label this vector as $y$. Then, due to the rank of
$A_{0}$, one has
\begin{equation}
  \label{eq:A-tilde-decomp-scaling}
  A_{0} \ = \ \mu^2 y y^{T} \;,
\end{equation}
for some scalar $\mu$, which is determined by comparing a column of $yy^{T}$ and
$A_{0}$. This gives that $\tilde{l}_{\infty} = \mu y$. The intersection of
$\tilde{l}_{\infty}$ with $l_{\infty}$ turns out to be given by
\begin{equation}
  \label{eq:intersection-new-old-infinity}
  \tilde{l}_{\infty} \times l_{\infty} \ = \
  \begin{pmatrix}
    0 \\
    1 \\
    1 
  \end{pmatrix}
  \ = \ J \;,
\end{equation}
which means that the transformation leaves $J$ invariant. Due to this fixed
point, the transformation is more restricted than a full projective
transformation. It is an example of a \textit{perspectivity} between two lines,
see \cref{app:rp1}.

It is interesting to see what happens to the points $p = \left[ S_w:S_n:1
\right] \in l_{\infty}$ if the transformation of $A$ is viewed as acting on the
points $p$ instead. From \cref{eq:primal-conic-transf}, this transformation
corresponds to $Q_{0}^{-1}p$, which gives
\begin{equation}
  \label{eq:Q-tilde-inv-saturations}
  Q_{0}^{-1}p = \left[ 1 - 2S_w:0:1 \right] \ = \ \left[ S_n - S_w:0:1 \right]
\end{equation}
after rescaling. This means that the affine coordinates are $\left( S_n - S_w, 0
\right)$ on $\tilde{l}_{\infty}$ as opposed to $\left( S_w, S_n \right)$ on $l_{\infty}$.
Note that in this view, the conic is considered fixed, while the points are
transformed.

Through the previous sections, only a few obvious points and lines follow
directly from the analysis in \cref{sec:rea,sec:spaces}: the homogeneous point
\begin{align}
  \label{eq:our-points}
  p \left( \vec{A} \right) \ = \ \left[ A_w:A_n:A_p \right]
\end{align}
with independent areas, and the lines
\begin{align}
  \label{eq:lines-l}
  l \ =& \ \left[ v_w:v_n:-v \right]  \\
  l_{\infty} \ =& \ \left[ 1:1:-1 \right] \label{eq:lines-l-infty} \;.
\end{align}
The intersection $l \times l_{\infty}$ was defined in \cref{eq:cross-prod-homogeneous},
which is henceforth labeled $L$.

The primal/dual pair $\left( A, B \right)$ determines all properties of the
Cayley-Klein geometry. A result of this structure is the maps $p \mapsto Ap$ and $m \mapsto
Bm$ for points $p$ and lines $m$, which defined polarities with respect to
$A,B$. These are important manifestations of the duality introduced in
\cref{sec:affine-space-points}. The map $Ap$ (if $\neq 0$) produces a line $p^{\ast}$
called the \textit{polar} of $p$ with respect to the conic $A$. Dually, the map
$Bl = l^TB$ produces a point called the \textit{pole} $l^{*}$ of $l$ with
respect to $B$ \footnote{ When a conic $\mathcal{C}$ is non-degenerate, the pole
  determines the polar and opposite. However, this is not the case when $\mathcal{C}$ is
  degenerate. }. One can use the pole of a line to construct orthogonal lines
through specified points, where ``orthogonal'' is defined with respect to the
specific geometry defined by the conics $\left( A,B \right)$. In the present
case, this is hyperbolic orthogonality: hyperbolic orthogonal lines are
reflections about the asymptotes of a hyperbola, whose asymptotes are defined by
$I,J$. For instance, the point $l^{\ast}$ can be joined with an arbitrary point
$p$ to define a line orthogonal to $l$ going through $p$, with respect to $B$.

Given $\left( A,B \right)$, poles and polars introduce more lines and points
into the framework, and connect these objects to a notion of ``orthogonal'' sets
of velocities, discussed in \cref{sec:O-choice-polar-pole}. Following the
reasoning above, one can introduce
\begin{equation}
  \label{eq:l-ast}
  l^{\ast}_{0} \ = \ B_{0} l \ = \
  \begin{pmatrix}
    1 & 0 & 0 \\
    0 & -1 & 0 \\
    0 & 0 & 0
  \end{pmatrix}
  \begin{pmatrix}
    v_w \\
    v_n \\
    -v
  \end{pmatrix} \ = \ 
  \begin{pmatrix}
    v_w \\
    -v_n \\
    0
  \end{pmatrix} \;,
\end{equation}
the pole of $l$ defined with respect to the transformed conic $B_{0}$.

On $\tilde{l}_{\infty}$, pick $O = \left[0:0:1\right]$ as a reference point. In
fact, in affine coordinates, $O$ is the midpoint of $I_{0}, J$, which defines a
barycentric frame in the sense of \cref{sec:affine-transf-vm}. After
homogenization, $\left( I_{0}, J, O \right)$ becomes a projective frame
$\tilde{\Delta}$ on $\tilde{l}_{\infty}$.

Any line dual to $O$ is of the form $\left[u:w:0 \right]$, where $u,w$ are
arbitrary, and intersects $l_{\infty}$ in $\left[ w:-u:w - u \right]$. Hence,
any line whose last coordinate is zero is incident to $O$. This is exactly the
case in the line described by \cref{eq:Q-alg-line}. If $A_p$ is substituted for
$A_w+A_n$ in this equation, one can interpret the result as the line coordinates
$\left[\left( v_w - v \right):(v_n - v):0 \right]$ in $\mathbb{P}^{2}$. One gets
the same intersection point with $l_{\infty}$ as for $l$, however, one has an
intermediate representation given by two velocity-differences. If these line
coordinates are taken as the starting point, the corresponding point coordinates
would be either $\left[ A_w:A_n:\alpha \right]$ with $\alpha$ arbitrary, or they
are $\left[ 0:0:1 \right]$ up to scaling \footnote{ It is clear that none of
  these coordinates correspond to any physical configuration of the areas or
  velocities, only the intersection with $l_{\infty}$ does.}. Every line which that
can be expressed in terms of only two of the areas after imposing $F=0$, or
equivalently with two velocities (here ${v_i - v}$), is incident to $O$. The
first set of coordinates contains a free parameter $\alpha$, so the latter is picked
as a reference.

One can now obtain a line $m$ by joining $l^{\ast}_{0}$ to $O$,
\begin{equation}
  \label{eq:line-m}
  m \ = \ l^{\ast}_{0} \times O =
  \begin{pmatrix}
    v_n \\
    v_w \\
    0
  \end{pmatrix} \;.
\end{equation}
The point $M \in l_{\infty}$ is defined by intersecting $m$ with $l_{\infty}$,
\begin{equation}
  \label{eq:m-l-infty-intersection}
  M \ = \ m \times l_{\infty} =
  \begin{pmatrix}
    v_w \\
    - v_n \\
    v_w - v_n
  \end{pmatrix} \;.
\end{equation}
One can interpret this as mapping the unit point of the projective frame
$\tilde{\Delta}$ to a corresponding point on $l_{\infty}$, with respect to the pole
$l^{\ast}_{0}$.

The initial set of objects, \cref{eq:our-points,eq:lines-l,eq:lines-l-infty},
have now been extended by the pole $l^{\ast}_{0}$, the reference point $O$, and the
line $m$. In the process, four points of interest on $l_{\infty}$ has been defined:
$I,J,L$ and $M$. In addition, the line $\tilde{l}_{\infty}$ was defined, obtained
from transforming the dual conic to the reduced form described by the dual
matrix $B_0 = \mathrm{diag}(+1, -1, 0)$. This line contains the points
$\tilde{I}, J, \tilde{L}, O$. Thus, two quadruples of points have been defined,
and a point $l^{\ast}_{0}$ not incident to any of the two lines containing these
points. One can formulate a projective map between these two quadruples of
points, which is done in the next section.

\subsection{Five-point conic \& parametric projective line}
\label{sec:five-point-conic}

\begin{figure}[h] \centering
  \includegraphics[width=\linewidth]{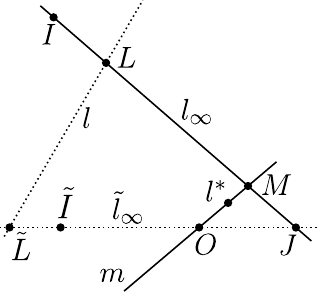}
  \caption{The point configuration $\left(I,J, L, O, l^{\ast} \right)$ defines a
    degenerate conic $l_{\infty} \cup m$. Moreover, there are two quadruples of points
    on two lines, $l_{\infty}$ and $\tilde{l}_{\infty}$. Note that the relative positions
    of the points and lines need not look like in the figure. This kind of
    configuration is often seen in demonstrations of \textit{Pappus theorem}, a
    special case of \textit{Pascal's theorem}
    \cite{richtergebertPerspectivesProjectiveGeometry2011}. }
  \label{fig:point-configuration-degenerate-conic}
\end{figure}

The considerations in the previous section translates into a setup where one can
compute the cross ratio of the points on the lines in terms of the saturations,
as will be shown in this section. The five points $(I,J, L, O, l^{\ast}_{0})$
define another degenerate conic in $\mathbb{P}^{2}$
\cite{richtergebertPerspectivesProjectiveGeometry2011}, see
\cref{fig:point-configuration-degenerate-conic}. The point $L$ depends on the
velocities, and $l^{\ast}_{0}$ depends on $L$, so both $L$ and $l^{\ast}_{0}$ move
dependently as functions of $S_w$.

\begin{figure}
  \centering
  \includegraphics[width=\linewidth]{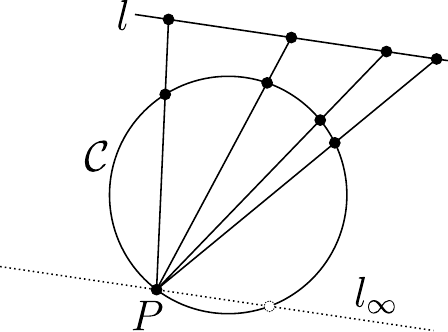}
  \caption{Isomorphism using stereographic projection between conic $\mathcal{C}$ and a
    projective line with finite points on $l$. The point $P$ is on $\mathcal{C}$, and the
    lines of the line bundle at $P$ intersect $\mathcal{C}$ in one point, and $l$ in one
    point, setting up a one-to-one correspondence. The blank point corresponds
    to the representation of the point at infinity of $l$. Note the relation
    holds even if $\mathcal{C}$ is degenerate. Adapted from
    \cite{richtergebertPerspectivesProjectiveGeometry2011}.}
  \label{fig:conic-line-isomorphism}
\end{figure}

In fact, five points of $\mathbb{P}^{2}$ in general position (meaning no three collinear)
defines a unique conic. If three of the points are collinear, the points still
define a conic, however, it can be reducible to a union of curves and is
possibly not unique. This conic is distinct from the conics $A,B,A_0, B_0$, and
is simply the union $l_{\infty} \cup m$. Conics in $\mathbb{P}^{2}$ are in fact isomorphic to
$\mathbb{P}^{1}$ \cite{richtergebertPerspectivesProjectiveGeometry2011}, see
\cref{fig:conic-line-isomorphism}.
One can then pick a point $O$ on the conic and compute the cross ratio using
\cref{eq:cross-ratio-projective-plane}. $\left[ M,L,I,J \right]_{O}$ is then
computed as
\begin{align}
  \label{eq:cross-ratio-our-conic}
  \left[ M,L,I,J \right]_O \ = \ \frac{[OMI][OLJ]}{[OLI][OMJ]} \ =& \ \frac{\left( v_n - v \right)}{\left( v_w - v \right)} \frac{v_n}{v_w} \nonumber \\
  =& \frac{S_w}{S_n} \tau \;,
\end{align}
where $\tau \equiv - v_n/v_w \equiv -\mathcal{S}$, with $\mathcal{S}$ the slip-ratio. Denote $\left[ M,L,I,J
\right]_O = k$. The cross-ratio can be used to define angles between lines in
the pseudo-Euclidean space
\cite{richtergebertPerspectivesProjectiveGeometry2011}. In the case of a
pseudo-Euclidean geometry, there is no distance measurement along a line, while
one has a hyperbolic measure of angles. The hyperbolic angle $\eta$ between two
lines $m,l$ with intersections $M,L$ with $l_{\infty}$ is defined as
\begin{equation}
  \label{eq:hyperbolic-angle}
  \eta \ = \ -\frac{1}{2} \ln{\left( k \right)} 
\end{equation}
where the prefactor of $-1/2$ is conventional. In this paper, $\eta$ is not used,
only the cross-ratio $k$. However, the interpretation of the angle is helpful:
it is the angle between two frames moving relative to each other, parametrized
by $S_w$. One has to be careful about the meaning of ``frame'', since there are
no observers involved. In this paper, a frame refers to a potentially dependent
set of elements that can be used as a reference to obtain coordinates, hence
different from a basis. An example of this is a projective frame, as has been
discussed earlier. The coordinates are the ones that are defined with respect to
the frame \footnote{In special relativity, this would be the coordinates with
  respect to some chosen coordinate system as measured by an observer undergoing
  motion.}, and the angle $\eta$ can be used to relate these coordinates
\footnote{This is the principle behind what physicists call \textit{Lorentz
    transformations} \cite{misnerGravitation1973} of Minkowski space.}.

Using \cref{eq:vw-transf,eq:vn-transf}, one can solve
\cref{eq:cross-ratio-our-conic} for $v_m$ to obtain
\begin{equation}
  \label{eq:vm-solved-from-k}
  v_m \ = \ v_m \left( S_w; k \right) \ = \ \frac{ k \hat{v}_w S_n - \hat{v}_nS_w}{kS_n^2 + S_w^{2}} \;,
\end{equation}
where $v_m$ now depends on the function $k = k \left( S_w \right)$, the
cross-ratio. Note that when $k = S_w/S_n$,~\cref{eq:vm-solved-from-k} reduces to
$\hat{v}_w - \hat{v}_n = v^{\prime}$.\textbf{~\Cref{eq:vm-solved-from-k} is the first
  major result in this paper.} To make contact with earlier work on the theory,
\cref{eq:vm-solved-from-k} can be written in a form more reminiscent of
\cref{eq:vm-derivative-form,eq:vm-constitutive},
\begin{equation}
  \label{eq:vm-k-derivative-form}
  v_m(S_w ; k) \ \equiv \ v^{\prime} + \beta v \;,
\end{equation}
where
\begin{equation}
  \label{eq:beta-function}
  \beta \ \equiv \ \frac{kS_n - S_w}{k S^2_n + S_w^{2}} \;.
\end{equation}
The procedure gives an expression for the term $v_n - v_w$ in
\cref{eq:vm-derivative-form}. If $\beta v$ in \cref{eq:vm-k-derivative-form} is
replaced by a linear fit of $\beta v$ as a function of $v^{\prime}$, one obtains a
relation of the form of \cref{eq:vm-constitutive}.

It is the parametrization of $k$ that gives the parametrization of $v_m$. To
obtain this parametrization, the cross ratio $k$ must be described in terms of a
projective map between two sets of saturations, $\left( S_w, S_n \right) \mapsto
\left( \tilde{S}_w, \tilde{S}_n \right)$, viewed as homogeneous coordinates.

Instead of defining the cross-ratio in terms of the degenerate conic $m \cup
l_{\infty}$, consider instead the two lines $l_{\infty}, \tilde{l}_{\infty}$, which are related
by a projective transformation as per the discussion in
\cref{sec:transf-degenerate-conic}. This transformation had a fixed point $J$
which was unaffected by the transformation. A projective map $f$ between two
lines $l_{\infty}, \tilde{l}_{\infty}$ that preserves their intersection, here $J =
l_{\infty} \times \tilde{l}_{\infty}$, is by definition a specialized projective
map called a \textit{perspectivity}, see \cref{app:rp1}. In terms of homogeneous
coordinates on the two lines, a perspectivity has the property that it can be
expressed as $\left[ S_w:S_n\right] \mapsto \left[ \tau S_w:S_n \right]$ for
some number $\tau$, possibly a function. One can now define a projective map
between the two lines as a point-projection from $l^{\ast}_{0}$. Thus, the
points of $l_{\infty}$ are related to those on $\tilde{l}_{\infty}$ by
intersection with the lines from $l_{\infty}$ incident to the lines.

Recall from~\cref{sec:frames} the parametric description of a projective line,
\cref{eq:parametric-projective-line}. In this description, for two points
$p\left( e_{1} \right), p \left( e_2 \right) \in \mathbb{P}^2$, one can
interpret $\lambda, \mu$ as homogeneous coordinates on this projective line:
simply pick $p\left( e_{1} \right) = I$, $p \left(e_{2} \right) = J$, $\lambda =
S_w$ and $\mu = S_n$ to obtain a parametrization of the projective line
$l_{\infty}$. This is expressed as
\begin{equation}
  \label{eq:l-infty-parametrized}
  l_{\infty} \ = \ p \left( S_w e_1 + S_n e_2 \right) \ = \ p \left( \frac{S_w}{S_n} e_1 + e_2 \right) \;.
\end{equation}
The lines $l_{\infty}, \tilde{l}_{\infty}$ can be viewed as the same line under
a change of frame, i.e. one views $\tilde{l}_{\infty}$ as $l_{\infty}$ by
picking a different basis $e^{´}_{1}, e^{´}_2$. This is a valid viewpoint of the
action of any projective map $f$. Such a change of frame is given by linear
combinations of the points $e_1, e_2$, represented by a $2\times2$-matrix of
coefficients $k_{ij}$. Fixing $J$ means that the representative vector $e_2$ can
only transform as $e_2 \mapsto k_{22} e_2 $ for some number $k_{22}$, possibly a
function. The requirement is that $J$ is fixed for all values of the parameter
$S_w$. This restricts the transformation in $e_1$ to be $e_1 \mapsto k_{11}
e_{1}$, since if the transformation matrix contained off-diagonal elements, the
vector $e_2$ would not transform as required, hence $J$ would not be fixed. One
then has
\begin{align}
  \label{eq:perspectivity-frame-description}
  p \left( S_w e_1 + S_n e_2 \right) \ \mapsto & \  p \left( S_w k_{11} e_1 + S_n k_{22} e_2 \right) \nonumber \\
  =& \ p \left( \frac{S_w k_{11}}{S_n k_{22}} e_1 + e_2 \right) \nonumber \\
  \equiv& \  p \left( \tilde{\tau} \frac{S_w}{S_n} e_1 + e_2 \right) \;,
\end{align}
where $\tilde{\tau} = k_{11}/k_{22}$. \Cref{eq:perspectivity-frame-description}
defines a map $\left[ S_w:S_n \right] \mapsto \left[ \tilde{\tau} S_w : S_n
\right]$. It then follows from~\cref{sec:cross-ratio}, in particular
\cref{eq:cross-ratio-hom-coords,eq:cross-ratio-hom-coords-demonstration}, that
$\tilde{\tau} S_w S^{-1}_n$ is the ratio of homogeneous coordinates of a point
\begin{equation}
  \label{eq:point-X}
  X \ = \ p \left(\tilde{\tau} S_w e_1 + S_n e_{2} \right) \;.
\end{equation}
with respect to the frame
\begin{align}
  \label{eq:cross-ratio-from-frame-I}
  I \ =& \ p \left( e_{1} \right) \\
  J \ =& \ p \left( e_{2} \right) \label{eq:cross-ratio-from-frame-J} \\
  U \ =& \ p \left(e_1 + e_{2} \right) \label{eq:cross-ratio-from-frame-U} \;. 
\end{align}
This ratio is by definition the cross ratio \cite{bergerGeometry1994}. One can
therefore set
\begin{equation}
  \label{eq:k-cross-ratio-defined}
  k \ \equiv \ \frac{S_w k_{11}}{S_n k_{22}} \equiv \tau\frac{S_w}{S_n} \;, 
\end{equation}
with $\tau = -\mathcal{S}$ as in \cref{eq:cross-ratio-our-conic}.

\subsection{Projection onto saturations }
\label{sec:second-projection}

Since the function $\tau$ only enters via homogeneous coordinates, it is
arbitrary whether it is absorbed into $S_w$, $S_n$, or distributed in some way
between the two. However, there is one unique way of performing this assignment
which allows for interpreting the transformed ratio of homogeneous
coordinates,~\cref{eq:k-cross-ratio-defined}, as a ratio of saturations. This
can be done in a way that leaves the cross-ratio invariant.

For now, absorb $\tau$ into $S_n$ by defining
\begin{equation}
  \label{eq:Sn-tau}
  S_{n,\tau} \equiv S_n \tau^{-1} \;.
\end{equation}
The ordinary saturations satisfy $S_w + S_n = 1$, while in general, $S_w +
S_{n,\tau} \neq 1$. To define a pair of homogeneous coordinates that sum to
unity, rewrite the ratio $S_w/S_{n, \tau}$ as $\underline{S}_w/\underline{S}_n$,
where $\underline{S}_w + \underline{S}_n = 1$. One then seeks the map $g: \left[
  S_w:S_{n,\tau} \right] \mapsto \left[
  \underline{S}_w:\underline{S}_{n}\right]$.

Recall that $\left[ S_w:S_{n,\tau} \right]$ are homogeneous coordinates of a
projective point on the line $l_{\infty}$ after a change of frame. The points on
$l_{\infty}$ are all defined as the coordinates where the two first components
sum to the third.
Now define $\omega = S_w + S_{n, \tau}$, such that $\left[ S_w:S_{n,
    \tau}:\omega \right]\in l_{\infty}$. In affine coordinates, $\left( S_w
  \omega^{-1}, S_{n, \tau} \omega^{-1} \right)$ determines a curve in this
plane, while $l_{\infty}$ has the points $\left( S_w, S_n \right)$.

It is now possible to simply project the points of the curve $\omega^{-1} \left(
  S_w, S_{n,\tau} \right)$ onto $l_{\infty}$ to obtain a pair $\left(
  \underline{S}_w, \underline{S}_n \right)$. One then need to specify a method
of projection between the points, which is here taken to be projection from a
point with a point of projection $p^{\ast} = \left( S_w^{\ast}, S_n^{\ast}
\right)$ \footnote{ The point of projection $p^{\ast}$ between the two lines
  could be placed ``at infinity'' in the projective plane, or at a finite
  position. The former is excluded here, as it only results in affine
  transformations between the lines \cite{stillwellFourPillarsGeometry2005}.}.
The points $ \omega^{-1}\left( S_w, S_{n,\tau} \right)$ are then projected onto
$l_{\infty}$ via the lines through $p^{\ast}$ and $\omega^{-1}\left( S_w,
  S_{n,\tau} \right)$. In homogeneous coordinates, this projection is defined as
(see \cref{app:rep-maps})
\begin{align}
  & \begin{pmatrix}
    1 & 0 & -S_w^{\ast} \omega^{-1} \\
    0 & 1 & -S_n^{\ast}  \omega^{-1} \\
    1 & 1 & -(S^{\ast}_w +S_n^{\ast}) \omega^{-1} 
  \end{pmatrix}
    \begin{pmatrix}
      S_w \\
      S_{n,\tau} \\
      \omega
    \end{pmatrix} \nonumber \\
  & \mapsto \left( \frac{S_w - S_w^{\ast}}{S_w + S_{n,\tau} - S_w^{\ast} - S_n^{\ast}},
    \frac{S_{n,\tau} - S_n^{\ast}}{S_w + S_{n,\tau} - S_w^{\ast} - S_n^{\ast}} \right) \nonumber \\
  & \equiv \ \left( \underline{S}_w, \underline{S}_{n} \right)\label{eq:underlined-saturations} \;.
\end{align}
The homogeneous coordinates $\left[ \underline{S}_w:\underline{S}_n:1 \right]$
also describes points of $l_{\infty}$. Going back to viewing $\underline{S}_{w},
\underline{S}_n$ as describing a point on the projective line as in
\cref{eq:point-X}, the homogeneous coordinates are now $\left[ \underline{S}_w:
  \underline{S}_n \right]$. Since only a projection between two lines has been
performed , the cross-ratio $k$ is invariant under $\left[ S_w:S_{n,\tau}
\right] \mapsto \left[ \underline{S}_{w}:\underline{S}_n \right]$, so that
\begin{equation}
  \label{eq:invariant-cross-ratio-projection}
  k = \tau \frac{S_w}{S_n} \ = \ \frac{\underline{S}_w}{\underline{S}_{n}} \;.
\end{equation}
The point $p^{\ast}$ could possibly be defined as the irreducible saturations of
wetting- and non-wetting fluid, but no attempt to verify this systematically is
performed in this article. With a few exceptions (see \cref{sec:noisy-data}),
$p^{\ast} = \left( 0,0 \right)$ is always used.

Summed up, one has the perspectivity map $f: [S_w:S_n] \mapsto \left[ S_w:
  S_{n,\tau} \right]$ from the before, and the second perspectivity $g: \left[
  S_w : S_{n,\tau} \right] \mapsto \left[ \underline{S}_w:\underline{S}_n
\right]$ defined in this section. This defines the composite transformation $h:
\left[ S_w:S_n \right] \mapsto \left[ \underline{S}_w:\underline{S}_n \right]$
as
\begin{equation}
  \label{eq:perps-comp}
  h = g \circ f \;,
\end{equation}
which by definition is a homography as it is a composition of perspectivities.

The discussion in this section and the previous one verifies that the
cross-ratio of four points on the degenerate conic $l_{\infty} \times m$ can be
interpreted in terms of a projective map between two lines. The same cross-ratio
then appears as the homogeneous coordinates of a point on the line with respect
to some projective frame.

In the next section, the transformation $\left[ S_w:S_n \right] \mapsto \left[
  \underline{S}_w:\underline{S}_n \right]$ is formulated as a linear fractional
transformation, see~\cref{app:rp1}. From this, the set $\tilde{S}_w,
\tilde{S}_n$ is obtained, which are the \textit{parametrized} saturations. The
parameters are the three parameters of the linear fractional transformation.

The relation between the coordinates $\left\{ \tilde{S}_i \right\}$ and $\left\{
  \underline{S}_{i} \right\}$ is the following: when $\left\{ \underline{S}_i
\right\}$ is found, the homography $h: \left[ S_w:S_n \right] \mapsto \left[
  \underline{S}_w:\underline{S}_{n} \right]$ is computed in the form of
eq.~\eqref{eq:lin-hom-frac}, with $t = S_w/S_n$. From the obtained homography,
one defines $\left\{ \tilde{S}_i \right\}$. The ratio $\tilde{S}_w/\tilde{S}_n$
is thus an approximation to the ratio $k=\underline{S}_w/\underline{S}_n$,
meaning the true cross-ratio $k$ is approximated by some function $k_e =
\tilde{S}_w /\tilde{S}_{n}$.


  
\section{Computation of $v_m$ from data, hydraulic conductivity}
\label{sec:vm-numeric}

This section outlines how to obtain the estimation $k_{e}$. No particular curve
fitting of the velocity function data will be performed, which means that the
data can contain noise \footnote{Cubic splines were used between data points
  such that the data series could be treated as a function, but this function is
  only evaluated at the nodes. To eliminate large oscillations close to the
  saturation boundaries, natural splines were used. This means that the second
  derivative $v^{\prime\prime}$ at the saturation endpoints is set to zero.}.

$\left( v,v_w, v_n, S_w \right)$ are available as lists of data, and $v^{\prime}$ is
computed from $v$ and $S_w$. From this, $v_m$ is computed using
\cref{eq:vm-derivative-form}, and $\hat{v}_{i}$ as
\cite{hansenRelationsSeepageVelocities2018}
\begin{align*}
  \hat{v}_w \ =& \ v + S_n v^{\prime} \;, \\ 
  \hat{v}_n \ =& \ v - S_w v^{\prime} \;.
\end{align*}

Setting $\tau = -v_n/v_w$, one has
\begin{equation}
  \label{eq:Sn-tilde}
  S_{n,\tau} \ = \ \frac{S_n}{\tau} \;.
\end{equation}
The projection $g: \left[ S_w : S_{n,\tau} \right] \mapsto \left[
  \underline{S}_w:\underline{S}_n \right]$ in \cref{eq:underlined-saturations}
is then computed with $p^{\ast} = \left( 0,0 \right)$. In some cases, $p^{\ast}$ is
non-zero, which will be noted in the figure.

There are now two saturation ranges given by $\left( S_w, S_n \right)$ and
$\left( \underline{S}_w, \underline{S}_n \right)$, which are related by a 1d
homography from the discussion in \cref{sec:five-point-conic}. The details of
homography estimation between two ranges of points is found elsewhere
\cite{luoReviewHomographyEstimation2023}. The map between the ranges, noting
that the vectors are defined up to scaling as usual, is
\begin{equation}
  \label{eq:hom-matrix}
  \begin{bmatrix}
    \underline{S}_w^{e} \\
    \underline{S}_n^{
    e}
  \end{bmatrix} \ = \ 
  \mathbb{M}
  \begin{bmatrix}
    S_w \\
    S_n
  \end{bmatrix} =
  \begin{pmatrix}
    a & b \\
    c & d
  \end{pmatrix}
  \begin{bmatrix}
    S_w \\
    S_n
  \end{bmatrix} \;,
\end{equation}
for constants $a,b,c,d$, and the superscript $e$ signifying that the quantities
are obtained from a homography estimation procedure with ranges given by the
points $\left(S_w, S_n \right)$, $\left( \underline{S}_w, \underline{S}_n
\right)$ as input. The matrix $\mathbb{M}$ is defined up to a scalar. By
dividing all entries by $d$, the number of parameters is reduced to three. The
same notation is kept for $a,b,c$, is equivalent to setting $d=1$.

The projective transformation in~\cref{eq:hom-matrix} gives the affine
coordinate
\begin{equation}
  \label{eq:projectivized-hom}
  \frac{\underline{S}_w^{e}}{\underline{S}_n^{e}} \ = \ \frac{aS_w + bS_{n}}{cS_w + S_n}  \;.
\end{equation}
Eq. \eqref{eq:projectivized-hom} is a linear fractional transformation in the
parameter $S_w/S_n \equiv t$. The cross ratio estimate is then
\begin{equation}
  \label{eq:tilde-ratio-defined}
  k_{e} \ = \ \frac{a t + b }{c t + 1} \;.
\end{equation}
$k$ equals the ratio $\underline{S}_w/\underline{S}_n$. Now, $k_e$ is defined as
the ratio $\tilde{S}_w / \tilde{S}_n$, where $\tilde{S}_w + \tilde{S}_n = 1$.
$\tilde{S}_w$ is obtained from $k_{e}$ as
\begin{equation}
  \label{eq:send-ratio}
  1 - \left( 1 - k_e \right)^{-1} \equiv \tilde{S}_{w} \;.
\end{equation}
where it is used that $\left\{ \tilde{S}_i \right\}$ sum to unity \footnote{An
  extra minus sign in front of $k_e$ is used here due to absorbing a minus sign
  in $\tau$ in \cref{eq:cross-ratio-our-conic}. }.

$k_{e}$ can now be used as input in \cref{eq:vm-solved-from-k} to compute an
estimate of $v_m$, parametrized by $a, b, c$. Note that the definition in
\cref{eq:vm-solved-from-k} is exactly $v_{m}$ if one uses $k$ as in
\cref{eq:k-cross-ratio-defined} . In practice, $c \approx 0$ in many cases, leaving
only the parameters $a,b$. This is related to the fixed point $J$ of the
transformation. This fixed point is not enforced in the homography estimation,
see \cref{sec:noisy-data}.

\subsection{Relative permeability, hydraulic conductivity}
\label{sec:relperm-hydr}

It is straightforward to define constitutive relations for $\left\{ v_i
\right\}$ to see how physical quantities enter the geometric considerations of
the preceding sections. Here, relative permeability theory is used as an
example, in which the constitutive relations for the seepage velocities $\left\{
  v_i \right\}$ are defined as \cite{wyckoffFlowGasLiquid1936}
\begin{equation}
  \label{eq:relperm-constitutive}
  v_i \ = \ \frac{K k_{ri}}{\phi S_i \mu_{i}} \left| \nabla P \right| \;,
\end{equation}
where $K$ is the absolute (intrinsic) permeability, $\nabla P$ is the pressure
gradient, $\phi$ the porosity and for fluid $i=w,n$, $k_{ri}$ is the relative
permeability and $\mu_i$ is the dynamic viscosity. $v$ is then computed as
\begin{equation}
  \label{eq:v-relperm}
  v \ = \ \mu_w v_0 \left[ \frac{k_{rw}}{\mu_w} + \frac{k_{rn}}{\mu_n} \right] \;,
\end{equation}
with
\begin{equation}
  \label{eq:v0}
  v_0 \ = \ \frac{K \left|  \nabla P \right|}{\mu_w \phi} \;.
\end{equation}
$v_m$ is then
\begin{equation}
  \label{eq:vm-relperm}
  v_m \ = \ v^{\prime} + \mu_w v_0 \left[ \frac{k_{rn}}{\mu_n} - \frac{k_{rw}}{\mu_w} \right] \;.
\end{equation}

Using \cref{eq:vm-k-derivative-form,eq:vm-relperm,eq:v-relperm} and defining the
hydraulic conductivity $\mathcal{K}_i$ of fluid $i$ as $\mathcal{K}_i = Kk_{ri}/\mu_i$, one finds that
\begin{equation}
  \label{eq:beta-v-hydr}
  \beta v = \mu_w v_0 \left( \frac{k_{rn}}{\mu_n} - \frac{k_{rw}}{\mu_w} \right) = \frac{\mu_wv_0}{K} \left(\mathcal{K}_n - \mathcal{K}_{w}  \right) \;.
\end{equation}
From \cref{eq:v-relperm}, the relation between the cross-ratio $k$ and $\left\{
  \mathcal{K}_i \right\}$ is then
\begin{equation}
  \label{eq:beta-hydr}
  \beta = \left.  \left( \frac{k S_n - S_w}{kS_n^2 + S_w^2 } \right) \right|_{r} =  \frac{\mathcal{K}_n - \mathcal{K}_{w}}{ \mathcal{K}_n + \mathcal{K}_{w}} \;,
\end{equation}
where the subscript $r$ signifies that the quantities in the function are those
defined in terms of the relative permeability data. Thus, the estimation of
$v_m$ is described entirely as a function of the hydraulic conductivities
$\left\{ \mathcal{K}_i \right\}$.

However, the presence of irreducible saturations must be also be taken into
account if one uses \cref{eq:beta-hydr}. Denote the irreducible
wetting-saturation and the residual non-wetting saturation after imbibition by
$S_w^{i}$ and $S_n^{i}$ respectively (for a particular value of $\Delta P$)
\footnote{ These in general depend on whether the data describes a drainage- or
  imbibition process.}. $v_m$ is found from applying $\partial_{S_w}$ to $v = S_{w,r}
v_w \left( S_{w,r} \right) + S_{n,r} v_n \left( S_{w,r} \right)$, where
$S_{w,r}$ is the saturation in \cref{eq:saturation-scaling} and $\left\{ v_i
\right\}$ are the velocities in \cref{eq:relperm-constitutive}. The relation
between $S_w$ and $S_{w,r}$ is \cref{eq:saturation-scaling}, with $a_{r}=\left(
  1 - S_w^{\ast} - S_n^{\ast} \right)^{-1}$ and $b_{r} = - S_w^{\ast} \left( 1 - S_w^{\ast}
  - S_n^{\ast} \right)^{-1}$. $S_{w}$ in the partial derivative $\partial_{S_w}$ comes
from the thermodynamic velocities, where the entire range $S_w \in \left[ 0,1
\right]$ is considered \footnote{ This is so because the partial derivatives
  that define $\left\{ \hat{v}_i \right\}$ alters the saturation via a
  unphysical addition or subtraction of fluid that can always be performed. The
  physical saturations can only be obtained via physical flow processes, and are
  subject to the restrictions of the porous medium, which means that there are
  history-dependent irreducible saturations which restricts the domain.}. Since
the derivative $\partial_{S_w}$ enters in $v_m$ (see~\cref{eq:co-moving-convex-comb})
and $v^{\prime}$ in \cref{eq:vm-relperm-hydr}, one has to apply the chain rule in
changing variables from $S_w$ to $aS_w + b$ in these terms, which introduces a
factor of $a_{r}$. There is no derivative in the term $\beta v$ if the hydraulic
conductivities are used as in \cref{eq:beta-hydr}, hence no scaling by $a_r$.
$v_m$ is therefore in this case given by
\begin{equation}
  \label{eq:vm-relperm-hydr}
  v_m = v^{\prime} + \frac{1}{a_{r}}\left( \frac{\mathcal{K}_n - \mathcal{K}_{w}}{ \mathcal{K}_n + \mathcal{K}_{w}}  \right)v \;,
\end{equation}
where all terms was multiplied by $1/a_{r}$.

\section{Results}
\label{sec:results}
\begin{figure*}
  \includegraphics[width=.32\textwidth]{"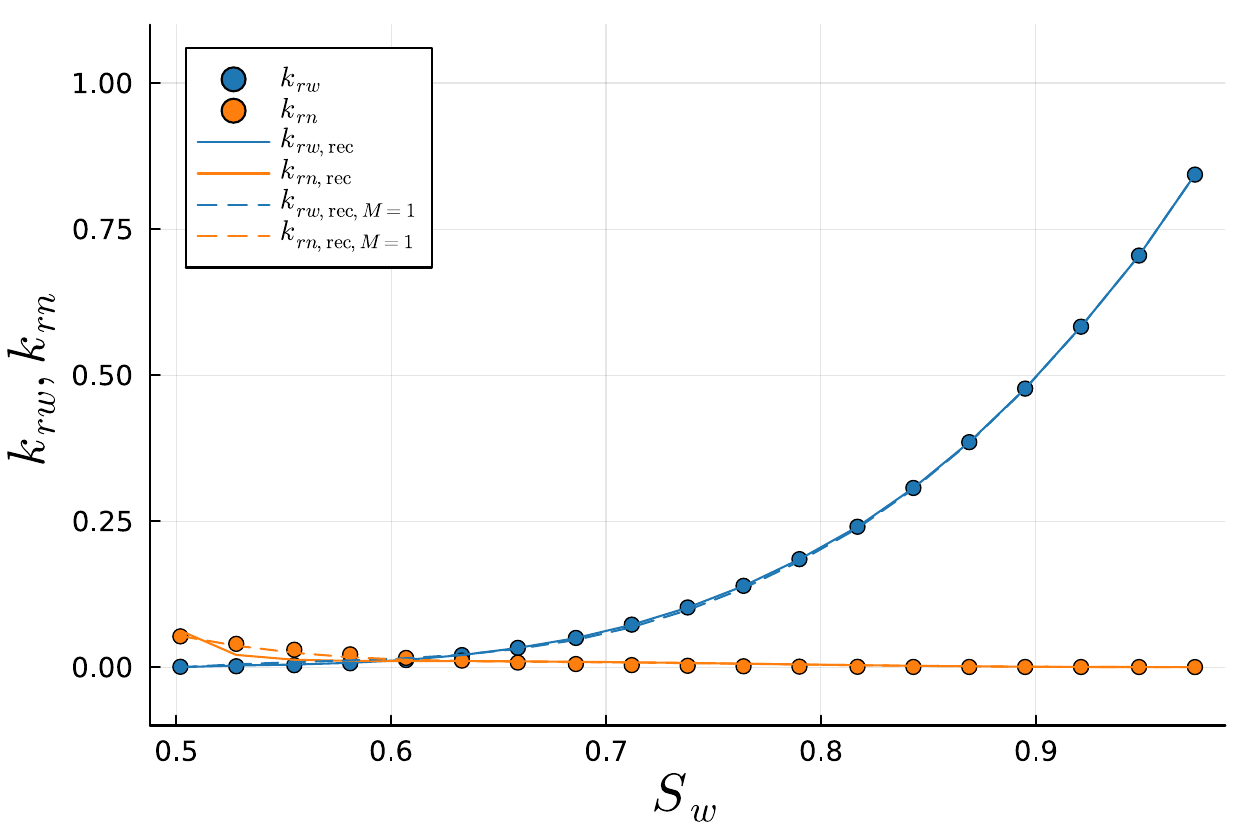"}
  \includegraphics[width=.32\textwidth]{"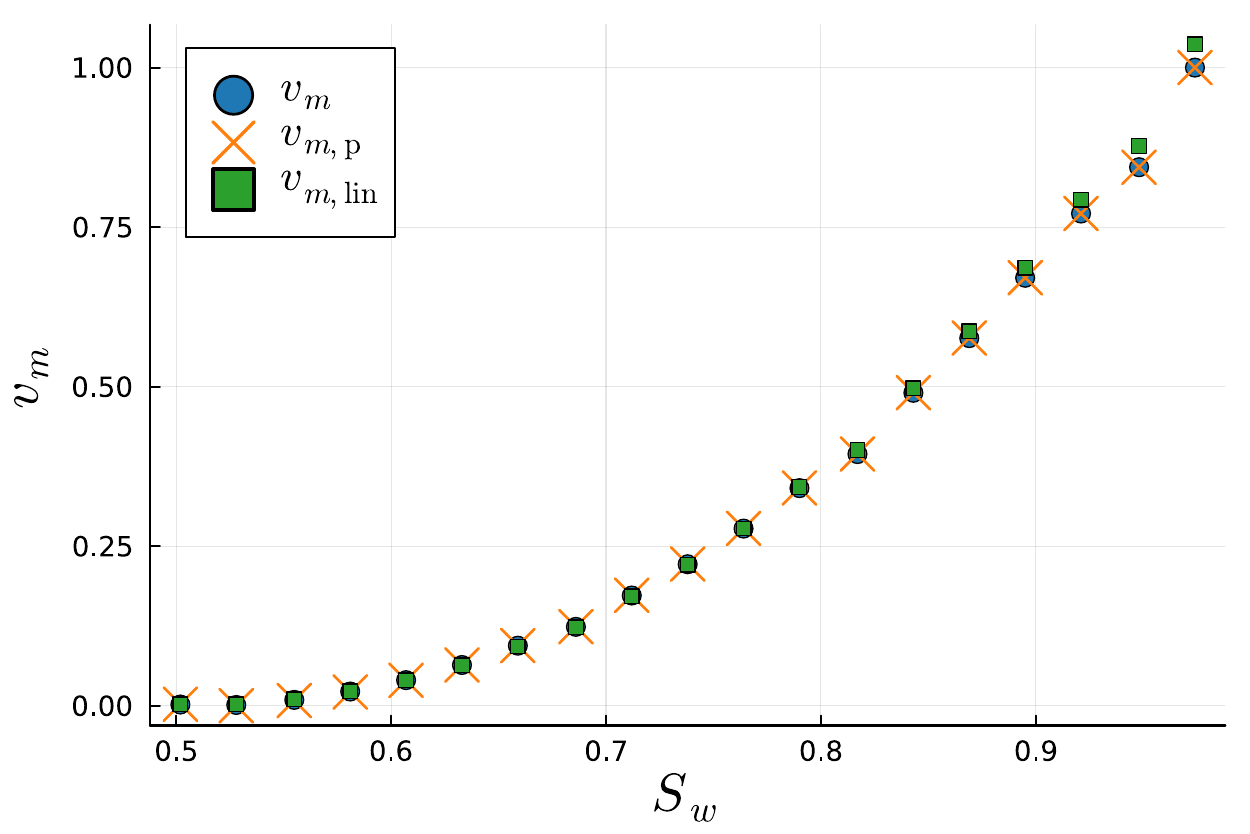"}
  \includegraphics[width=.32\textwidth]{"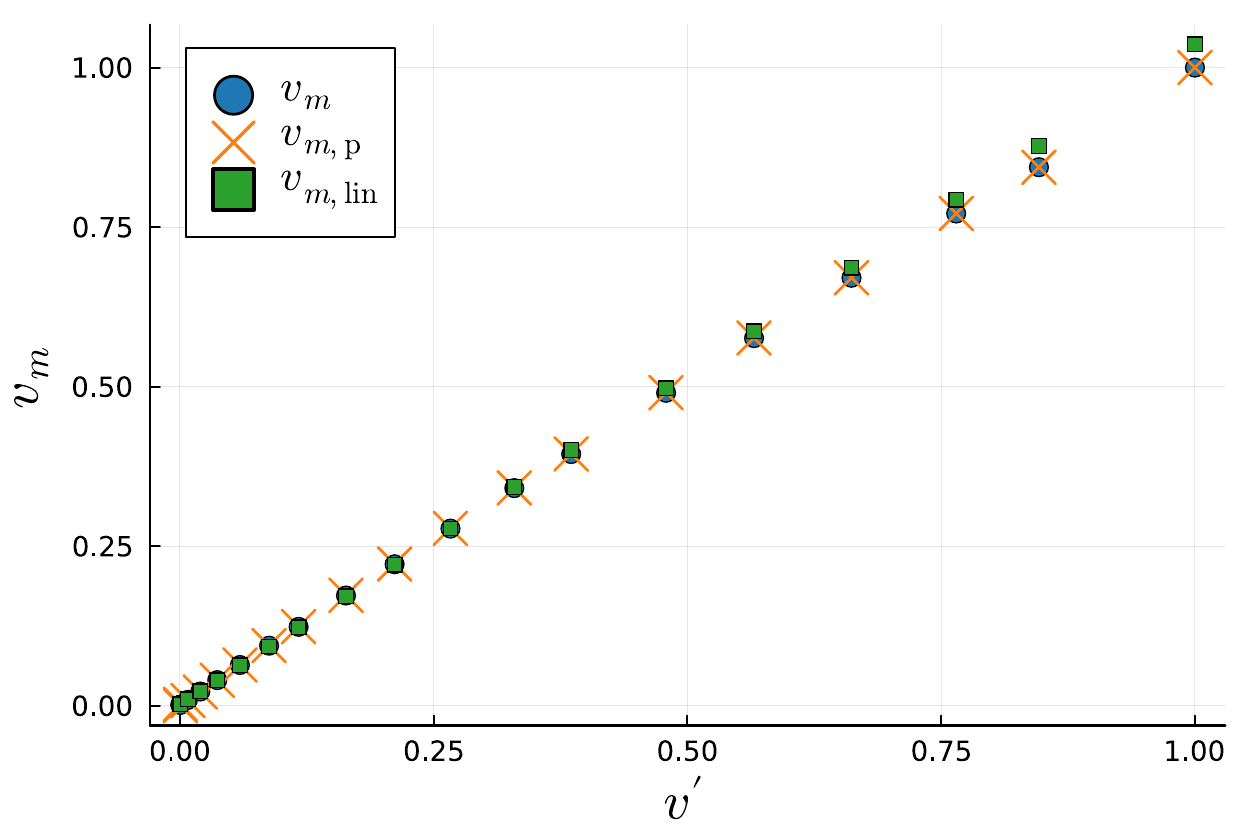"} \par
  \vspace{5pt}
  \includegraphics[width=.32\textwidth]{"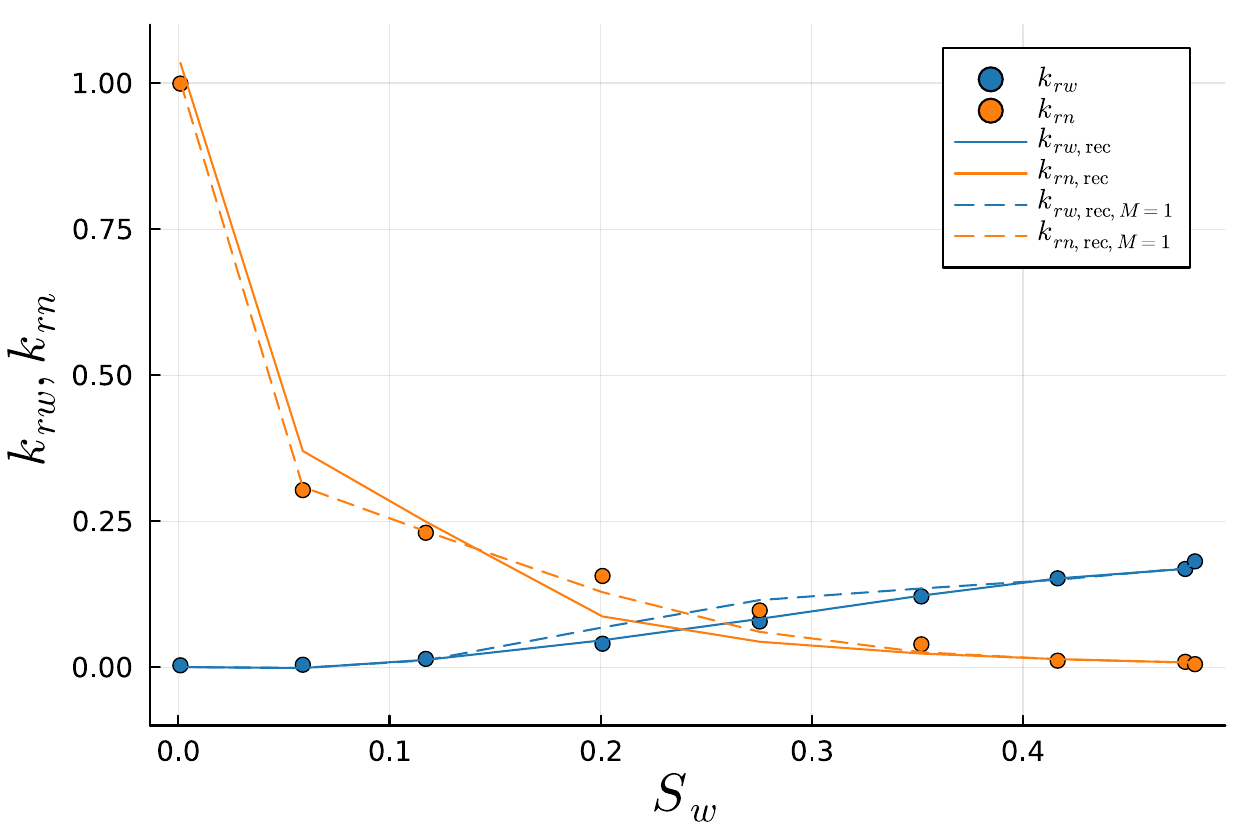 "}
  \includegraphics[width=.32\textwidth]{"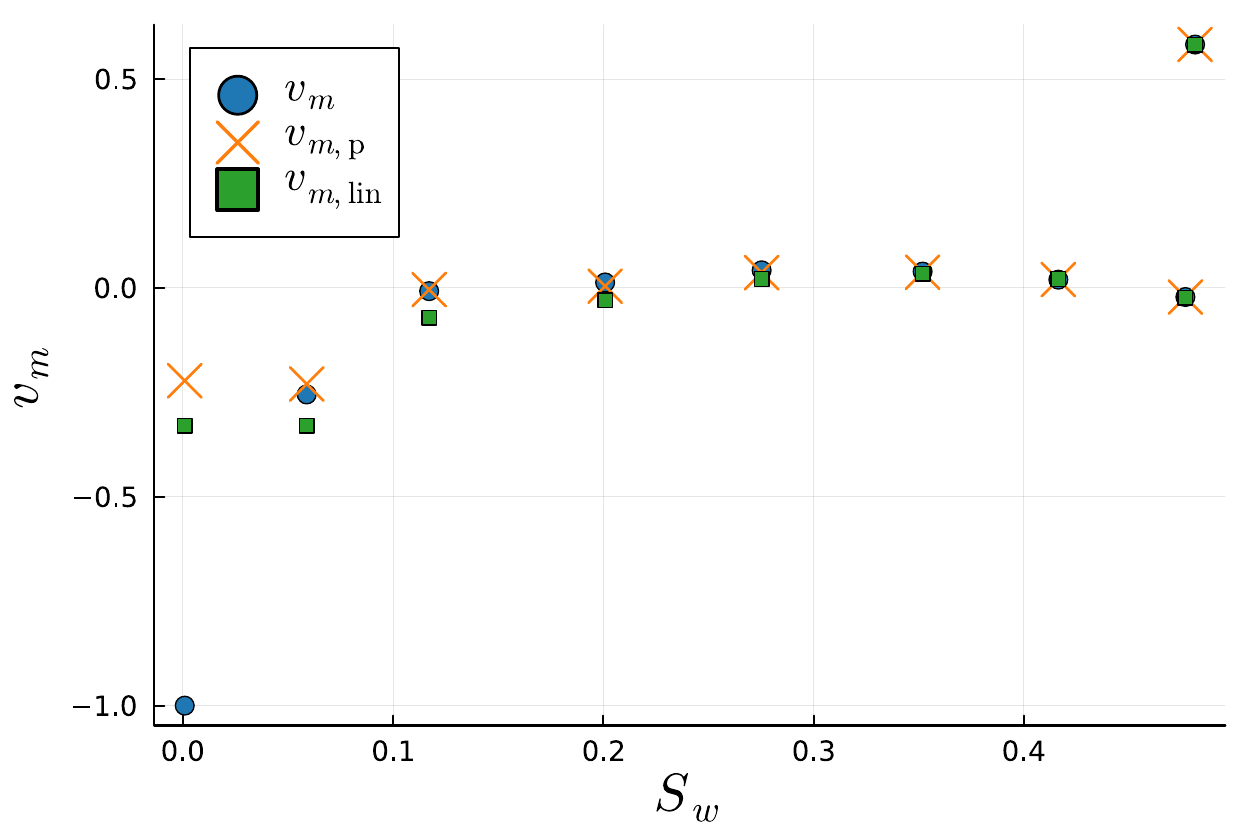"}
  \includegraphics[width=.32\textwidth]{"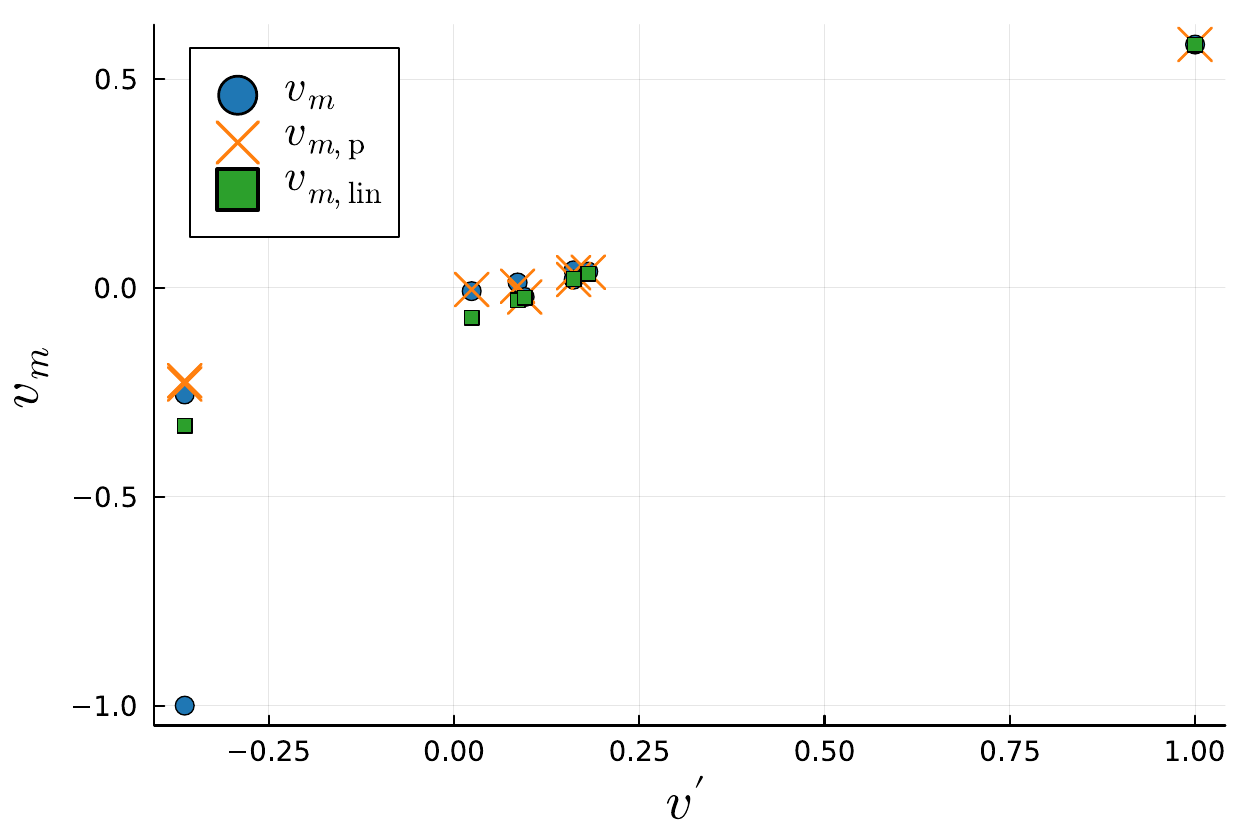"} \par
  \vspace{5pt}
  \includegraphics[width=.32\textwidth]{"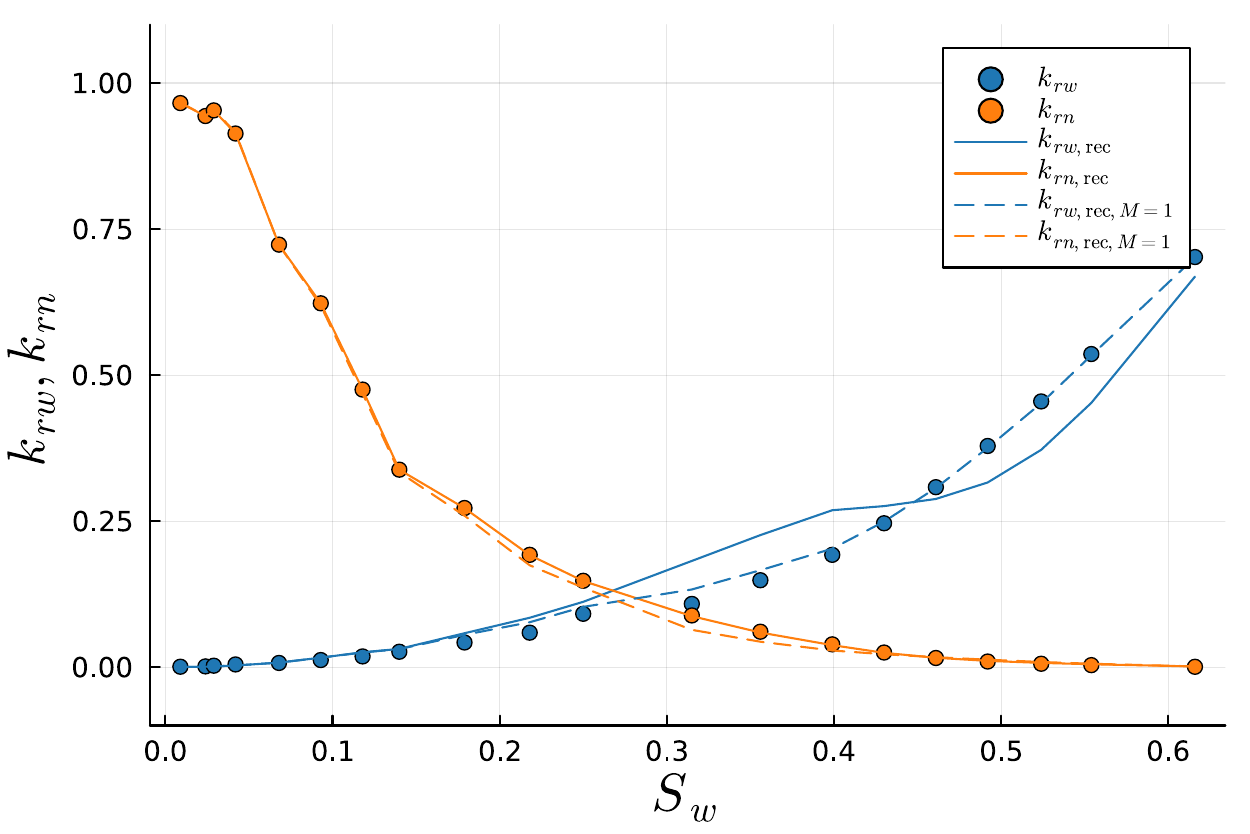"}
  \includegraphics[width=.32\textwidth]{"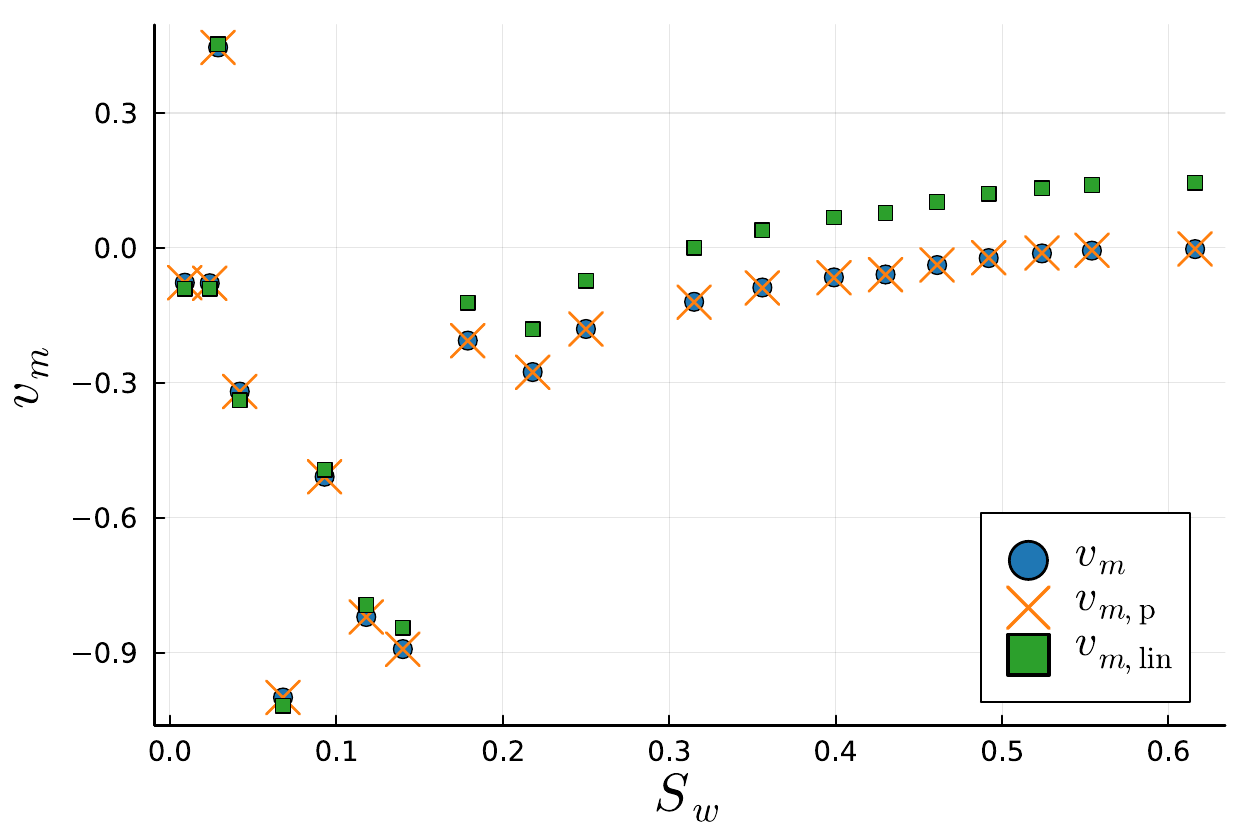"}
  \includegraphics[width=.32\textwidth]{"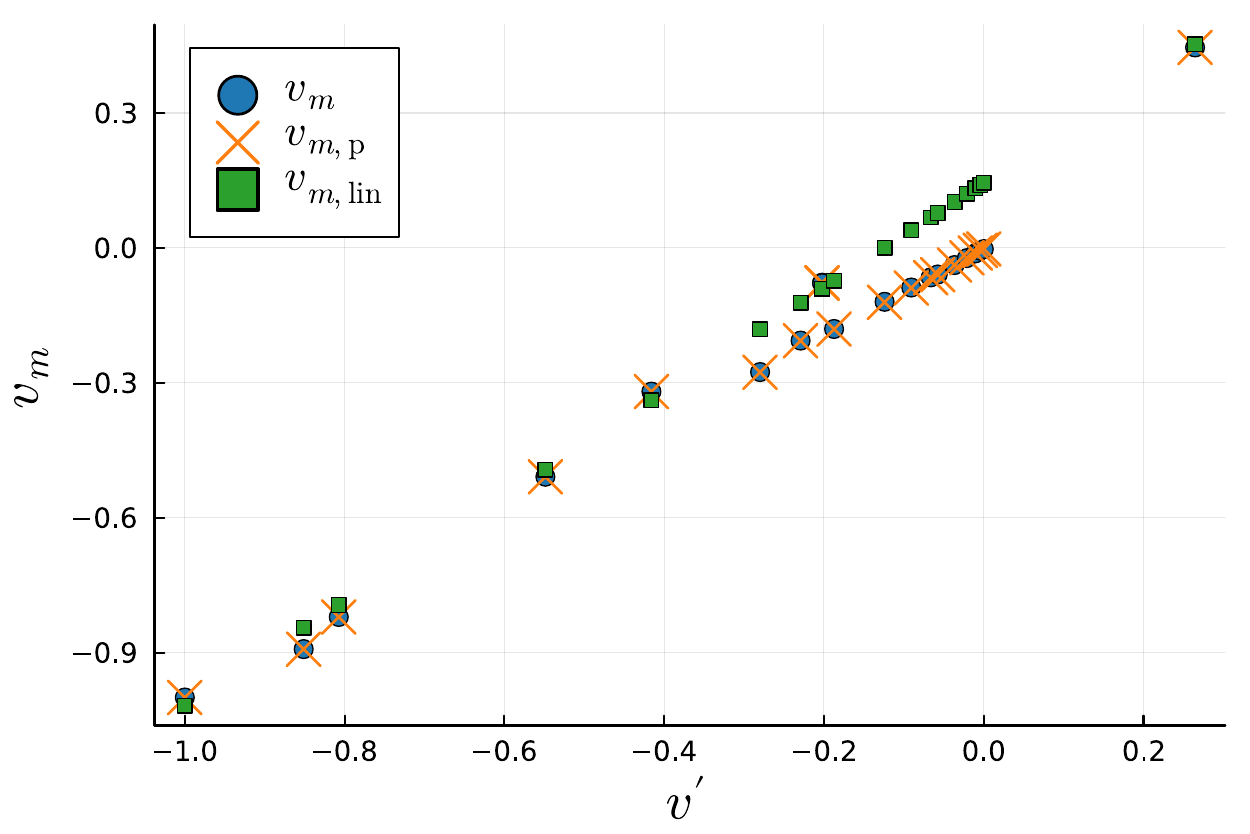"} \par
  \vspace{5pt}
  \includegraphics[width=.32\textwidth]{"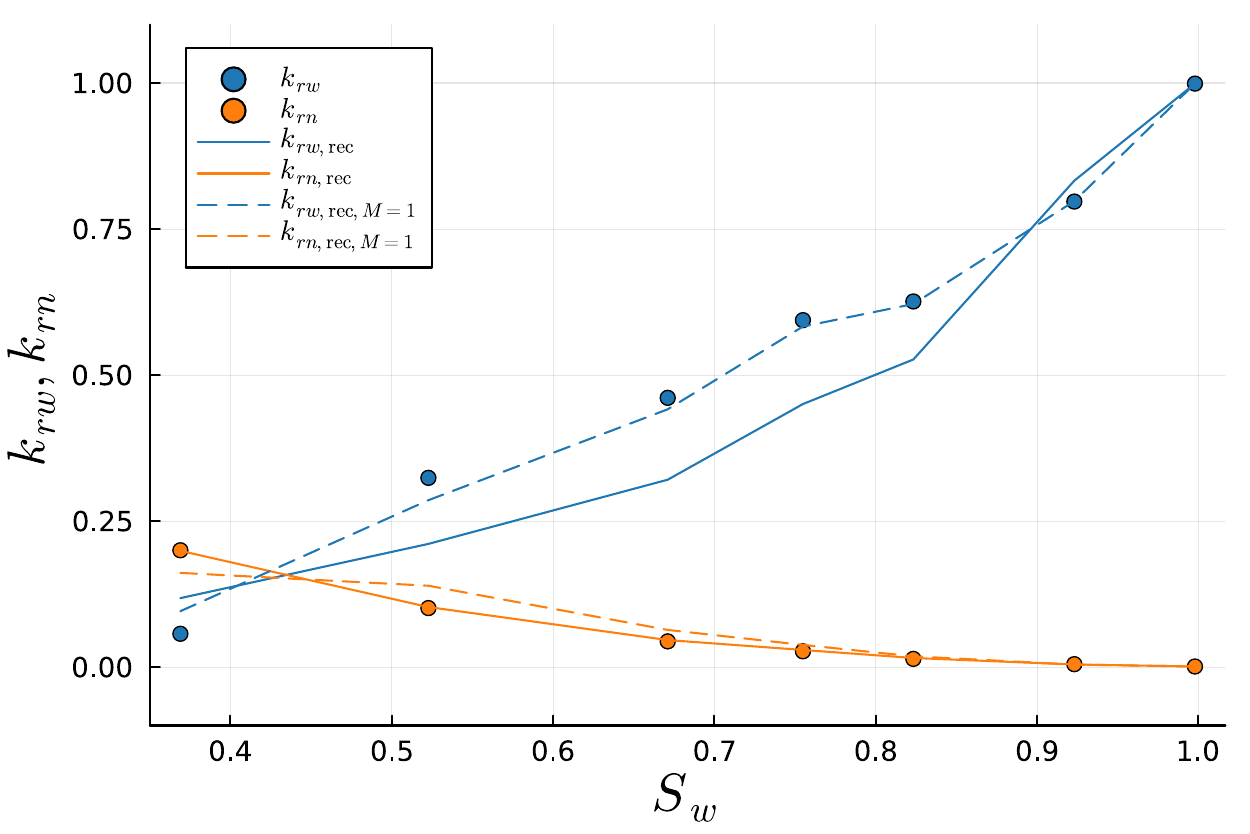 "}
  \includegraphics[width=.32\textwidth]{"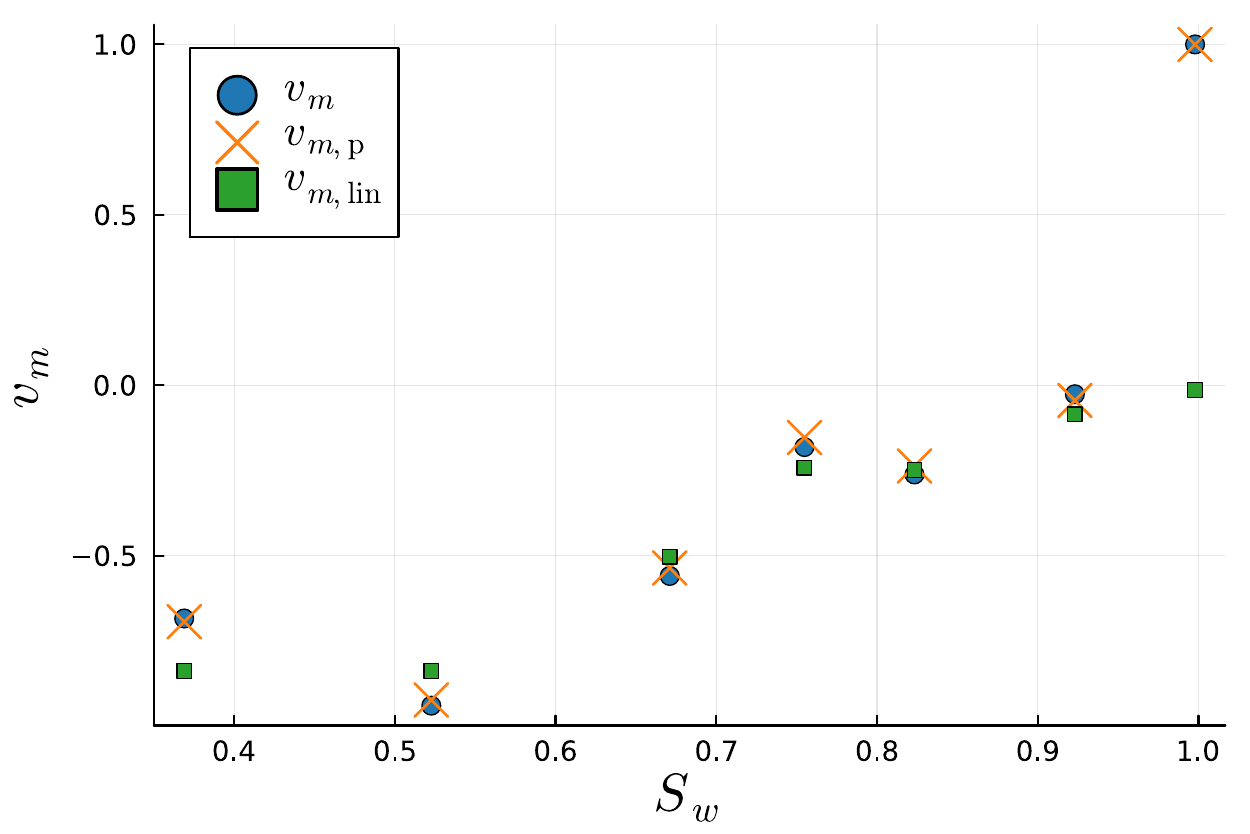"}
  \includegraphics[width=.32\textwidth]{"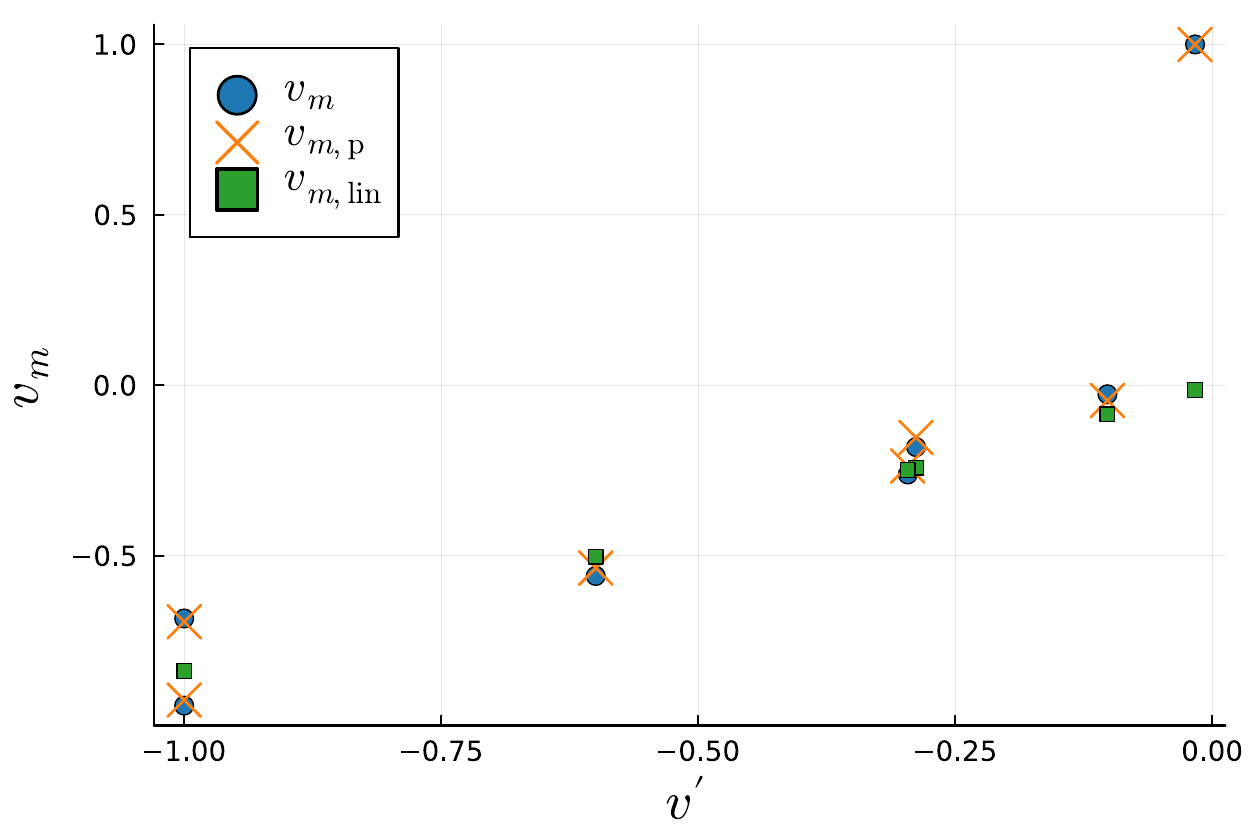"} \par
  \vspace{5pt}
  \includegraphics[width=.32\textwidth]{"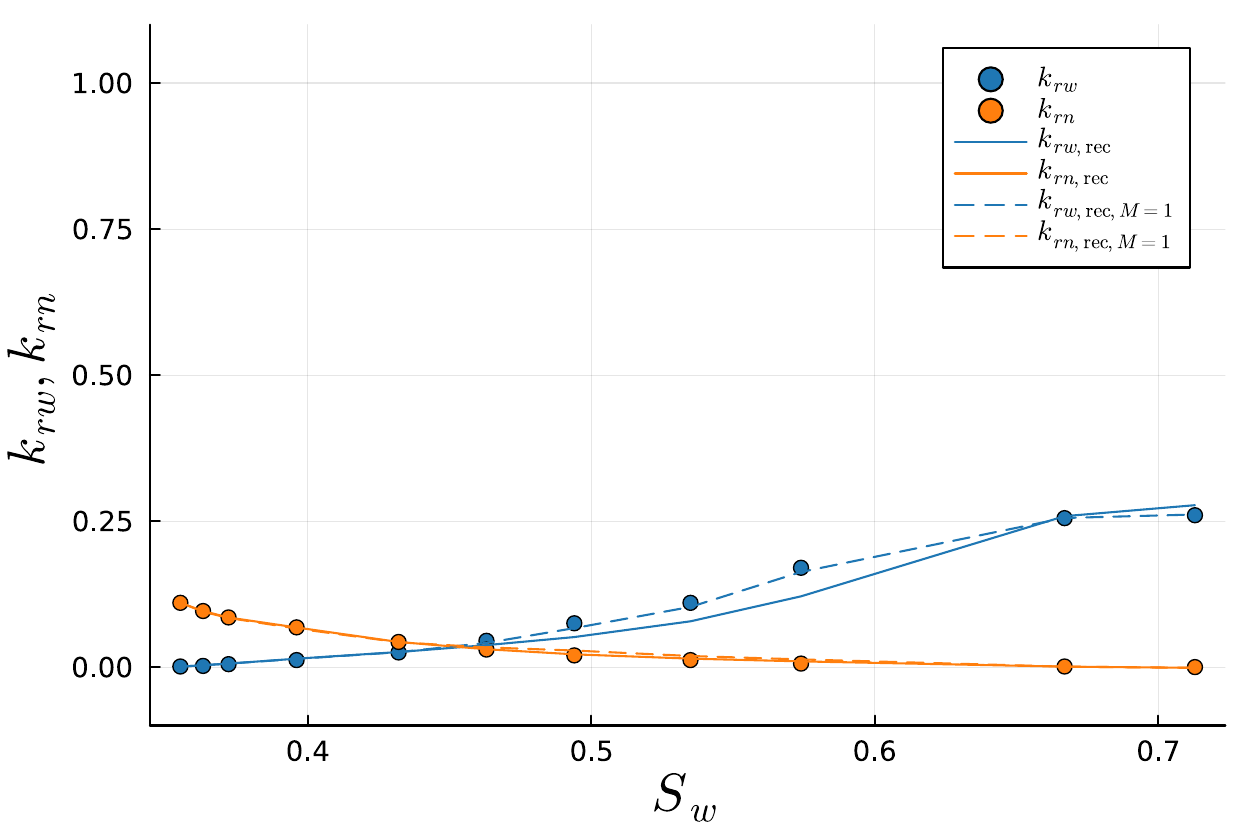"}
  \includegraphics[width=.32\textwidth]{"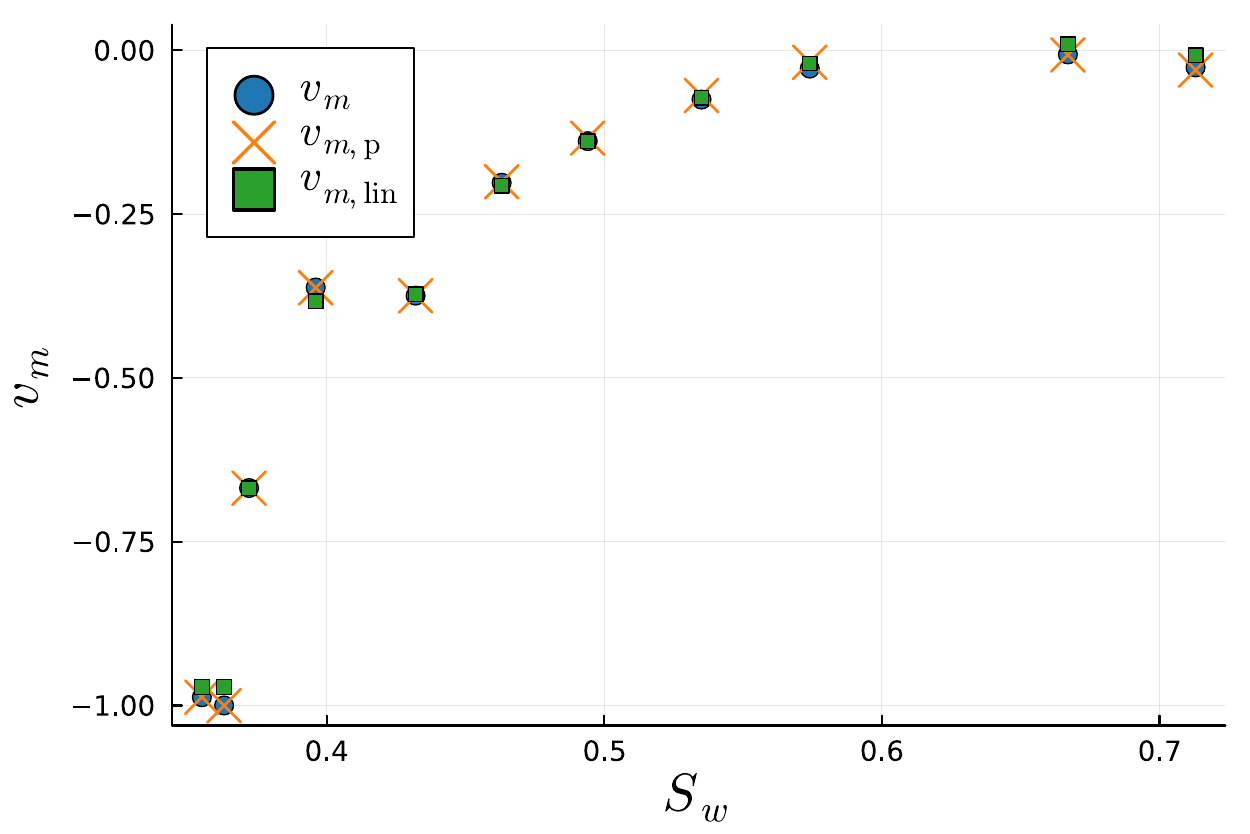"}
  \includegraphics[width=.32\textwidth]{"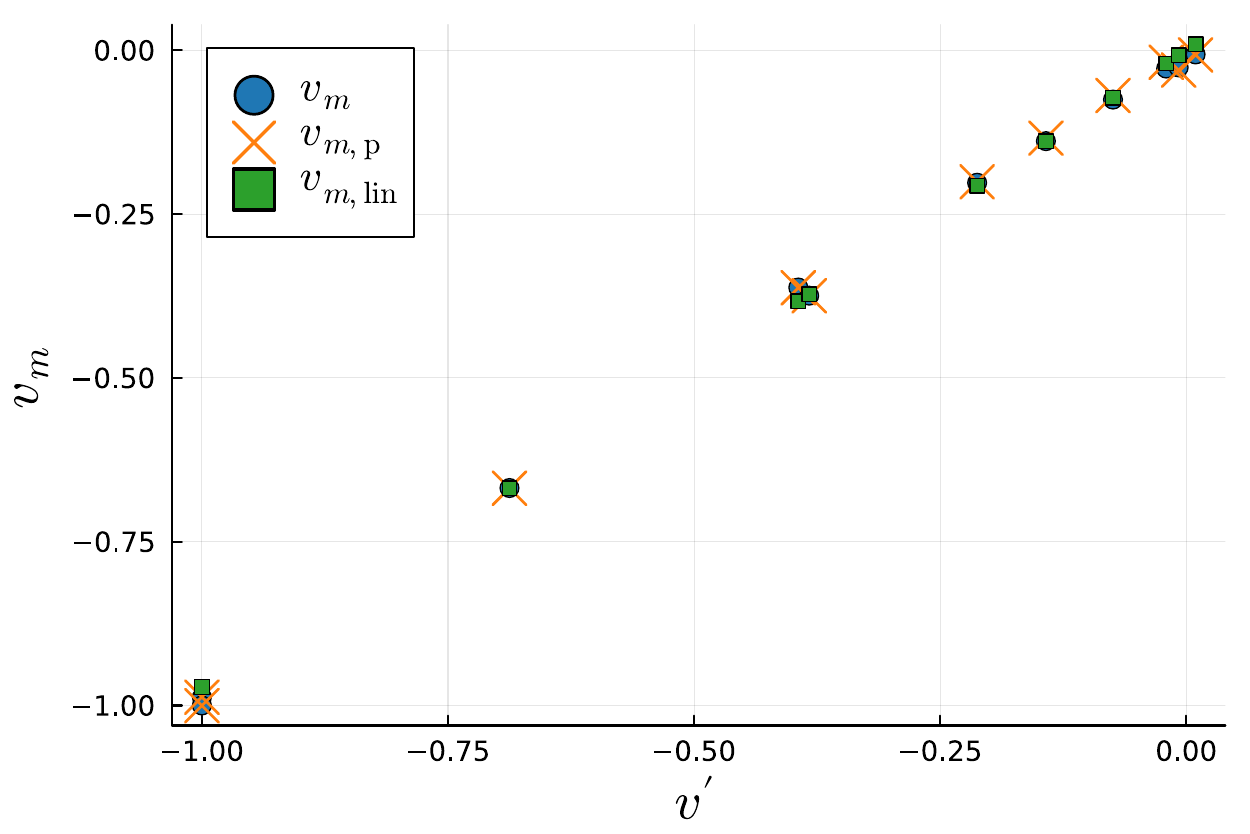"}
  \caption{Selected relative permeability data series, where data from the same
    datasets were considered in \cite{royCoMovingVelocityImmiscible2022}.
    $k_{rw}, k_{rn}$ are the relative permeabilities from the data, while
    $k_{rw,\mathrm{rec}}, k_{rn,\mathrm{rec}}$ are the recreated data from the
    procedure in this paper. $\left\{ k_{ri,\mathrm{rec}, M=1} \right\}$ are the
    same recreated velocities with $M=1$ imposed from the start. Each row in the
    figure represents a data series, all primary drainage or imbibition. There
    can be discrepancies in the recreated curves even though the estimation in
    \cref{sec:vm-numeric} reproduces $v_m$ exactly, see \cref{sec:noisy-data}.
    First row: table 3a in
    \cite{bennionRelativePermeabilityCharacteristics2005}, $\mathrm{Ca} \sim
    10^{-6} - 10^{-7}$, $M \sim 15$. Second row: fig. 4 ($0.2$
    \si{\centi\meter\cubed\per\minute}) in
    \cite{virnovskyImplementationMultirateTechnique1998}, $\mathrm{Ca} \sim
    10^{-2}$, $M \sim 12$. Third row: fig. 4 in
    \cite{oakThreePhaseRelativePermeability1990} (water/gas), $\mathrm{Ca}\sim
    10^{-5}$, $M \sim 0.02$. Fourth row: fig. 9 (run 18) in
    \cite{fulcherEffectCapillaryNumber1985}, $\mathrm{Ca} \sim 10^{-2} -
    10^{-3}$, $0.02$. Fifth row: table 3. in
    \cite{reynoldsCharacterizingFlowBehavior2015}, $\mathrm{Ca} \sim 10^{1}$, $M
    \sim 0.08$.\label{fig:relperm} }
\end{figure*}

The theory presented in this paper is now demonstrated on selected datasets from
the relative permeability literature, and data from a 2d Dynamic Pore Network
Model (DNM). A similar analysis of relative permeability dataset were performed
in \cite{royCoMovingVelocityImmiscible2022}. Here, no thorough analysis is
performed, and serves only as an example of recreating relative permeability
curves using the theory. The relative permeability data
\cite{bennionRelativePermeabilityCharacteristics2005,
  fulcherEffectCapillaryNumber1985,
  virnovskyImplementationMultirateTechnique1998,oakThreePhaseRelativePermeability1990,
  reynoldsCharacterizingFlowBehavior2015} is used to define $\left\{ v_i
\right\}$ as in \cref{sec:relperm-hydr}, from which $v_m$ can be computed via
\cref{eq:vm-relperm} or \cref{eq:vm-relperm-hydr} (the former is used here). If
values for a physical parameter was missing from the references, an estimated
value was used in its place. The computed $\left\{ v_i
\right\}_{\mathrm{relperm}}$ were then used in the procedure described in
\cref{sec:vm-numeric}, resulting in $v_m \left( S_w ; k_e
\right)_{\mathrm{relperm}}$ with $k_e$ the estimated cross-ratio. One then
defines $\left\{ v_i \right\}$ from \cref{eq:vw-transf,eq:vn-transf}, with $v_m$
replaced by the estimate. The resulting velocities are then converted to
relative permeabilities using \cref{eq:relperm-constitutive}. The result is
$k_{rw, \mathrm{rec}}$ and $ k_{rn, \mathrm{rec}}$. The quantities $\left\{
  k_{rw, \mathrm{rec}, M=1} \right\}$ in \cref{fig:relperm} are the recreated
relative permeabilities where $\mu_n = \mu_w$ is set from the beginning. This
has been done to illustrate the $M$-sensitivity of the recreated curves, as $M$
appears linearly in the cross-ratio function $k$ when the velocities are
replaced by relative permeabilities, see \cref{sec:noisy-data}. Note that in the
graphs, the linear approximation of $v_m (v^{\prime})$ uses only two parameters,
while three parameters is used for $v_{m,p}$ in \cref{fig:relperm}.

The datasets were selected to showcase a range of capillary numbers
$\mathrm{Ca}$. The definition for $\mathrm{Ca}$ used in this paper is
\cite{alzubaidiImpactWettabilityComoving2024}
\begin{equation}
  \label{eq:Ca-relperm}
  \mathrm{Ca} \ = \ \frac{\mu_w v_w + \mu_n v_n}{\gamma} \;, 
\end{equation}
and the one for the viscosity ratio $M$ is $\mu_n/\mu_{w}$. $p^{\ast} = \left(
  0,0 \right)$ has been used in all relative permeability datasets, even though
adjusting $p^{\ast}$ might give a better fit in some cases. The results are
shown in \cref{fig:relperm}, where $v_m$ is plotted on a scale relative to the
maximum value of $\left| v_m \right|$, as the scale of $v_m$ is uncertain due to the
estimated physical parameters in these datasets.

\begin{figure*}[t]
  \includegraphics[width=.32\textwidth]{"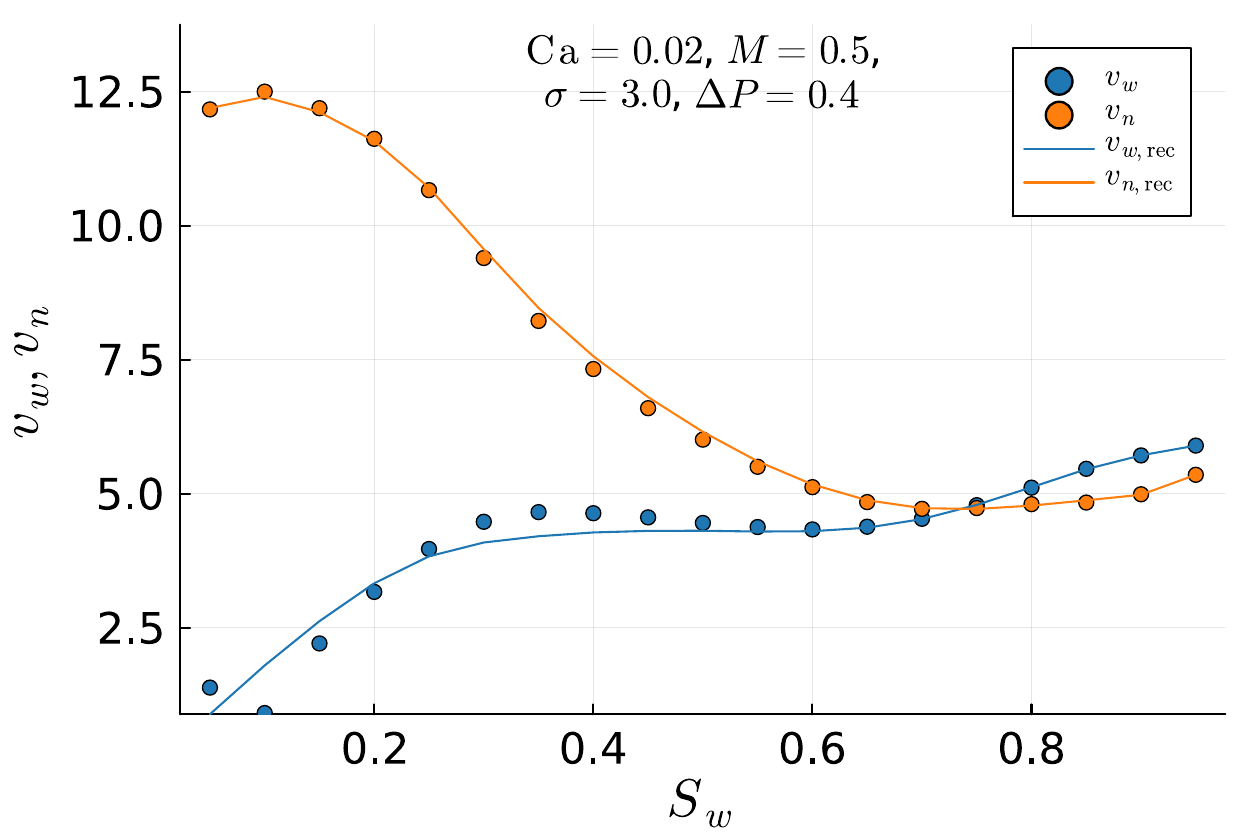"}
  \includegraphics[width=.32\textwidth]{"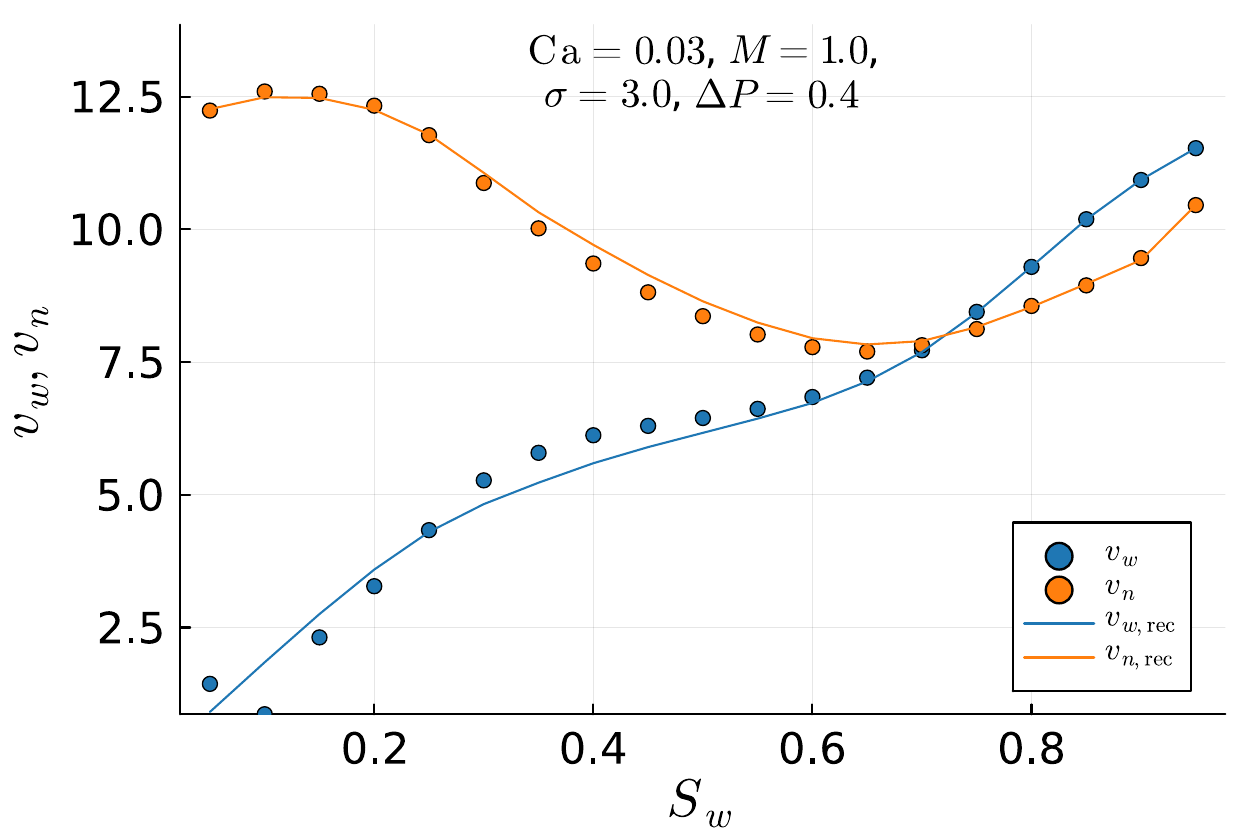"}
  \includegraphics[width=.32\textwidth]{"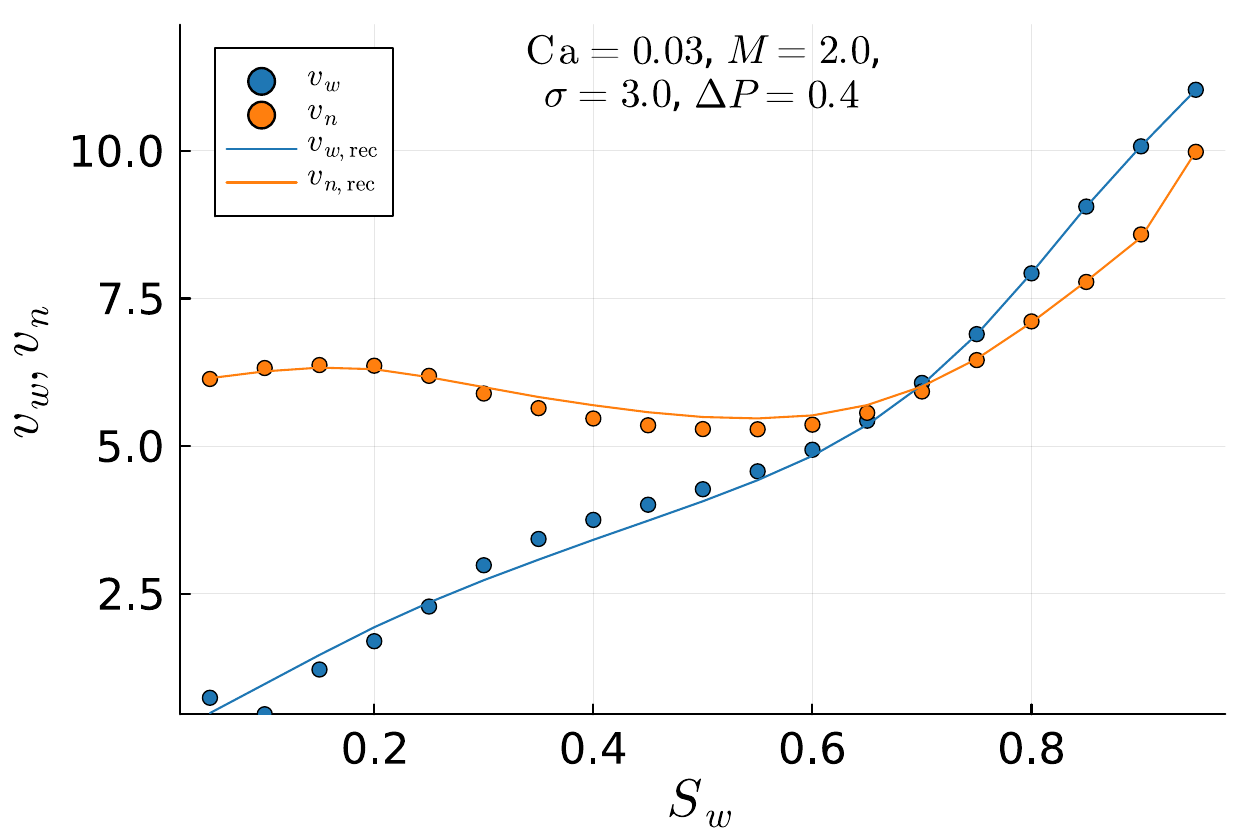"} \par
  \vspace{10pt}
  \includegraphics[width=.32\textwidth]{"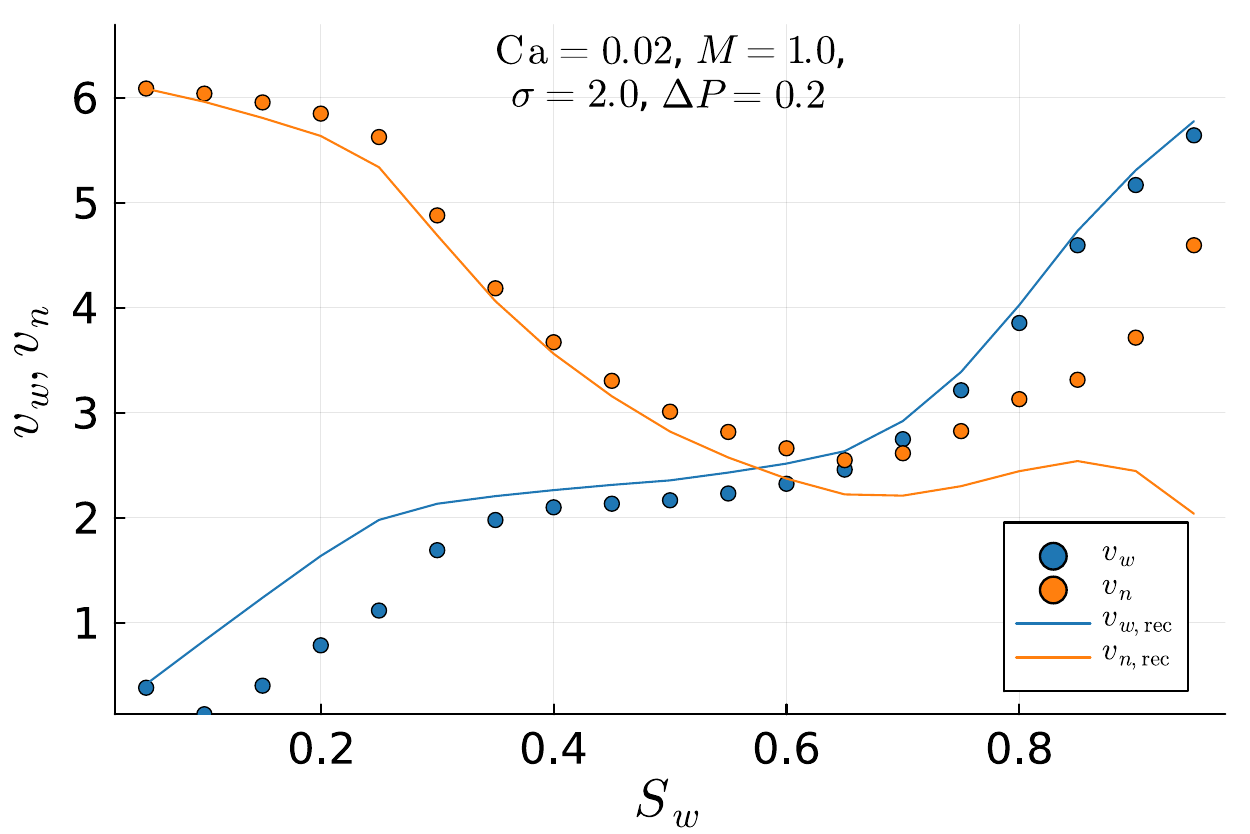"}
  \includegraphics[width=.32\textwidth]{"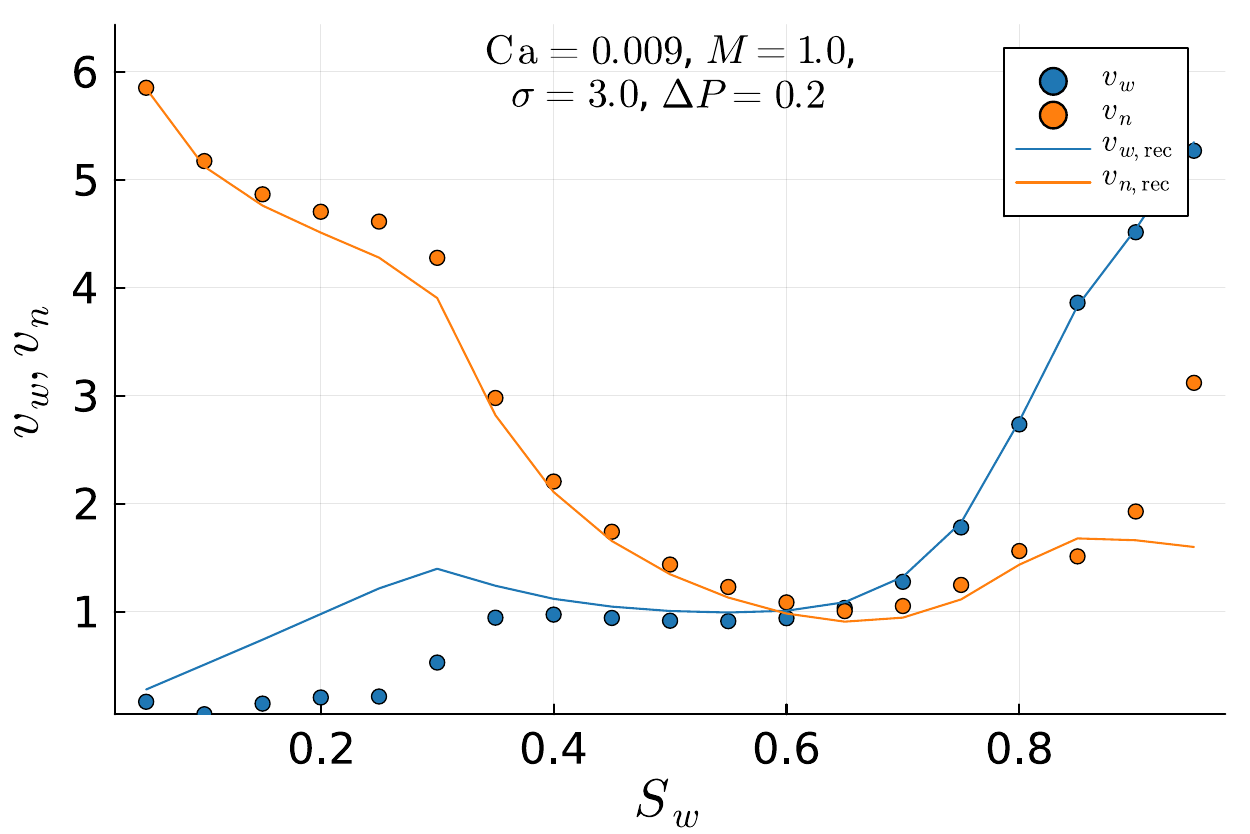"}
  \includegraphics[width=.32\textwidth]{"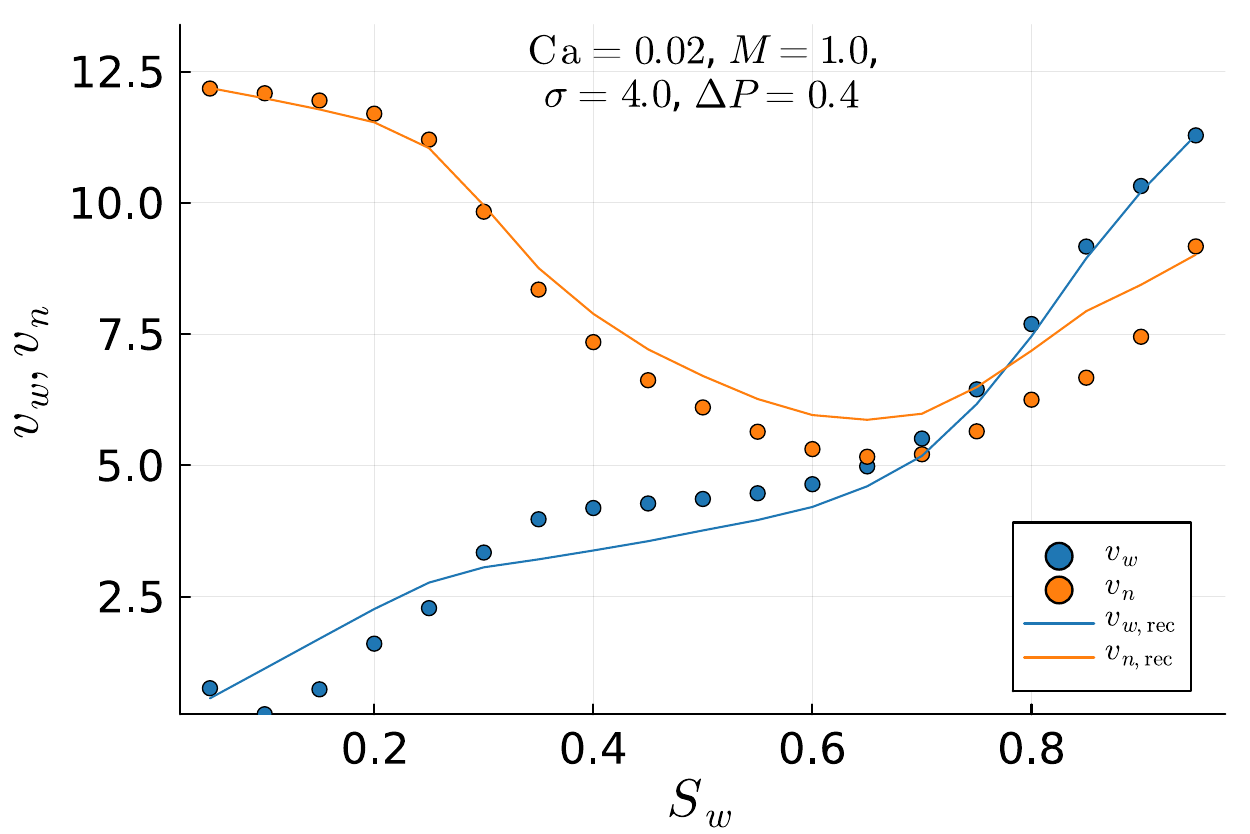"} \par
  \vspace{10pt}
  \includegraphics[width=.32\textwidth]{"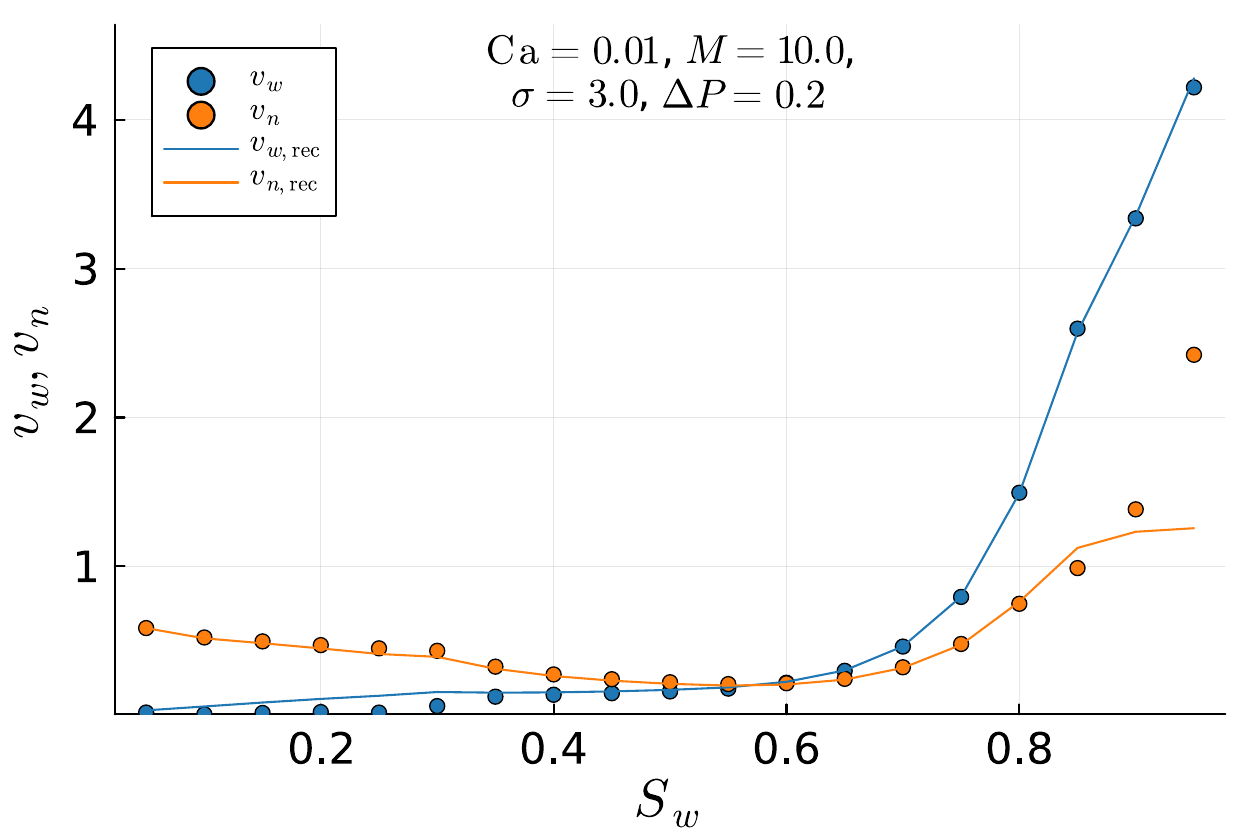"}
  \includegraphics[width=.32\textwidth]{"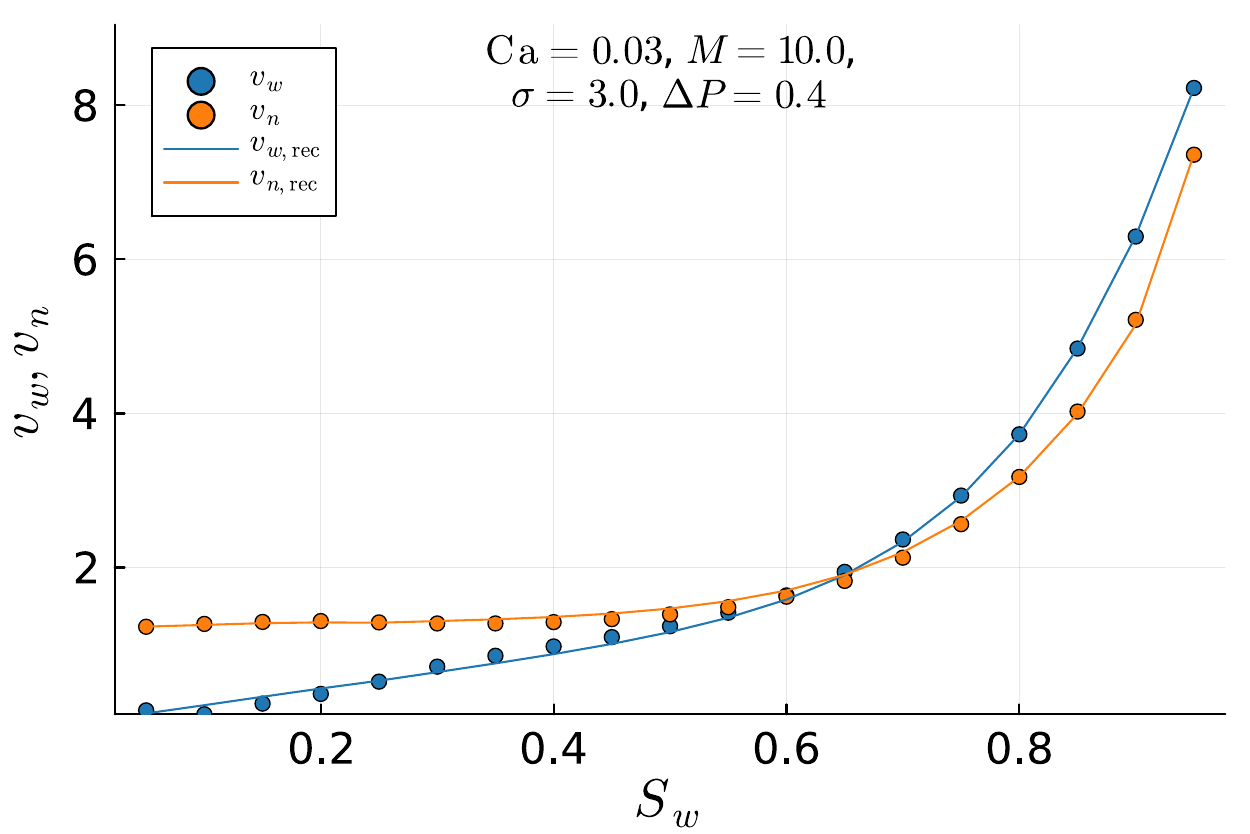"}
  \includegraphics[width=.32\textwidth]{"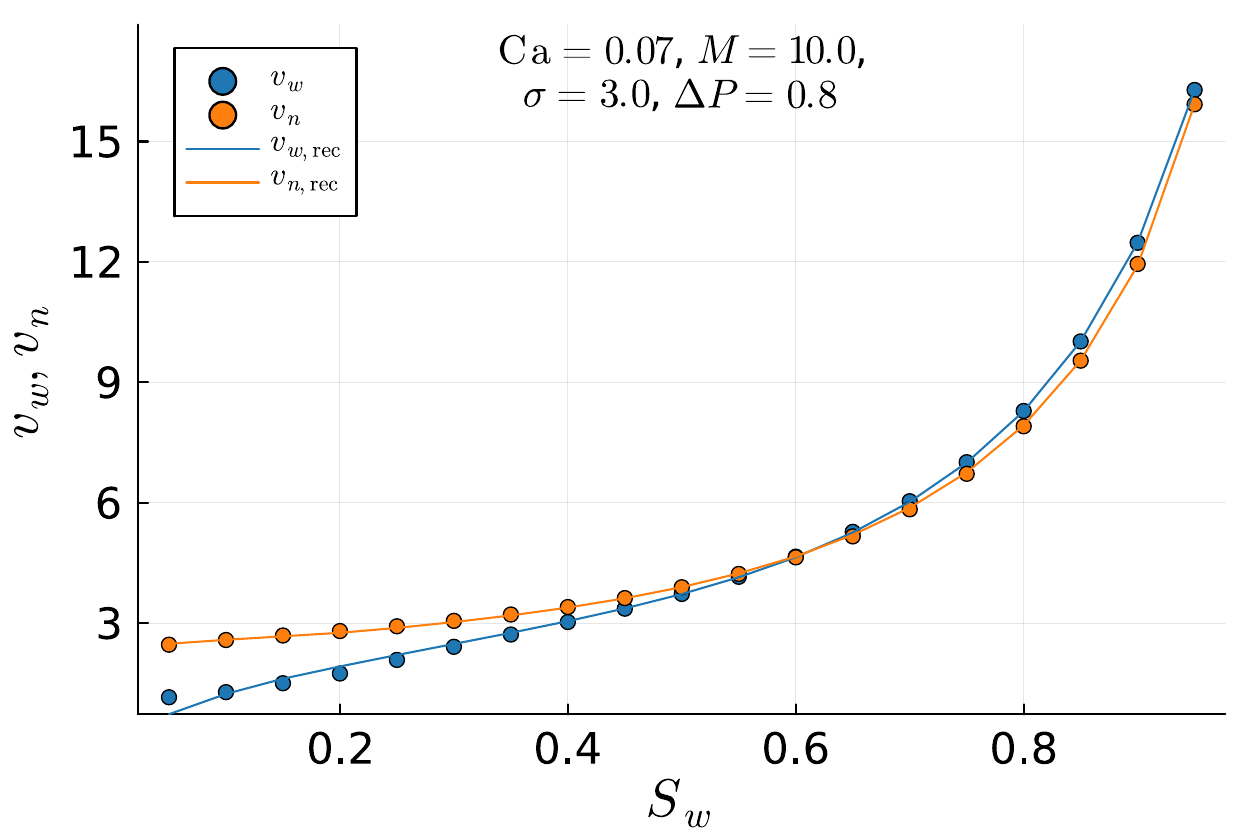"} \par
  \vspace{10pt}
  \includegraphics[width=.32\textwidth]{"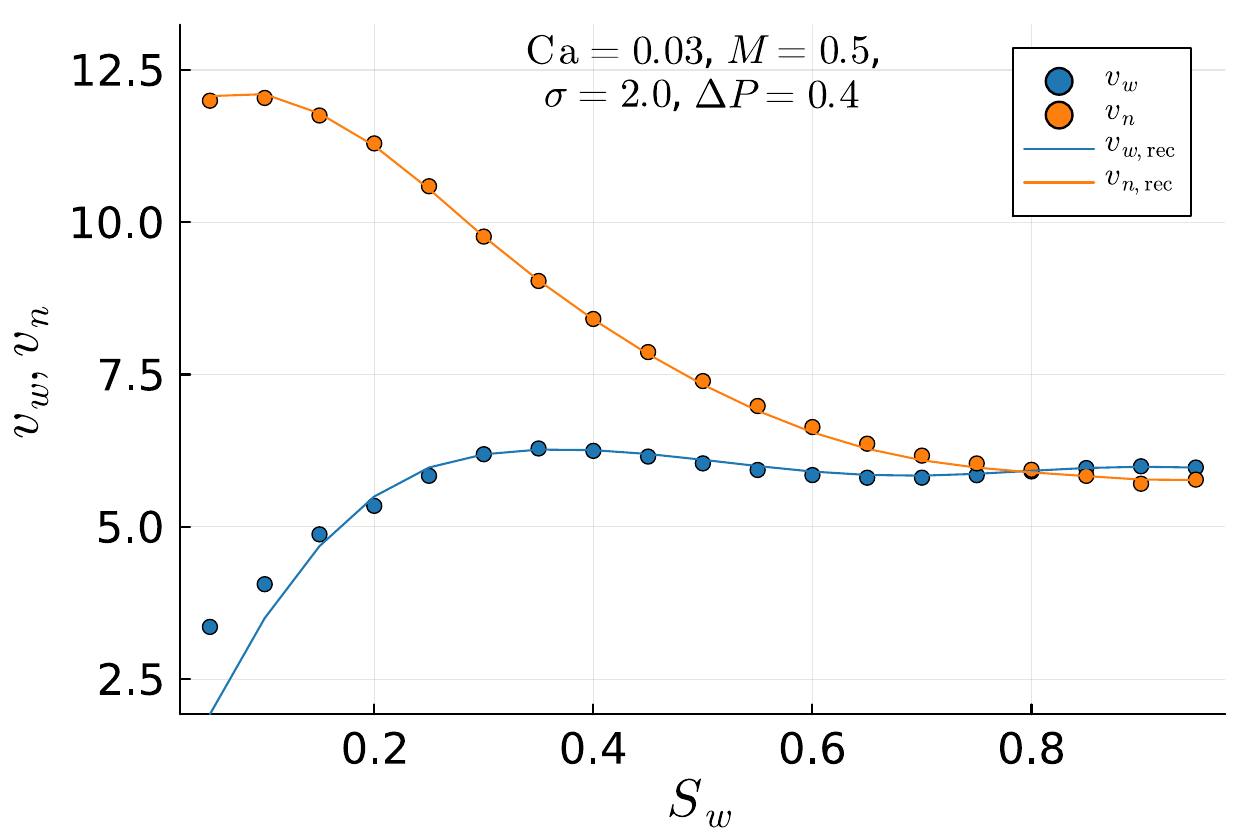"}
  \includegraphics[width=.32\textwidth]{"img/rec-M-0.5-T-3.0-P-0.4.pdf"}
  \includegraphics[width=.32\textwidth]{"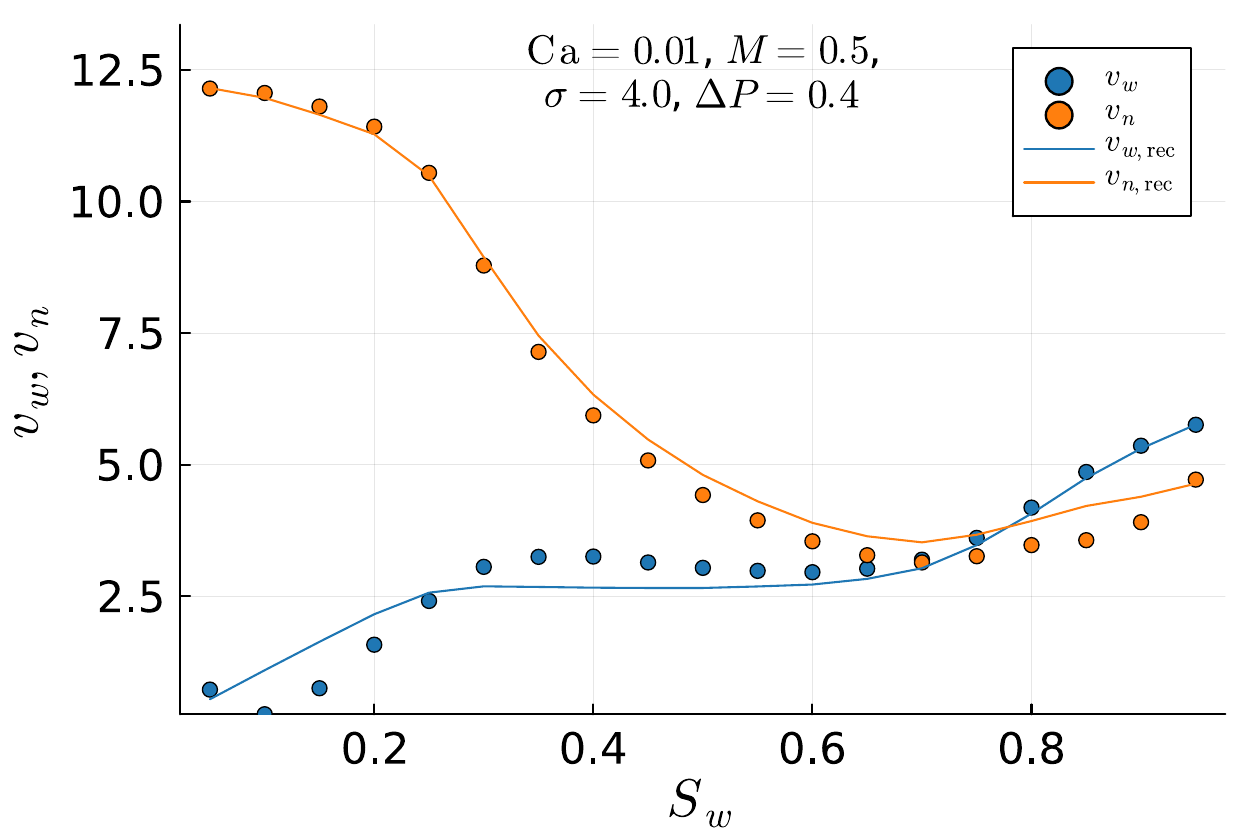"} \par
  \caption{Recreated seepage velocities (DNM) using the approximated form of
    $v_m = v_m \left(S_w ; k_{e} \right)$ in the definitions
    \cref{eq:vw-transf,eq:vn-transf}. The plots have been selected to highlight
    the cases where the seepage velocities have difficult functional forms.
    First row: varying $M$. Second row: varying $\sigma$, the dataset for
    $\Delta P =0.2$ does not exist, so $\Delta O = 0.4$ is used instead. Third
    row: varying $\Delta P$. Fourth row: varying $\sigma$. For larger values of
    $\Delta P$, the fits align almost exactly with the data. Typically, for
    small $\Delta P$, the $v_n$-curve is not well approximated unless
    $S_i^{\ast}$ is adjusted, as demonstrated by the second row, where
    $S^{\ast}_n = 0.08$. The transition regions into single-phase flow is not
    well-fitted, as expected. As with the rest of the results, no adjustment for
    the saturation range has been performed.\label{fig:rec-seepage-dnm} }
\end{figure*}

\begin{figure*}[t]
  \includegraphics[width=.24\textwidth]{"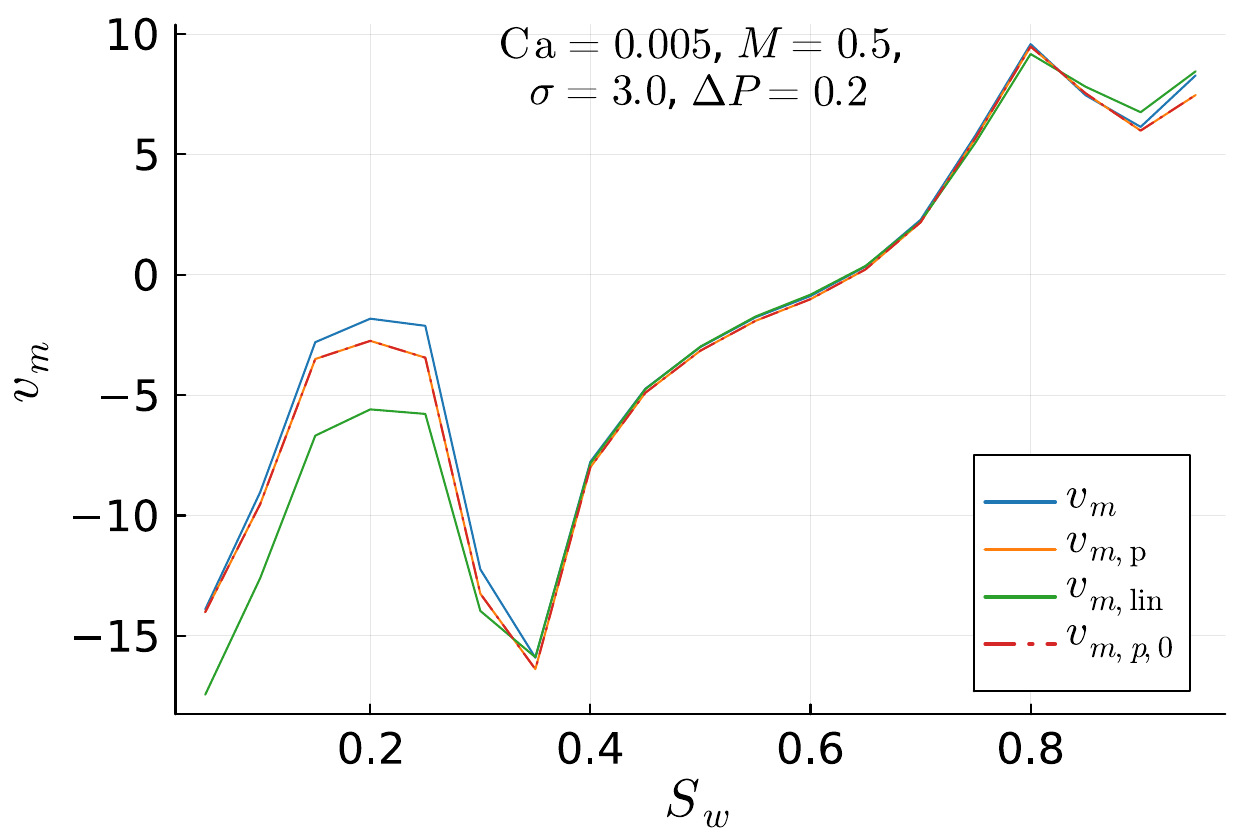"}
  \includegraphics[width=.24\textwidth]{"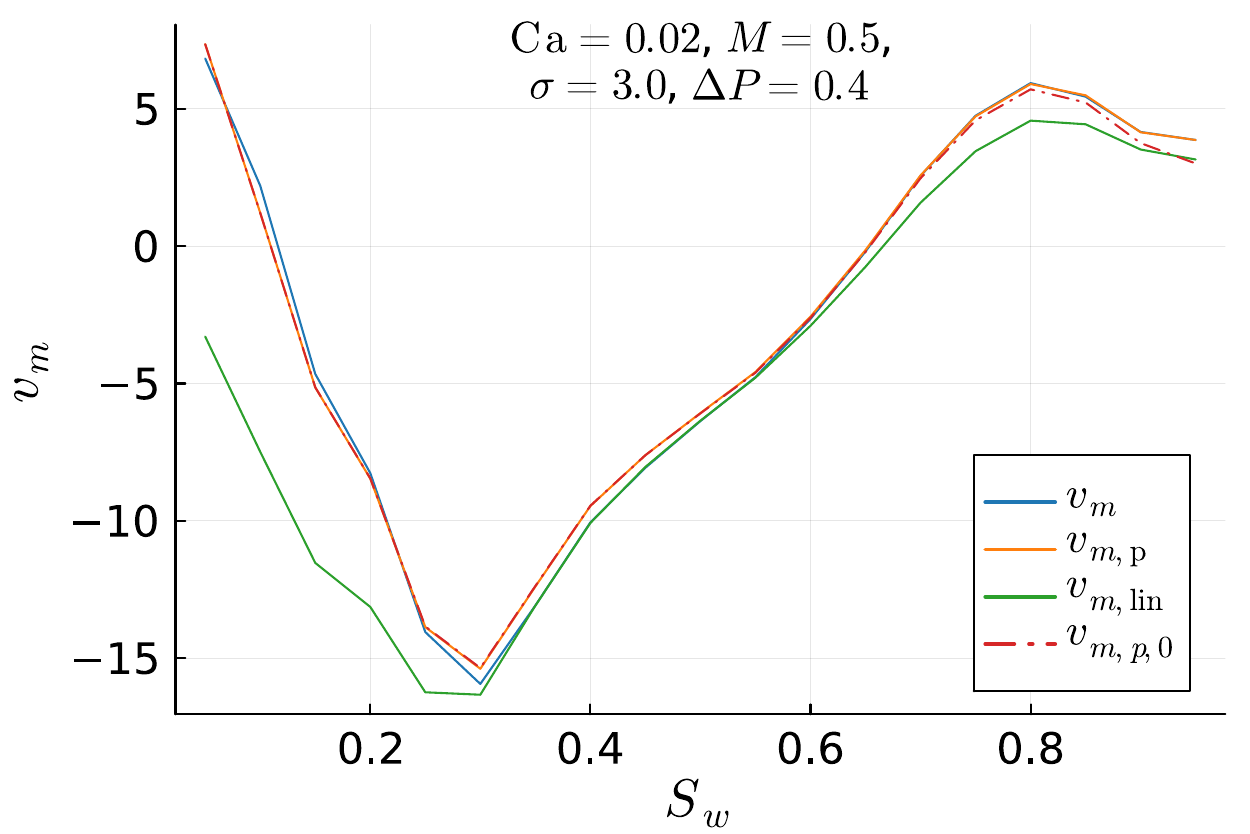"}
  \includegraphics[width=.24\textwidth]{"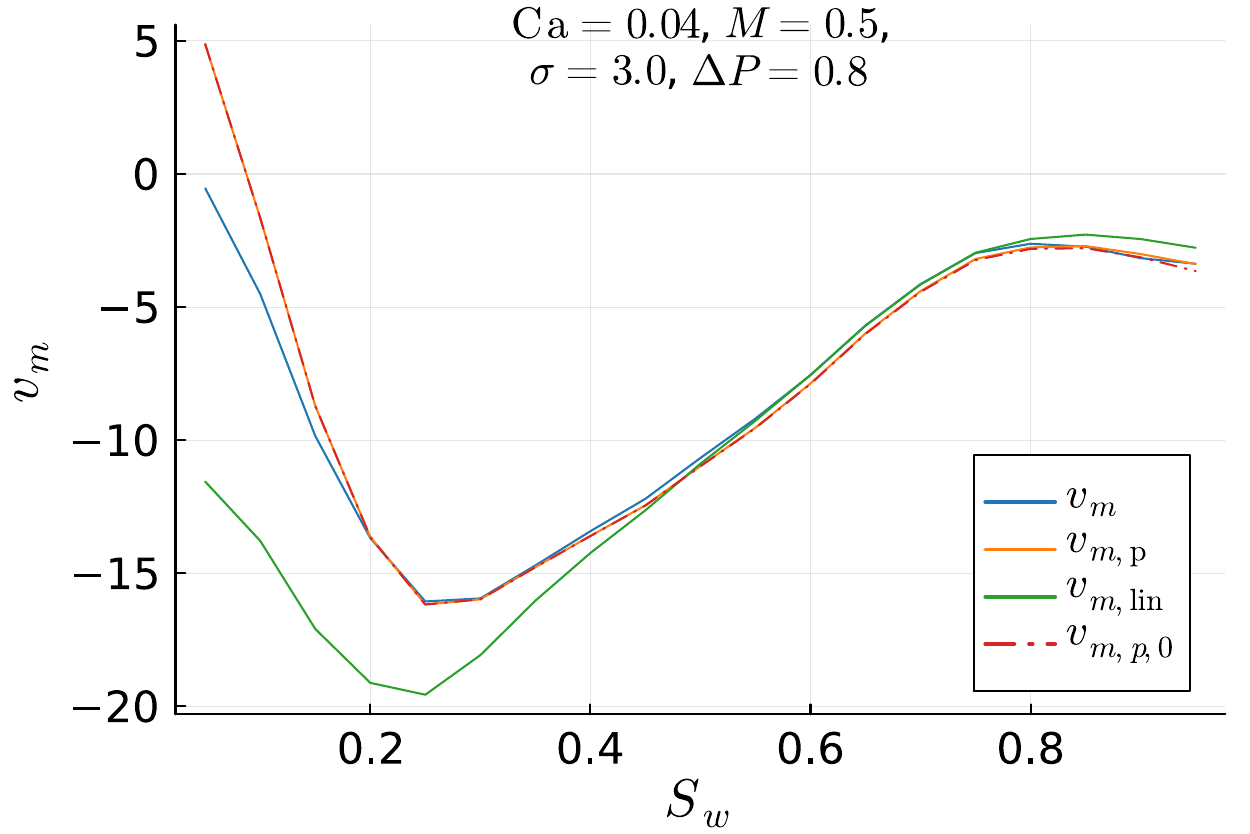"}
  \includegraphics[width=.24\textwidth]{"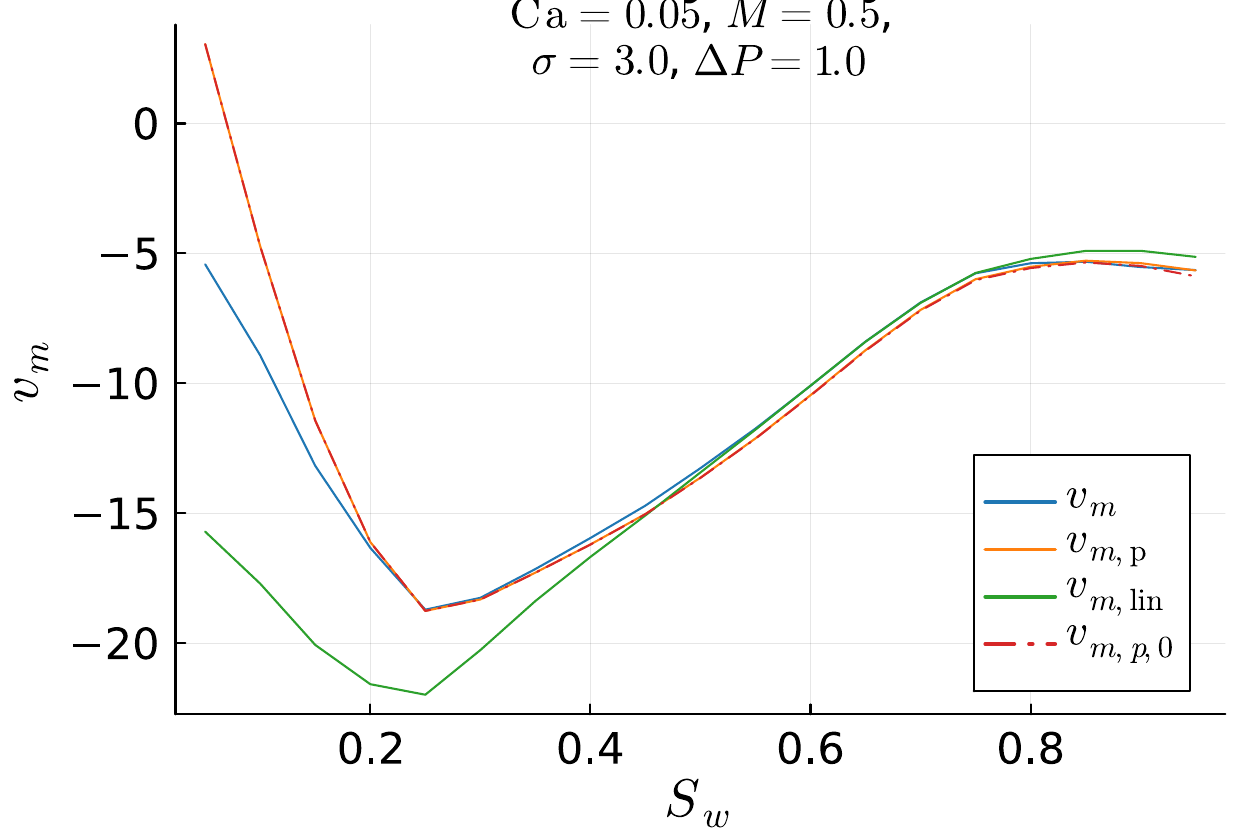"} \par
  \vspace{10pt}
  \includegraphics[width=.24\textwidth]{"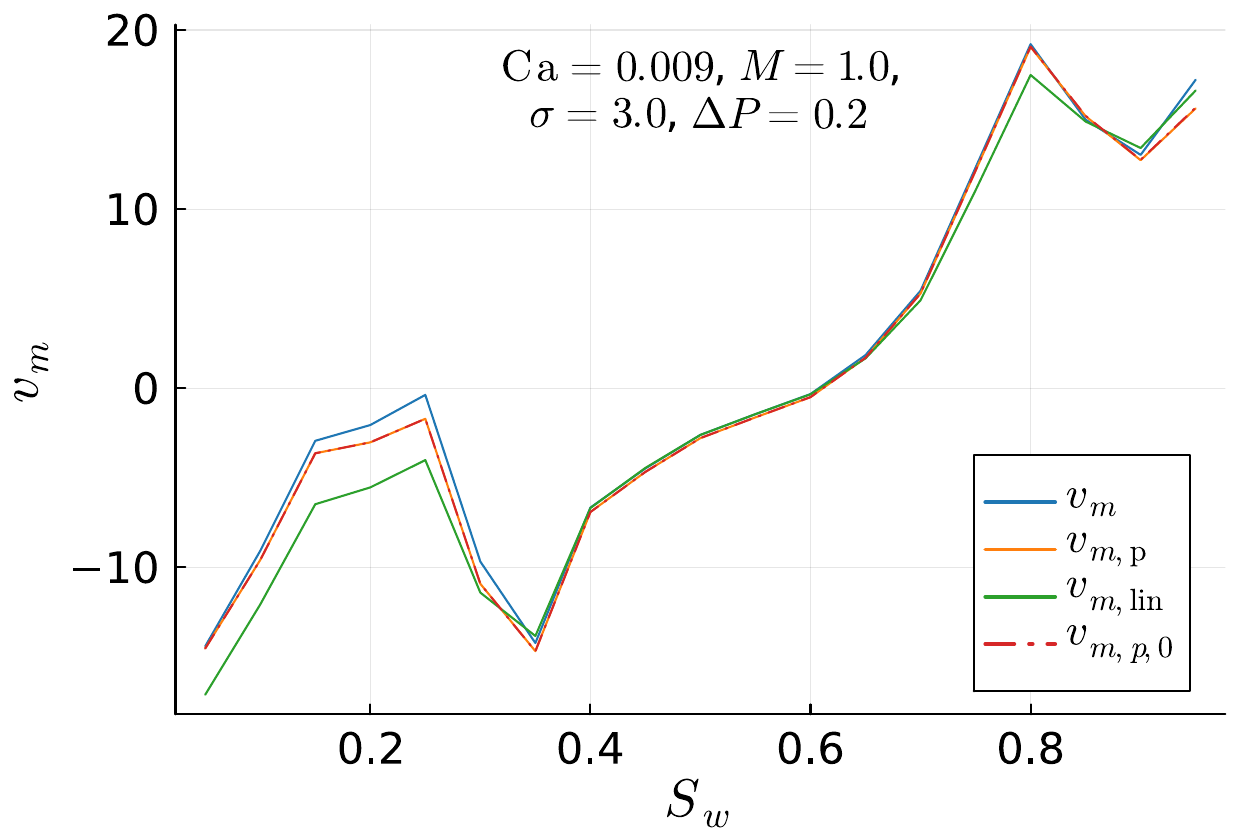"}
  \includegraphics[width=.24\textwidth]{"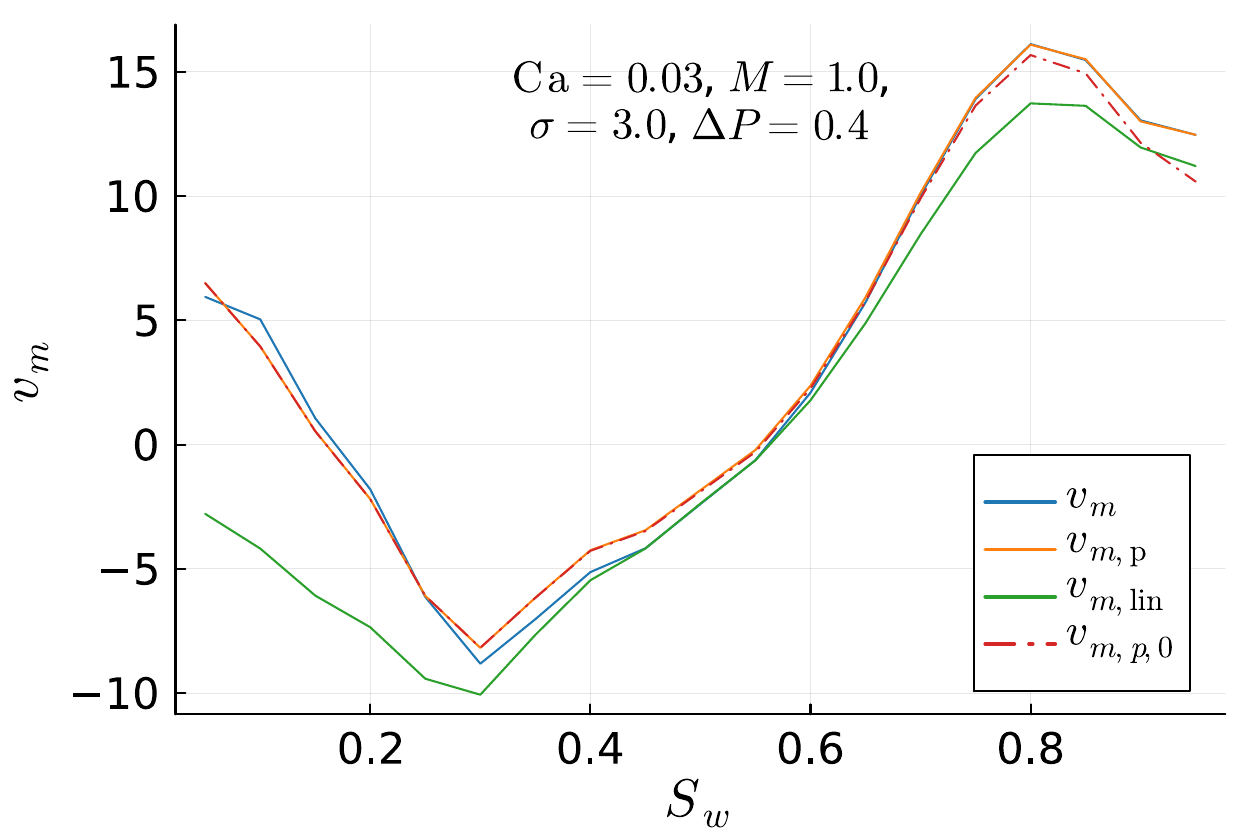"}
  \includegraphics[width=.24\textwidth]{"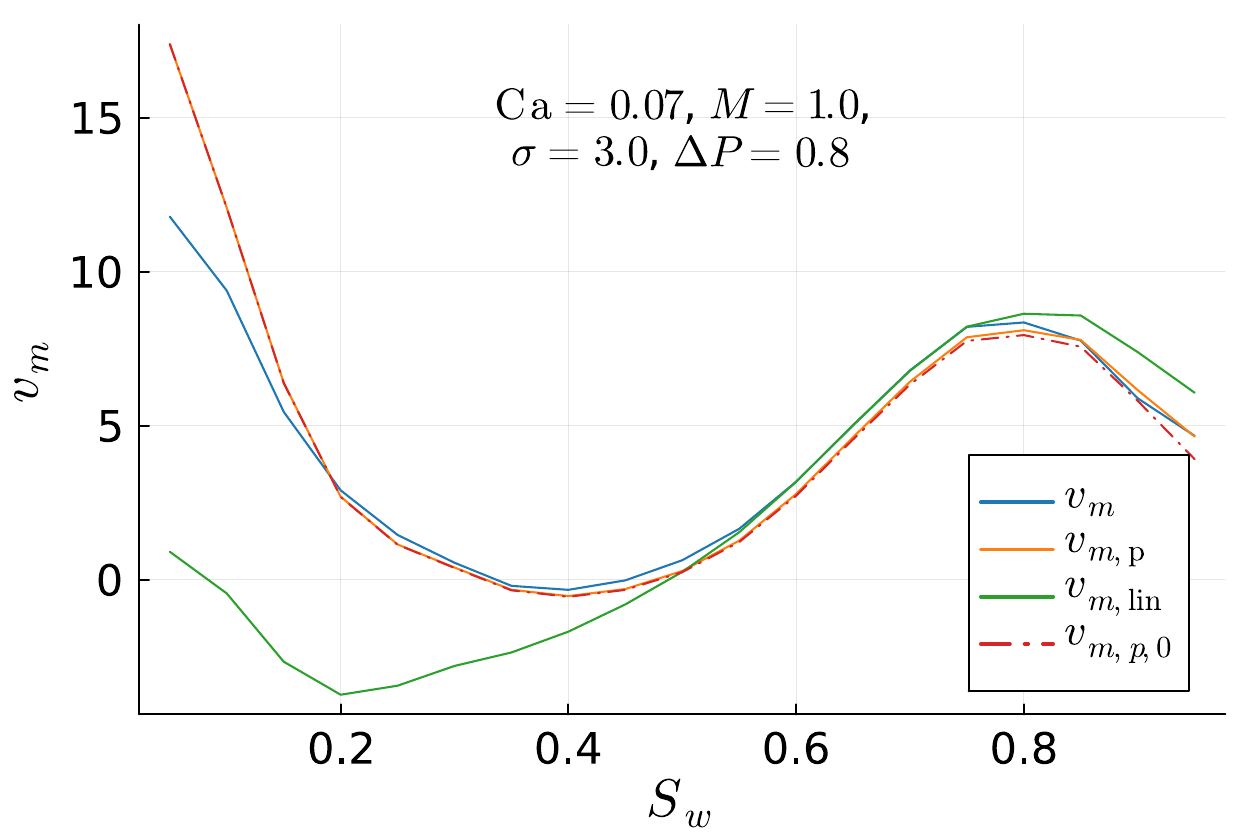"}
  \includegraphics[width=.24\textwidth]{"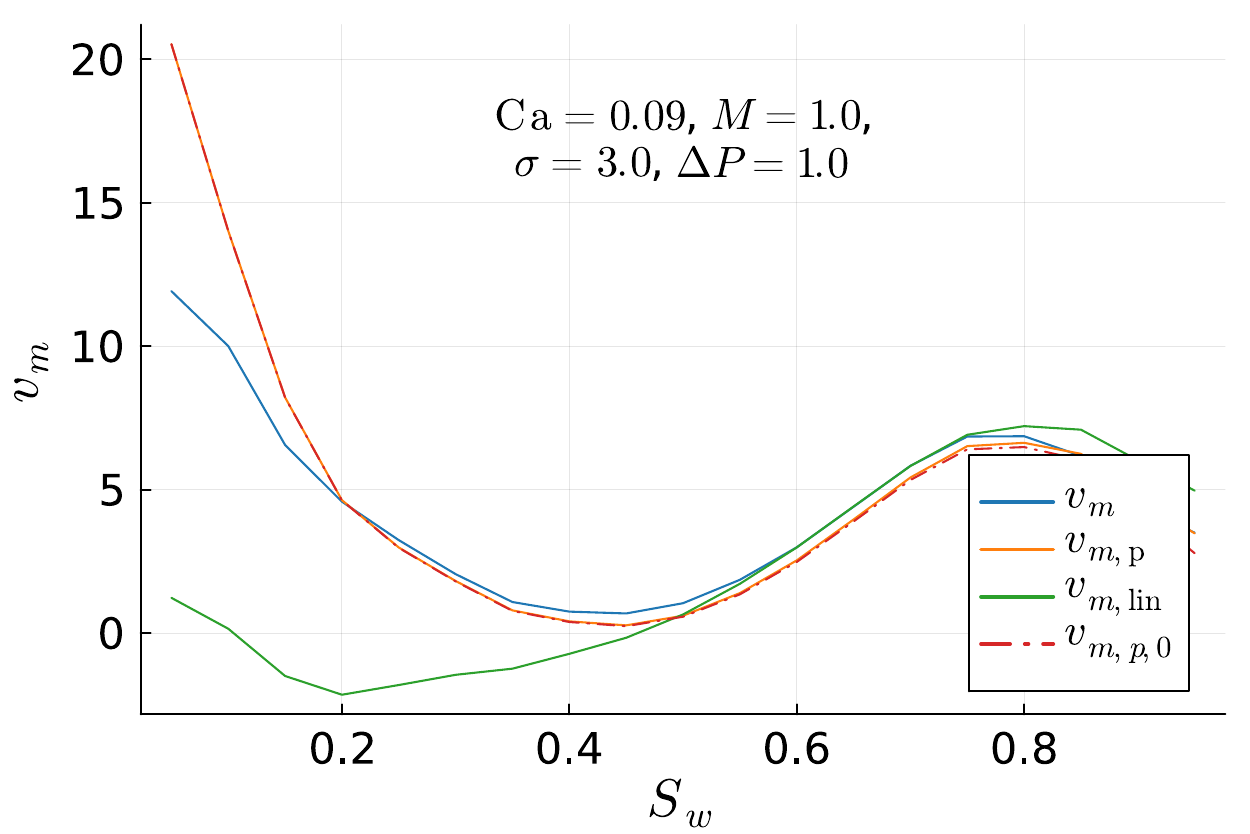"} \par
  \vspace{10pt}
  \includegraphics[width=.24\textwidth]{"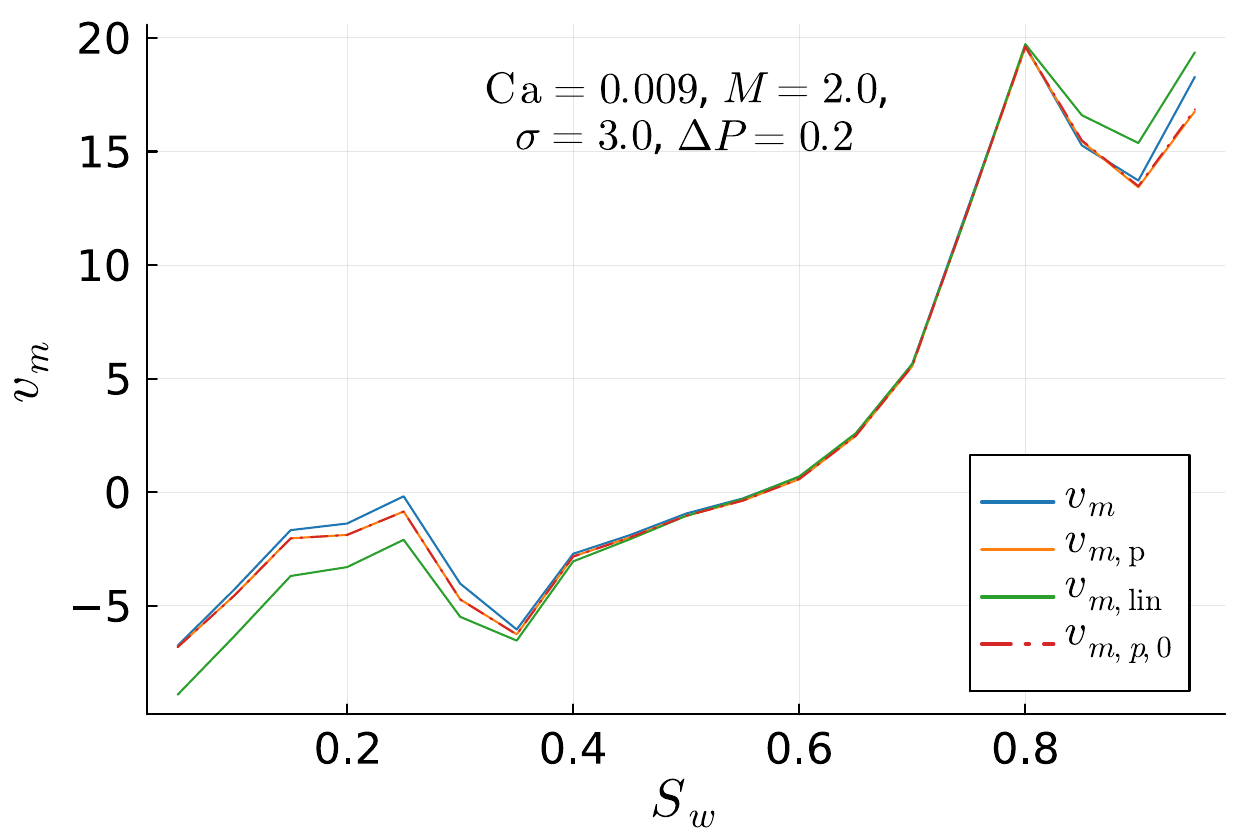"}
  \includegraphics[width=.24\textwidth]{"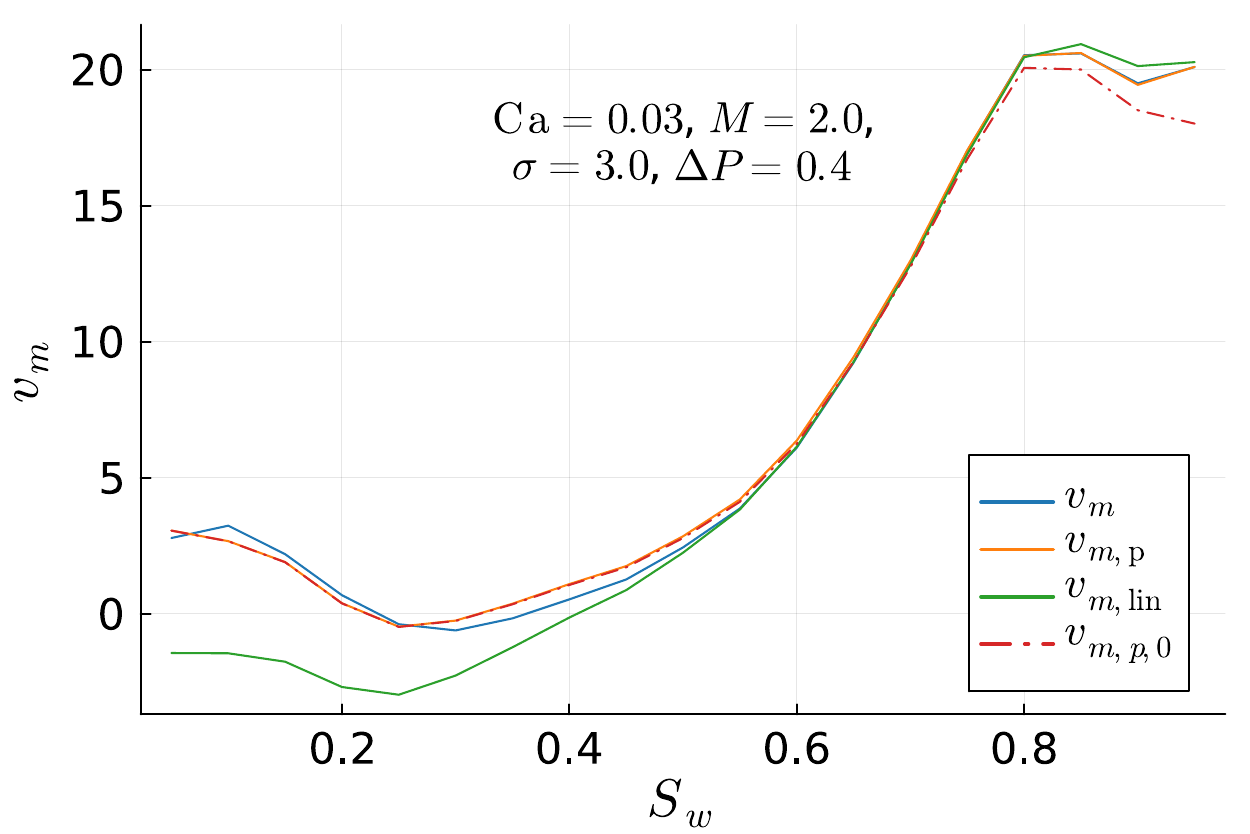"}
  \includegraphics[width=.24\textwidth]{"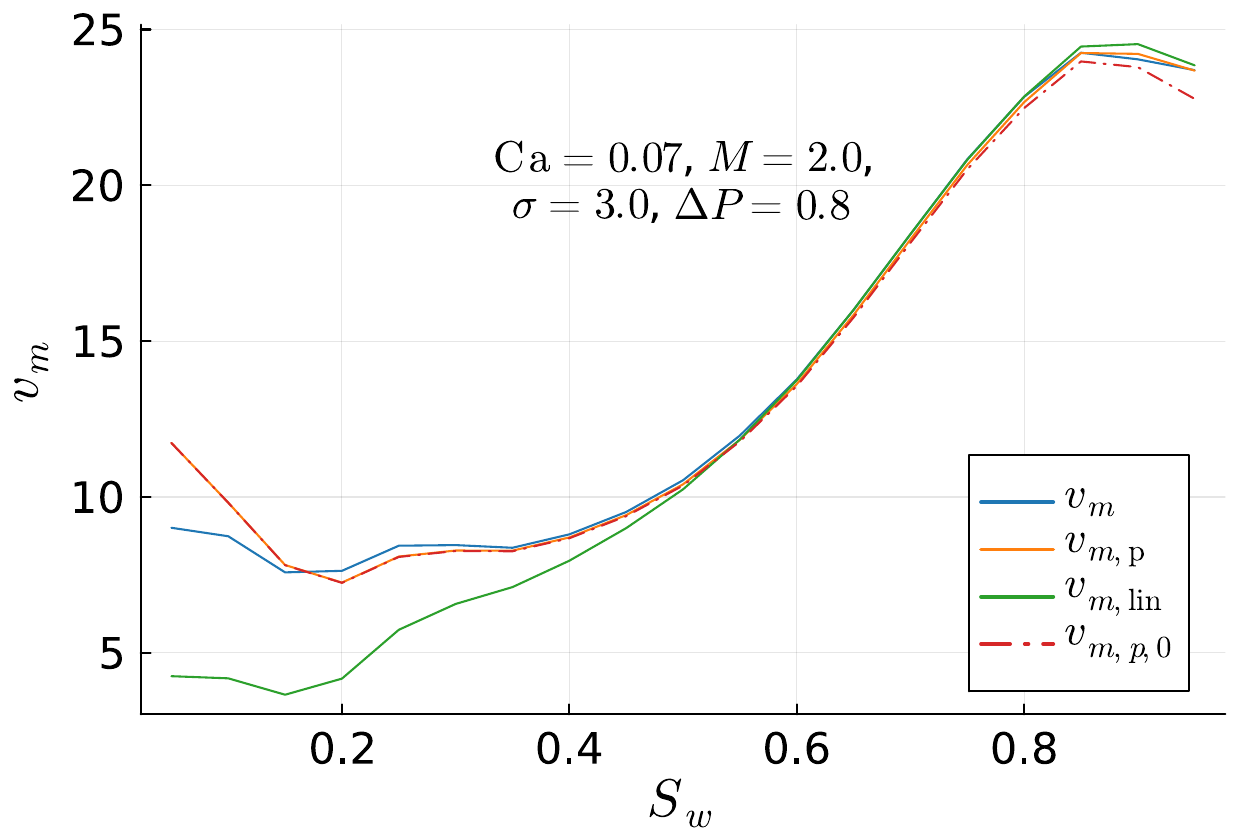"}
  \includegraphics[width=.24\textwidth]{"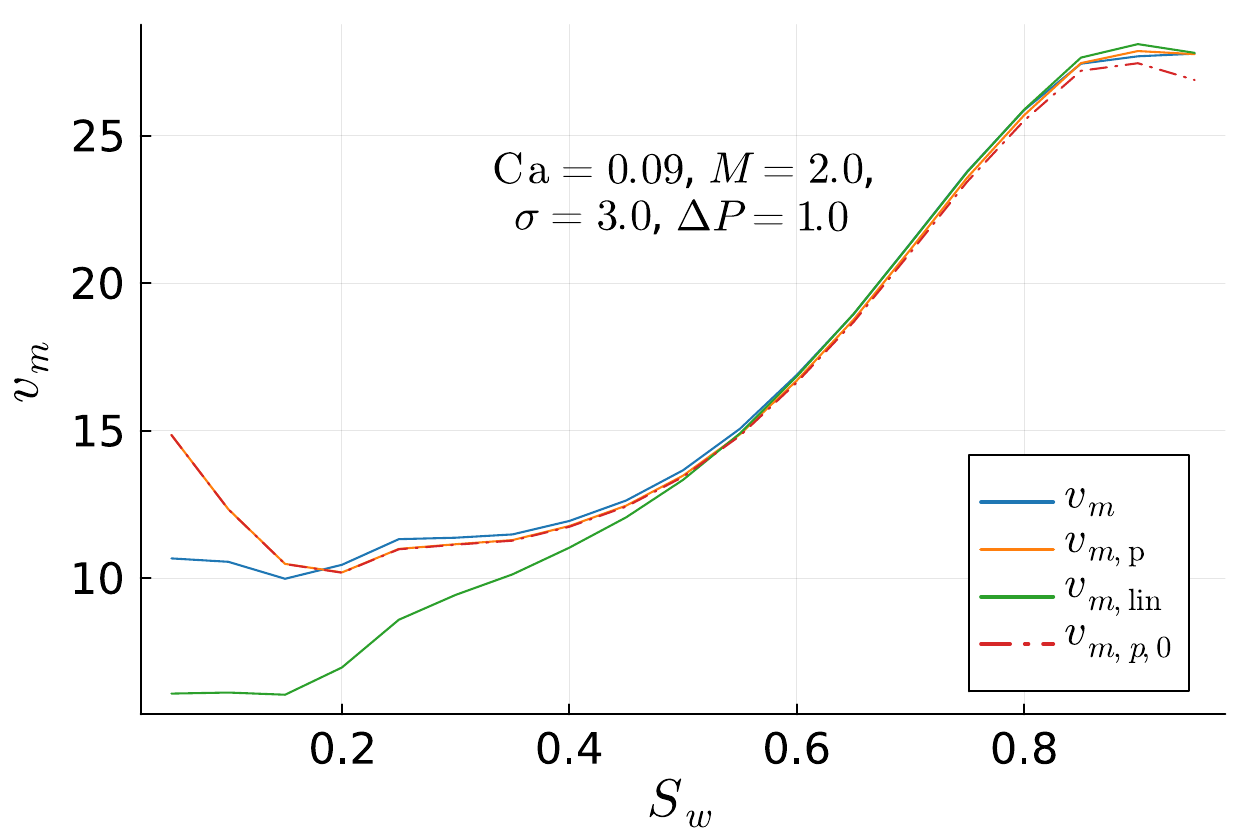"} \par
  \vspace{10pt}
  \includegraphics[width=.24\textwidth]{"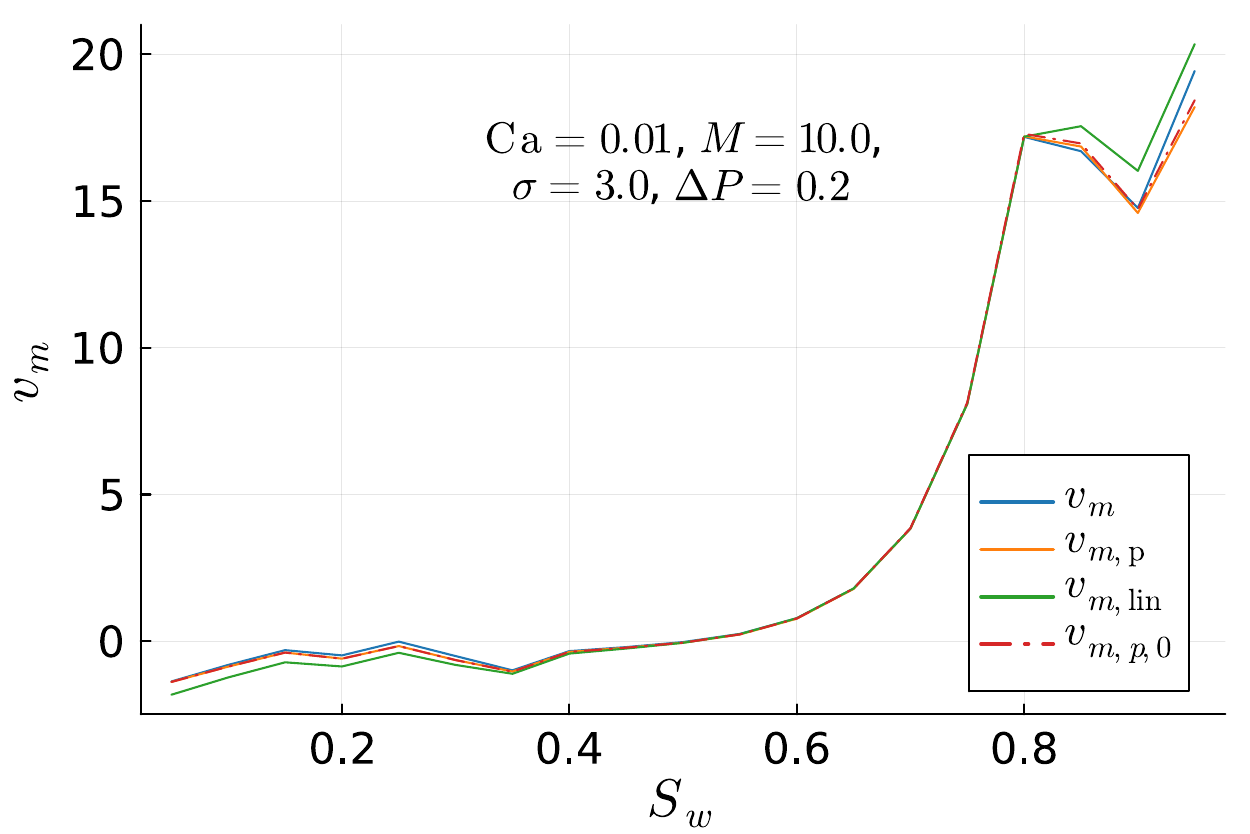"}
  \includegraphics[width=.24\textwidth]{"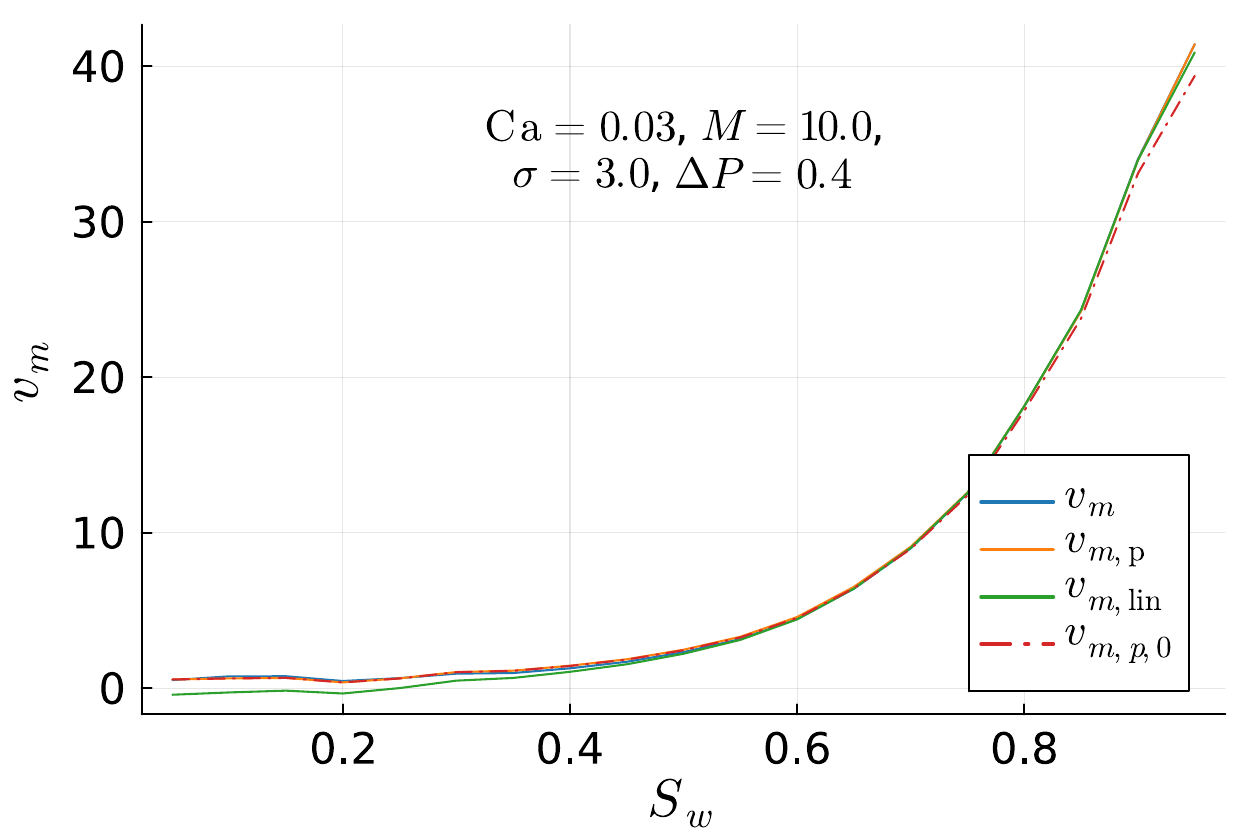"}
  \includegraphics[width=.24\textwidth]{"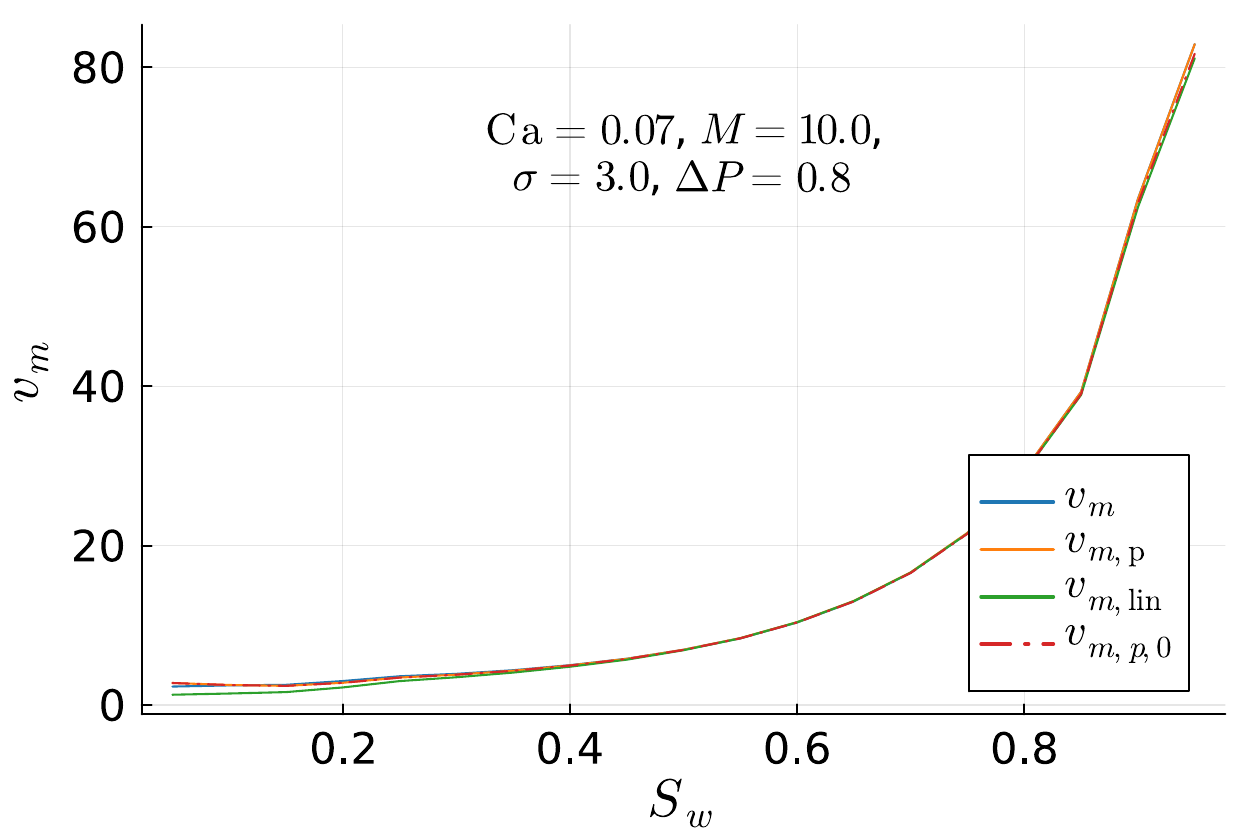"}
  \includegraphics[width=.24\textwidth]{"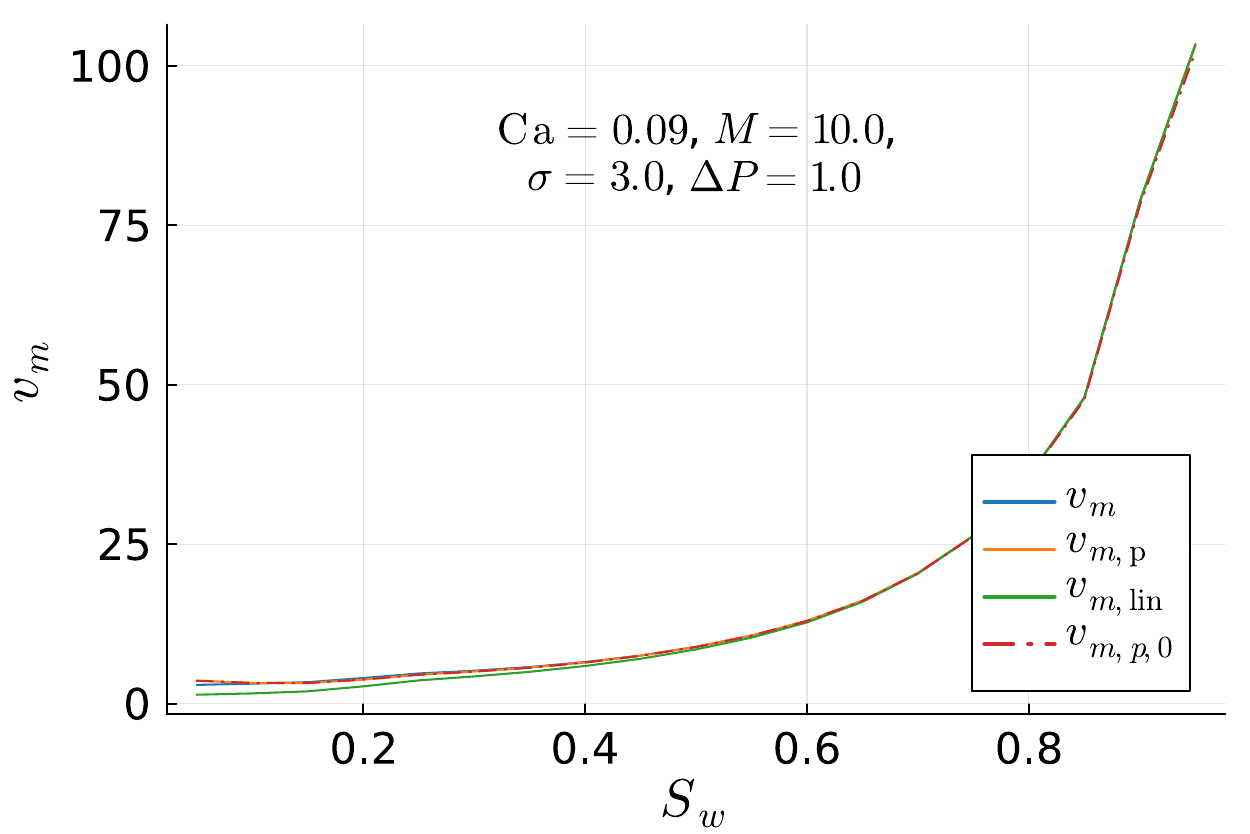"}
  \caption{$v_m(S_w)$ (DNM) for a range of $M$- and $\Delta P$-values with
    $\sigma = 3.0$, varying with column. In the leftmost column, $S_n^{\ast} =
    0.1$ is used to get better fits. The linear approximation has been performed
    on the largest linear region of $v_m \left( v^{\prime} \right)$ within a
    reasonable tolerance.\label{fig:vm-sw-dnm} }
\end{figure*}

\begin{figure*}[t]
  \includegraphics[width=.32\textwidth]{"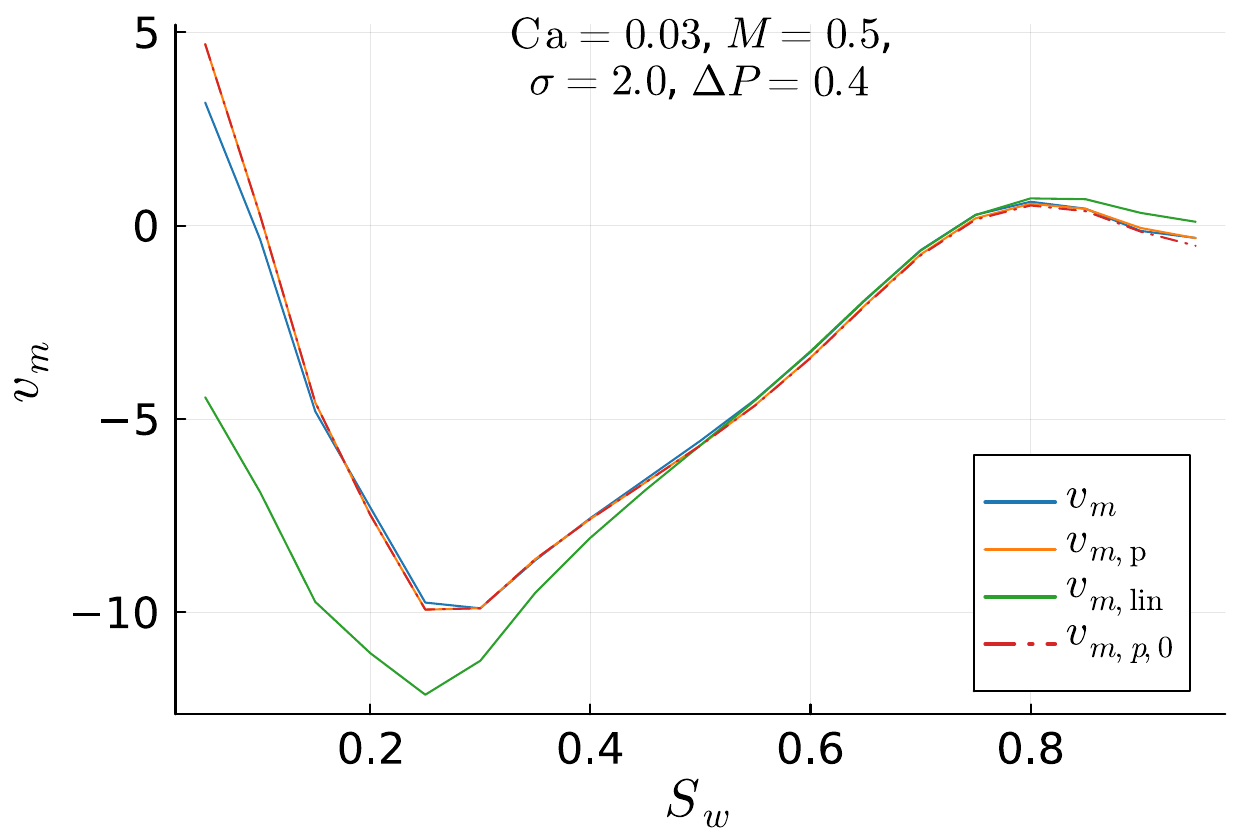"}
  \includegraphics[width=.32\textwidth]{"img/vmSw-M-0.5-T-3.0-P-0.4.pdf"}
  \includegraphics[width=.32\textwidth]{"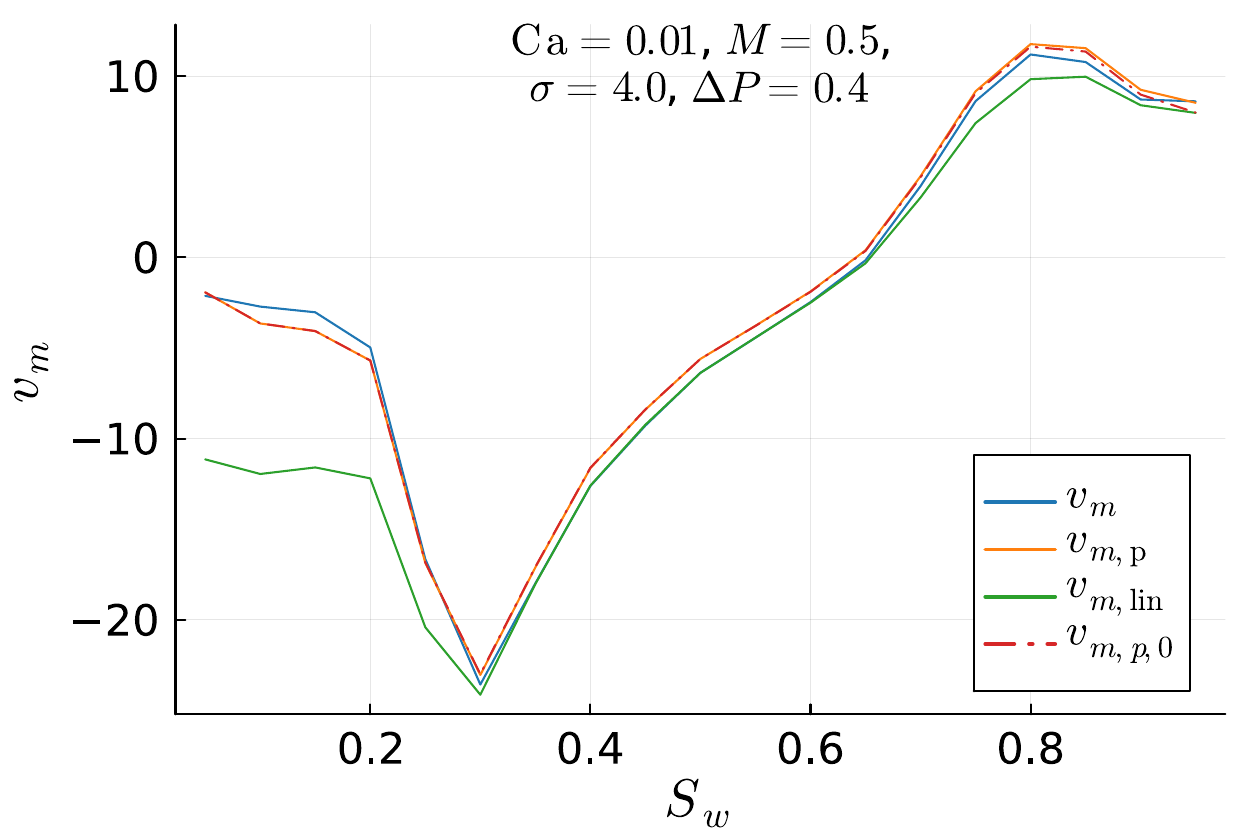"} \par
  \vspace{10pt}
  \includegraphics[width=.32\textwidth]{"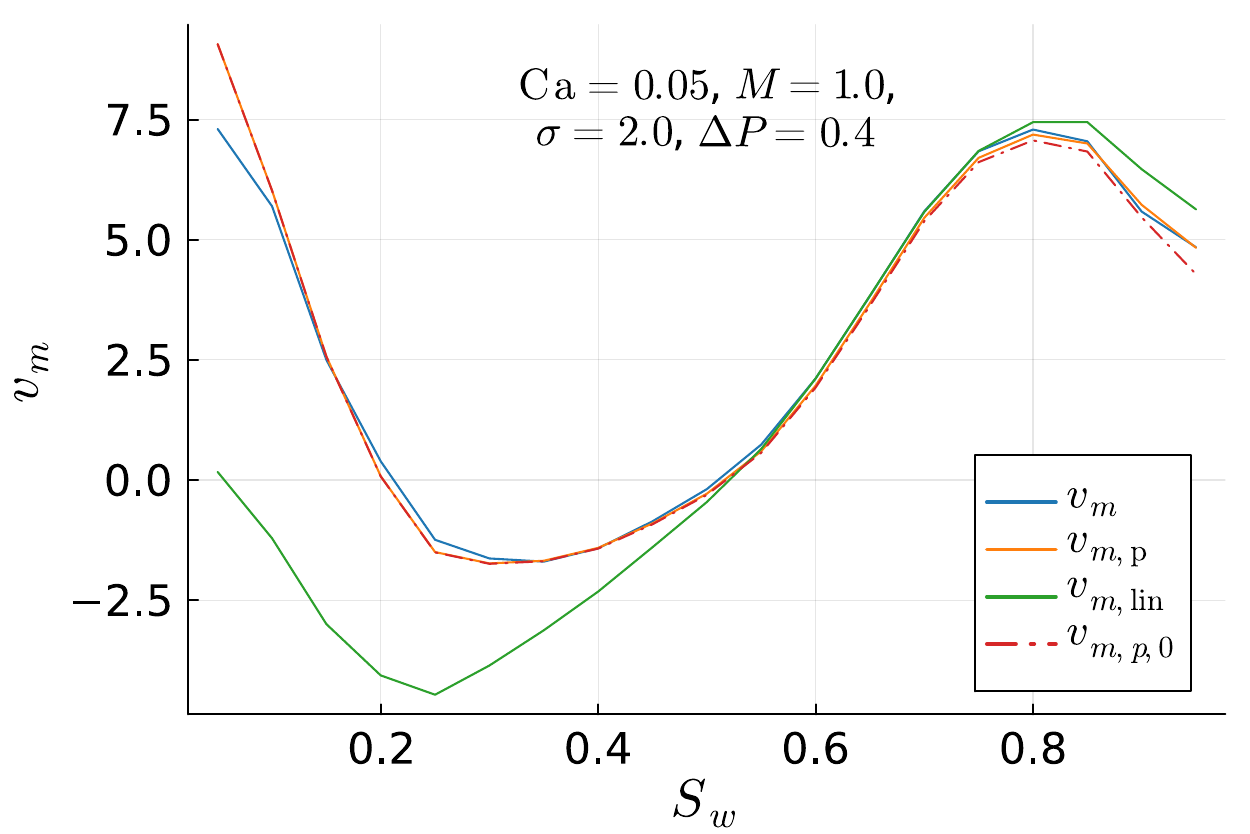"}
  \includegraphics[width=.32\textwidth]{"img/vmSw-M-1.0-T-3.0-P-0.4.pdf"}
  \includegraphics[width=.32\textwidth]{"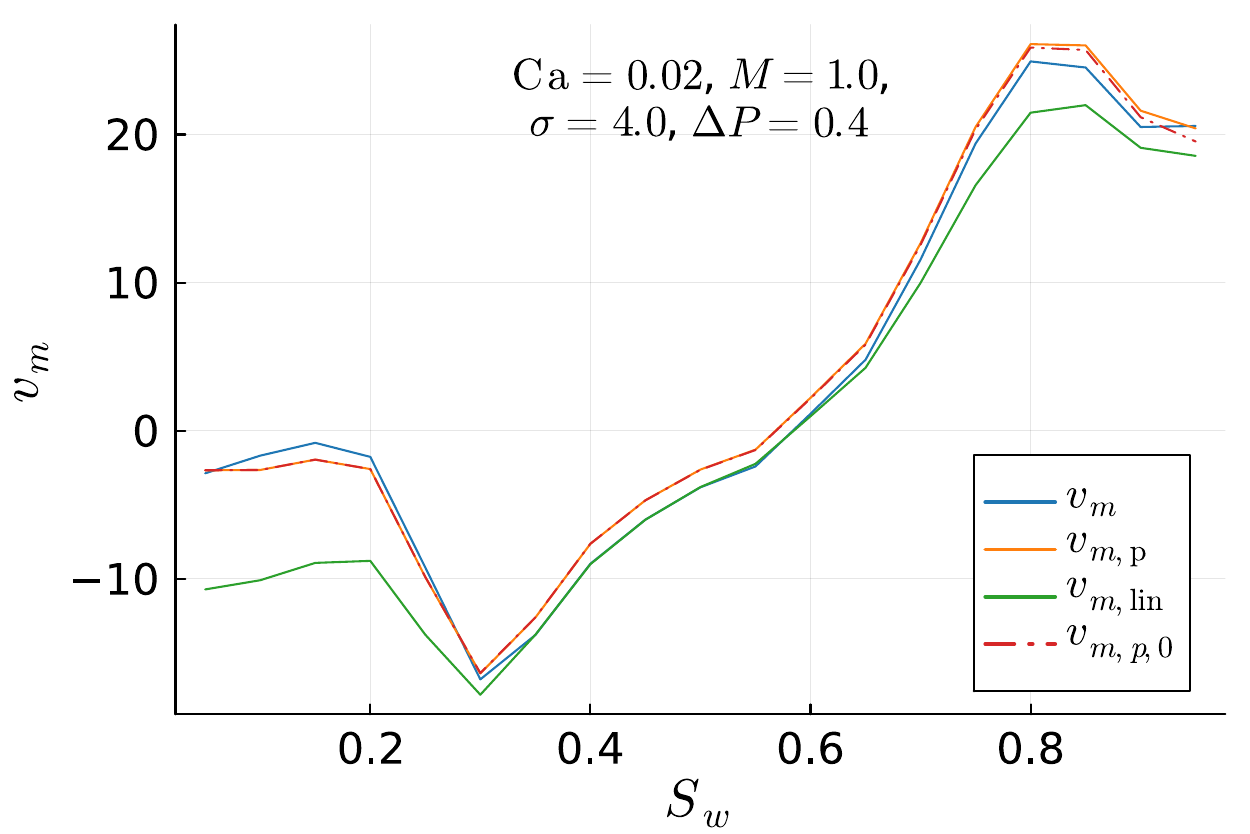"} \par
  \vspace{10pt}
  \includegraphics[width=.32\textwidth]{"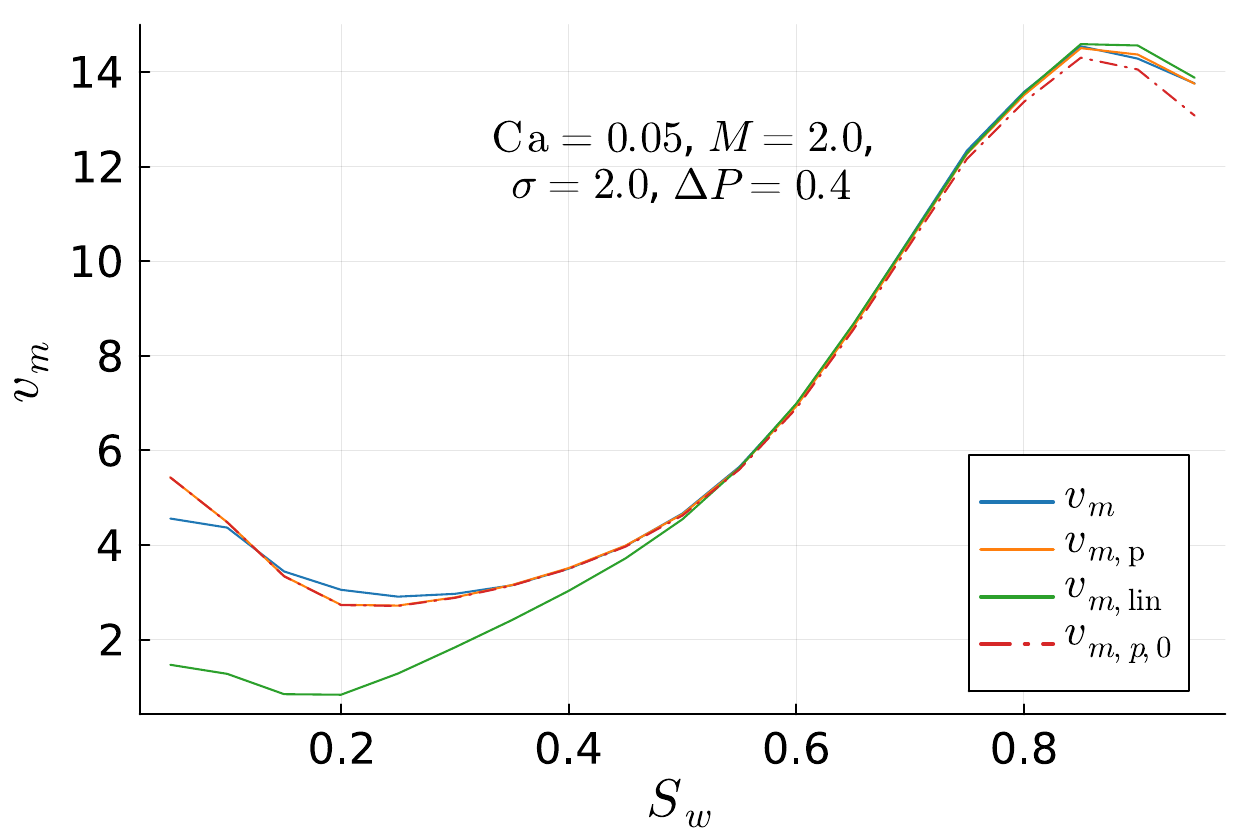"}
  \includegraphics[width=.32\textwidth]{"img/vmSw-M-2.0-T-3.0-P-0.4.pdf"}
  \includegraphics[width=.32\textwidth]{"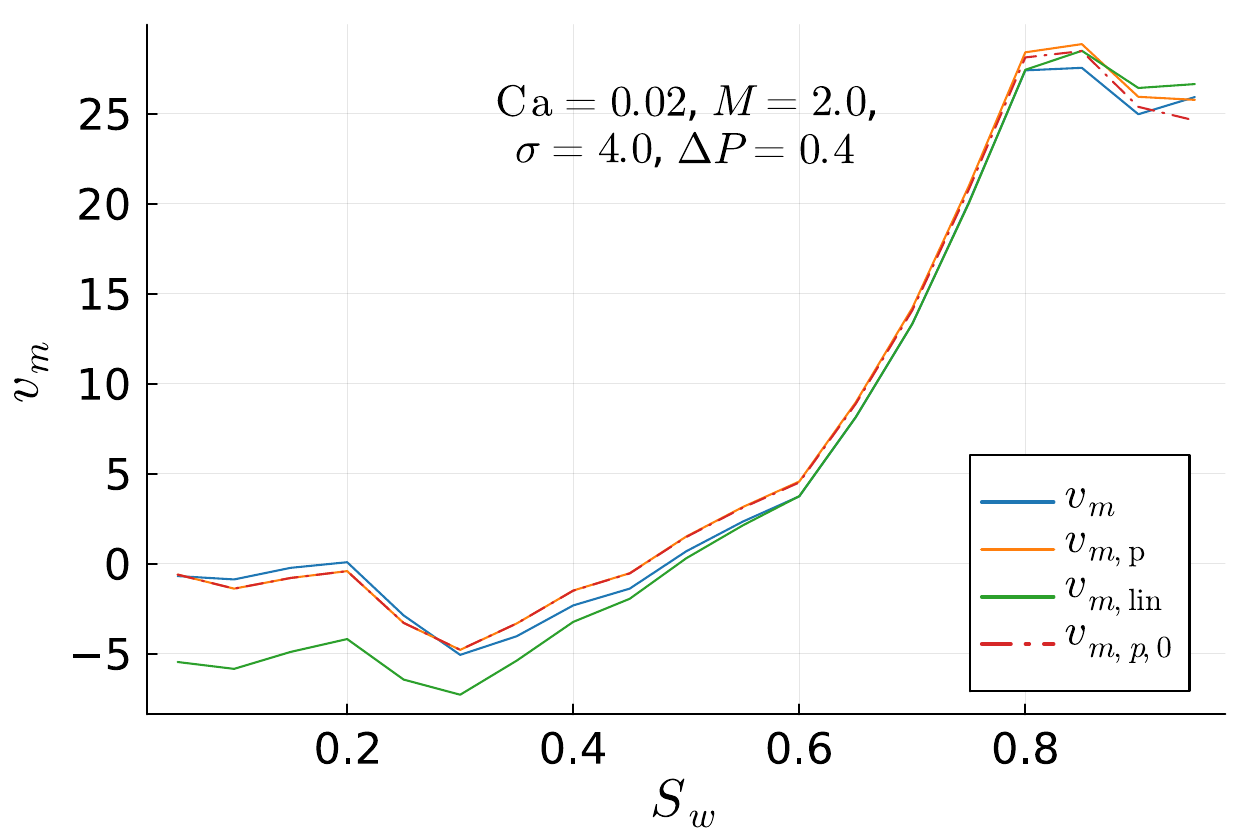"}
  \caption{Variation in $\sigma$ and $M$ for fixed $\Delta P = 0.4$, varying
    with column. \label{fig:surface-tension-variation} }
\end{figure*}

\begin{figure*}[t]
  \includegraphics[width=.24\textwidth]{"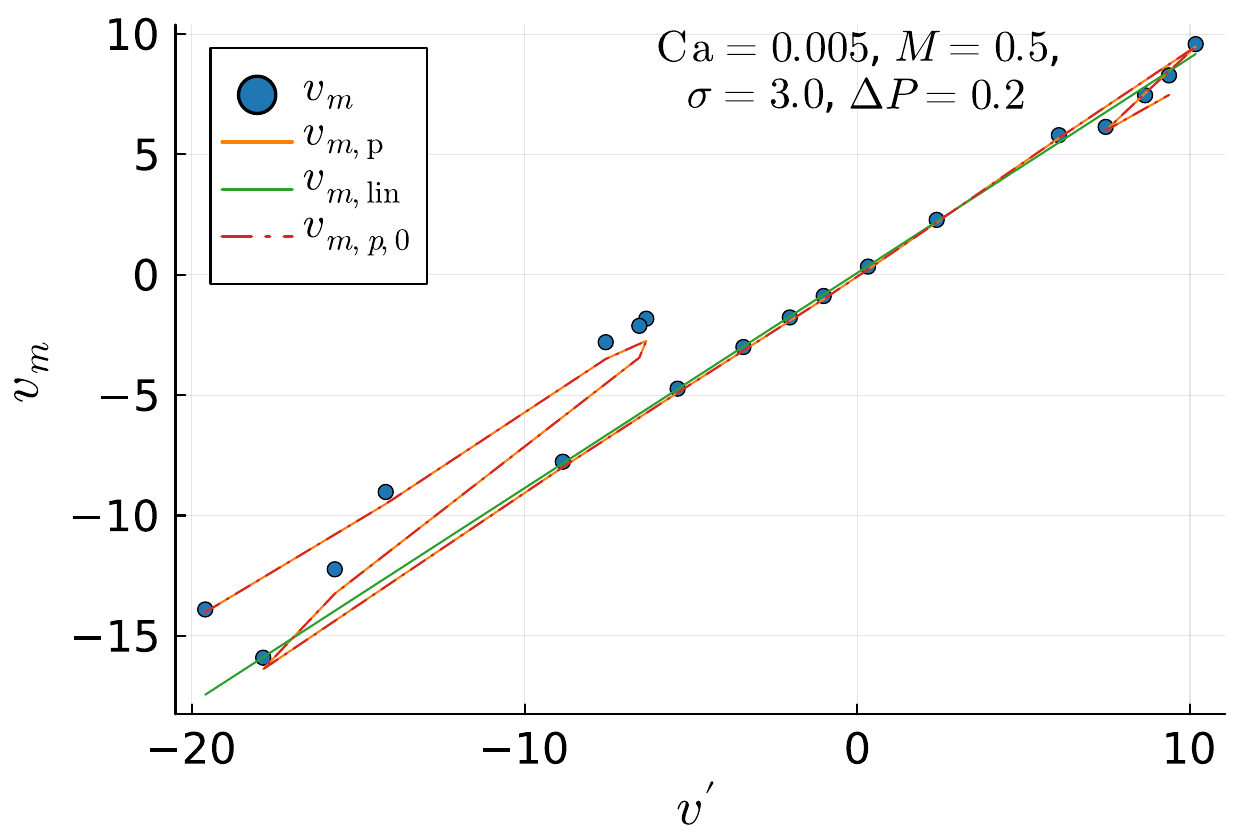"}
  \includegraphics[width=.24\textwidth]{"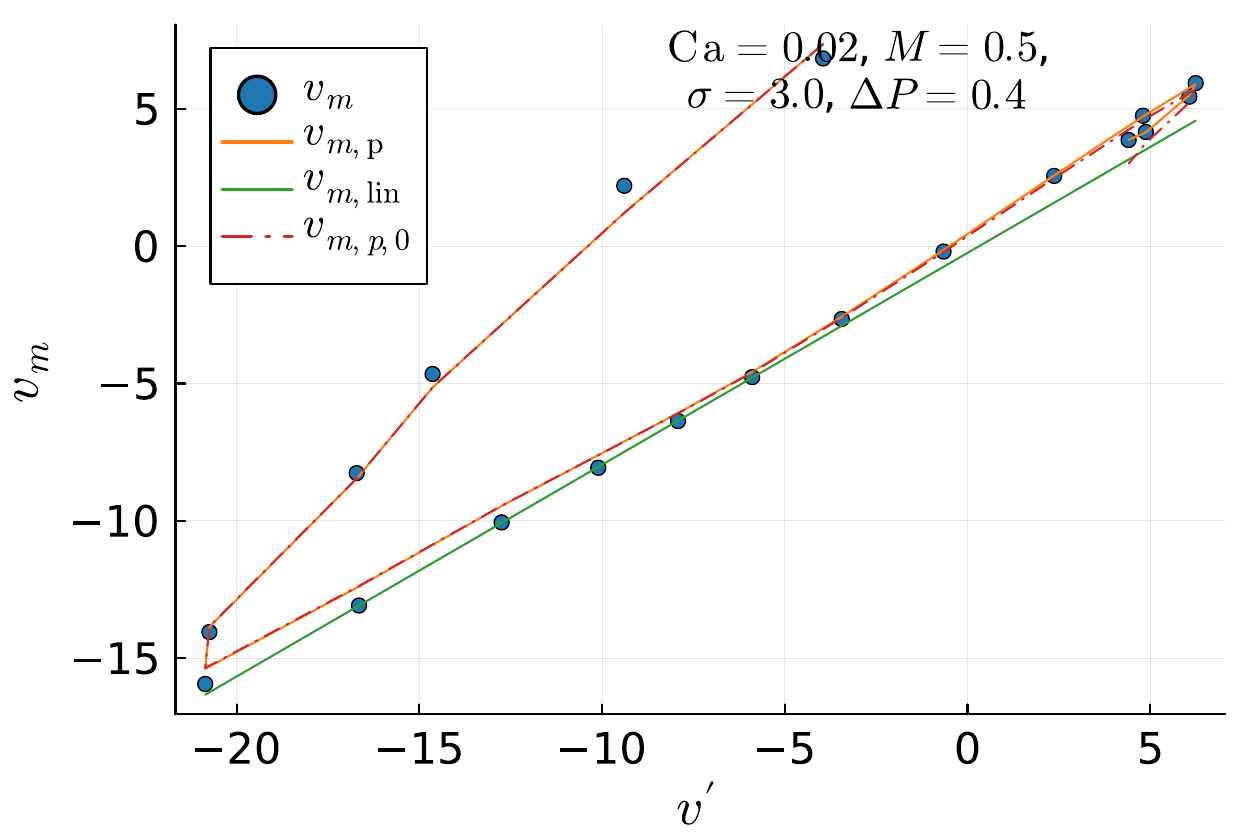"}
  \includegraphics[width=.24\textwidth]{"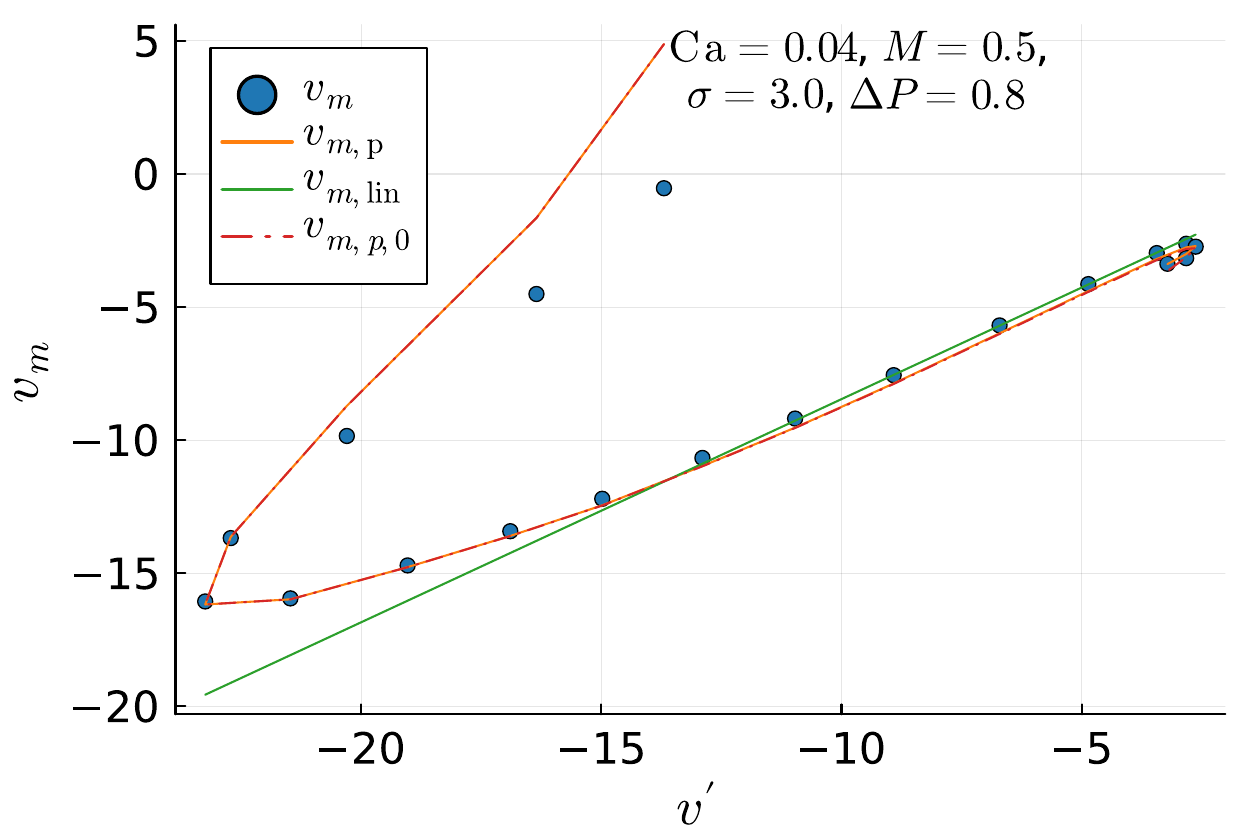"}
  \includegraphics[width=.24\textwidth]{"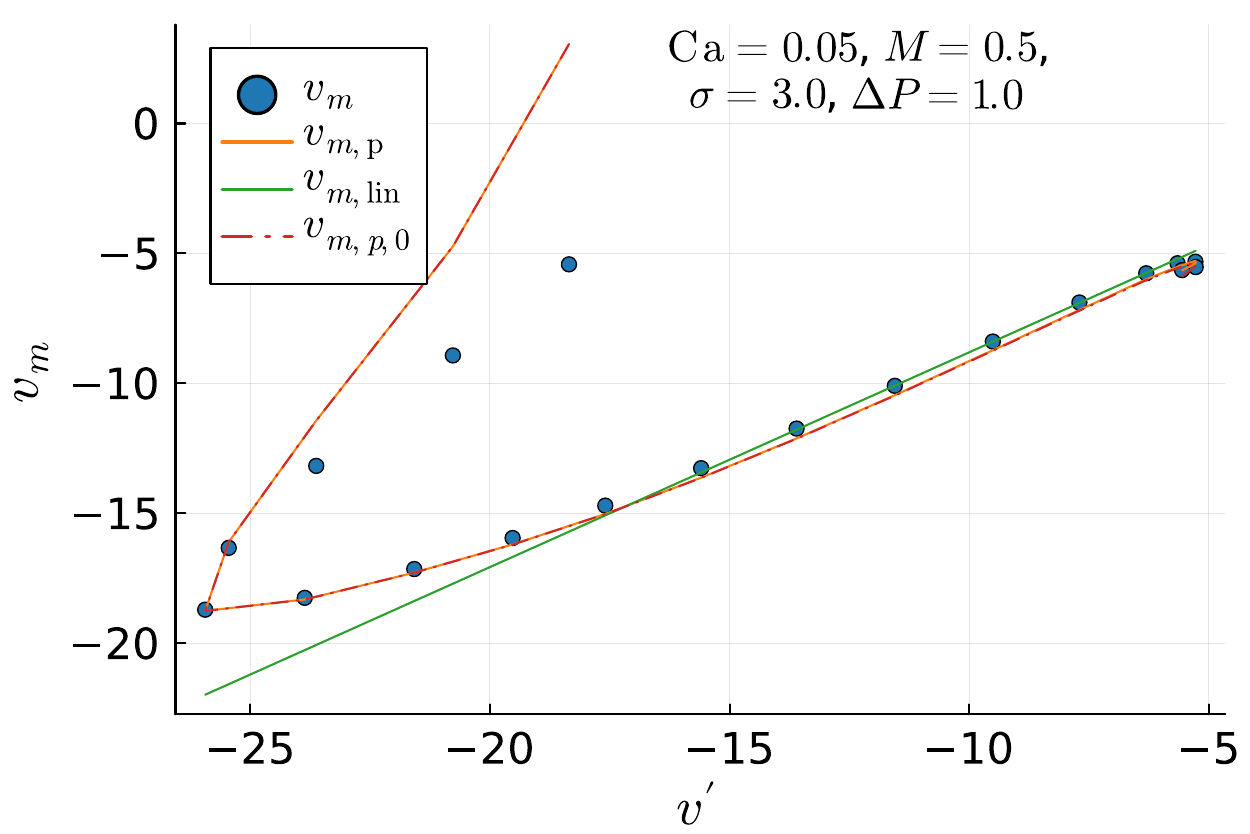"} \par
  \vspace{10pt}
  \includegraphics[width=.24\textwidth]{"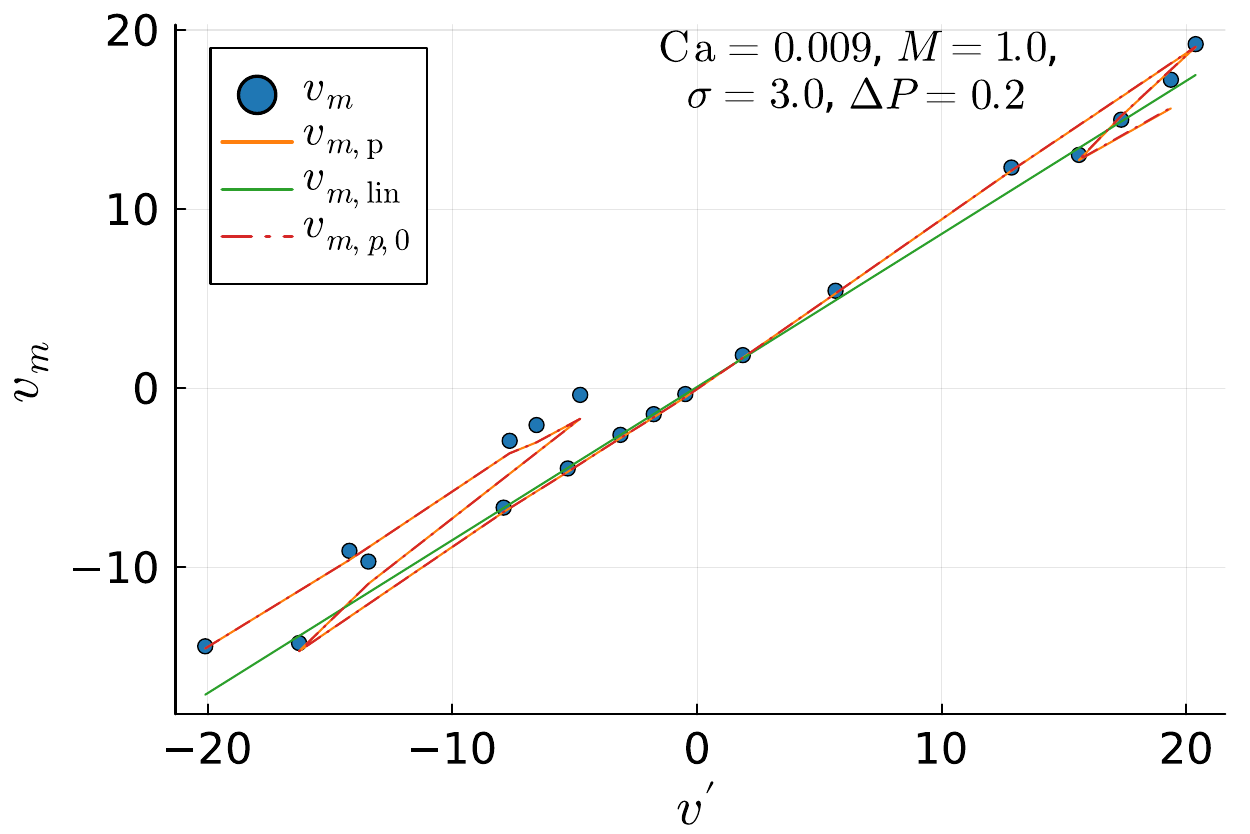"}
  \includegraphics[width=.24\textwidth]{"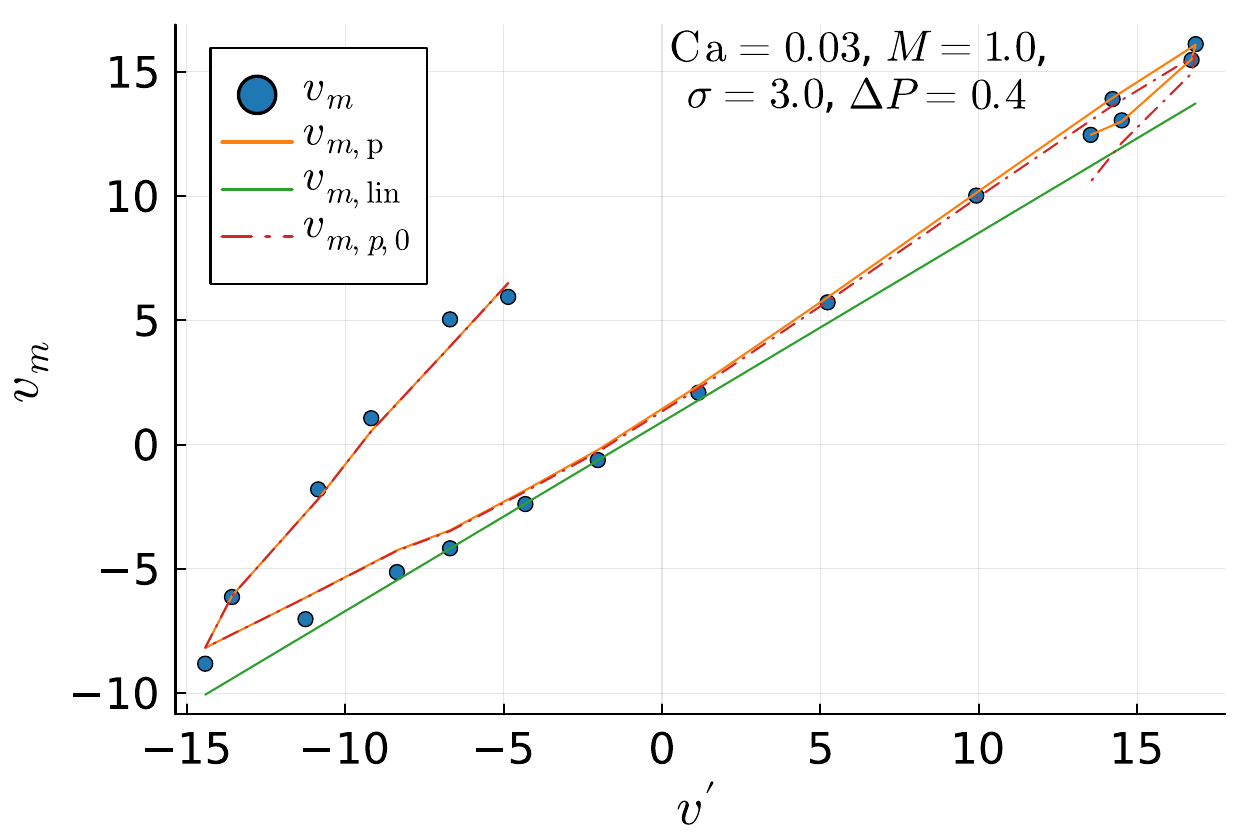"}
  \includegraphics[width=.24\textwidth]{"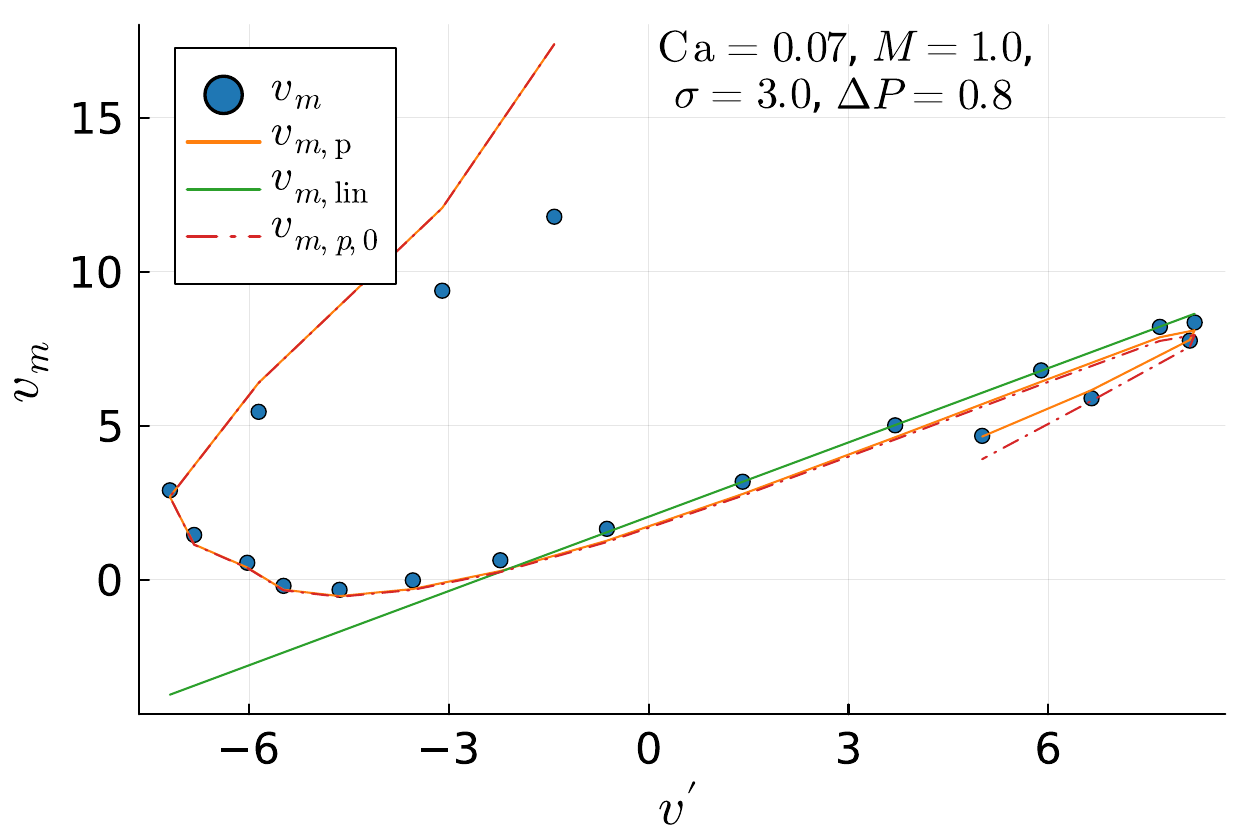"}
  \includegraphics[width=.24\textwidth]{"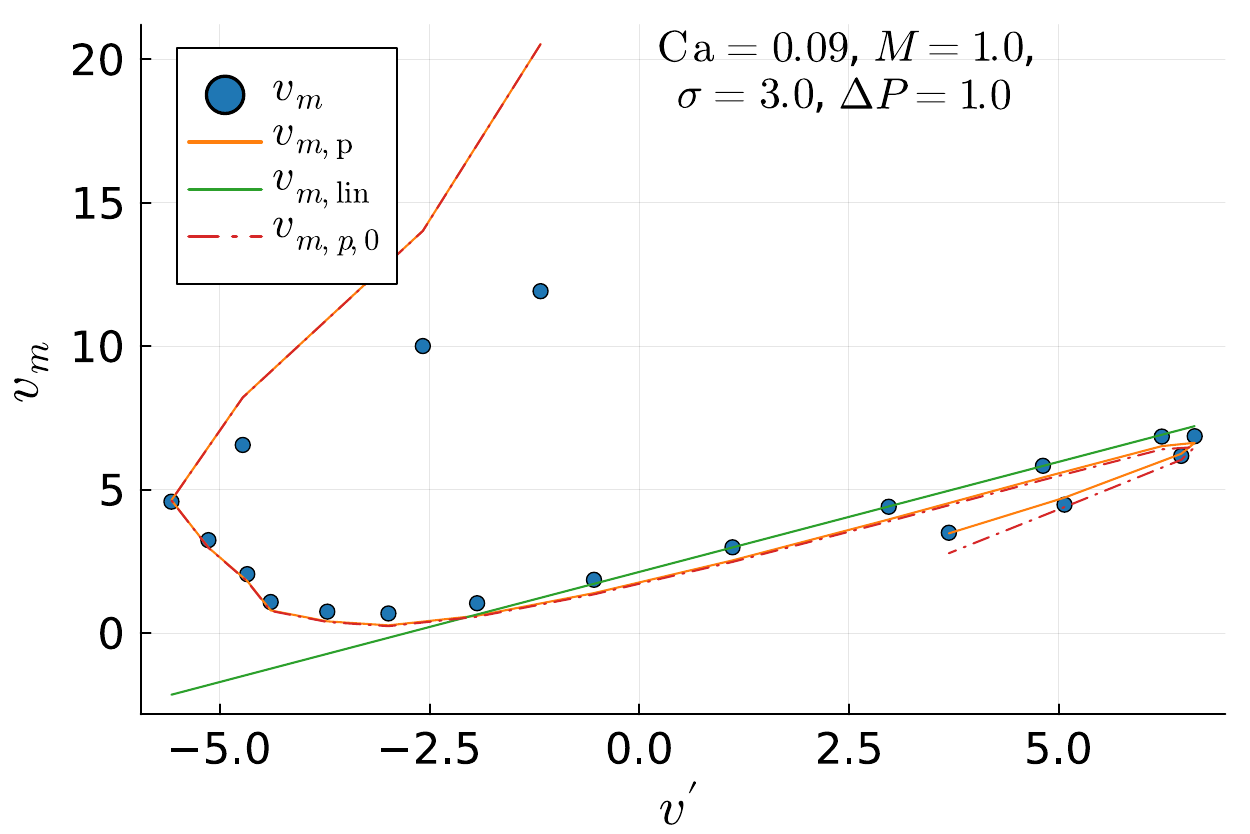"} \par
  \vspace{10pt}
  \includegraphics[width=.24\textwidth]{"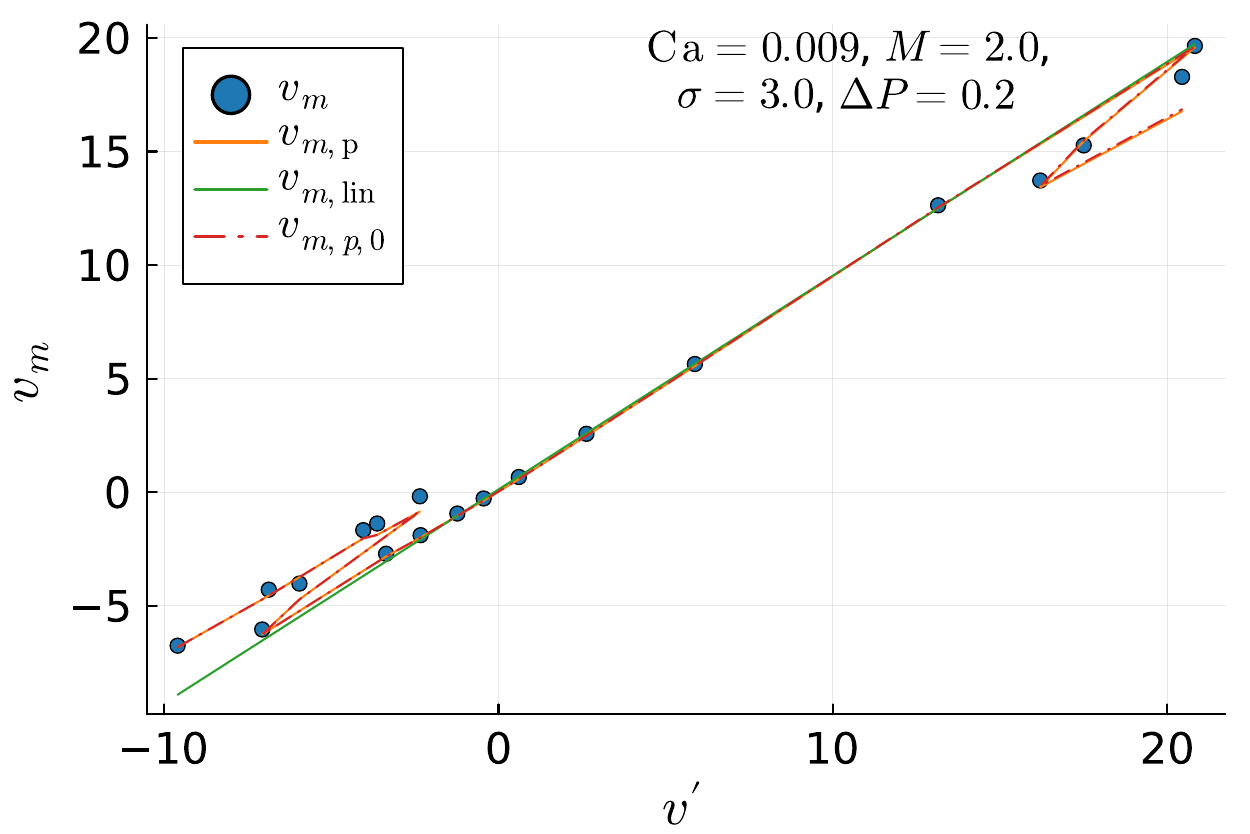"}
  \includegraphics[width=.24\textwidth]{"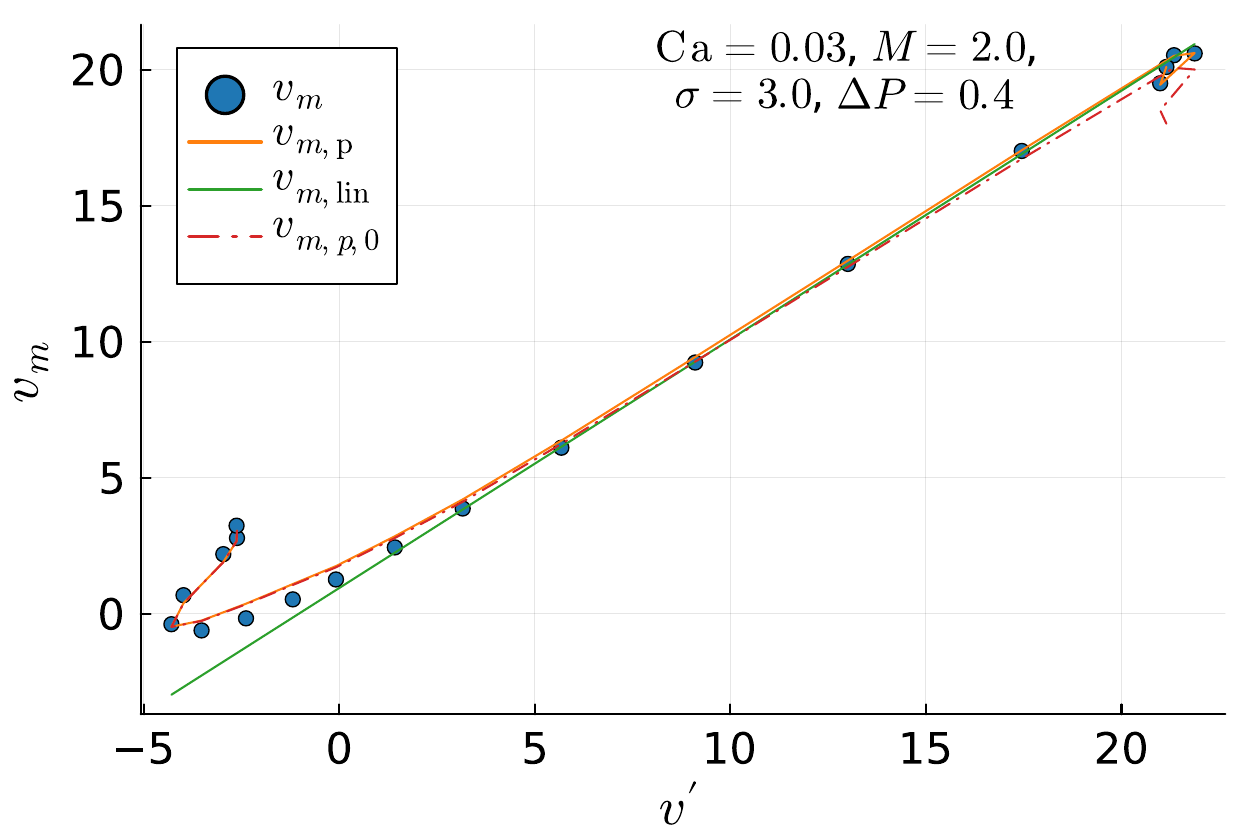"}
  \includegraphics[width=.24\textwidth]{"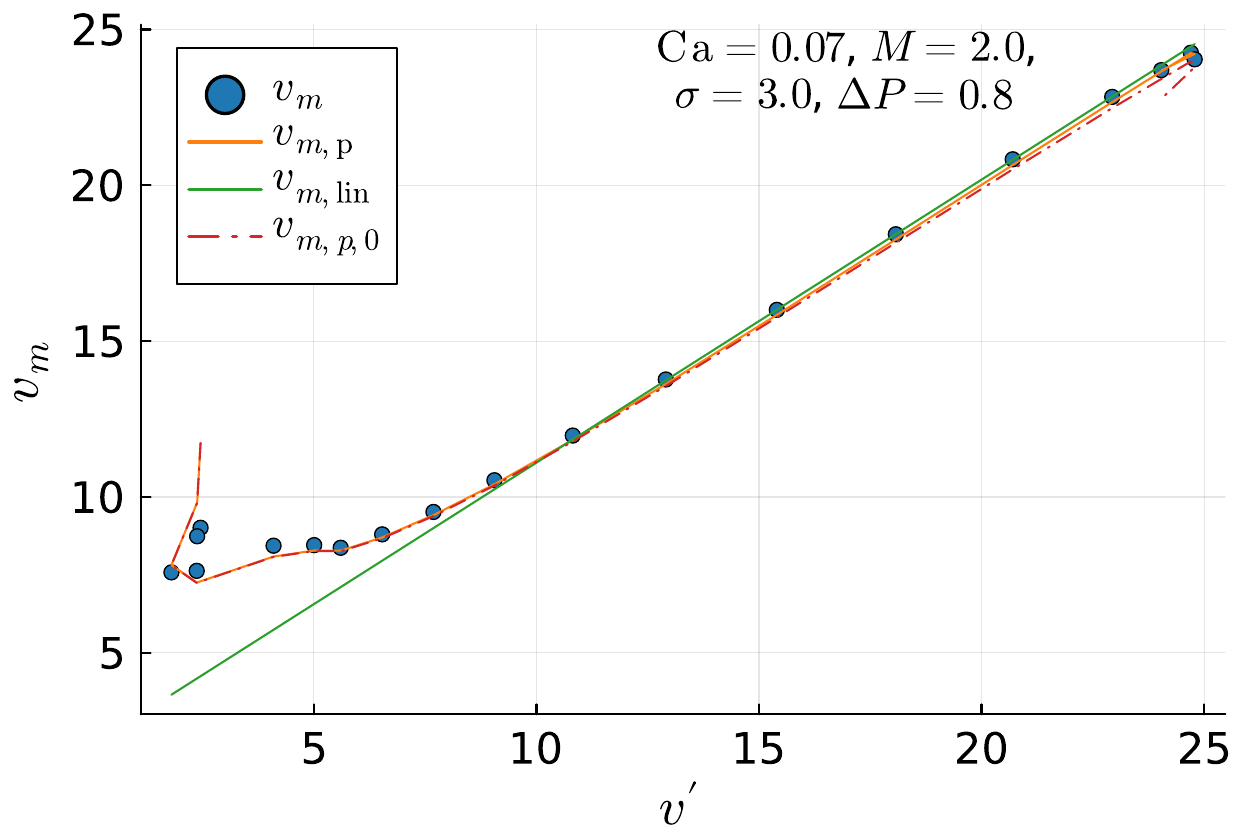"}
  \includegraphics[width=.24\textwidth]{"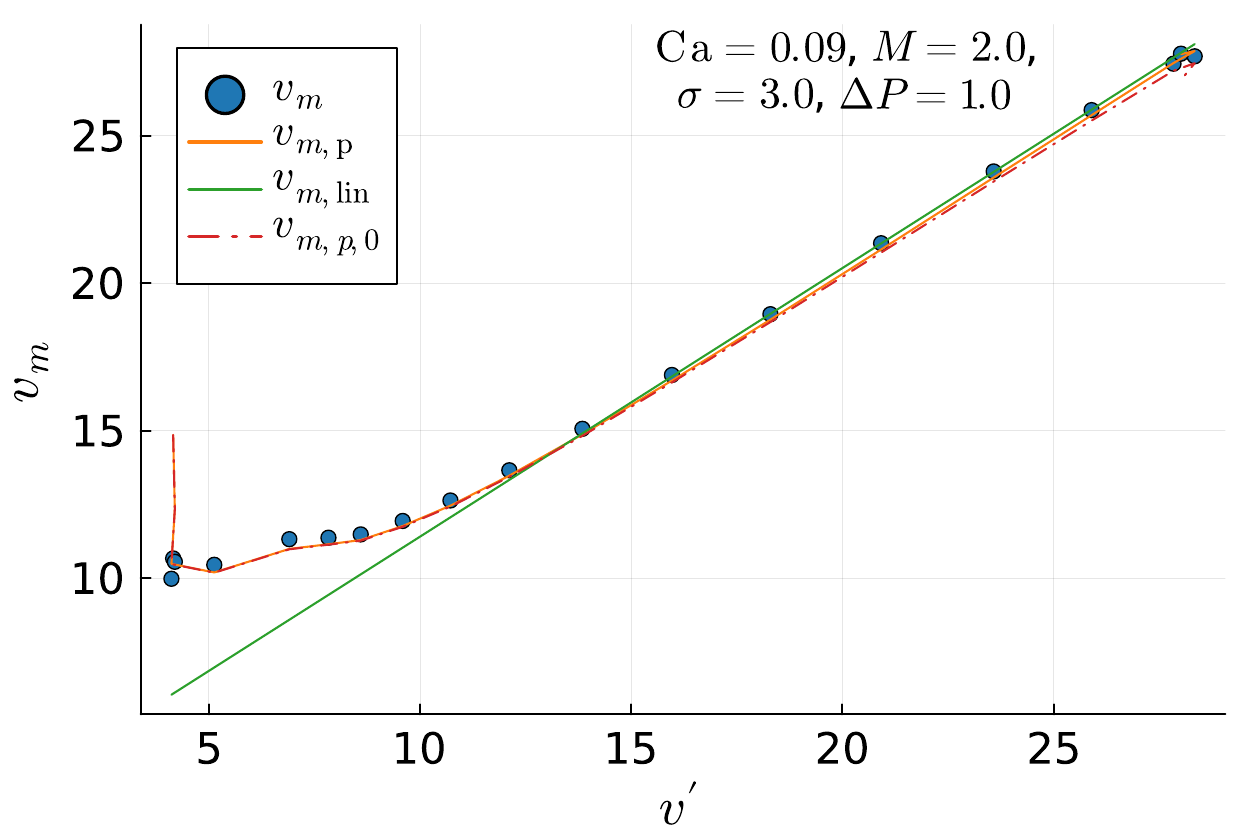"} \par
  \vspace{10pt}
  \includegraphics[width=.24\textwidth]{"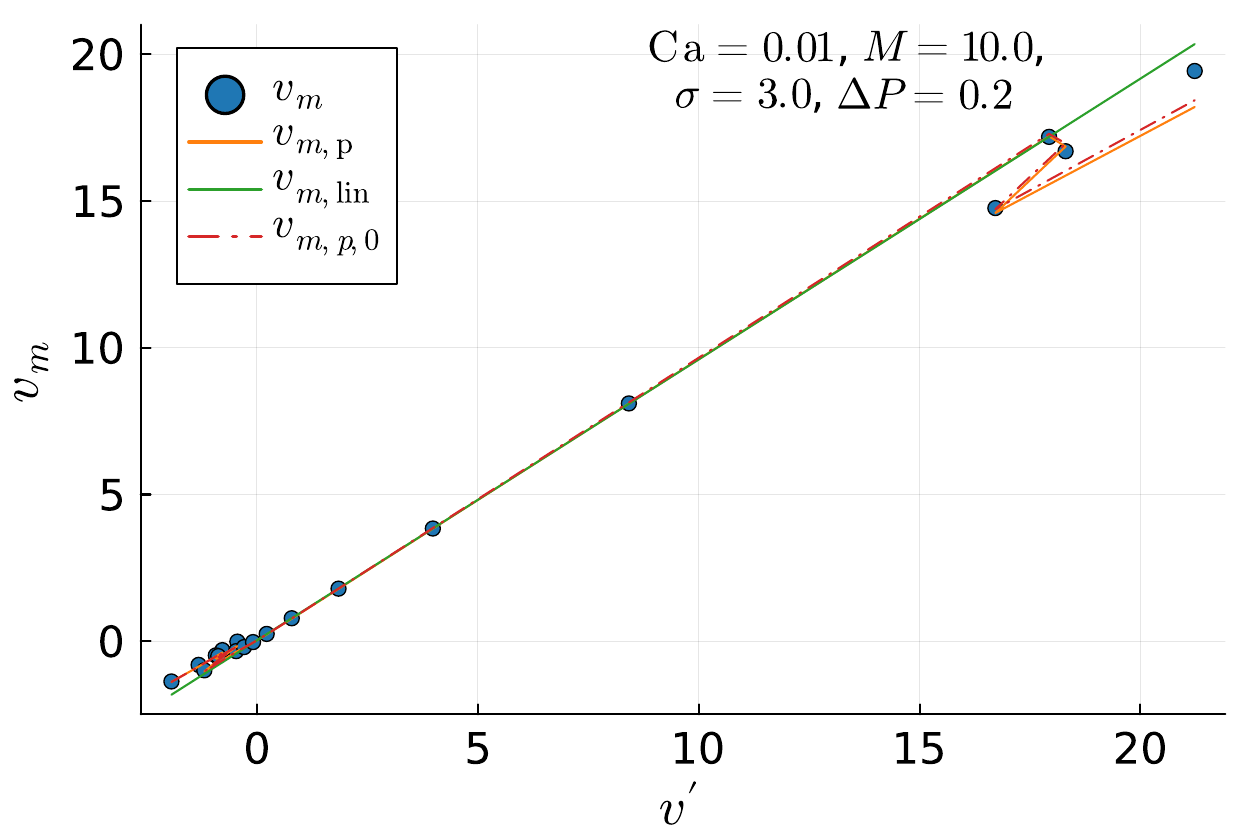"}
  \includegraphics[width=.24\textwidth]{"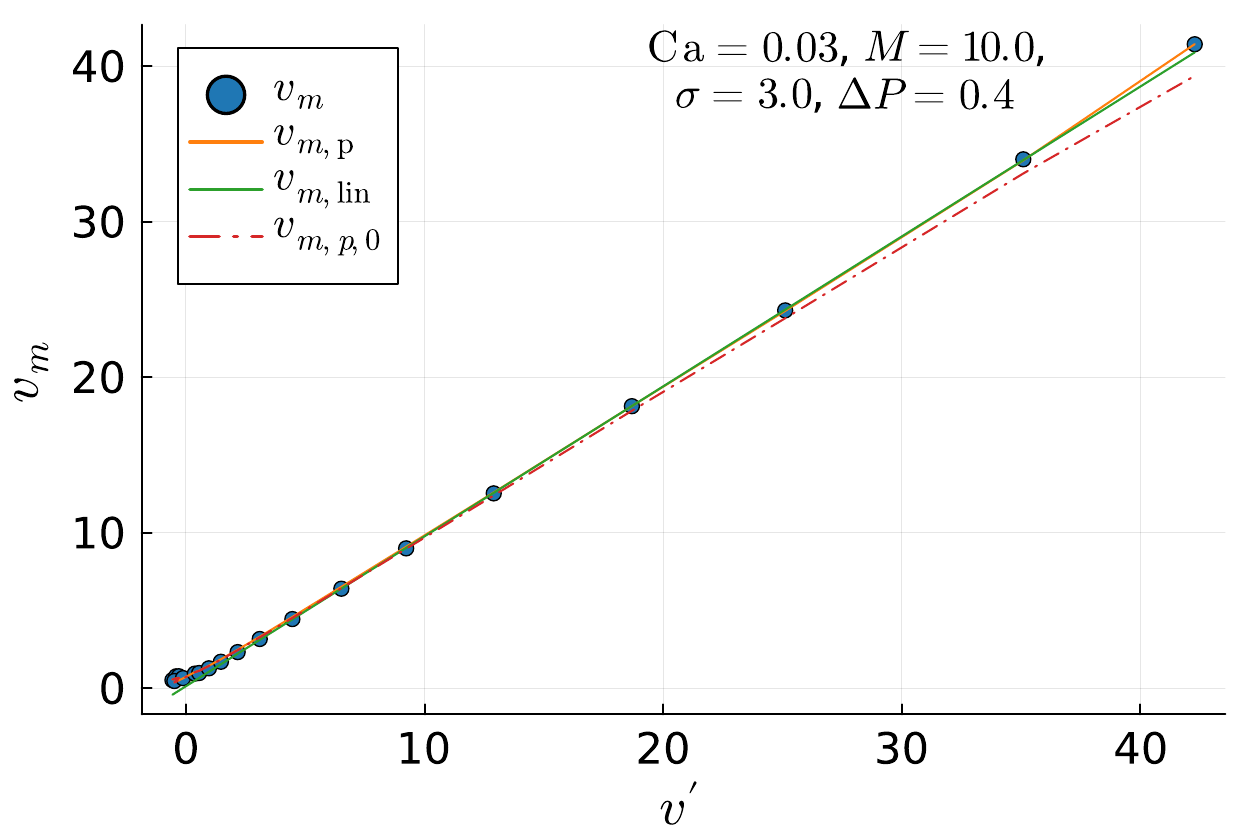"}
  \includegraphics[width=.24\textwidth]{"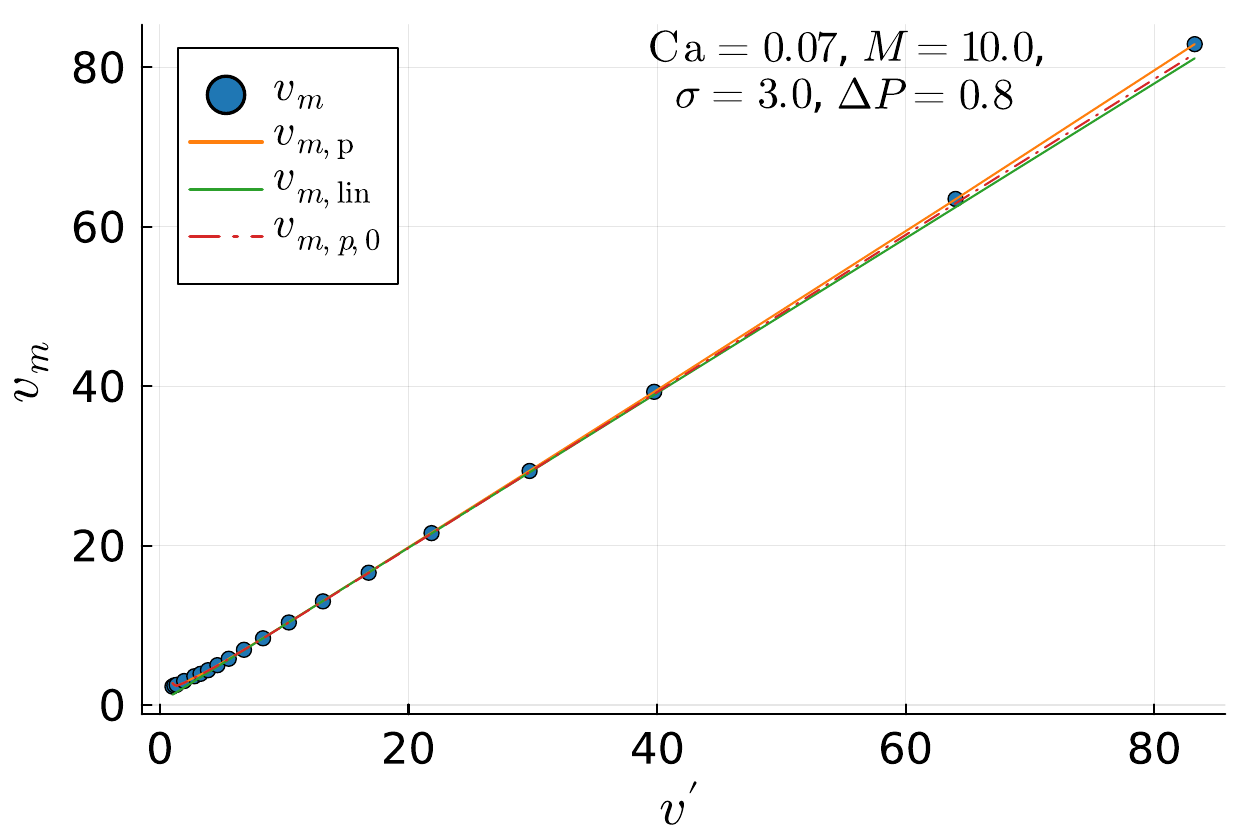"}
  \includegraphics[width=.24\textwidth]{"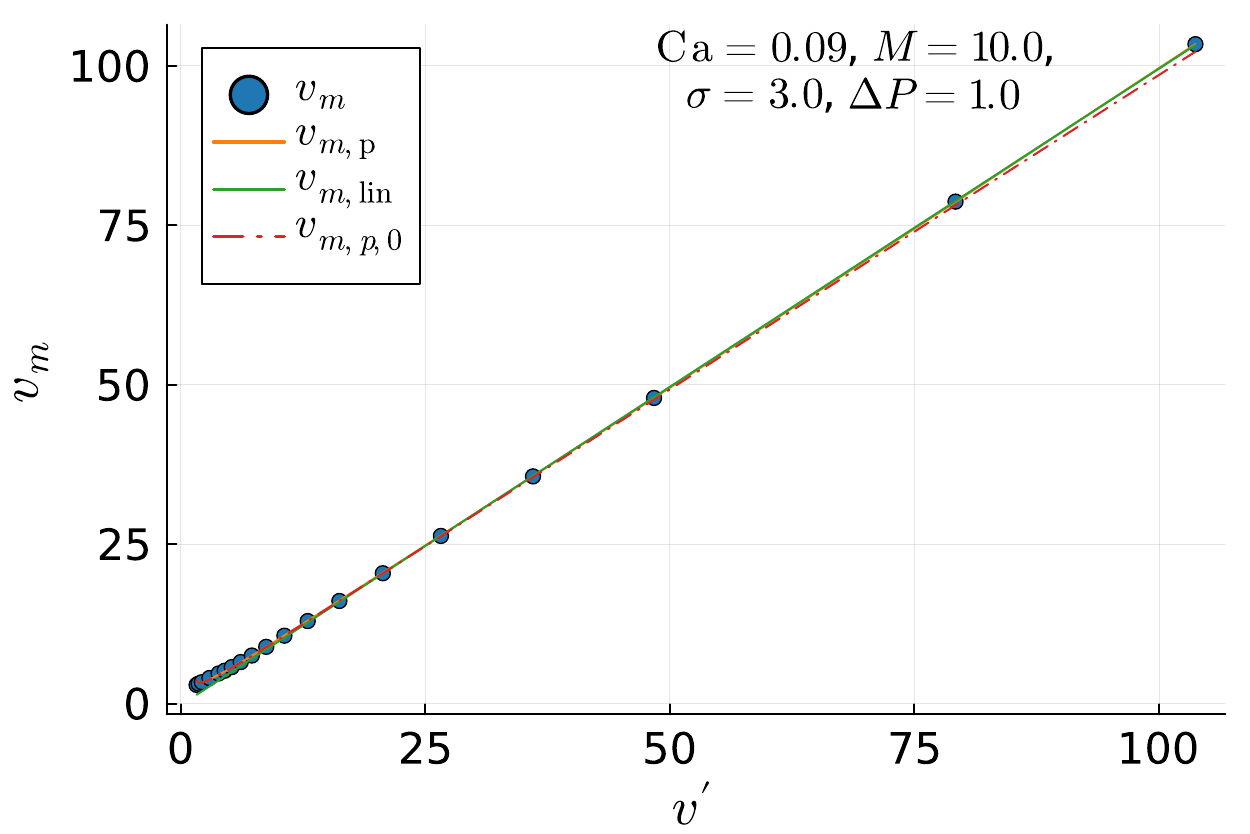"}
    \caption{$v_m(v^{\prime})$ (DNM) for a range of $M$- and $\Delta P$-values with $\sigma =
      3.0$, varying with column.
      In the leftmost column, $S_n^{\ast} = 0.1$ is used to get better fits. The
      linear approximation has been performed on the largest linear region of
      $v_m \left( v^{\prime} \right)$ within a reasonable tolerance. The
      ``slingshot-effect'' due to the non-monoticity of $v^{\prime}(S_w)$ is evident.
      \label{fig:vm-vd-dnm} }
  \end{figure*}

\begin{figure*}[t]
    \includegraphics[width=.32\textwidth]{"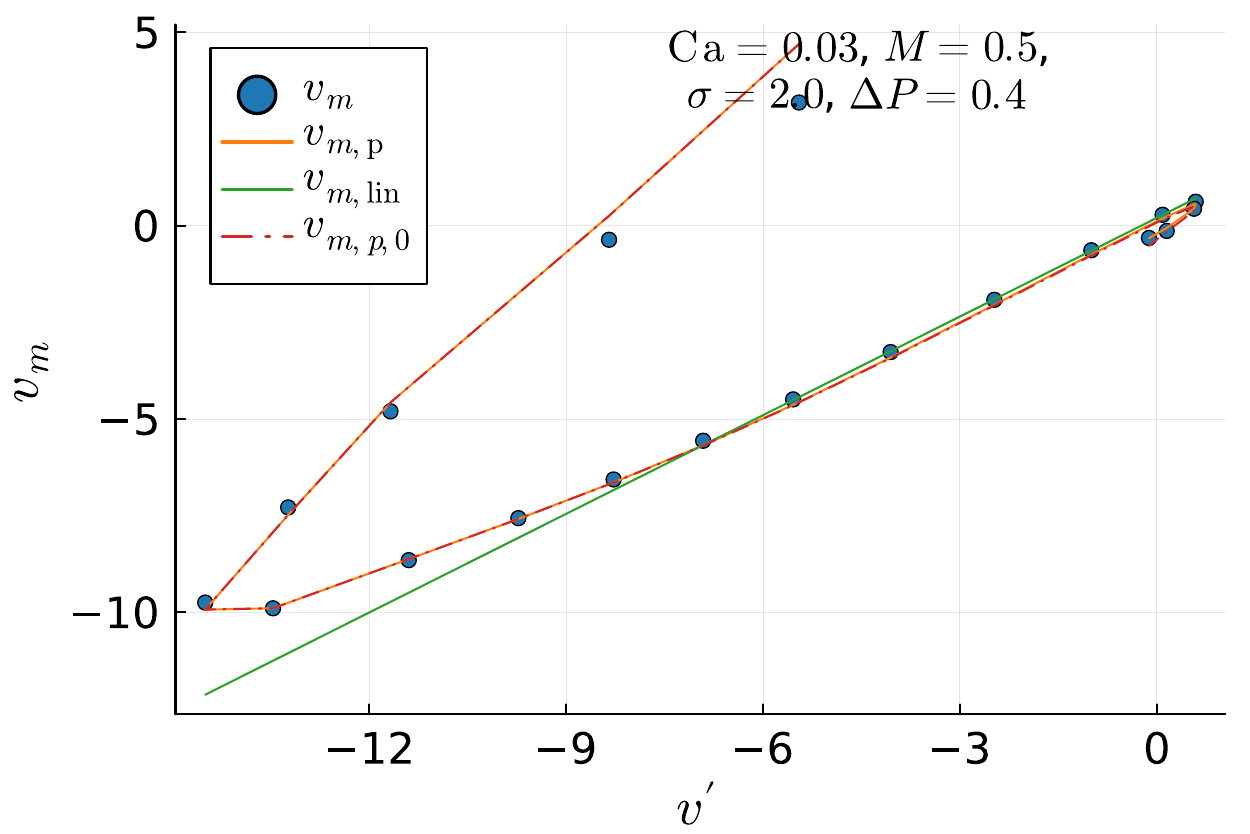"}
    \includegraphics[width=.32\textwidth]{"img/vmvd-M-0.5-T-3.0-P-0.4.pdf"}
    \includegraphics[width=.32\textwidth]{"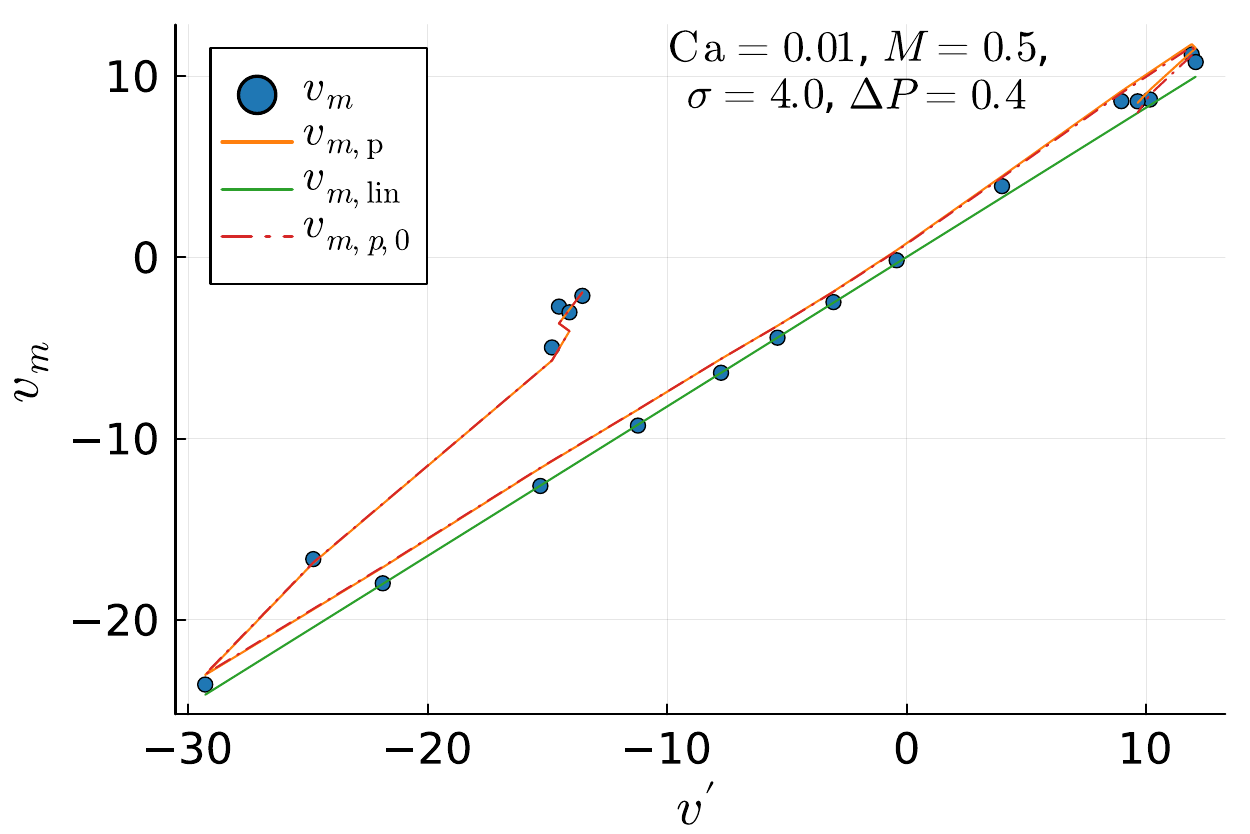"} \par
    \vspace{10pt}
    \includegraphics[width=.32\textwidth]{"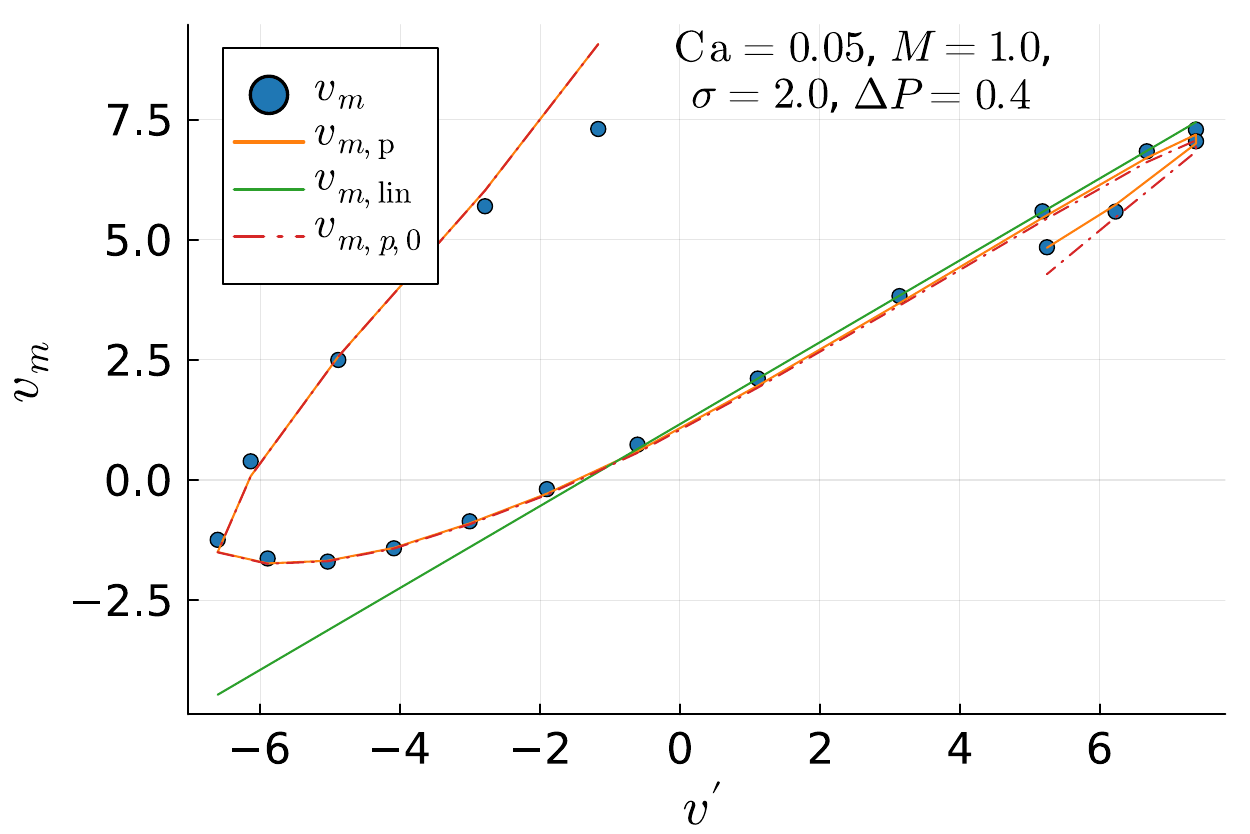"}
    \includegraphics[width=.32\textwidth]{"img/vmvd-M-1.0-T-3.0-P-0.4.pdf"}
    \includegraphics[width=.32\textwidth]{"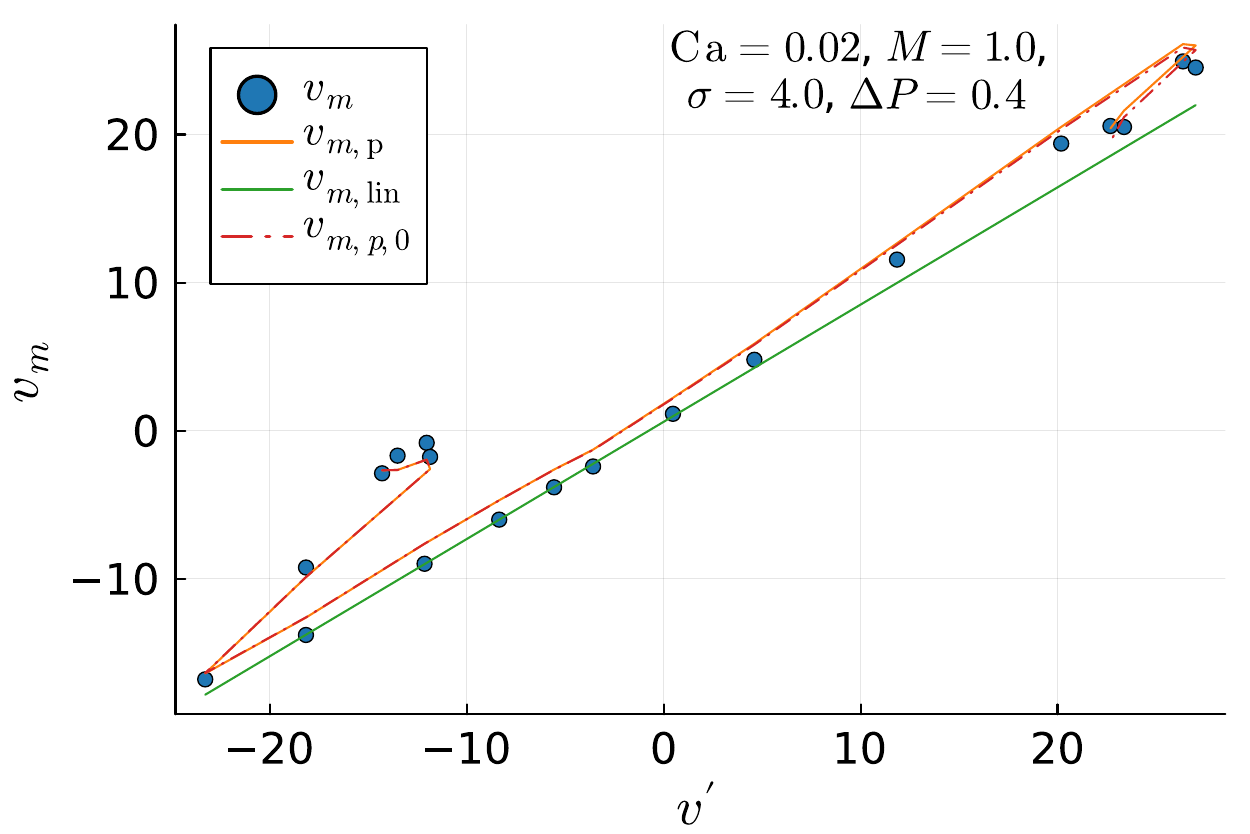"} \par
    \vspace{10pt}
    \includegraphics[width=.32\textwidth]{"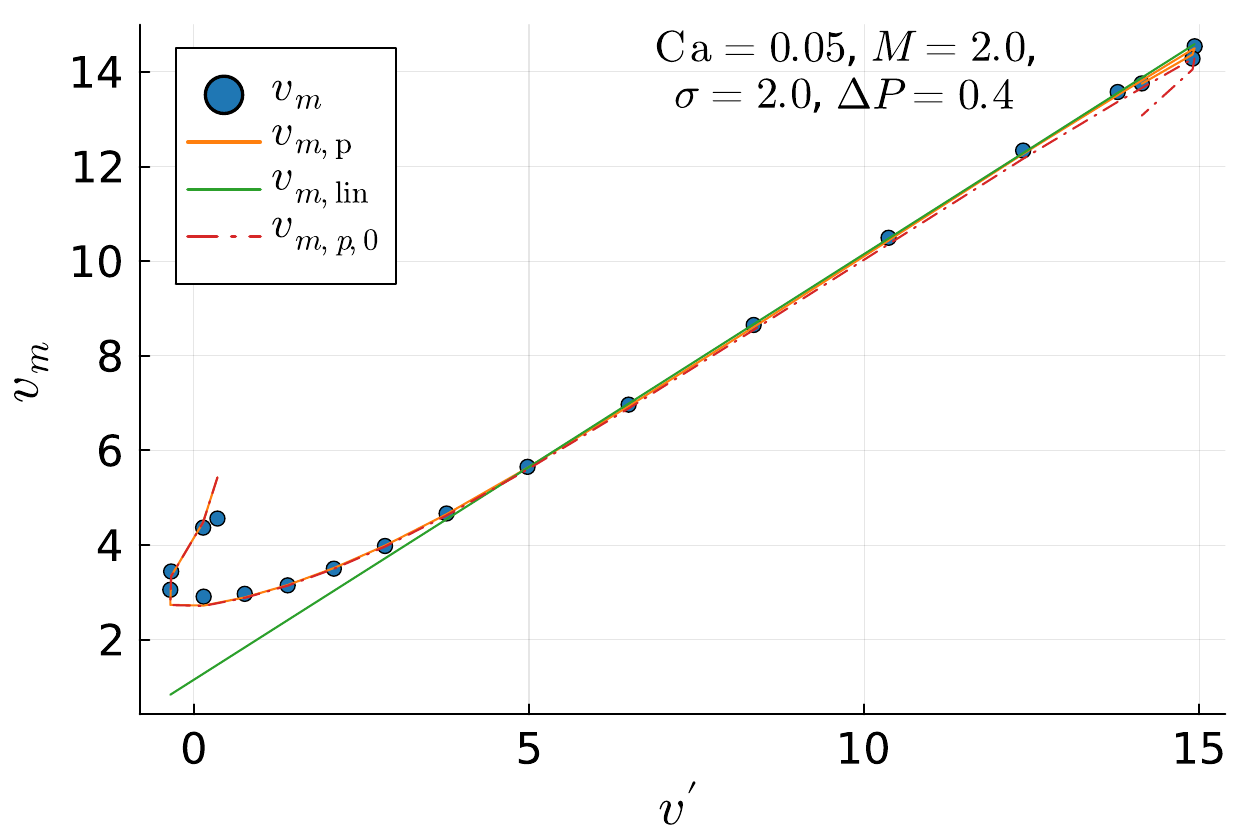"}
    \includegraphics[width=.32\textwidth]{"img/vmvd-M-2.0-T-3.0-P-0.4.pdf"}
    \includegraphics[width=.32\textwidth]{"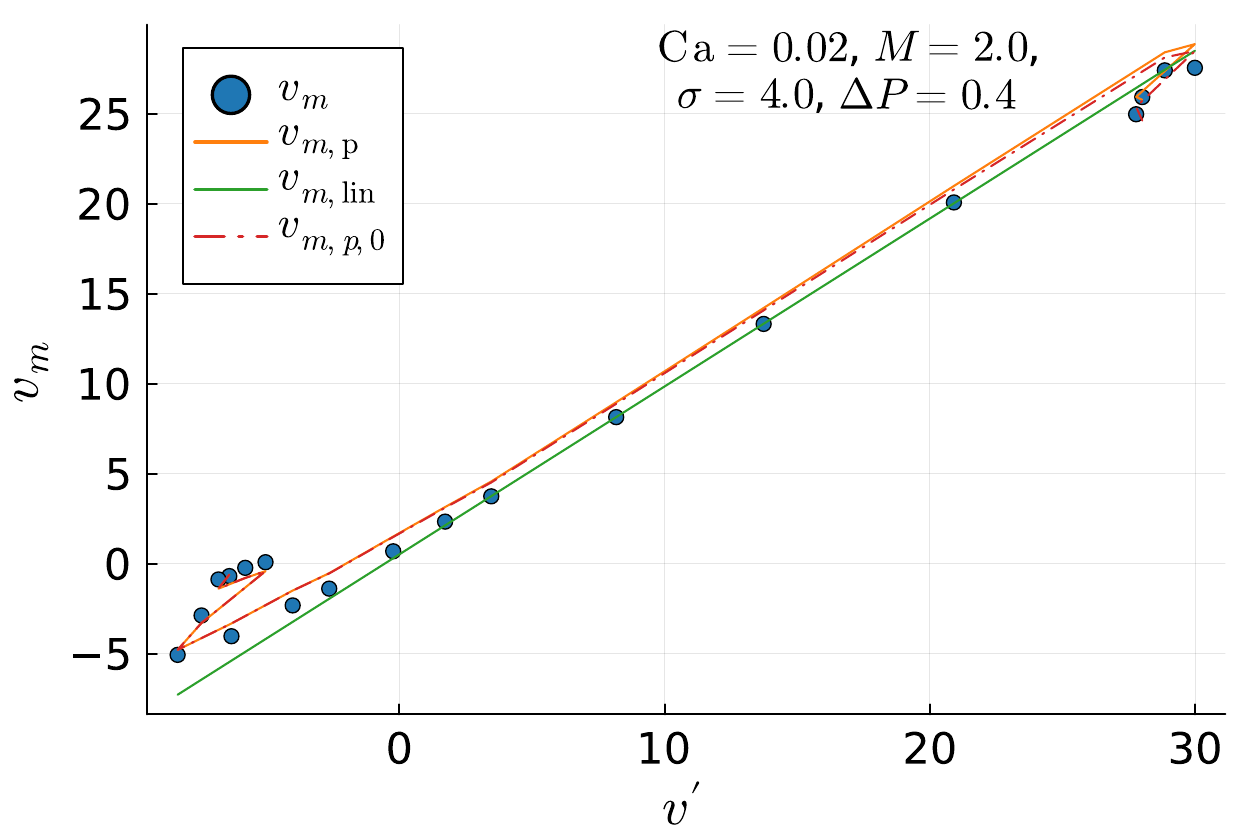"}
    \caption{$v_m \left( v^{\prime} \right)$ (DNM), with varying $\sigma$ and
      $M$ for fixed $\Delta P = 0.4$. \label{fig:surface-tension-variation-vd}.
    }
  \end{figure*}

\Cref{fig:rec-seepage-dnm,fig:vm-sw-dnm,fig:surface-tension-variation,fig:vm-vd-dnm,fig:surface-tension-variation-vd}
uses 2d dynamic network model data. This data has been published elsewhere
\cite{sinhaFluidMeniscusAlgorithms2021}, and the same conventions as there are
used here. Data for for different values of the viscosity ratio $M = \mu_n/\mu_{w} =
(0.05, 1, 2, 10)$, pressure gradient $\Delta P = (0.2, 0.4, 0.8, 1.0)$
\si{\mega\pascal\per\meter} and interfacial tension $\sigma = (2,3,4)$
\si{\dyne\per\milli\meter} is presented. $\mu = 0.1$ \si{\pascal\second} has been
used, and the velocities are in units of \si{\milli\meter\per\second}. Due to
the numerics behind the numerical network model, the regions of the graphs close
to the edges of the saturation domain can contain unphysical predictions and
should be disregarded.

In the linear approximation for $v_m \left( v_{d} \right)$, $v_{m,
  \mathrm{lin}}$, the largest linear region of the graph is used for the fit.
The linear fit in terms of $\beta v$ with the parameter $k_e$ as mentioned in
\cref{sec:five-point-conic} is not plotted, as they are nearly identical to the
linear fit of to $v_m \left( v_d \right)$. In all figures, $v_m$ is computed
from the definition in \cref{eq:vm-derivative-form}, $v_{m, \mathrm{lin}}$ is
the linear approximation in \cref{eq:vm-constitutive}, $v_{m,p}$ is computed
from \cref{eq:vm-solved-from-k} with $k_e$ as estimated in
\cref{sec:vm-numeric}, and $v_{m, p, 0}$ is the same as $v_{m,p}$ with the
parameter $c$ set to zero.

  \begin{figure*}
    \includegraphics[width=.32\textwidth]{"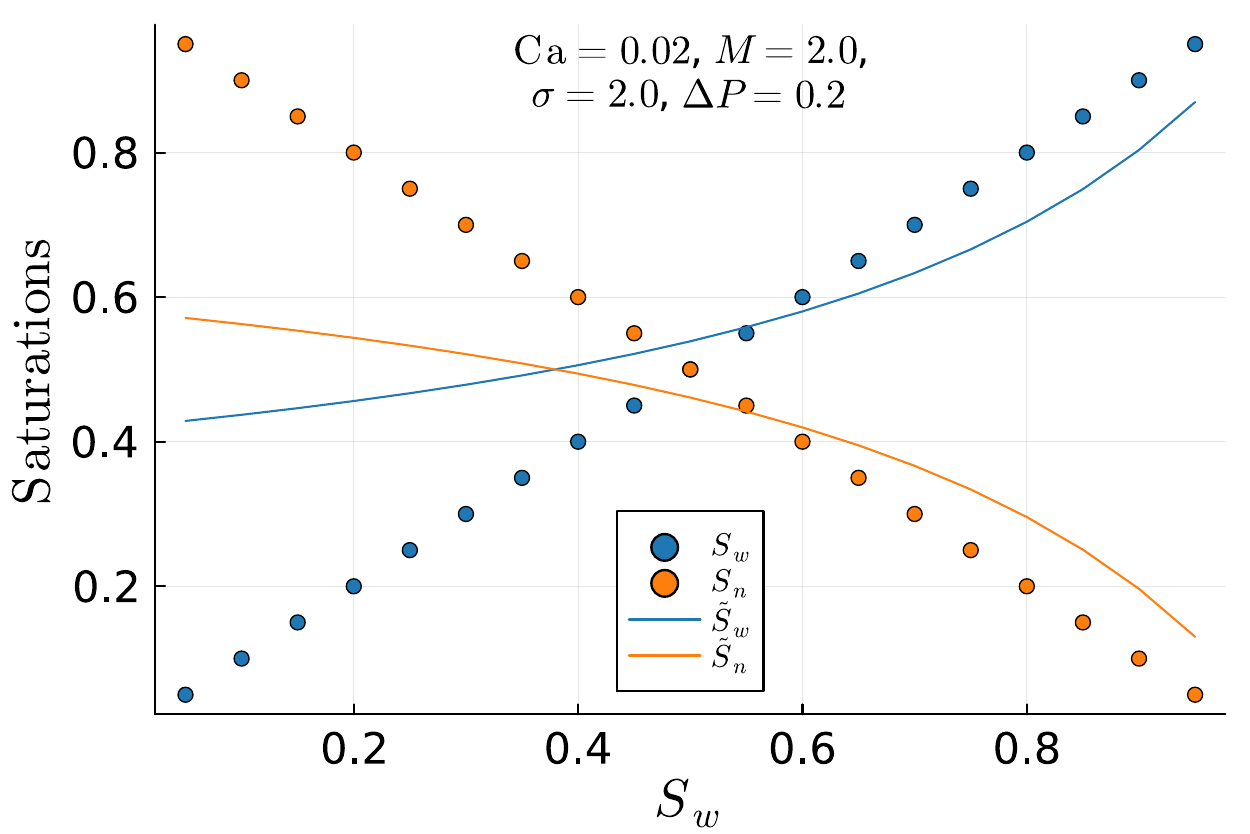"}
    \includegraphics[width=.32\textwidth]{"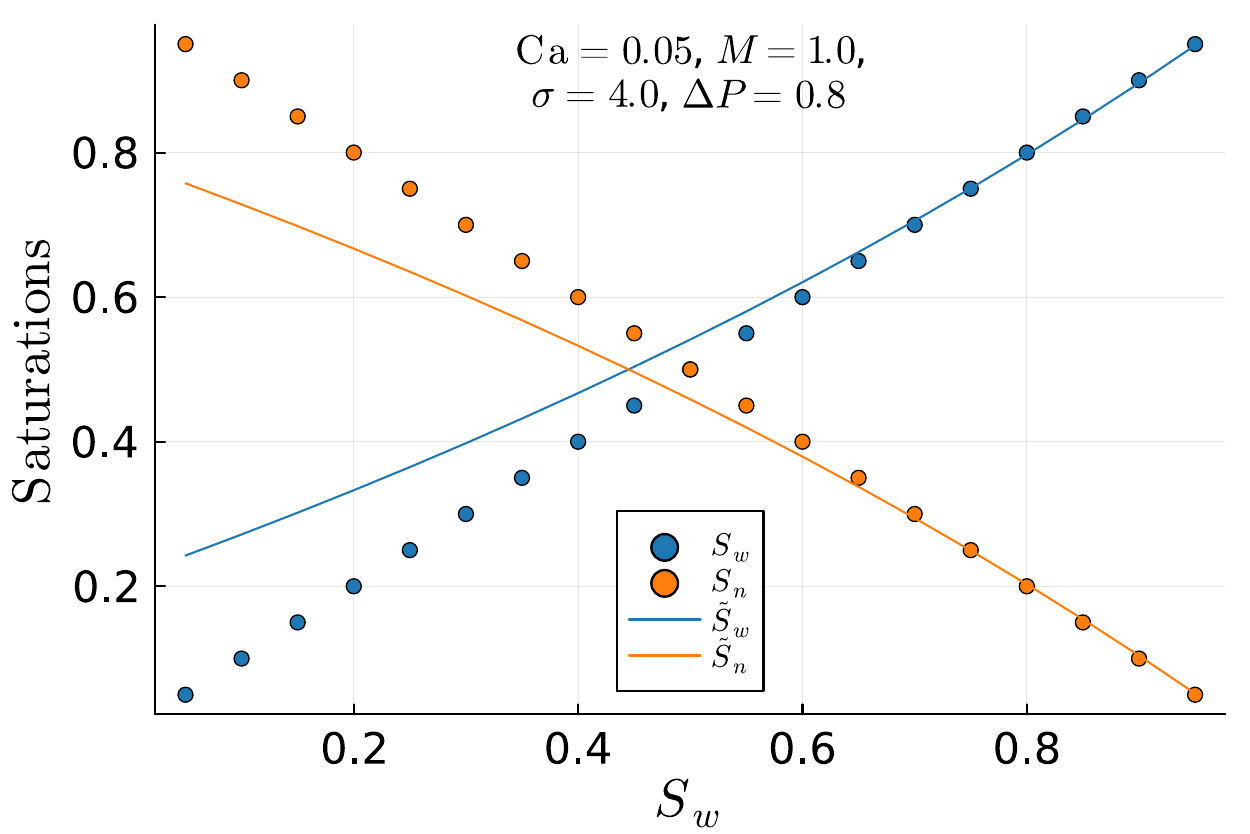"}
    \includegraphics[width=.32\textwidth]{"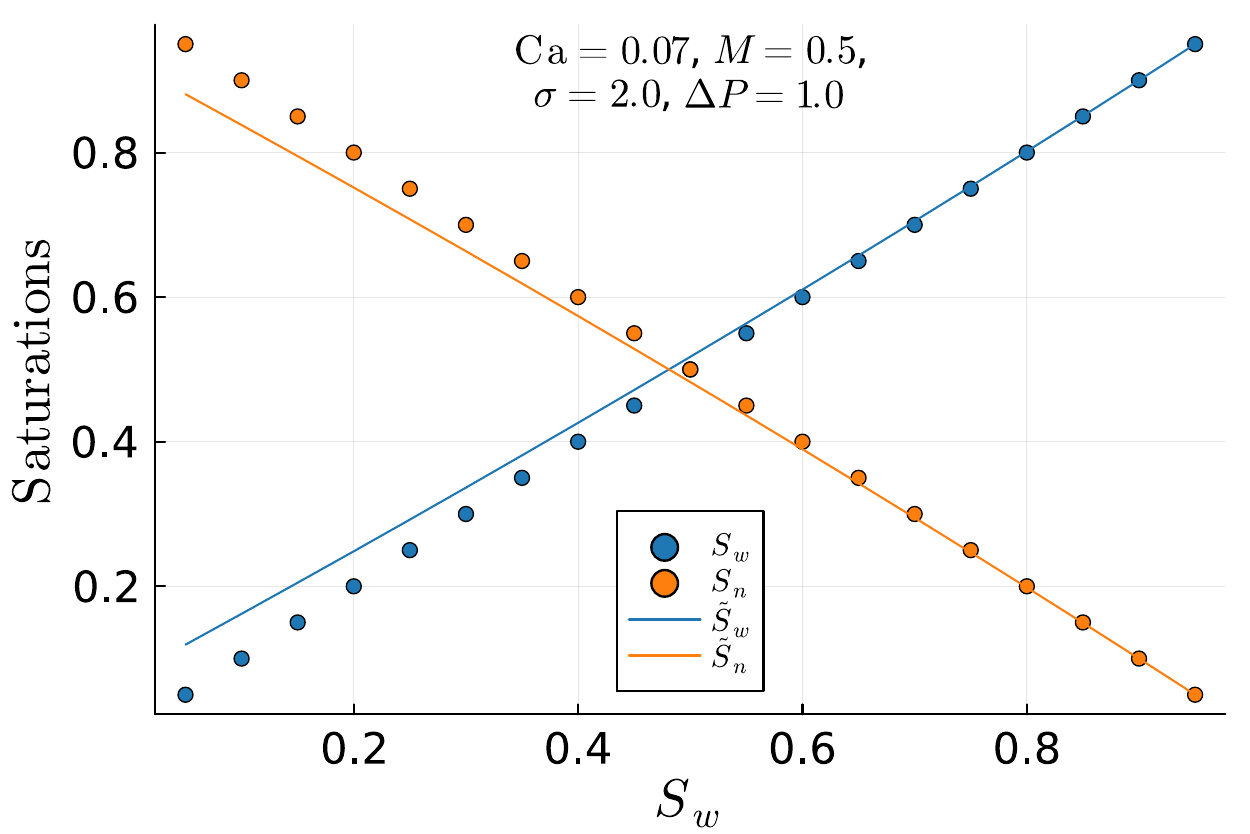"} \par
    \vspace{10pt}
    \includegraphics[width=.32\textwidth]{"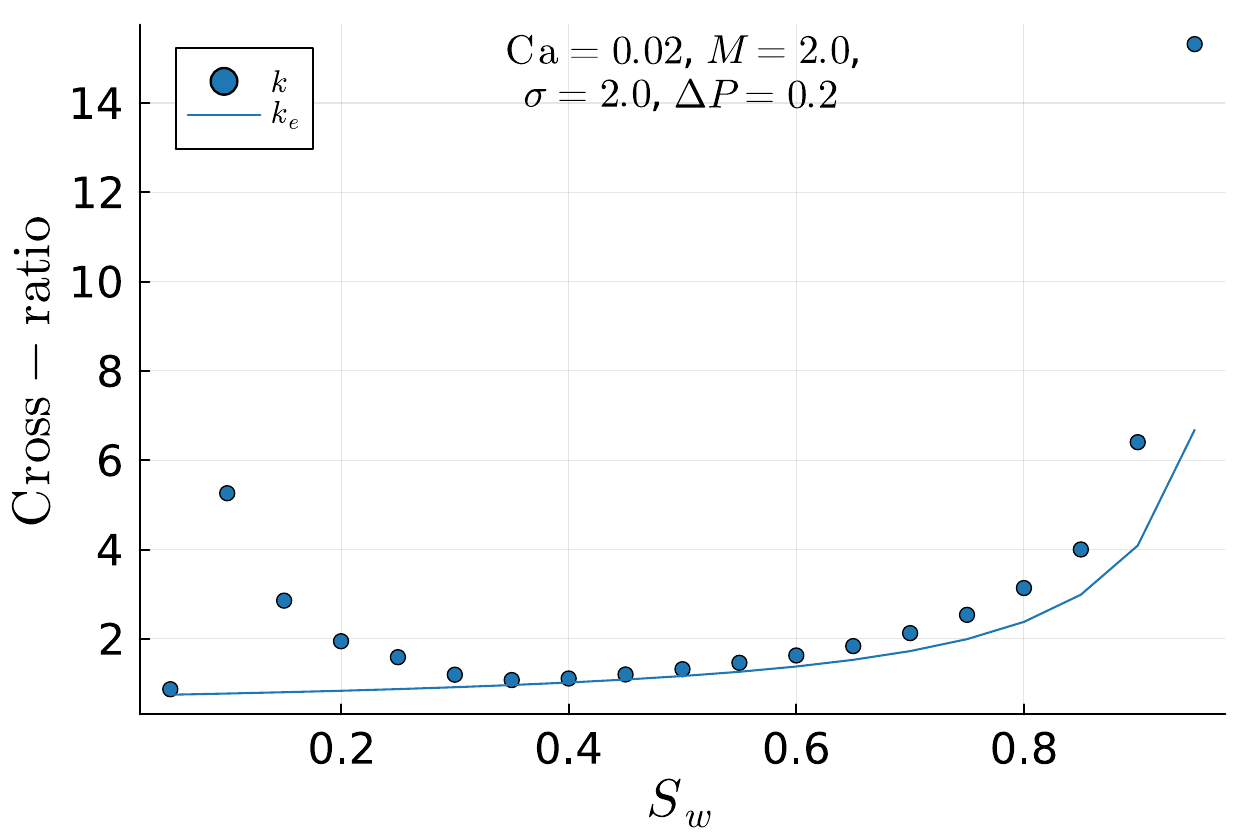"}
    \includegraphics[width=.32\textwidth]{"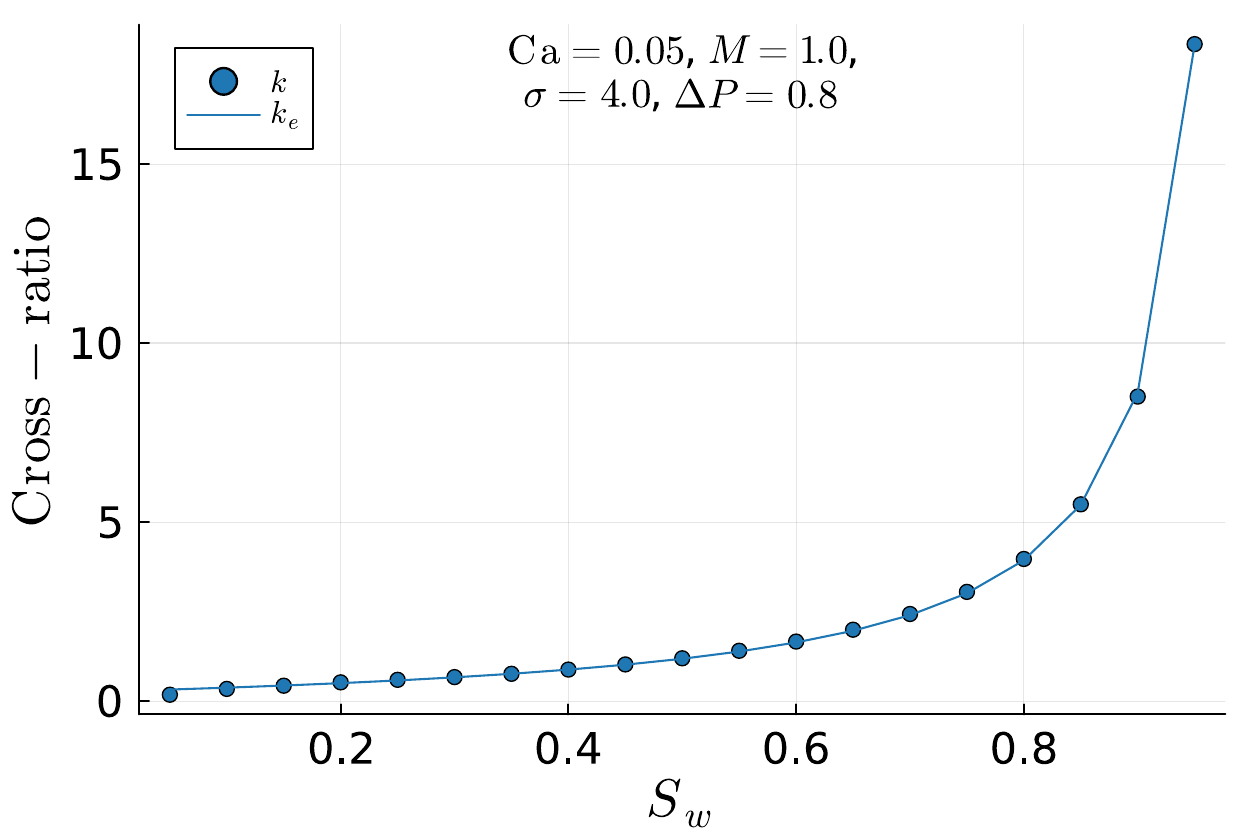"}
    \includegraphics[width=.32\textwidth]{"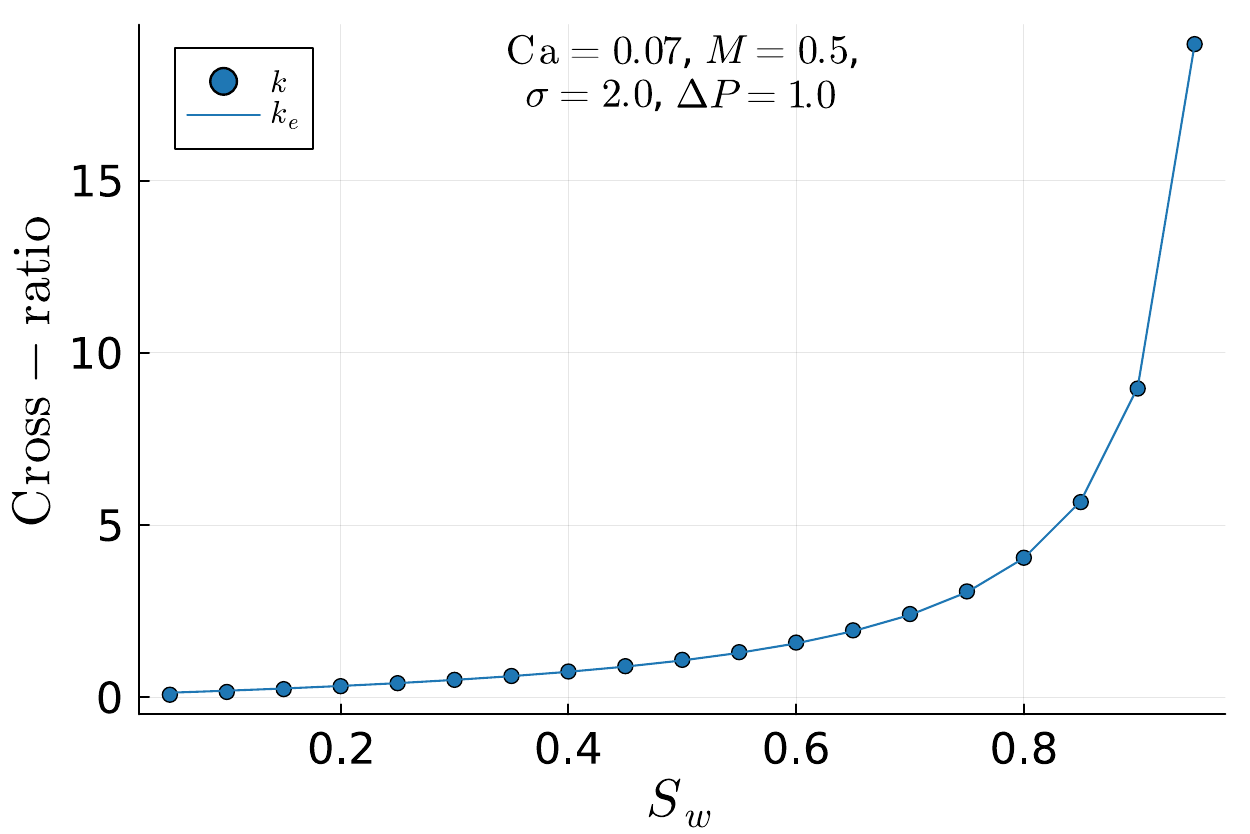"} \par
    \vspace{10pt}
    \includegraphics[width=.32\textwidth]{"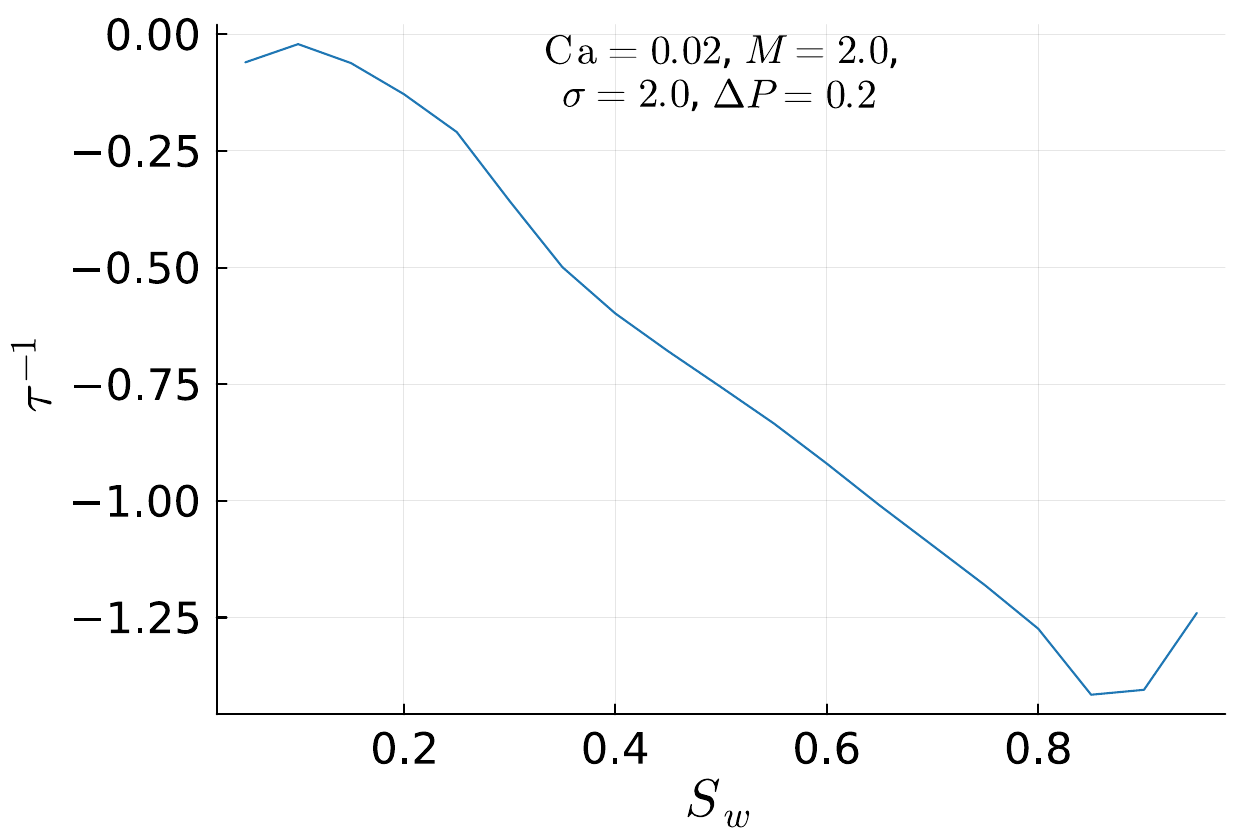"}
    \includegraphics[width=.32\textwidth]{"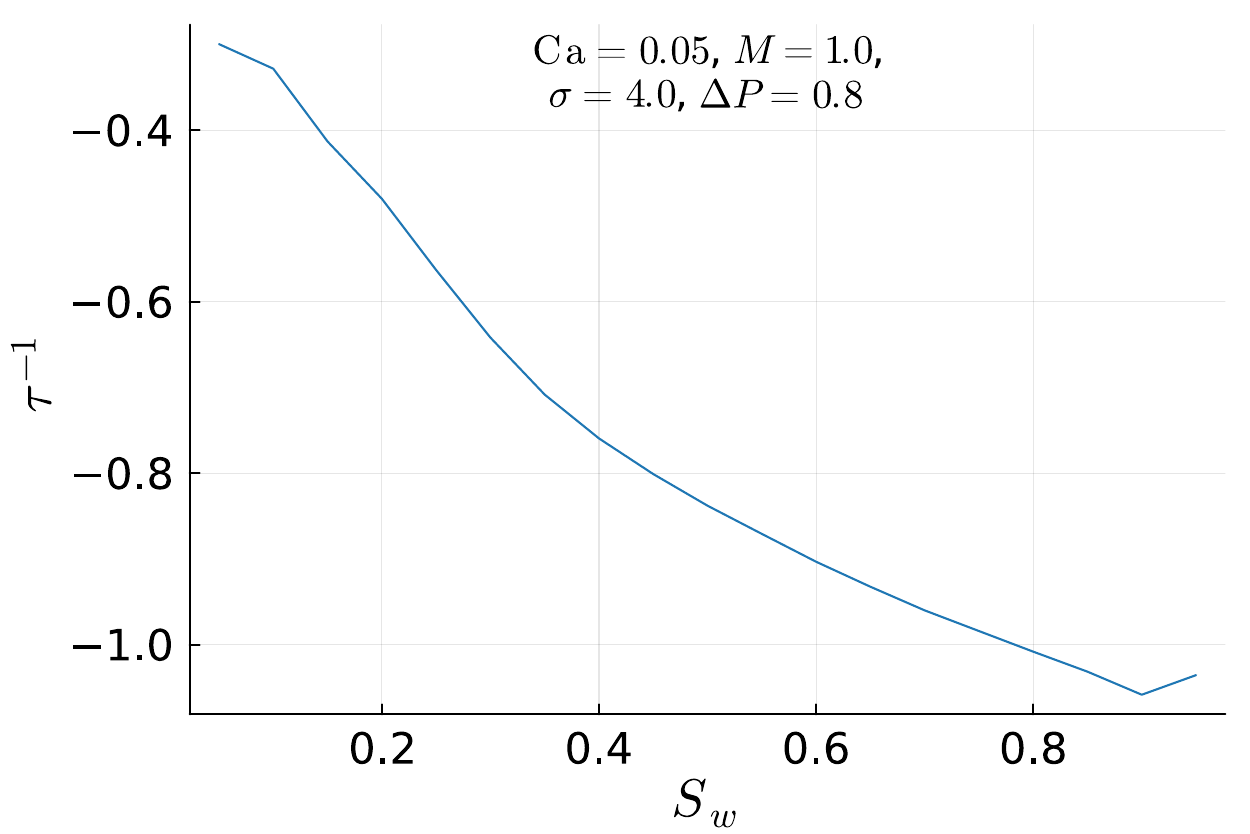"}
    \includegraphics[width=.32\textwidth]{"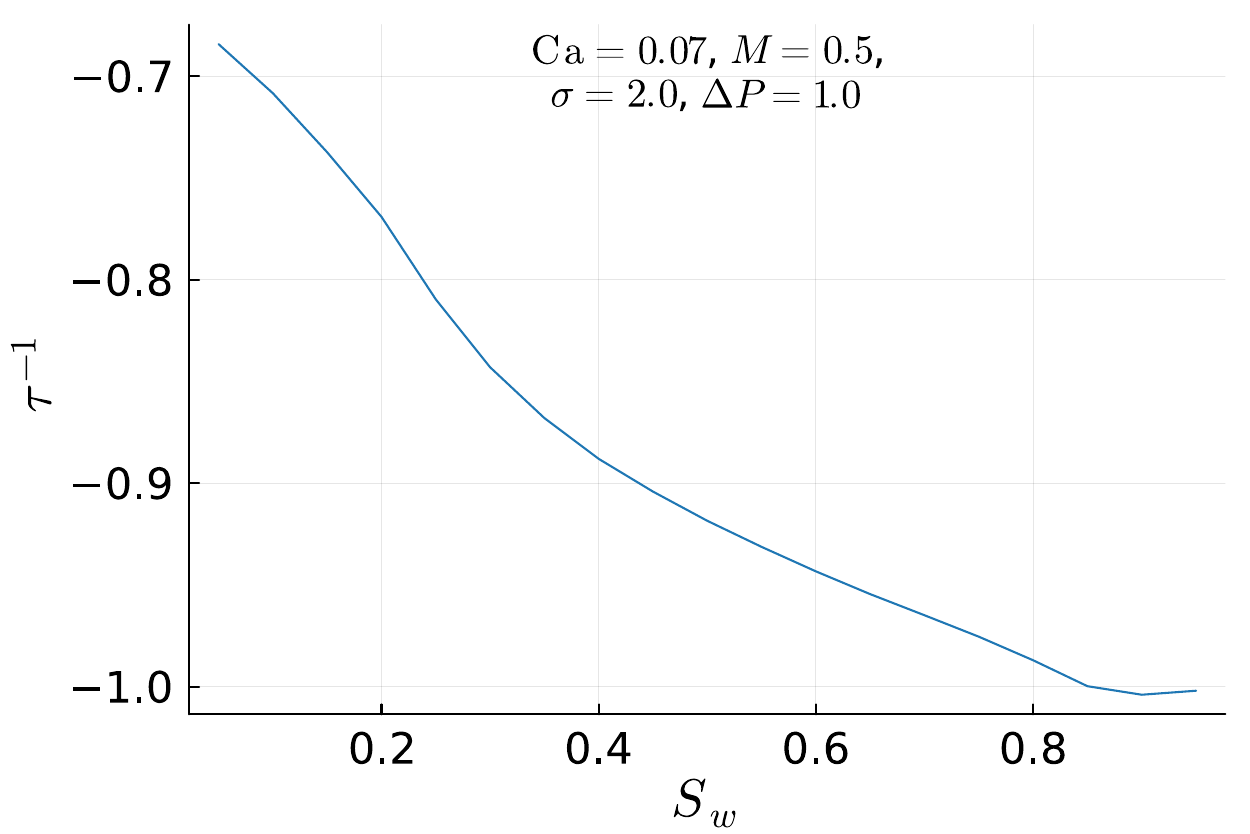"}

    \caption{First row: $\tilde{S}_w, \tilde{S}_n$. Second row: the cross-ratio
      function $k$ and its estimate $k_{e}$. Third row: $\tau^{-1} = -v_w/v_n$.
      All quantities are plotted as functions of $S_w$ for the same values of
      the parameters using the DNM-data. }
  \end{figure*}

  \section{Discussion}
  \label{sec:discussion}

  \subsection{Orthogonality and equivalent configurations}
  \label{sec:O-choice-polar-pole}

  In \cite{pedersenParameterizationsImmiscibleTwophase2023b}, Euclidean
  orthogonality was used to characterize a vector which represented the
  co-moving velocity, which in that framework was orthogonal to a vector of
  areas. A similar idea is applied here. Polarity is introduced in
  \cref{sec:degenerate-quadratic-form} to define orthogonal directions of $l$
  with respect to $B_0$. This represents a direction in which one is free to
  move in the configuration space, at the same time representing the same line
  $l$. Since the geometry is pseudo-Euclidean, one can apply an analogy with
  Minkowski space. The relations between the wetting- and non-wetting quantities
  are the same as between the time- and space-axes in $1+1$-dimensional
  Minkowski space. The signature of $B, B_{0}$ is due to the existence of
  saturation limits $0,1$, which by duality is equivalent to limits on the
  physical velocities. These limits are defined from the points $I,J$. However,
  note that the velocities in this paper are not kinematic velocities, but
  rather the duals of the areas.
  
  The choice of dual conic corresponds to determining the location of the points
  $I, J$. In terms of $A_0$, the points of the line $l_{\infty}$ with affine
  coordinates $\left(S_w, S_n \right)$ was mapped onto the points $\left(1 -
    2S_{w}, 0 \right)$. In the transformation $v_{i} \leftrightarrow \hat{v}_i$,
  one encounters the two quantities $S_{n} v_m$ and $-S_w v_m$. The factors
  $S_n, -S_w$ each span a part of the range $\left( 1 - 2S_w \right) \in \left[
    -1, 1 \right]$. One cannot from this conclude that the map $\left( S_w, S_n
  \right) \mapsto \left( 1 - 2S_w, 0 \right)$ is related to the definition of
  $v_m$, but the same procedure of grouping together pairs of measurable
  quantities define both $v_m$ and $\left( 1 - 2S_w \right)$. $B_0$ represents
  the simplest possible form of a transformation with this effect.

  Analogous to special relativity, the line $m$ through $O$ orthogonal to $l$
  contains all ``events'' that are ``simultaneous'' with respect to the line
  $l$. An ``event'' is simply a point $\in \mathbb{P}^{2}$. ``Simultaneous'' or
  simply orthogonal is analogous to $m$ containing ``inaccessible''
  configurations of velocities with respect to $l$ \footnote{ One could think of
    $l$ and $m$ as the axes of a $1+1$ Minkowski diagram with respect to some
    (potentially boosted) observer. If $l$ represented a timelike direction,
    then $m$ would represent a spacelike direction (and opposite).}. Said
  differently, if $l$ represents a physical configuration of velocities, then
  $m$ represent configurations that has no physical realization. Of course,
  there are infinitely many such configurations. With respect to a chosen fixed
  conic and event, a unique \footnote{This is not entirely true, because the
    infinite line of the degenerate conic is orthogonal to every other finite
    line in the pseudo-Euclidean plane
    \cite{richtergebertPerspectivesProjectiveGeometry2011}. It is here assumed
    that the orthogonal line is a finite one.} such ``inaccessible''
  configuration is defined up to scaling. The hyperbolic angle $\eta$ of
  \cref{sec:five-point-conic} can then be seen as quantifying the difference
  between the physical velocities encoded in $l$ and the unphysical
  configurations encoded in $m$. This implies the existence of two frames
  separated by $\eta$.

  What events are simultaneous depends on $B_0$ and the choice of event, which
  here is taken to be $O \in \mathbb{P}^{2}$. In this article, if a line
  described by velocities is incident to some point, it means that one can
  partition the velocities described by that line in some manner depending on
  the point, see \cref{sec:transf-degenerate-conic}. For instance, in
  \cref{eq:Q-alg-line}, the velocities can be distributed such that they are
  coefficients of $A_w$ and $A_{n}$ only, which is the result of picking $O$ as
  an event.

  As $l = l \left( S_w \right)$ varies, the points that are simultaneous with
  $O$ lie on $m$. $m$ then contains all possible partitions equivalent with the
  one of $O$ that corresponds to the observed velocities $v_w, v_n, v$. $O$ is a
  convenient partitioning for the problem in this paper; every line incident to
  $O$ has a representation in which $v$ can be absorbed into the seepage
  velocities. The phase velocities are then defined relative to $v$. This is
  exactly how the \textit{drift velocities} in two-phase flow are defined
  \cite{zuberAverageVolumetricConcentration1965}. Thus, one can encode the fact
  that the average or ``center of mass'' velocity $v$ is ignored in the
  analysis: only velocities relative to $v$ matter. Since this is all defined
  with respect to $B_0$, which implies picking a frame where $v$ is dropped from
  the dependent set of velocities and the seepage velocities are given different
  signs, $v$ simply disappears from the partitioning described by $O$. In fact,
  one can consider any point $O^{\prime}$ giving some arbitrary partition of the
  velocities. With respect to $\left( A_0, B_0 \right)$, the partitioning is the
  same with $v$ set to zero and a minus sign added in front of $v_n$.

  Note that these equivalent partitions of the line $m$ are not physically
  measurable velocities: they are generally inaccessible due to the hyperbolic
  orthogonality. However, since $m$ is defined with respect to $B_0$ and not
  $B$, the intersection $M$ can land within the segment defined by $I,J$ on
  $l_{\infty}$. With respect to the conic $B_{0}$, the orthogonal direction of $l$
  through $O$, $m$, is the line $\left( v_n, v_w, 0 \right)^{T}$, as has already
  been shown. The transformation of $B$ into the standard form $B_0$ allowed for
  defining $A \mapsto A_0$. The action of $B_0$ on the only finite line of $\mathbb{P}^{2}$
  that was introduced from the velocity relations in \cref{sec:rea}, namely $l$,
  produced the pole $l^{\ast}_{0} = \left( v_w, -v_n, 0 \right)$. The join $O \times
  l^{\ast}_{0}$ defined $m$, which is orthogonal to $l$ with respect to the conic
  $B_0$ because $l^{T}B_0m = 0$. The intersection of $m$ with $l_{\infty}$ gave the
  point $M$. The point $M \in l_{\infty}$ is thus a representation of the orthogonal
  direction of $l$ with respect to $B_0$. This encodes a frame where the average
  velocity is $0$ and the fluids move in two different directions.

  As mentioned earlier, the geometry on $\mathbb{P}^{2} \setminus \mathcal{C}$ (where $\mathcal{C}$ is either one of
  the conics) is pseudo-Euclidean. On these sets, one then must have all the
  features of such a geometry, including an analogy of Lorentz-transformations
  \cite{misnerGravitation1973}. These transformation mix the velocities of the
  two fluids in the same way as space and time is intertwined via Lorentz-boosts
  in Minkowski spacetime. In this paper, the projective viewpoint is kept, so
  these types of transformations are not used. This is simply because only the
  physical case of interest is where one has area conservation. Thus, all point-
  and line-configurations have to be represented on $l_{\infty}$, and all
  configurations are obtained from reasonable joins of points or intersections
  of lines. What the pseudo-Euclidean geometry provides in this context is a
  means of expressing equivalent partitions of the velocities with respect to
  some frame. This can be seen as a formalization of the manipulation in
  \cref{eq:Q-alg-line}. The frame is being implied by the conic, and do not
  represent the frames encountered in special relativity; here, they are
  projective frames, where the representative points/lines are given by the
  columns of the transformation matrices
  in~\cref{sec:degenerate-quadratic-form}.

  \subsection{Alternative geometric formulations, thermodynamics}
  \label{sec:alternative-geo-form}

  The usage of projective spaces is not prevalent in porous media research, but
  it \textit{is} in geometric formulations of both equilibrium- and
  non-equilibrium thermodynamics
  \cite{vanderschaftLiouvilleGeometryClassical2021,
    vanderschaftLiouvilleGeometryClassical2021,
    vanderschaftGeometryThermodynamicProcesses2018,
    balianHamiltonianStructureThermodynamics2001,
    arnoldMathematicalMethodsClassical1978}. However, the projective viewpoint
  in this work is not equivalent to one applied in thermodynamics. There, one
  does not choose the extensive variables as done here, i.e. $(A_w, A_n, A_p)$
  with dual variables given by the velocities, and then intersect with the
  surface defined by $A_w + A_n - A_p$. Instead, one exchanges $A_p$ for $Q$,
  which has no thermodynamic conjugate variable. Equilibrium properties of the
  system are obtained by defining the state function $Q = Q \left( A_w, A_n
  \right)$, which gives the intensive, conjugate variables $\left\{ \hat{v}_i
  \right\}$ and their equations of state, \cref{eq:vw-therm,eq:vn-therm}.
  Extensivity is used to scale all extensive properties by the
  \textit{dependent} variable $A_p$, the total amount of area in the REA. By
  only including two pore areas as variables, this paper is working with an
  incomplete set of thermodynamic variables. Most importantly, $Q$ should also
  be a function of the \textit{configurational entropy} or \textit{cluster
    entropy} $\Sigma$ \cite{hansenStatisticalMechanicsFramework2023}, in
  addition to $A_w, A_n$. $\Sigma$ enters the statistical mechanics of the
  problem via a maximum entropy principle
  \cite{jaynesInformationTheoryStatistical1957}. Homogeneity of $Q$ under
  scaling of $\Sigma$ is a topic that is outside the scope of this article.
  
  In a thermodynamic parlance, $v_m$ relates partial quantities to specific
  quantities, $\left\{ \hat{v}_i \right\}$ and $\left\{ v_i \right\}$
  respectively. Eq.~\eqref{eq:v-both-definitions} can be stated as (normalized
  by $A_p$)
  \begin{equation}
    \label{eq:mixing}
    S_w \left( \hat{v}_w - v_w \right) + S_n \left( \hat{v}_n - v_n \right) = 0 \;.
  \end{equation}
  The velocities $\left\{ v_i \right\}$ are ratios of extensive quantities
  pertaining to one of the fluids, i.e. a ``pure'' substance. Thus,
  \cref{eq:mixing} is completely analogous to how one define properties of
  mixing in thermodynamics
  \cite{callenThermodynamicsIntroductionThermostatistics1985}. \Cref{eq:mixing}
  states that there is no change in pore velocity $v$ due to mixing of the two
  fluids, i.e $\Delta v_{\mathrm{mix}} = 0$. Extending this analogy,
  \cref{eq:vw-transf,eq:vn-transf} define \textit{excess properties} of the
  fluid mixture with total velocity $v$, so that $S_n v_m$, $-S_wv_m$ are
  \textit{excess partial velocities} of the wetting- and non-wetting fluids
  respectively. They can be understood as the deviation from a non-interacting
  flow where the two fluids do not interact, analogous to a ideal solution in
  thermodynamics. Note that the above analogy is only possible to state given
  that the other intensive quantities of the system is kept constant, which is
  not controlled in this paper. For instance, if the cluster entropy $\Sigma$
  was included in the formalism, one would need to define how its conjugate
  variable, the \textit {agiture}
  \cite{hansenStatisticalMechanicsFramework2023}, is held fixed in this system.

  The formalism presented in this work is quite general, and do not hinge on
  special assumptions about two-phase flow in porous media; every system where
  \cref{eq:Q-homogeneity} is fulfilled can be treated in the same framework.
  This paper treats the simplest possible case of two extensive variables. This
  was possible because the seepage velocities are functions of $S_w$ only,
  meaning they could be placed on a single affine line. More variables requires
  a consideration of the geometric configurations they form. For instance, do
  they lie in a special configuration such as along a line, a common plane
  (where the cross ratio can become complex
  \cite{richtergebertPerspectivesProjectiveGeometry2011}), or along some other
  type of curve or within some subset. A simple use case like the one presented
  here might not be found in such a system, and other methods might then be more
  fruitful.

  \subsection{Affine slip ratio, Legendre transformation}
  \label{sec:slip-ratio}

  It is clear that it is the functional form of $\tau = -v_n/v_{w}$ that
  fundamentally allow for an effective parametrization of $v_n - v_w$. $v_w$ and
  $v_n$ individually can be complicated functions, however, the two are related
  through the average flow velocity $v$. Equivalently, one can know the relative
  velocity $v_n - v_w$, or any of the drift velocities $\left\{ v_i - v
  \right\}$. The relative velocity $v_n - v_w$ often has a simpler functional
  form than $v_w, v_{n}$ individually.

  $\tau$ turns out to be very well approximated by an affine function in
  $\left\{ S_i \right\}$, or in some cases by an affine function in $\left\{
    S_i^{-1} \right\}$, in some saturation range $\in \left[ 0,1 \right]$. Both
  gives the same conclusion in determining why the parametrization works, so
  only the former case is treated here. Hence, this means that $\tau$ is
  modelled as $\tau \sim A S_w + B$ for some numbers $A$, $B$. This means that
  $S_w \tau \sim A S_w^{2} + B S_w$, which the equation of a parabola, a conic
  section. As explained in \cref{sec:five-point-conic}, a conic $\mathcal{C}$
  can be seen as a representation of a projective line. Equivalently, one has
  that $S_n \tau^{-1} \sim S_n/(A S_w + B) $, which has the form of a hyperbola.
  This is clearer if one writes
  \begin{equation}
    \label{eq:hyperbola}
    S_n \tau^{-1} = \frac{S_n}{A S_w + B} = \frac{1 + A^{-1}B}{AS_w + B} - A^{-1} \;,
  \end{equation}
  which is an affinely transformed hyperbola. A point of projection $p$ is then
  selected as in sec. \ref{sec:second-projection}, and the points of the conic
  are projected onto the line defined by $\left( S_w, S_n \right)$ via lines
  through $p$, see \cref{fig:homography-L} for an example between two lines (the
  concept here is the same). For this to work, the point of projection discussed
  in sec.~\ref{sec:second-projection} strictly speaking has to be placed on the
  conic $\mathcal{C}$ \cite{richtergebertPerspectivesProjectiveGeometry2011}.
  All of the lines of the line bundle at a point $p \in \mathcal{C}$ intersect
  the conic in one other point, and a arbitrary line in one point, which creates
  a correspondence between these two points. If $p$ is not on the conic, this is
  not the case.

  There is a connection between conics and the Legendre transform, which can be
  formulated in terms of a pole/polar-duality
  \cite{richtergebertPerspectivesProjectiveGeometry2011,
    pappasProjectiveQuadricsPoles1996}. The thermodynamic conjugate variable
  duality in thermodynamics is equivalent to interchanging a point on a
  parametrized curve (or the graph of a function) by the tangent hyperplane at
  that point, described by the gradient at the same point. The pole in this
  context is just the chosen point on the curve, and the polar line is the
  tangent line at that point. For instance, one can find that $\hat{v}_n$ is the
  Legendre transform of $v$ with respect to $S_w$,
  \begin{equation}
    \label{eq:legendre-vnhat}
    \hat{v}_n = v - S_w \frac{d v}{dS_w} \;.
  \end{equation}
  Here, $S_w$ is the parameter, $v(S_w)$ is the curve, and $\frac{d v}{d S_w} =
  v^{\prime}$ is the dual or \textit{conjugate} variable to $S_w$ \footnote{Note
    that one here has to require $v$ to be convex. If not, one must either
    consider the transformation locally, or generalize to the Legendre-Fenchel
    transformation.}. Similarly, one can define $\hat{v}_w$ as the Legendre
  transformation of $v$ with respect to $S_n$.

  It is clear that the actual physics of the problem has to be related to the
  affine approximation of $\tau = - \mathcal{S} = - v_n / v_{w}$, which is
  equivalent to asking about the functional form of the slip-ratio
  $\mathcal{S}$. \textbf{Hence, the problem of the functional form of $v_m$ is
    related to that of the functional form of $\mathcal{S}$}. The slip ratio can
  be related to different constitutive relations in the two-phase flow
  literature depending on the system. One example is dispersed/bubbly flow
  \cite{croweMultiphaseFlowHandbook2005}, where the relation between
  $\mathcal{S}$ and the Stokes number $\mathrm{Stk}$ is approximately
  \begin{equation}
    \label{eq:slip-stokes}
    \mathcal{S} \approx \frac{1}{1 + \mathrm{Stk}} \;.
  \end{equation}

  \subsection{Other interpretations of $v_m$}
  \label{sec:interpretation-vm}

  The approach taken in this article was to only parametrize the term $v_n -
  v_w$ in eq. \eqref{eq:vm-derivative-form}. The velocities $\left\{ \hat{v}_i
  \right\}$, $\left\{ v_i \right\}$ enter the projective formalism as components
  of a hyperplane, i.e. a projective line. $v_m$ appeared in intersections of
  these lines, where it is scaled out. Hence, at each point of $\mathbb{P}^2$,
  we have a map that picks out a vector with an overall factor of $v_m$, so it
  is a map $\mathbb{P}^2 \rightarrow V = \mathbb{R}^{3}$ to the vector space
  $V$. Such a map is called a \textit{section $s$ of a tautological line bundle}
  $E$ \cite{leeIntroductionSmoothManifolds2012,
    spivakComprehensiveIntroductionDifferential1999}. The moniker
  ``tautological'' refers to the fact that the fiber at each point $x \in
  \mathbb{P}^{2}$ is the vector space (or rather line) $W \subset V$
  corresponding to the point $x = p \left( W \right)$ itself. If one denotes the
  homogeneous coordinates of the line $W$ as $\left[ X_0:X_1:X_2 \right]$, the
  sections $s \left( W \right) = \vec{v}$ over $\mathbb{P}^{2}$ has the form
  \begin{equation}
    \label{eq:section-line-bundle}
    s \left( \left[ X_0:X_1:X_2 \right] \right) = g \left( X_0,X_1,X_2 \right) \left[ X_0:X_1:X_2 \right]
  \end{equation}
  However, this view is not what defines $v_m$, but rather just classifies $v_m$
  in this context.

  One can say, intuitively, that $v_m$ scales the vector corresponding to the
  direction of infinity. Since the points on $l_{\infty}$ looks like $\left[
    S_w:S_n:1 \right] = \left( S_w, S_n \right)$ up to scaling, or just $S_w$
  since $S_w, S_n$ are dependent, the tangent vector basis of $l_{\infty}$ is
  just $\partial_{S_w}$. A general tangent vector at each point of $l_{\infty}$
  looks like $f \left( S_w \right) \partial_{S_w}$ for some non-zero (possibly
  constant) function $f(S_w)$. A tangent vector to the line at infinity is then
  transversal to the direction indicated by the points on the line with
  direction $\left( S_w, S_n \right)$.

  It is clear that the relations that govern the theory are not those of
  thermodynamics, since the seepage velocities are not thermodynamic variables.
  Thus, in an effort to parametrize $v_m$, a kinematic view was applied, wherein
  the cross-ratio $k$ was central. It has been shown how the Cayley-Klein
  geometries produces metric spaces, where $k$ defines a notion of distance and
  angle, hence a metric. This metric is what completely determines $v_m$. One
  way of extending on the idea of a metric \footnote{This metric is different to
    the ones encountered in e.g. contact geometry
    \cite{grmelaContactGeometryMesoscopic2014}.} is to do so within a
  differential geometric context, outlined in \cite{pedersen_geometric_2025}.
  The Cayley-Klein geometries can produce metric spaces, e.g. Euclidean or
  pseudo-Euclidean, but this metric is not defined on the conic representing the
  infinite objects.

  \subsection{Noisy data, relative permeability data}
  \label{sec:noisy-data}

  For DNM-datasets with low $\Delta P$ and/or noisy data, the homography estimation
  might fail in finding parameters that produces a satisfying fit. In the
  derivation of the routine, $J$ was a fixed point of the map $\left[ S_w:S_n
  \right] \mapsto \left[ \tau S_w:S_n \right]$. When $S_w \rightarrow 1$, $S_w/S_n \rightarrow \infty$. This
  behaviour should be preserved in $\tau S_w/S_n$, which means that in terms of the
  saturations $S_{w}, S_n$ as affine coordinates, $S_w = 1$ is mapped to $\infty$.
  This can be checked by verifying that
  \begin{equation}
    \label{eq:infty-to-infty}
    \left( -\frac{d}{c} \right)^{\ast} \approx \left( \frac{a}{c} \right)^{\ast} \approx 1 \;.
  \end{equation}
  where the asterisk means that $-d/c, a/c$ are permuted as in
  \cref{eq:send-ratio} to correspond with saturations. Adjusting $p^{\ast}$, often
  by small values in the saturations ($\sim 0.1$), can give substantially better
  fits. Since the homography is computed by setting up a correspondence between
  $\left[ \tau S_w:S_n \right]$ and $\left[ S_w:S_n \right]$ via point-projection
  through some $p^{\ast}$, a source of difficulty in the estimation routine comes
  from certain points or ranges of points switches order when the projection is
  performed, or when many points are mapped to a single point. This can happen
  for instance when $\tau$ contains large jumps. It is especially clear from the
  relative permeability data that the presence of irreducible saturations can
  significantly affect the goodness of fit, and the test
  \cref{eq:infty-to-infty} might not be a good indicator in this case. If these
  irreducible saturations should be handled in general by incorporating them in
  the geometry (i.e. in the points $I,J$ in \cref{eq:I-J}), in the point of
  projection $p^{\ast}$, in the coordinate change in \cref{sec:relperm-hydr}, or
  some combination of these three is currently unclear, and requires a more
  rigorous analysis.


  In the recreated relative permeability data, note that the fit of $v_{m,p}$ to
  $v_m$ can be almost exact, while the recreated relative permeability curves do
  not necessarily overlap the data completely \footnote{Given the average
    velocity, the co-moving velocity \textit{exactly} reproduces the seepage
    velocities from the thermodynamic velocities.}. Thus, the source of the
  discrepancy cannot lie in the procedure presented in this paper, but must be
  attributed to the definition of the velocity from the relative permeability
  itself (\cref{eq:relperm-constitutive}), or with the values of the physical
  parameters during the experimental procedure in the datasets. In the relative
  permeability data, the pressure drop was not always held constant, while in
  \cref{fig:relperm}, an average pressure was used. However, the value of the
  pressure itself does not enter the computation of the cross-ratio estimate
  $k_e$, as only ratios of velocities matter. What does matter is the ratio $M$,
  since
  \begin{equation}
    \label{eq:relperm-ratio}
    \mathcal{S} \ = \ \frac{v_n}{v_w} \ = \ \frac{k_{rn}}{k_{rw}}\frac{S_w\mu_w}{S_n\mu_{n}} 
  \end{equation}
  from the constitutive relation in \cref{eq:relperm-constitutive}. If $M$ is
  not constant as a function of $S_w$, this has an impact on
  \cref{eq:relperm-ratio} and hence the recreated curves. The effect of altering
  the viscosity is illustrated in \cref{fig:relperm}.

  Interestingly, \cref{eq:beta-hydr} have several possible interpretations.
  First of, $\beta$ might be similar to a reflection coefficient of some abstract
  wave interacting with a medium. The conductivities $\left\{ \mathcal{K}_i \right\}$ are
  then the inverses of the corresponding hydraulic resisitivities. These are
  special cases of impedances, which are often used in expressing reflection
  coefficients. Here, $\left\{ \mathcal{K}_i \right\}$ are functions of $S_w$, so the
  reflection is parametrized by $S_w$. However, this implies some wave equation
  which at present is not possible to define sensibly, so the hydraulic analogy
  is dubious to apply without more information. Instead, the identification $v_n
  - v_w = \beta v$ allow for interpreting \cref{eq:vm-relperm-hydr} as a form of
  balance equation, where $v$ is the flow velocity. Since steady state flow is
  assumed, time-dependence do not enter, so \cref{eq:vm-k-derivative-form} and
  \cref{eq:vm-relperm-hydr} are really steady flow advection equations,
  potentially with source/sink terms depending on how one chooses to label the
  terms. For instance, one could rewrite \cref{eq:vm-k-derivative-form} as
  \begin{align}
    \label{eq:advect-eq}
    \psi v_m - \partial_{S_w} \left( \psi v \right) \ =& \ \psi v_m - \psi v^{\prime} - \psi^{\prime} v = 0 \nonumber \\
    =& \ v_m - v^{\prime} - \frac{\psi^{\prime}}{\psi} v \;,
  \end{align}
  where a prime indicates a derivative with respect to $S_w$ and $\psi$ would
  correspond to a type of density. One could then set
  \begin{equation}
    \label{eq:beta-log-derivative}
    \beta = \frac{kS_n - S_w}{kS_n^2 + S_w^2} =\frac{ \psi^{\prime}}{\psi }= \partial_{S_w}\left( \ln{\psi} \right)
  \end{equation}
  by comparing with \cref{eq:vm-k-derivative-form}. $\beta v$ is then analogous to
  an advection term, while $v_m$ must be a source/sink-term, however, whether
  this is a sensible way of viewing \cref{eq:vm-k-derivative-form} requires
  further analysis.

  \section{Conclusion}
  \label{sec:conclusion}

  In this article, the principles of a novel geometric view of immiscible
  two-phase flow in porous media was presented, and a constitutive relation for
  the co-moving velocity was obtained in this framework. This relation expressed
  $v_m$ in terms of the protectively invariant cross-ratio, and a procedure for
  parametrizing this function was outlined. The formalism builds upon the
  investigations in \cite{pedersenParameterizationsImmiscibleTwophase2023b,
    pedersen_geometric_2025}.

  Relations between pore areas and velocities were re-interpreted as elements of
  a vector space up to scaling, and the projective plane was introduced as the
  ambient space of these objects. The dependency between the pore-areas and the
  saturation limits were used to define a degenerate primal/dual pair of conics
  in the projective plane. This pair defined a pseudo-Euclidean Cayley-Klein
  model. A transformation to a simpler dual conic was introduced, motivated by
  removing $v$ from our dependent set of velocities and distinguishing the
  seepage velocities by their signs. It was then showed that the infinite points
  of the model geometry, a line defined by the seepage velocities and its
  orthogonal line with respect to the transformed conic defined a simple
  cross-ratio function where the slip-ratio of velocities entered.

  Using the points and lines of the projective plane, a projective
  transformation in the homogeneous coordinates between the points of two lines
  was introduced. To arrive at these homogeneous coordinates, a
  perspectivity was defined in \cref{eq:Sn-tilde}, explained in
  \cref{sec:cross-ratio}. The homogeneous coordinates then mapped to
  $\left[S_w:S_{n,\tau} \right]$. This perspectivity was composed with a
  projection via lines through the point $p^{\ast}$ to obtain the homogeneous
  coordinates $\left[ \underline{S}_w:\underline{S}_n \right]$, which summed to
  unity. To show that the map $\left[ S_w:S_n \right] \mapsto \left[
    \underline{S}_w:\underline{S}_n \right]$ is well described by a homography,
  the parameters of the homography was estimated from DNM- and relative
  permeability data. This defined a pair of estimated saturations $\tilde{S}_w, \tilde{S}_{n}$, which were used to
  parametrize $v_m$ through \cref{eq:vm-solved-from-k}. The estimates provided
  good overall fits for most of the available datasets, with possibilities of
  adjustments and further simplification by setting one of the parameters to
  zero. This latter situation leaves just two parameters, which in most cases
  provided almost exactly the same results as with three for the DNM-data. A linear approximation
  in $\beta v$ as a function of $k_{e}$ is also possible, but gives almost identical
  graphs to the linear fit of $v_m \left( v^{\prime} \right)$. Moreover, for two
  parameters, the fit in terms of $k_e$ with provides an overall better fit. The
  estimation procedure keeps track of the order of points, and allows for
  curvature in the fits of $v_m \left( v^{\prime} \right)$. Thus, the
  ``slingshot-effect'' often seen in these graphs are covered by this procedure.
  The general procedure should be applicable for effective descriptions of a
  two-component system modeled as extensive in a pair of macroscopic variables.
  Possible avenues of exploration is then to investigate other similar systems
  with the same procedure. Furthermore, the set of thermodynamic variables
  considered in this effective description is incomplete. It would be
  interesting to see how the effective theory is modified upon introduction of
  more variables like the configurational entropy.

\bigskip

\textbf{Acknowledgments}

The author thanks Alex Hansen and Santanu Sinha for helpful discussions and
input, and Santanu Sinha for providing the pore network model data.

\textbf{Funding}

This work was partly supported by the Research Council of Norway through its
Centres of Excellence funding scheme, project number 262644.

\textbf{Data availability}

The dynamical pore network model data is used in \doi{10.3389/fphy.2020.548497},
and available from the authors upon request. A script used for the computations
in this article can be found at \url{https://github.com/hakkped/VmProjective.jl}.


\bibliographystyle{ieeetr}
\bibliography{main.bib}

\begin{thebibliography}{10}

\bibitem{bearDynamicsFluidsPorous1988}
J.~Bear, {\em Dynamics of Fluids in Porous Media}.
\newblock Dover Books on Physics and Chemistry, {New York}: {Dover}, 1988.

\bibitem{bluntMultiphaseFlowPermeable2017}
M.~J. Blunt, {\em Multiphase {{Flow}} in {{Permeable Media}}: {{A Pore-Scale
  Perspective}}}.
\newblock {Cambridge}: {Cambridge University Press}, 2017.

\bibitem{federPhysicsFlowPorous2022}
J.~Feder, E.~G. Flekk{\o}y, and A.~Hansen, {\em Physics of {{Flow}} in {{Porous
  Media}}}.
\newblock {Cambridge}: {Cambridge University Press}, 2022.

\bibitem{andersonMoreDifferent1972}
P.~W. Anderson, ``More {{Is Different}},'' {\em Science (New York, N.Y.)},
  vol.~177, pp.~393--396, Aug. 1972.

\bibitem{wyckoffFlowGasLiquid1936}
R.~D. Wyckoff and H.~G. Botset, ``The {{Flow}} of {{Gas}}-liquid {{Mixtures
  Through Unconsolidated Sands}},'' {\em Physics}, vol.~7, no.~9, pp.~325--345,
  1936.

\bibitem{leverettCapillaryBehaviorPorous1941}
M.~Leverett, ``Capillary {{Behavior}} in {{Porous Solids}},'' {\em Transactions
  of the AIME}, vol.~142, pp.~152--169, Dec. 1941.

\bibitem{leverettFlowOilWaterMixtures1939}
M.~C. Leverett, ``Flow of {{Oil-Water Mixtures Through Unconsolidated
  Sands}},'' {\em Transactions of the AIME}, vol.~132, no.~01, pp.~149--171,
  1939.

\bibitem{grayIntroductionThermodynamicallyConstrained2014}
W.~G. Gray and C.~T. Miller, {\em Introduction to the {{Thermodynamically
  Constrained Averaging Theory}} for {{Porous Medium Systems}}}.
\newblock Advances in {{Geophysical}} and {{Environmental Mechanics}} and
  {{Mathematics}}, {Cham}: {Springer International Publishing}, 2014.

\bibitem{hassanizadehMechanicsThermodynamicsMultiphase1990}
S.~M. Hassanizadeh and W.~G. Gray, ``Mechanics and thermodynamics of multiphase
  flow in porous media including interphase boundaries,'' {\em Advances in
  Water Resources}, vol.~13, pp.~169--186, Dec. 1990.

\bibitem{kjelstrupNonisothermalTransportMultiphase2018}
S.~Kjelstrup, D.~Bedeaux, A.~Hansen, B.~Hafskjold, and O.~Galteland,
  ``Non-isothermal {{Transport}} of {{Multi-phase Fluids}} in {{Porous Media}}.
  {{The Entropy Production}},'' {\em Frontiers in Physics}, vol.~6, p.~126,
  Nov. 2018.

\bibitem{kjelstrupNonisothermalTransportMultiphase2019}
S.~Kjelstrup, D.~Bedeaux, A.~Hansen, B.~Hafskjold, and O.~Galteland,
  ``Non-isothermal {{Transport}} of {{Multi-phase Fluids}} in {{Porous Media}}.
  {{Constitutive Equations}},'' {\em Frontiers in Physics}, vol.~6, p.~150,
  Jan. 2019.

\bibitem{valavanidesMechanisticModelSteadyState1998}
M.~S. Valavanides, G.~N. Constantinides, and A.~C. Payatakes, ``Mechanistic
  {{Model}} of {{Steady-State Two-Phase Flow}} in {{Porous Media Based}} on
  {{Ganglion Dynamics}},'' {\em Transport in Porous Media}, vol.~30,
  pp.~267--299, Mar. 1998.

\bibitem{valavanidesReviewSteadyStateTwoPhase2018}
M.~S. Valavanides, ``Review of {{Steady-State Two-Phase Flow}} in {{Porous
  Media}}: {{Independent Variables}}, {{Universal Energy Efficiency Map}},
  {{Critical Flow Conditions}}, {{Effective Characterization}} of {{Flow}} and
  {{Pore Network}},'' {\em Transport in Porous Media}, vol.~123, pp.~45--99,
  May 2018.

\bibitem{valavanidesOilFragmentationInterfacial2018}
M.~S. Valavanides, ``Oil {{Fragmentation}}, {{Interfacial Surface Transport}}
  and {{Flow Structure Maps}} for {{Two-Phase Flow}} in {{Model Pore
  Networks}}. {{Predictions Based}} on {{Extensive}}, {{{\emph{DeProF}}}}
  {{Model Simulations}},'' {\em Oil \& Gas Sciences and Technology
  \textendash{} Revue d'IFP Energies nouvelles}, vol.~73, p.~6, 2018.

\bibitem{mcclureGeometricStateFunction2018}
J.~E. McClure, R.~T. Armstrong, M.~A. Berrill, S.~Schl{\"u}ter, S.~Berg, W.~G.
  Gray, and C.~T. Miller, ``Geometric state function for two-fluid flow in
  porous media,'' {\em Physical Review Fluids}, vol.~3, p.~084306, Aug. 2018.

\bibitem{armstrongPorousMediaCharacterization2019}
R.~T. Armstrong, J.~E. McClure, V.~Robins, Z.~Liu, C.~H. Arns, S.~Schl{\"u}ter,
  and S.~Berg, ``Porous {{Media Characterization Using Minkowski Functionals}}:
  {{Theories}}, {{Applications}} and {{Future Directions}},'' {\em Transport in
  Porous Media}, vol.~130, pp.~305--335, Oct. 2019.

\bibitem{schroder-turkMinkowskiTensorsAnisotropic2013}
G.~E. {Schr{\"o}der-Turk}, W.~Mickel, S.~C. Kapfer, F.~M. Schaller,
  B.~Breidenbach, D.~Hug, and K.~Mecke, ``Minkowski tensors of anisotropic
  spatial structure,'' {\em New Journal of Physics}, vol.~15, p.~083028, Aug.
  2013.

\bibitem{hansenRelationsSeepageVelocities2018}
A.~Hansen, S.~Sinha, D.~Bedeaux, S.~Kjelstrup, M.~A. Gjennestad, and
  M.~Vassvik, ``Relations {{Between Seepage Velocities}} in {{Immiscible}},
  {{Incompressible Two-Phase Flow}} in {{Porous Media}},'' {\em Transport in
  Porous Media}, vol.~125, pp.~565--587, Dec. 2018.

\bibitem{royFlowAreaRelationsImmiscible2020}
S.~Roy, S.~Sinha, and A.~Hansen, ``Flow-{{Area Relations}} in {{Immiscible
  Two-Phase Flow}} in {{Porous Media}},'' {\em Frontiers in Physics}, vol.~8,
  p.~4, Jan. 2020.

\bibitem{royCoMovingVelocityImmiscible2022}
S.~Roy, H.~Pedersen, S.~Sinha, and A.~Hansen, ``The co-moving velocity in
  immiscible two-phase flow in porous media,'' {\em Transport in Porous Media},
  vol.~143, no.~1, pp.~69--102, 2022.

\bibitem{hansenStatisticalMechanicsFramework2023}
A.~Hansen, E.~G. Flekk{\o}y, S.~Sinha, and P.~A. Slotte, ``A statistical
  mechanics framework for immiscible and incompressible two-phase flow in
  porous media,'' {\em Advances in Water Resources}, vol.~171, p.~104336, Jan.
  2023.

\bibitem{jaynesInformationTheoryStatistical1957}
E.~T. Jaynes, ``Information {{Theory}} and {{Statistical Mechanics}},'' {\em
  Physical Review}, vol.~106, pp.~620--630, May 1957.

\bibitem{leeIntroductionSmoothManifolds2012}
J.~M. Lee, {\em Introduction to {{Smooth Manifolds}}}, vol.~218 of {\em
  Graduate {{Texts}} in {{Mathematics}}}.
\newblock {New York}: {Springer New York}, 2012.

\bibitem{milnorCharacteristicClasses1974}
J.~W. Milnor and J.~D. Stasheff, {\em Characteristic Classes}.
\newblock No.~no. 76 in Annals of Mathematics Studies, {Princeton, N.J}:
  {Princeton University Press}, 1974.

\bibitem{hansenFiberBundleModel2015}
A.~Hansen, P.~C. Hemmer, and S.~Pradhan, eds., {\em The {{Fiber Bundle Model}}:
  {{Modeling Failure}} in {{Materials}}}.
\newblock {Wiley}, first~ed., Oct. 2015.

\bibitem{arnoldMathematicalMethodsClassical1978}
V.~I. Arnold, {\em Mathematical {{Methods}} of {{Classical Mechanics}}},
  vol.~60 of {\em Graduate {{Texts}} in {{Mathematics}}}.
\newblock {New York}: {Springer New York}, 1978.

\bibitem{bravettiContactGeometryThermodynamics2019}
A.~Bravetti, ``Contact geometry and thermodynamics,'' {\em International
  Journal of Geometric Methods in Modern Physics}, vol.~16, p.~1940003, Feb.
  2019.

\bibitem{bergerGeometry1994}
M.~Berger, {\em Geometry}.
\newblock Universitext, {Berlin ; New York}: {Springer-Verlag}, corr. 2nd
  print~ed., 1994.

\bibitem{pedersenParameterizationsImmiscibleTwophase2023b}
H.~Pedersen and A.~Hansen, ``Parameterizations of immiscible two-phase flow in
  porous media,'' {\em Frontiers in Physics}, vol.~11, Feb. 2023.

\bibitem{garcia-arizaGeometricApproachConcept2019}
M.~{\'A}. {Garc{\'i}a-Ariza}, ``A {{Geometric Approach}} to the {{Concept}} of
  {{Extensivity}} in {{Thermodynamics}},'' {\em Symmetry, Integrability and
  Geometry: Methods and Applications}, Mar. 2019.

\bibitem{pedersen_geometric_2025}
H.~Pedersen and A.~Hansen, ``Geometric structure of parameter space in
  immiscible two-phase flow in porous media,''

\bibitem{vanderschaftLiouvilleGeometryClassical2021}
A.~{van der Schaft}, ``Liouville geometry of classical thermodynamics,'' {\em
  Journal of Geometry and Physics}, vol.~170, p.~104365.

\bibitem{misnerGravitation1973}
C.~W. Misner, K.~S. Thorne, and J.~A. Wheeler, {\em Gravitation}.
\newblock {San Francisco}: {W. H. Freeman}, 1973.

\bibitem{balianHamiltonianStructureThermodynamics2001}
R.~Balian and P.~Valentin, ``Hamiltonian structure of thermodynamics with
  gauge,'' {\em The European Physical Journal B - Condensed Matter and Complex
  Systems}, vol.~21, pp.~269--282, May 2001.

\bibitem{casas-alveroAnalyticProjectiveGeometry2014}
E.~{Casas-Alvero}, ``Analytic {{Projective Geometry}}.''
  https://ems.press/books/etb/211, May 2014.

\bibitem{richtergebertPerspectivesProjectiveGeometry2011}
J.~{Richter-Gebert}, {\em Perspectives on {{Projective Geometry}}: {{A Guided
  Tour Through Real}} and {{Complex Geometry}}}.
\newblock {Berlin, Heidelberg}: {Springer}, 2011.

\bibitem{stillwellFourPillarsGeometry2005}
J.~Stillwell, {\em The {{Four Pillars}} of {{Geometry}}}.
\newblock Undergraduate {{Texts}} in {{Mathematics}}, {New York, NY}:
  {Springer}, 2005.

\bibitem{sinhaFluidMeniscusAlgorithms2021}
S.~Sinha, M.~A. Gjennestad, M.~Vassvik, and A.~Hansen, ``Fluid {{Meniscus
  Algorithms}} for {{Dynamic Pore-Network Modeling}} of {{Immiscible Two-Phase
  Flow}} in {{Porous Media}},'' {\em Frontiers in Physics}, vol.~8, 2021.

\bibitem{bennionRelativePermeabilityCharacteristics2005}
B.~Bennion and S.~Bachu, ``Relative {{Permeability Characteristics}} for
  {{Supercritical CO2 Displacing Water}} in a {{Variety}} of {{Potential
  Sequestration Zones}},'' in {\em {{SPE Annual Technical Conference}} and
  {{Exhibition}}}, ({Dallas, Texas}), {Society of Petroleum Engineers}, 2005.

\bibitem{virnovskyImplementationMultirateTechnique1998}
G.~Virnovsky, K.~Vatne, S.~SkjaeveLand, and A.~Lohne, ``Implementation of
  {{Multirate Technique}} to {{Measure Relative Permeabilities Accounting}},''
  {\em All Days}, pp.~SPE--49321--MS, Sept. 1998.

\bibitem{oakThreePhaseRelativePermeability1990}
M.~J. Oak, L.~E. Baker, and D.~C. Thomas, ``Three-{{Phase Relative
  Permeability}},'' p.~8.

\bibitem{reynoldsCharacterizingFlowBehavior2015}
C.~A. Reynolds and S.~Krevor, ``Characterizing {{Flow Behavior}} for {{Gas
  Injection}}: {{Relative Permeability}} of {{Co2-brine}} and {{N2-water}} in
  {{Heterogeneous Rocks}},'' {\em Water Resources Research}, vol.~51, no.~12,
  pp.~9464--9489, 2015.

\bibitem{fulcherEffectCapillaryNumber1985}
R.~Fulcher, T.~Ertekin, and C.~Stahl, ``Effect of {{Capillary Number}} and
  {{Its Constituents}} on {{Two-Phase Relative Permeability Curves}},'' {\em
  Journal of Petroleum Technology}, vol.~37, pp.~249--260, Feb. 1985.

\bibitem{hansenThermodynamicslikeFormalismImmiscible2025}
A.~Hansen and S.~Sinha, ``Thermodynamics-like {{Formalism}} for {{Immiscible}}
  and {{Incompressible Two-Phase Flow}} in {{Porous Media}},'' {\em Entropy},
  vol.~27, p.~121, Feb. 2025.

\bibitem{hansen2024linearity}
A.~Hansen, ``Linearity of the co-moving velocity,'' vol.~151, no.~13,
  pp.~2477--2489.

\bibitem{gallierGeometricMethodsApplications2011}
J.~Gallier, {\em Geometric {{Methods}} and {{Applications}}: {{For Computer
  Science}} and {{Engineering}}}, vol.~38 of {\em Texts in {{Applied
  Mathematics}}}.
\newblock {New York}: {Springer New York}, 2011.

\bibitem{crampinApplicableDifferentialGeometry1987}
M.~Crampin and F.~A.~E. Pirani, {\em Applicable {{Differential Geometry}}}.
\newblock London {{Mathematical Society Lecture Note Series}}, {Cambridge}:
  {Cambridge University Press}, 1987.

\bibitem{rockafellarConvexAnalysis1997}
R.~T. Rockafellar, {\em Convex Analysis}.
\newblock Princeton {{Landmarks}} in Mathematics and Physics, {Princeton, NJ}:
  {Princeton Univ. Press}, 10. print. and 1. paperb. print~ed., 1997.

\bibitem{arnoldOrdinaryDifferentialEquations1992}
V.~I. Arnol'd and V.~I. Arnol'd, {\em Ordinary Differential Equations}.
\newblock Springer Textbook, {Berlin Heidelberg}: {Springer}, 1992.

\bibitem{callenThermodynamicsIntroductionThermostatistics1985}
H.~B. Callen and H.~B. Callen, {\em Thermodynamics and an Introduction to
  Thermostatistics}.
\newblock {New York}: {Wiley}, 2nd ed~ed., 1985.

\bibitem{touchetteWhenQuantityAdditive2002}
H.~Touchette, ``When is a quantity additive, and when is it extensive?,'' {\em
  Physica A}, 2002.

\bibitem{oppenheimThermodynamicsLongrangeInteractions}
J.~Oppenheim, ``Thermodynamics with long-range interactions: {{From Ising}}
  models to black holes,'' {\em Physical Review E}, vol.~68, no.~1, p.~016108,
  2003.

\bibitem{fyhnLocalStatisticsImmiscible2023}
H.~Fyhn, S.~Sinha, and A.~Hansen, ``Local statistics of immiscible and
  incompressible two-phase flow in porous media,'' {\em Physica A: Statistical
  Mechanics and its Applications}, vol.~616, p.~128626, Apr. 2023.

\bibitem{belgiornoQuasiHomogeneousThermodynamicsBlack2005}
F.~Belgiorno, ``Quasi-{{Homogeneous Thermodynamics}} and {{Black Holes}},''
  tech. rep., Dec. 2005.

\bibitem{bravettiZerothLawQuasihomogeneous2017}
A.~Bravetti, C.~Gruber, C.~S. {Lopez-Monsalvo}, and F.~Nettel, ``The zeroth law
  in quasi-homogeneous thermodynamics and black holes,'' {\em Physics Letters
  B}, vol.~774, pp.~417--424, Nov. 2017.

\bibitem{anderssonComplexConvexityAnalytic2004}
M.~Andersson, R.~Sigurdsson, and M.~Passare, {\em Complex {{Convexity}} and
  {{Analytic Functionals}}}.
\newblock {Basel}: {Birkh\"auser Basel}, 2004.

\bibitem{perrinAlgebraicGeometry2008}
D.~Perrin, {\em Algebraic {{Geometry}}}.
\newblock {London}: {Springer London}, 2008.

\bibitem{touchetteLegendreFenchelTransformsNutshell}
H.~Touchette, ``Legendre-{{Fenchel}} transforms in a nutshell,''

\bibitem{zuberAverageVolumetricConcentration1965}
N.~Zuber and J.~A. Findlay, ``Average {{Volumetric Concentration}} in
  {{Two-Phase Flow Systems}},'' {\em Journal of Heat Transfer}, vol.~87,
  pp.~453--468, Nov. 1965.

\bibitem{luoReviewHomographyEstimation2023}
Y.~Luo, X.~Wang, Y.~Liao, Q.~Fu, C.~Shu, Y.~Wu, and Y.~He, ``A {{Review}} of
  {{Homography Estimation}}: {{Advances}} and {{Challenges}},'' {\em
  Electronics}, vol.~12, p.~4977, Jan. 2023.

\bibitem{alzubaidiImpactWettabilityComoving2024}
F.~Alzubaidi, J.~E. McClure, H.~Pedersen, A.~Hansen, C.~F. Berg, P.~Mostaghimi,
  and R.~T. Armstrong, ``The {{Impact}} of {{Wettability}} on the {{Co-moving
  Velocity}} of {{Two-Fluid Flow}} in {{Porous Media}},'' {\em Transport in
  Porous Media}, vol.~151, pp.~1967--1982, Sept. 2024.

\bibitem{vanderschaftGeometryThermodynamicProcesses2018}
A.~{van der Schaft} and B.~Maschke, ``Geometry of {{Thermodynamic
  Processes}},'' {\em Entropy}, vol.~20, p.~925, Dec. 2018.

\bibitem{pappasProjectiveQuadricsPoles1996}
R.~C. Pappas, ``Projective {{Quadrics}}, {{Poles}}, {{Polars}}, and {{Legendre
  Transformations}},'' in {\em Clifford ({{Geometric}}) {{Algebras}}} (W.~E.
  Baylis, ed.), ({Boston, MA}), pp.~441--448, {Birkh\"auser}, 1996.

\bibitem{croweMultiphaseFlowHandbook2005}
C.~T. Crowe and C.~T. Crowe, eds., {\em Multiphase {{Flow Handbook}}}.
\newblock {Boca Raton}: {CRC Press}, Sept. 2005.

\bibitem{spivakComprehensiveIntroductionDifferential1999}
M.~Spivak, {\em A Comprehensive Introduction to Differential Geometry}.
\newblock {Houston, Tex}: {Publish or Perish, Inc}, 3rd ed~ed., 1999.

\bibitem{grmelaContactGeometryMesoscopic2014}
M.~Grmela, ``Contact {{Geometry}} of {{Mesoscopic Thermodynamics}} and
  {{Dynamics}},'' {\em Entropy. An International and Interdisciplinary Journal
  of Entropy and Information Studies}, vol.~16, pp.~1652--1686, Mar. 2014.

\end{thebibliography}

\appendix

\section{Representation of affine \& projective maps}
\label{app:rep-maps}

The representation of affine and projective maps is straightforward using
homogeneous coordinates, and acts as matrix multiplication on the vector of
homogeneous coordinates. If one has a vector $\vec{x}$ of dimension $n$ in the
vector space $\mathcal{V}$ associated to an
affine space $\mathcal{A}$, one can
consider the homogeneous coordinates $\left( \vec{x}, 1 \right)$ as points of $\mathbb{A}^{n+1}$. An affine
transformation $T_{\mathrm{aff}}$ of  $\mathcal{A}$ then acts as a matrix of dimensions $(n+1) \times
(n+1)$ as
\begin{align}
  \label{eq:matrix-transform-general}
  \begin{pmatrix}
    \vec{y} \\
    1 
  \end{pmatrix}
  \ =& \ T_{\mathrm{aff}}
       \begin{pmatrix}
         \vec{x} \\
         1 
       \end{pmatrix} \\
  =&  \begin{bmatrix}
    \begin{array}{ c | c }
      M & \vec{b} \\ \hline
      0 \cdots 0 & 1 \\ 
    \end{array} 
  \end{bmatrix}
     \begin{pmatrix}
       \vec{x} \\
       1 
     \end{pmatrix} \;,
\end{align}
where $M$ is a $(n \times n)$-matrix (the linear part of the transformation),
$\vec{x}$ and $\vec{y}$ are vectors, and $\vec{b}$ is a translation
vector. The inverse of the affine transformation $T_{\mathrm{aff}}$, denoted
$T_{\mathrm{aff}}^{-1} $ is given by
\cite{gallierGeometricMethodsApplications2011}
\begin{equation}
  \label{eq:inverse-affine}
  \left( T_{\mathrm{aff}}^{-1} \right)_{ij} \ = \  \begin{bmatrix}
    \begin{array}{ c | c }
      M^{-1} & -M^{-1}\vec{b} \\ \hline
      0 \cdots 0 & 1 \\ 
    \end{array} 
  \end{bmatrix} \;.
\end{equation}

Affine transformations form a Lie group, the \textit{affine group}
$\mathrm{Aff}\left(n, K\right)$, with $n$ the dimension of the underlying affine
space and $K$ some field, here taken to be $\mathbb{R}$. The group $\mathrm{Aff} \left(n,
  K \right)$ is a semi-direct product,
\begin{equation}
  \label{eq:aff-semi-direct}
  \mathrm{Aff} \left(\mathcal{V} \right) \ = \ \mathcal{V} \rtimes \mathrm{GL}\left( \mathcal{V} \right) 
\end{equation}
\noindent where $\mathcal{V}$ is the vector space on which $\mathcal{A}$ is modelled, and
$\mathrm{GL}\left( \mathcal{V} \right)$ is the \textit{general linear group} over the
vector space $\mathcal{V}$. This means that affine transformations consist of both linear
transformations given by the action of elements of $\mathrm{GL}\left( \mathcal{V} \right)$
and translations by actions of vectors in $\mathcal{V}$ on points of $\mathcal{A}$. One often writes
$\mathrm{Aff}\left(n, K\right)$ to specify the dimension $n$ of the
representation of the group elements over the field $K$.

\textit{Projective transformations} or \textit{homographies} are automorphisms
of projective space. Given an injective linear map of some
vector space $E$ of finite dimension, one can construct a homography acting on the
projectivization $\mathbb{P} \left( E \right)$ of said vector space. Consider such a map $g$ of $E$. $g$ induces a transformation of $\mathbb{P}(E)$ \footnote{In general, if $g$ is a linear map but not
  necessarily injective, and $M$, $N$ are two vector spaces with $g: M \rightarrow N$ and
  the projective map $\mathbb{P} \left( g \right)$ associated with $g$, one
  has $\mathbb{P}\left( g \right): \mathbb{P}(M) \setminus \mathrm{ker}\left( g \right) \rightarrow \mathbb{P}(N)$. This is
  a \textit{partial projective map}, but not a projective transformation. One
  example can be projection of a curve onto a subspace.}. The set of
homographies of the projective space $\mathbb{P}(E)$ is called the \textit{projective
  linear group}, denoted $PGL(E)$. When $E$ is a (real) vector space of
dimension $n$, one often uses the notation $PGL(n,\mathbb{R})$ \footnote{ $PGL(2, \mathbb{R})$ is well known among physicists
  and engineers: the homographies in this group can be represented as
  \textit{linear fractional transformations}. In the context of complex numbers,
  these transformations
  are better known as \textit{Möbius transformations}.}.

Homographies
act on homogeneous coordinates, introduced in sec.
\ref{sec:projective-space}. They can be formulated in the same way as affine
transformations, \cref{eq:matrix-transform-general}, but with the last row in
the transformation matrix in
general non-zero. As an example, consider $n=3$. One then has that a projective transformation
$T_{\mathrm{proj}}$ acts on a vector of homogeneous coordinates $\left( x,y,z
\right)$ as
\begin{align}
  \label{eq:matrix-transform-general-projective}
  \begin{pmatrix}
    x^{\prime} \\
    y^{\prime} \\
    z^{\prime} 
  \end{pmatrix}
  \ =& \ T_{\mathrm{proj}}
  \begin{pmatrix}
    x \\
    y \\
    z 
  \end{pmatrix}
  =&  \begin{bmatrix}
\begingroup
\renewcommand*{\arraystretch}{1.5}
    \begin{array}{ c | c }
      M & \vec{b} \\ \hline
      \vec{a}^{T} & 1 \\ 
    \end{array}
\endgroup
  \end{bmatrix}
  \begin{pmatrix}
    x \\
    y \\
    z 
  \end{pmatrix} \;.
\end{align}

\subsection{Homographies of $\mathbb{P}^{1}$}
\label{app:rp1}

A homography of $\mathbb{P}^1$ corresponds to the case $n=2$ in \cref{eq:matrix-transform-general-projective}. Consider the homogeneous coordinates $\left[ A_w: A_n \right]$ on $l = \mathbb{P}^{1}$.
Setting $A_n \neq 0$ is equivalent to considering $\left[ A_w/A_{n}:1
\right] \mapsto A_w/A_{n}$. This is a
coordinate on the affine line $\mathbb{A}^{1} \cong \mathbb{R}$. One can set $A_w/A_n = t \sim S_w/S_n $ by dividing out a
common scale factor.

A homography $G \in PGL(2, \mathbb{R})$ is expressed in homogeneous coordinates $\left[
  A_w:A_n \right]$ as \cite{bergerGeometry1994,
  gallierGeometricMethodsApplications2011}
\begin{align}
  \label{eq:fract-lin-hom-coord-general}
  G \left[ A_w:A_n \right] \ =& \
                                \begin{pmatrix}
                                  a  & b \\
                                  c & d \\
                                \end{pmatrix}
                                \left[ A_w:A_n \right] \nonumber \\
  =& \ \left[(a A_w + b A_n):(c A_w + d A_n)  \right] \nonumber \\
  =& \ \left[ \tilde{A}_w:\tilde{A}_n \right] \;,
\end{align}
for scalars $a,b,c,d$, where the matrix $G$ is just a linear map of the
underlying $2d$ vector space.~\Cref{eq:fract-lin-hom-coord-general} is
equivalent to \eqref{eq:matrix-transform-general-projective}.

In affine coordinates, the map in eq.~\eqref{eq:fract-lin-hom-coord-general} is
simply
\begin{equation}
  \label{eq:lin-frac-transf-1}
   G\left[ A_w:A_n \right] \mapsto \frac{a A_w + b A_n}{c A_w + d A_n} \ = \  \frac{a
    t + b }{c t + d } \ = \ \frac{\tilde{A}_w}{\tilde{A}_n} \;.
\end{equation}
 One then has a transformation
\begin{equation}
  \label{eq:lin-hom-frac}
  t \ \mapsto \ \frac{a
    t + b }{c t + d } \equiv h(t) \;,
\end{equation}
of $t$, a
\textit{linear fractional transformation} in the affine parameter $t$.



\begin{figure}[htb!]
  \centering
  \includegraphics[width=\linewidth]{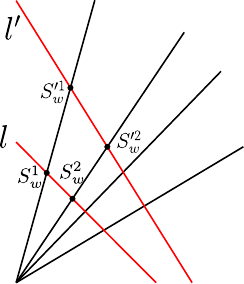}
  \caption{ A perspectivity, which is a type of homography. A homography maps
    lines to lines, while the distance between points are not preserved in
    general. However, the cross-ratio is preserved
    \cite{stillwellFourPillarsGeometry2005}. \label{fig:homography-L} }
\end{figure}

Fig.~\ref{fig:homography-L} illustrates the effect of a particular linear
fractional transformation relating the points of two lines $l, l^{\prime}$ of
$\mathbb{R}^{2}$. Let for instance $l$ be the points that satisfies $A_w + A_n =
\mathrm{const}$, meaning the points of $l$ are given by $\left( S_w, S_n
\right)$ up to scaling. The line $l^{\prime}$ can be any arbitrary affine line. The
points of $l$, $l^{\prime}$ are related by scaling along the radial lines through the
origin $(A_w, A_n) = \left( 0,0 \right)$. To find the intersection point $p^{\prime}
\in l^{\prime}$ of a line $m$ from the origin that goes through a given point $p =
S^1_w \in l$ , one needs to know the slope of $m$. This slope can be found by
dividing the coordinates of $p$. This is equivalent to a linear fractional
transformation $h$ in $t$. The image of the points of $l$ under $h$ can be
imagined as the points of $l$ projected onto $l^{\prime}$ ``as seen from'' the
origin. The location of the projected points in terms of $S_w$ is exactly given
by a function of the form $h \left( t \right)$. The projection in fig.
\ref{fig:homography-L} is a special type of homography called a
\textit{perspectivity} \cite{stillwellFourPillarsGeometry2005}. A homography of
the projective line, which has the general form of eq. \eqref{eq:lin-hom-frac}
in an affine parameter $t$, can be an arbitrary composition of such
perspectivities.


A perspectivity has a particular form expressed in homogeneous coordinates. For homogeneous coordinates $\left[ x:y \right]$ and $\left[\tilde{x}:\tilde{y}
\right]$, the homography that relates them is a perspectivity if one can write
it as \cite{richtergebertPerspectivesProjectiveGeometry2011}
\begin{equation}
  \label{eq:perspectivity-hom-coord}
  \left[ \tilde{x}:\tilde{y} \right]  \ = \  \left[\tau x : y \right] \;,
\end{equation}
for some scalar $\tau$. \Cref{eq:perspectivity-hom-coord} is then
\begin{equation}
  \label{eq:saturations-perspectivity}
  \frac{\tilde{x}}{\tilde{y}} \ = \ \tau \frac{x}{y} 
\end{equation}
in affine coordinates. If these are homogeneous coordinates of two parametrized projective
lines of a projective plane $\mathbb{P}^2$, the relation still holds if the map is a perspectivity.

\end{document}